\documentclass[12pt]{article}

\usepackage{cancel}

\usepackage{latexsym}
\usepackage{epsfig,amssymb,euscript,slashed}

\usepackage{xcolor}
\usepackage{colortbl}
\definecolor{dark-red}{rgb}{0.80,0.12,0.12} 
\definecolor{dark-green}{rgb}{0,0.60,0.30} 

\usepackage{hyperref}
\hypersetup{
    colorlinks=true,
    linkcolor=dark-red,
    filecolor=blue,      
    urlcolor=cyan,
    citecolor=dark-green
  }
  
\usepackage[font={small}]{caption}

\usepackage{amsmath}
\usepackage[nosort]{cite}
\usepackage{array,calc,epsfig}
\usepackage{bbm}
\usepackage{fancybox}
\usepackage{mdframed}
\usepackage{mathrsfs}

\usepackage[left=1.5cm,right=1.5cm,top=3.5cm,bottom=3.5cm]{geometry}

\def\myell{l}

\numberwithin{equation}{section} 
\usepackage[]{graphicx}
\usepackage{xcolor}
\usepackage{tikz}

\begin{document}
\font\cmss=cmss10 \font\cmsss=cmss10 at 7pt


\hfill
	\vspace{18pt}
	\begin{center}
		{\Large 
			\textbf{Four-point correlators with BPS bound states\\ in AdS$_3$ and AdS$_5$ }}
		
	\end{center}

	\vspace{8pt}
	\begin{center}
		{\textsl{Francesco Aprile}}
		\vspace{.2cm}
				
		\textit{\small Departamento de F\'\i sica Te\'orica \& IPARCOS,  Facultad de Ciencias F\'\i sicas, \\Universidad Complutense de Madrid, 28040 Madrid, Spain} \\ 
		
		\vspace{.5cm}
		
		{\textsl{Stefano Giusto}}
		\vspace{.2cm}

		\textit{\small Dipartimento di Fisica,  Universit\`a di Genova, Via Dodecaneso 33, 16146, Genoa, Italy} \\  \vspace{6pt}
		
		\textit{\small I.N.F.N. Sezione di Genova,
			Via Dodecaneso 33, 16146, Genoa, Italy}\\
			
			\vspace{.5cm}

	{\textsl{Rodolfo Russo}}
		\vspace{.2cm}
				
		\textit{\small School of Mathematical Sciences, Queen Mary University of London, \\Mile End Road, London, E1 4NS, United Kingdom} \\ 
					
	\end{center}

	\vspace{12pt}
	
	\begin{center}
		\textbf{Abstract}
	\end{center}
	
	\vspace{4pt} {\small
		\noindent 
We consider heavy-heavy-light-light (HHLL) correlators in AdS/CFT, focussing on the D1D5 CFT$_2$ 
and the ${\cal N}= 4$ super Yang-Mills theory.  Out of the lightest $1/2$-BPS operator in the spectrum, 
$O$, we construct a particular heavy operator $O_H$ given by a coherent superposition of multi-particle 
operators $O^n$, and study the HHLL correlator. When $n$ is of order of the central charge, 
we show that the bulk equation that computes our boundary HHLL correlators is always a Heun equation.
		By assuming that the form of the correlator can be continued to the regime where $n$ is ${\mathcal O}(1)$,
		we first reproduce the known single-particle four-point correlators for $n=1$ and then predict new results 
		for the multi-particle correlators $\langle O^n O^n O O\rangle$. Explicit expressions can be
		written entirely in terms of $n$-loop ladder integrals and their derivatives, and we provide 
		them for $n=2$ and $n=3$ both in position and in Mellin space.
		Focussing on the AdS$_5$ case, we study the OPE expansion of these multi-particle correlators 
		and show that several consistency relations with known CFT data are non-trivially  satisfied. 
		Finally, we extract new CFT data for double and triple-particle long operators.
		\vspace{1cm}

		\thispagestyle{empty}

		\setcounter{footnote}{0}
		\setcounter{page}{0}

		
		\baselineskip=17pt
		\parskip=5pt
		
		\newpage

{
  \hypersetup{linkcolor=black}
  \tableofcontents
}

\section{Introduction} \label{sec:intro}

In the AdS/CFT correspondence, 
scattering amplitudes and holographic correlators 
are a preferred set of observables that embody the holographic principle \cite{
Maldacena:1997re, Witten:1998qj, Gubser:1998bc,Heemskerk:2009pn}.
Over the years there has been tremendous progress in computing 
holographic correlators in the regime of semiclassical gravity, i.e.~when
the Planck and the string lengths are small in units of the radius of AdS. This is especially true for highly supersymmetric theories that have a top-down construction in string theory, since then some entries of the AdS/CFT dictionary are well understood and the bulk description is most tractable. 
All these developments have focussed mainly on the study of 
the so called single-particle operators, which are the operators 
dual to the ``elementary'' fields on the gravity side.
However, the spectrum of BPS local operators is much richer than 
the single-particle fields, as it must contain BPS multi-particle operators 
that can be obtained by taking the Operator Product Expansions (OPE) 
of the single-particle fields mentioned above. The existence of BPS 
multi-particle operators is therefore generic,
and the study of correlation functions with multi-particle operators 
is crucial to progress towards a complete understanding of the holographic principle.

In this article we take a concrete step towards a systematic study of 
four-point correlators with multi-particle insertions, building on the previous work done in \cite{Ceplak:2021wzz,Aprile:2024lwy}. 
Although our approach could be applied to all holographic CFTs that are dual to string theory on an AdS background,
we will focus on two canonical examples of AdS/CFT pairs in type IIB string theory, 
namely, the four dimensional ${\cal N}=4$ SYM,  dual to string theory on AdS$_5\times$S$^5$, and
the so-called D1D5 system,
dual to string theory on  AdS$_3 \times$S$^3\times{\cal M}_4$ where ${\cal M}_4$ can be either $T^4$ or $K_3$.

To illustrate our method, let ${O}_\Delta$ be a $1/2$-BPS single-particle operator 
of dimension $\Delta$ and let ${O}_\Delta^n$ be the $n$-particle operator. We will consider a coherent superposition of operators 
defined by a convergent power series, for instance in the case of AdS$_5\times$S$^5$
\begin{equation}\label{introOperator5}
\ \ O_H = \sum_{n=0}^{\infty} \frac{1}{n!} \left(\frac{\alpha}{\sqrt{2}}\right)^n O_2^n= e^{\frac{\alpha}{\sqrt{2}}\,O_2}\;,
\end{equation}
where $\alpha$ is a continuous parameter satisfying $|\alpha|^2\in[0,\frac{1}{2})$ (for simplicity we assume it to be real and positive), and we choose the normalisation of ${O}_2$ in such a way that its norm squared is of order of the central charge, $c$.
The supergravity description of $O_H$
depends on the magnitude of $\alpha$ relative to the central charge of the CFT, which we will always take to be largest parameter, i.e.~$c\gg 1$.
In the small $\alpha$ regime, more precisely $\alpha^2\sim c^{-1}$, the operator $O_H$ is a small deformation of the identity. In particular, at order $O(\alpha)$ it describes 
one of the (Kaluza-Klein) linear perturbations of AdS$\times$S classified in 
\cite{Kim:1985ez} for AdS$_5$ and in \cite{Deger:1998nm} for AdS$_3$. 
However, when  $\alpha$ is a finite number of order one, the sum defining $O_H$ 
is peaked over values of $n$ of order $c$ and, hence, $O_H$ is effectively as heavy as the central charge $c$. Such operators backreact on the vacuum when inserted in correlation functions. 
In the simplest non-trivial case we have two conjugate heavy operators placed respectively at the far past and future. In the dual description they are described by a regular solution of supergravity that is asymptotically AdS, see~\cite{Lin:2004nb} and~\cite{Lunin:2001jy} for concrete examples in AdS$_5$ and AdS$_3$. It is also possible to provide a geometric description when the insertion points of the heavy operators are brought to finite values~\cite{Abajian:2023jye}.

The classification and construction of all $1/2$-BPS microstate geometries has been found explicitly both in AdS$_5\times$S$^5$ \cite{Lin:2004nb} 
and AdS$_3 \times$S$^3\times{\cal M}_4$ \cite{Lunin:2001jy,Kanitscheider:2007wq}. 
However, given a microstate geometry, it is non trivial to identify the corresponding coherent operator
by using Kaluza-Klein holographic methods~\cite{Kanitscheider:2006zf,Skenderis:2007yb,Giusto:2015dfa}, and viceversa. 
The delicate issue is to disentangle the mixing between
$1/2$-BPS operators with the same quantum numbers that might appear in the coherent sum.
Isolating which solution is dual to an operator $O_H$ of the form given in \eqref{introOperator5}
is therefore not straightforward. In order to simplify this problem it is useful to focus on solutions that 
belong to a consistent truncation, respectively to 3D \cite{Mayerson:2020tcl} and 
5D supergravity \cite{Cvetic:2000nc}. In the 5D case the only
 allowed Chiral Primary Operator (CPO) in the truncation is the lightest single-particle operator of the theory, $O_2$, and necessarily $O_H$ can only be made by $O_2^n$ operators, because mixing with operators built out of different Kaluza-Klein modes would violate the consistency conditions of the truncation \cite{DHoker:2000xhf}. Thus there is a unique 1/2-BPS geometry, sometimes called ``AdS bubble", that was obtained in various 
forms in the literature \cite{Chong:2004ce,Liu:2007xj,Giusto:2024trt},\footnote{In \cite{Chen:2007du} it was shown 
that the same geometry coincides with the LLM solution corresponding to an ellipsoidal bubble.}
but the precise identification of the dual $O_H$ was lacking. Our first result, that follows by generalizing the detailed analysis done in \cite{Giusto:2024trt}, is the precise determination of form of $O_H$, which was already anticipated in~\eqref{introOperator5}.

In the 3D supergravity, the lightest single-particle operator is not the graviton, but rather the CPO in the tensor multiplet, $O_1$. However in this case, the situation is more subtle since the truncation includes also $O_2$, one of the next-to-lightest CPOs. Because of this, it is indeed possible to find a family of supersymmetric solutions that agree at the linear order in $\alpha$, but have different nonlinear completions~\cite{Ganchev:2021iwy}. We will consider explicitly two geometries in this family that have a particular simple dual CFT interpretation: the first one is a solution found in~\cite{Kanitscheider:2007wq}, dubbed here as the KST geometry, while the other is the so-called ``Special Locus'' geometry first found in~\cite{Ganchev:2021pgs}. Again, a similar analysis to the one in 5D leads to the precise determination of form of $O_H$ for both cases. In particular, the operator $O_H$ dual to the Special Locus reads
\begin{equation}\label{introOperator3}
O_H   = \sum_{n=0}^{2N} \frac{1}{n!} \left(\frac{\alpha}{2\sqrt{2}}\right)^n O_1^n\;,
\end{equation} 
where $\alpha\in[0,\infty)$. We note that the sum in \eqref{introOperator3} truncates due to the AdS$_3$ 
stringy exclusion principle \cite{Maldacena:1998bw}, but otherwise the result for AdS$_3$ is analogous to the one in AdS$_5$.

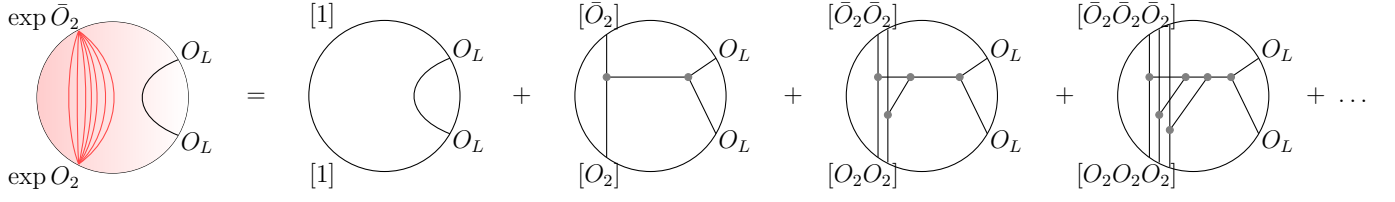
\begin{figure}[t!]
\begin{minipage}{\textwidth}
\begin{equation}\notag
\begin{array}{ccccc}
	\begin{tikzpicture}
	\def\cospisei{0.866};
	\def\sinpisei{0.5};
	\def\sinA{0.81};
	\def\cosA{-0.58};
	\def\sinB{0.88};
	\def\cosB{-0.45};
	\def\ptx{.5};
	\def\pty{.25};

	\draw[] (0,0) circle (1.cm);
	\draw (1.9,0)   node[scale=.8]  {$=$};

	\shade[left color=red!20,right color=white!20] (0,0) circle (1.cm);
	
	\draw (\cospisei,1*\sinpisei) .. controls (0.5*\cospisei-.2 ,0.5*\sinpisei) and (0.5*\cospisei-.2,-0.5*\sinpisei) .. (\cospisei,-1*\sinpisei);
	
	\foreach \s in {-.4,-.25,-.1,0,.1,.25,.4}
	\draw[red!72] (\cosB,1*\sinB) .. controls (0.5*\cosB+\s ,0.5*\sinB) and (0.5*\cosB+\s,-0.5*\sinB) .. (\cosB,-1*\sinB);

	\draw (\cosA-.3,\sinA+.25) node[scale=.8] {$\!\exp{{\bar O}_2}$};
	\draw (\cosA-.3,-\sinA-.25) node[scale=.8] {$\!\exp{{O}_2}$};
	\draw (\cospisei+.25, \sinpisei+.1)   node[scale=.8]  {${O}_L$};
	\draw (\cospisei+.25, -\sinpisei-.1)   node[scale=.8]  {${O}_L$};

	\end{tikzpicture}
&
	\begin{tikzpicture}
	\def\cospisei{0.866};
	\def\sinpisei{0.5};
	\def\sinA{0.81};
	\def\cosA{-0.58};
	\def\sinB{0.88};
	\def\cosB{-0.45};
	\def\ptx{.5};
	\def\pty{.25};

	\draw[] (0,0) circle (1.cm);
	\draw (-.7,0)   node[scale=.8]  {$~$};

	\draw (\cospisei,1*\sinpisei) .. controls (0.5*\cospisei-.2 ,0.5*\sinpisei) and (0.5*\cospisei-.2,-0.5*\sinpisei) .. (\cospisei,-1*\sinpisei);

	\draw (\cosA-.1,\sinA+.25) node[scale=.8] {$[1]\phantom{\,_2]}$};
	\draw (\cosA-.1,-\sinA-.25) node[scale=.8] {$[1]\phantom{\,_2]}$};
	\draw (\cospisei+.25, \sinpisei+.1)   node[scale=.8]  {${O}_L$};
	\draw (\cospisei+.25, -\sinpisei-.1)   node[scale=.8]  {${O}_L$};

	\end{tikzpicture}
&
\!\!\!\begin{tikzpicture}
	\def\cospisei{0.866};
	\def\sinpisei{0.5};
	\def\sinA{0.81};
	\def\cosA{-0.58};
	\def\sinB{0.88};
	\def\cosB{-0.45};
	\def\ptx{.5};
	\def\pty{.25};

	\draw[] (0,0) circle (1.cm);

	\draw (-1.7,0)   node[scale=.8]  {$+$};

	\draw[] (\cosA,\sinA)  --   (\cosA,-\sinA);

	\draw[] (\cosA, \pty) --  (\ptx,\pty);

	\draw[] (\cospisei,1*\sinpisei) -- (\ptx,\pty);
	\draw[] (\cospisei,-1*\sinpisei) -- (\ptx,\pty);

	\filldraw[gray] (\ptx,\pty) circle (1.3pt);
	\filldraw[gray] (\cosA,\pty) circle (1.3pt);

	\draw (\cosA-.1,\sinA+.25) node[scale=.8] {$[{\bar O}_2]$};
	\draw (\cosA-.1,-\sinA-.25) node[scale=.8] {$[{O}_2]$};
	\draw (\cospisei+.25, \sinpisei+.1)   node[scale=.8]  {${O}_L$};
	\draw (\cospisei+.25, -\sinpisei-.1)   node[scale=.8]  {${O}_L$};

	\end{tikzpicture}
&
	\!\!\begin{tikzpicture}
	\def\cospisei{0.866};
	\def\sinpisei{0.5};
	\def\sinA{0.81};
	\def\cosA{-0.58};
	\def\sinB{0.88};
	\def\cosB{-0.45};
	\def\ptx{.5};
	\def\pty{.25};

	\draw[] (0,0) circle (1.cm);
	\draw (-1.7,0)   node[scale=.8]  {$+$};

	\draw[] (\cosA,\sinA)  --   (\cosA,-\sinA);
	\draw[] (\cosB,\sinB)  --   (\cosB,-\sinB);

	\draw[] (\cosA, \pty) --  (\ptx,\pty);
	\draw[] (\cosB, -.25) --  (\cosB+.3, \pty);

	\draw[] (\cospisei,1*\sinpisei) -- (\ptx,\pty);
	\draw[] (\cospisei,-1*\sinpisei) -- (\ptx,\pty);

	\filldraw[gray] (\ptx,\pty) circle (1.3pt);
	\filldraw[gray] (\cosA,\pty) circle (1.3pt);

	\filldraw[gray] (\cosB, -.25)  circle (1.3pt);
	\filldraw[gray] (\cosB+.3, \pty) circle (1.3pt);

	\draw (\cosA-.2,\sinA+.25) node[scale=.8] {$[{\bar O}_2{\bar O}_2]$};
	\draw (\cosA-.2,-\sinA-.25) node[scale=.8] {$[{O}_2{O}_2]$};

	\draw (\cospisei+.25, \sinpisei+.1)   node[scale=.8]  {${O}_L$};
	\draw (\cospisei+.25, -\sinpisei-.1)   node[scale=.8]  {${O}_L$};

	\end{tikzpicture}
&
	\!\!\begin{tikzpicture}
	\def\cospisei{0.866};
	\def\sinpisei{0.5};
	\def\sinA{0.81};  
	\def\cosA{-0.58};
	\def\sinB{0.88};  
	\def\cosB{-0.45};
	\def\sinC{0.94}; 
	\def\cosC{-0.31};
	\def\ptx{.5};
	\def\pty{.25};

	\draw[] (0,0) circle (1.cm);
	\draw (-1.7,0)   node[scale=.8]  {$+$};
	\draw (+1.9,0)   node[scale=.8]  {$+\ \ldots$};

	\draw[] (\cosA,\sinA)  --   (\cosA,-\sinA);
	\draw[] (\cosB,\sinB)  --   (\cosB,-\sinB);
	\draw[] (\cosC,\sinC)  --   (\cosC,-\sinC);
	
	\draw[] (\cosA, \pty) --  (\ptx,\pty);
	\draw[] (\cosB, -.25) --  (\cosB+.35, \pty);
	\draw[] (\cosC, -.45) --  (\cosC+.5, \pty);

	\draw[] (\cospisei,1*\sinpisei) -- (\ptx,\pty);
	\draw[] (\cospisei,-1*\sinpisei) -- (\ptx,\pty);

	\filldraw[gray] (\ptx,\pty) circle (1.3pt);
	\filldraw[gray] (\cosA,\pty) circle (1.3pt);

	\filldraw[gray] (\cosB, -.25)  circle (1.3pt);
	\filldraw[gray] (\cosB+.35, \pty) circle (1.3pt);
	
	\filldraw[gray] (\cosC, -.45)  circle (1.3pt);
	\filldraw[gray] (\cosC+.5, \pty) circle (1.3pt);

	\draw (\cosA-.3,\sinA+.25) node[scale=.8] {$[{\bar O}_2{\bar O}_2{\bar O}_2]$};
	\draw (\cosA-.3,-\sinA-.25) node[scale=.8] {$[{O}_2{O}_2{O}_2]$};

	\draw (\cospisei+.25, \sinpisei+.1)   node[scale=.8]  {${O}_L$};
	\draw (\cospisei+.25, -\sinpisei-.1)   node[scale=.8]  {${O}_L$};

\end{tikzpicture}
\end{array}
\end{equation}
\end{minipage}
\caption{
Coherent geometries as generating functions for connected correlators with multi-particle insertions.
On the LHS the $\langle O_H \bar{O}_H O_L O_L\rangle$ correlator obtained by probing
 the heavy-heavy background with two $O_L$ insertions.
On the RHS the expansion in multi-particle correlators. The diagrams naturally organize 
into a ladder series and each bulk vertex counts for a $c^{-1/2}$ interaction. 
We view each of these diagrams as a limit of a higher-point diagram where 
the multi-particle operator (point-)splits into constituent CPOs. 
\label{fig1}}
\end{figure}

Having clarified the description of the heavy state, it is interesting to study its small deformations. Linear perturbations of a microstate geometry corresponding 
to a light single-particle operator of finite conformal dimension, $O_L$, are described by a 
wave function\footnote{Here $r$ denotes the bulk radial coordinate and $z$, $\bar z$ some coordinates on the AdS boundary.} $\Phi(r,z,\bar{z};\alpha)$, which in the simplest case is the solution of 
the Klein-Gordon equation in the microstate background. The asymptotic limit of this wave 
function, denoted by $\Phi_B(z,\bar{z};\alpha)$, computes
the Heavy-Heavy-Light-Light correlator (HHLL) according 
to the usual holographic dictionary,\footnote{See \cite{Galliani:2017jlg,Bombini:2017sge} for a first computation of this type in the AdS$_3$ context.} 
\begin{equation}
\Phi_B(z,\bar{z};\alpha) = \langle O_H(0) \bar{O}_H(\infty) O_L(1) O_L(z,\bar{z})\rangle  \,.
\end{equation}
The key idea of our approach now is to use this HHLL correlator as the generating function of 
four-point correlators with two single-particle and two $n$-particle operators: 
$\langle {O}^n \bar{O}^n  O_L O_L \rangle$. Formally this identification 
follows from the form of $O_H$, by expanding in $\alpha$, for instance in the case of the AdS$_5$ correlator we have
\begin{equation}\label{eq:corrperteta}
  \Phi_B(z,{\bar z};\alpha)  =  \sum_{n=0}^\infty \frac{\alpha^{2n}}{n!} \,\Phi_B^{(n)}(z,{\bar z})\;,\quad \mbox{and}\quad 
\Phi_B^{(n)}(z,{\bar z}) \sim \langle {O}^n \bar{O}^n  O_L O_L \rangle^c\,.
\end{equation}
Before continuing, it is necessary to clarify that the $\langle {O}^n \bar{O}^n O_L O_L \rangle^c$ 
above corresponds to a particular contribution in the full $\,\langle {O}^n \bar{O}^n O_L O_L \rangle$ correlator.
To explain this point  it is convenient to think about  $\,\langle {O}^n \bar{O}^n O_L O_L \rangle^c$ 
as the OPE limit of a $2n+2$-point function, where the first $n$ and the second $n$ points are taken to coincide. 
With this picture in mind, one can decompose $\,\langle {O}^n \bar{O}^nO_L O_L \rangle$ 
into a connected and a disconnected contribution, in the usual quantum field theory sense.  
Disconnected contributions appears at all orders. However, at order $c^{-i}$ for $i=0,1,\ldots,n-1$ the correlator
$\,\langle {O}^n \bar{O}^n O_L O_L \rangle$ only has disconnected components and the first 
connected contribution appears at order $c^{-n}$. This contribution is precisely the one given by the limit of a
\emph{higher point tree-level connected Witten diagrams}, as exemplified in Figure \ref{fig1}, and
it is precisely the one that is captured by our supergravity computation.
The first genuinely novel structures appear in six-point functions, corresponding to the four-point $n=2$ correlator $\,\langle {O}^2 \bar{O}^2 O_L O_L \rangle^c$ in the OPE limit . The non-trivial four-point Witten diagrams are precisely those dubbed ``Type II" in  \cite{Ma:2022ihn}. As pointed out in \cite{Ceplak:2021wzz} and \cite{Ma:2022ihn}, even though these diagrams are tree-level, they cannot be decomposed into sums of contact Witten diagrams. 

At this point one might worry that the equal sign in \eqref{eq:corrperteta} is formal because the 
microstate geometry is a valid description of the state for $\alpha$ of order unity in the large $c$ limit, 
while the multi-particle expansion of $O_H$ is valid in the small $\alpha$ limit, 
when $\alpha\sim c^{-1/2}$. While we expect that in the 1/2-BPS case the analytic 
continuation of $\Phi_B(\alpha)$ from one regime to the other is well defined, 
we will prove this expectation by showing that our results for $\langle {O}^n \bar{O}^n {O} O \rangle^c$ 
are fully consistent with the OPE properties required for a CFT correlator. There is a priori no such an input in the 
gravity computation, therefore this is an independent and highly non trivial test of our results.
This approach was first employed for $n=1$, see the discussion in 
\cite{Giusto:2018ovt,Giusto:2019pxc,Giusto:2020neo} for AdS$_3$ and \cite{Turton:2024afd} for AdS$_5$. 
Here we will consider explicitly the next two cases, focussing in particular on AdS$_5$,
\begin{equation}
\langle {O}_2^2 \bar{O}_2^2 \bar{O}_2 O_2 \rangle^c\qquad;\qquad\langle {O}_2^3 \bar{O}_2^3 \bar{O}_2 O_2 \rangle^c\,.
\end{equation}
The result for $\langle {O}_2^2 \bar{O}_2^2 \bar{O}_2 O_2 \rangle^c$ appeared in our 
previous work \cite{Aprile:2024lwy}, but the analysis we present now extends the one summarized in \cite{Aprile:2024lwy} and provides also the details of the supergravity derivation.

\subsection{Outline of results}

There are many interlocking features of our results and we would like to highlight 
some of them before presenting the outline of our paper, in order to provide guidance and facilitate the reading.

{\bf Heun equation.} The first problem we consider is to derive the HHLL holographic correlators for heavy operators that are described by coherent states of the type~\eqref{introOperator5} and~\eqref{introOperator3}.
The strategy we shall adopt is to work in the supergravity approximation and solve the wave equation of a scalar field (the dilaton in AdS$_5$) in the background generated by the heavy states. The perturbation we use, $\Phi(r,z,\bar{z};\alpha)$, is dual to a super-descendant, $O_L$, of the light single-particle CPO, namely, $O_2$ in \eqref{introOperator5} or $O_1$ in \eqref{introOperator3}: studying its asymptotic limit, $\Phi_B(z,\bar{z};\alpha)$, and then using a known Ward Identity, we obtain the function $\Psi_B(z,\bar{z};\alpha)$, which is the result for the HHLL correlator with CPOs.
This is advantageous because the wave equation of the scalar field is 
the simplest, i.e~the minimally coupled Klein-Gordon equation on the microstate background.
 It turns out that, after performing a Fourier transform in momentum space on the boundary, the problem reduces to a Heun equation both in AdS$_3$ and AdS$_5$\footnote{In the KST geometry the problem simplifies and the equation is a hypergeometric~\cite{Bombini:2017sge}.}, which can be solved systematically. 
As a result of our analysis, we will present a compact expression 
for all our HHLL correlators, that takes the following suggestive form,
\begin{equation}\label{Psi_introduction}
    \Psi_B(z,\bar{z};\alpha)  = \sum_{l=0}^{\infty} \int_{\mathbb{R}}\!\frac{d\omega}{2\pi i}\, 
    \mu^{(d)}_{l} \chi_{l}^{(d)}(z,\bar{z}) E_{\rm gravity}(\omega,l;\alpha)\;,
    \end{equation}
where the measure, $\mu^{(d)}_l$, and the Fourier factor, $\chi^{(d)}_{l}$, depend on the dimensionality of AdS$_{d+1}$, e.g. in AdS$_{5}$ they are
\begin{equation}\label{BMNdata}
\mu^{(4)}_{l}=\frac{l+1}{2}\qquad;\qquad \chi_{l}^{(4)}(z,\bar{z})=(z\bar{z})^{\frac{\omega-l}{2}} \left( \frac{z^{l+1}-\bar{z}^{l+1}}{z-\bar{z}} \right)\,,
\end{equation} 
and where
\begin{equation}\label{Egravity}
E_{\rm gravity}(\omega,l,\alpha)=\sum_{k=1}^\infty \left(\frac{1}{k+\frac{l}{2}+a(\omega,l;\alpha)}+\frac{1}{k+\frac{l}{2}-a(\omega,l;\alpha)}\right)\;;
\end{equation}
the function $a(\omega,l;\alpha)$ depends on the explicit form of the background geometry and encodes all the non-trivial dynamical information. In the small $\alpha$ expansion we have $a(\omega,l;\alpha) = \frac{\omega}{2} +\delta a(\omega,l;\alpha)$ and $\delta a$ starts at order  ${\cal O}(\alpha^2)$. An elegant approach to study the Heun equation is to map it to a problem in an ${\mathcal N}=2$ supersymmetric gauge theory~\cite{Aminov:2020yma} and, in this language, $a$ is the quantum Seiberg-Witten period.

Also, the integrand in \eqref{Psi_introduction} appears as a generalization of the known BMN-type integrals 
in the literature derived from integrability methods  \cite{Fleury:2016ykk,Basso:2017khq,Coronado:2018ypq}.

{\bf Gravitational ladder integrals.} 
In the context of correlators with heavy operators, it was found that the series of 
ladder integrals plays a role in perturbation theory \cite{Coronado:2018ypq,Brown:2024yvt}.
These are $n$-loop integrals and, up to overall factors, they have a representation which takes the same form 
as in \eqref{Psi_introduction}, but with the following different kernel
\begin{equation}\label{ladder_energy}
E_{\rm ladder}(\omega,l)=\frac{1}{\ \ (\frac{1+l+\omega}{2})^{n+1}( \frac{1+l-\omega}{2})^{n+1}} \,.
\end{equation} 
Since the $\alpha$ expansion of $E_{\rm gravity}$ generates integrands that generalize $E_{\rm ladder}$, 
we think of $\Psi_B(z,\bar{z})|_{\alpha^{2n}}$ as a gravitational generalization of the ladder series.
We can see how the generalization goes by performing the $\omega$ integration via Cauchy theorem by using the Feynman contour in both cases.
While with the ladder kernel in \eqref{ladder_energy} one finds a result whose building blocks 
are
\begin{equation}\label{buildingblockladder}
 \log^{m_1}(z\bar{z})  \sum_{l}  \frac{1}{(l+1)^{m_2}}\left(\frac{z^{l+1}-\bar{z}^{l+1}}{z-\bar{z}}\right) =  \log^{m_1}(z\bar{z})  \frac{{\rm Li}_{m_2}(z)-{\rm Li}_{m_2}(\bar{z}) }{z-\bar{z}}\,,
\end{equation}
the building blocks of the gravitational correlators have the form
\begin{equation}\label{buildingblockAdS}
 \log^{m_1}(z\bar{z}) \sum_{k,l} (z\bar{z})^k \frac{  p_{m_1,m_2}(k,l) }{( l+ 2k+m_3)^{m_2}}  \left( \frac{z^{l+1}-\bar{z}^{l+1}}{z-\bar{z}}\right)\,,
\end{equation}
with $m_1, m_2\in \mathbb{N}$, $m_3\in \mathbb{Z}$ and $p_{m_1,m_2}(k,l)$ some polynomials. 
As explained in \cite{Ceplak:2021wzz}, the gravitational building blocks in \eqref{buildingblockAdS} resum into rational functions 
of $z,\bar{z}$ times polylogs of the form ${\rm Li}_{L}(z)-{\rm Li}_{L}(\bar{z})$ with various degree.
In all worked examples, after assembling the building blocks, we find that the correlator can be written in terms of ladders and derivatives of ladders, as suggested by the structure of the integrand, and we conjecture this to be true at all orders.
The generalized structure we find in \eqref{buildingblockAdS}, and in particular the sum over $k$ that yields non trivial rational functions of $z,\bar{z}$, is crucial for showing
consistency of the correlator with the OPE at strong coupling, which is substantially different from perturbation theory.

Beyond the free term at $\alpha=0$, a first lesson we learn from $\Psi_B$ is that the first non-trivial contribution at $O(\alpha^2)$ is precisely the dynamical LLLL correlator, 
usually written in terms of $\overline{D}$ functions obtained from Witten diagram; for example, in AdS$_5$ one finds 
\begin{equation}\label{eq:exint}
\Psi_B(z,\bar{z})\Big|_{\alpha^{2}}=
-4 V^2 \bar{D}_{2422}\;,
\end{equation}
where $U=(1-z)(1-z)$, $V=z\bar{z}$, and 
\begin{equation}
\overline{D}_{2422}=(3+V\partial_V+U\partial_U)\partial_V\partial_U\bigg[\frac{  2  ({\rm Li}_2(z)-{\rm Li}_2(\bar{z} )) - \log(z \bar{z}) ({\rm Li}_1({z})-{\rm Li}_1(\bar{z} ))}{z-\bar{z}}\bigg]\,.
\end{equation}
Our result rewrites $\bar{D}_{2422}$ using a new integral representation of the type~\eqref{Psi_introduction}. 
Computing higher orders from \eqref{Psi_introduction}  is straightforward, and
we will study in detail correlators up to $\alpha^6$.

Note that, since we deduce the function $\Psi_B(z,\bar{z};\alpha)$ by integrating the Ward Identity that relates it to $\Phi_B(z,\bar{z};\alpha)$, the $\alpha$-expansion of \eqref{ladder_energy} gives the correlators with the CPOs $\langle O_{2}^n \bar{O}_2^{n} \bar{O}_2 O_2\rangle^c$ only up to rational terms that are annihilated by the differential operator implementing the Ward Identity. These rational terms, however, are easily fixed by using the free theory result. 

Since the Mellin transform of the ladder integrals is known~\cite{Aprile:2020luw}, we can easily provide the Mellin space expression for our multi-particle correlators. In these variables the results take a compact form with rational functions of the Mellin variables multipled by kernel with polygamma function of trascendentality up to $n-1$.

{\bf CFT analysis of $\langle O_{2}^n \bar{O}_2^{n} \bar{O}_2 O_2\rangle$.}
We have already mentioned that $\Psi_B(z,\bar z)|_{\alpha^{2n}}$ gives the connected contribution to the correlator at order $c^{-n}$.  This is not the only contribution at order $c^{-n}$, 
since disconnected contributions also appear at the same order. For instance, the disconnected terms
$\langle \bar{O}_2 O_2\rangle \times\langle O^{n-1}_2 \bar{O}^{n-1}_2 \bar{O}_2 O_2\rangle$ and $\langle \bar{O}^2_2 O^2_2\rangle \times\langle O^{n-2}_2 \bar{O}^{n-2}_2 \bar{O}_2 O_2\rangle$, where the four-point correlator is calculated at one and two loops respectively, must be included at the same order as  $\langle {O}_2^n \bar{O}_2^n \bar{O}_2 O_2\rangle^c$.
In general these quantum corrections have a different structure than those captured by our classical gravity computation. It is known that, starting from two loops, the quantum corrections to $\langle O_2 \bar{O}_2 \bar{O}_2 O_2\rangle$ do not follow the pattern of a ladder series, because their maximal logarithmic
discontinuity resums into a different kind of integrals, see \cite{Drummond:2012bg,Aprile:2018efk,Drummond:2022dxw}.

It is interesting at this point to ask, purely in CFT terms, which constraints and consistency conditions should be imposed on the connected component 
$\langle O_2^n \bar{O}_2^n \bar{O}_2 O_2\rangle^c$ of the full correlator $\langle  O_{2}^n \bar{O}_2^{n}  \bar{O}_2 O_2\rangle$.
We will consider two complementary approaches to this problem: the study of the
Regge limit, and the study of the OPE. In the first approach, and focusing on the AdS$_5$ correlator, we will the compare the result obtained upon taking the Regge limit on $\langle {O}_H \bar{O}_H \bar{O}_2 O_2\rangle$ in the small $\alpha$ expansion, with the semiclassical interpretation of the Regge limit in terms of a geodesic in the microstate background (computed independently from $\Psi_B$). We find that they match exactly as it happened in the case of AdS$_3$~\cite{Ceplak:2021wak}.
Concerning the OPE, we will start from $n=2,3$ and show that one can use knowledge 
about the double particle spectrum \cite{Aprile:2018efk,Aprile:2025nta} (protected and long) to infer
constraints on $\Psi_B|_{\alpha^{4}}$ and $\Psi_B|_{\alpha^{6}}$. This study  
will produce non trivial  checks about the gravity computation. Then, 
we will learn that 
in order to extend these checks and explain the pattern provided by $\Psi_B(\alpha)$ in 
the $\alpha$ expansion, there has to be new results about the scaling
in $N$ of the three-point couplings between external $1/2$-BPS operators 
and long operators, which do not follow in an obvious way from large-$N$ factorization.

\paragraph{Outline of the paper.}

To emphasize the 
similarities (but also the peculiarities)  between the gravity descriptions of the D1D5 CFT$_2$ 
and the $\mathcal{N}=4$ CFT$_4$, we will present results for both theories in a parallel manner. 
In Section~\ref{sec:ads3} we study the  D1D5 CFT$_2$ in AdS$_3$, considering the two microstate 
geometries of \cite{Kanitscheider:2007wq} and \cite{Ganchev:2021pgs}. In both cases we derive the exact coherent operators that are dual to these geometries and compute 
the corresponding HHLL correlators. These computations are very instructive and open up the way to the computation of the correlators in the $\mathcal{N}=4$ SYM. In the remaining sections we will focus on this 4D theory. In Section~\ref{sec:ads5} we study the microstate geometry in AdS$_5$ \cite{Chong:2004ce,Liu:2007xj,Giusto:2024trt}, derive the exact $1/2$-BPS heavy state to which it is  dual to in the 
$\mathcal{N}=4$ CFT$_4$, and compute the HHLL correlator.   
Then,  from the expansion in $\alpha$ of the wave function $\Psi_B(z,\bar{z};\alpha)$  we give explicit 
results for $\Psi|_{\alpha^2}$, $\Psi|_{\alpha^4}$, $\Psi|_{\alpha^6}$, in position space. In Section~\ref{sec:CFTgen} we reconstruct the full non-connected correlators 
with general R-charge vectors, using the input from $\Psi_B(z,\bar{z};\alpha)$, and we discuss the structure of the results, both in position space 
and in Mellin space. The result in Mellin space -- expressed in the variables, $s$ and $t$, conjugated 
to the cross rations, $U$ and $V$ -- offers a complementary perspective on $\Psi_B(z,\bar{z})|_{\alpha^{2n}}$, also providing a bridge for the CFT analysis. 
In Section~\ref{sec:OPE} we review basic knowledge about the spectrum of ${\cal N}=4$ SYM at strong coupling, 
we derive results on various OPE channels, then perform some checks and obtain new predictions.
In Section~\ref{sec:discussion}, we will give a final discussion, 
and conclude with an outlook.

Some technical results are collected in the appendices.  Appendix~\ref{app:heun} 
reviews the Heun connection problem, which we used to solve the wave equation and
compute the HHLL correlators in the form of \eqref{Psi_introduction}. In Appendix~\ref{app:bpsbubble} we show the equivalence between the AdS bubble presented
in \cite{Liu:2007xj}, the solution presented in \cite{Giusto:2024trt},
and their 10D uplift. The matching between the Regge limit of the HHLL correlator and the result from the geodesic approximation is shown in Appendix~\ref{app:ReggeLimit}. Appendix~\ref{N4blocks} gives a pedagogical introduction to the ${\cal N}=4$ superconformal block 
expansion of four-point functions of ${1}/{2}$-BPS operators, and collects a number of useful results/computations 
that we used in the OPE analysis of the $\langle O_2^n \bar{O}_2^n  \bar{O}_2{O}_2\rangle$ correlators. 
Appendix \ref{apptriplep} studies the {new} CFT data for triple-particle correlators that can be extracted from our results.
Appendix \ref{appLadder} computes the Mellin amplitude of some ladder 
integrals and shows  how this knowledge has been used to study our holographic correlators.

\section{AdS$_3$/CFT$_2$ multi-particle correlators}
\label{sec:ads3}

In this section we review some key properties of the D1D5 CFT which are relevant for the holographic multi-particle correlators of the type~\eqref{eq:corrperteta}. In subsection~\ref{ssec:ads3notation} below, we briefly summarise the $1/2$-BPS spectrum of the theory, paying particular attention to the definition of the multi-particle states. In doing so, it is useful to describe the CFT both in the free limit, where it reduces to a collection of free elementary 2D bosons and fermions, and in the leading gravity regime, where it is captured by generalised free fields in AdS$_3$. In subsection~\ref{ssec:ads3HHLL} we focus on two closely related classes of heavy $1/2$-BPS multi-particle operators whose dual description in terms of asymptotitcally AdS supergravity solutions is known. The quadratic fluctuations around such solutions capture the holographic HHLL 4-point functions. Then in subsection~\ref{ssec:ads3LLLL} we show how to interpret these results as correlators involving two {\em light} multi-particle operators obtaining various examples of such correlators in AdS$_3$. Finally we show how these results can be written in a compact form both in configuration and in Mellin space. The formalism developed in this discussion applies directly also to ${\cal N}=4$ SYM as it will be shown in the next sections.

\subsection{The $1/2$-BPS sector of the spectrum}
\label{ssec:ads3notation}

The $1/2$-BPS sector of the spectrum is captured by the supergravity approximation which, in the case at hand, involves a 6D ${\mathcal N} = (2, 0)$ theory coupled to $n_f$ tensor multiplets.\footnote{We have $n_f=5$ if the original 10D theory is compactified on a $T^4$ and $n_f=21$ when it is compactified on $K_3$. For the purposes of this section, we will not need to treat the two cases separately.} The supergravity fields arising from the Kaluza-Klein compactification on S$^3$ are in one-to-one correspondence with a subclass of local operators in the CFT language which are dubbed as the single-particle states. In this section we will focus on the lightest operators of this type\footnote{There are $n_f$ such operators with conformal dimension $(h,\bar{h})=(1/2,1/2)$. At the classical supergravity level, there is a $SO(n_f)$ flavor symmetry rotating them, so we can choose any. In the $T^4$ theory, there are even lighter states with dimensions $(h,\bar{h})=(1/2,0)$ or $(h,\bar{h})=(0,1/2)$ which are related to the fermionic partners of the $U(1)$ affine symmetries appearing in the affine super-Virasoro algebra relevant for that case. We will not need to consider these operators.} and the $1/2$-BPS multi-particle states obtained by taking the $n^{\rm th}$ power of the same single-particle seed. As already mentioned, there is a sub-locus of the D1D5 superconformal moduli space where the CFT can be described in terms of free fields: this requires to switch off some couplings that are the analogue of the 't~Hooft coupling in ${\mathcal N}=4$ SYM and, in the $K_3$ case, choose the geometric moduli so that $K_3=T^4/Z_2$. Then, on the CFT side, we can describe a subset of the lightest single-particle operators as bilinears in the elementary fermions, for instance:
\begin{equation}
  \label{eq:O1ads3}
  O = \sum_{r=1}^N \psi_{(r)}^{+ A} \tilde\psi_{(r)}^{+ B}\epsilon_{AB}\equiv \sum_{r=1}^N  O_{(r)} \;,
\end{equation}
where  $A$, $B$ are $SU(2)$ indices related\footnote{The elementary bosons $X_{A \dot{A}}$ have also $\dot{A}$-indices of the other $SU(2)$ from the $SO(4)$ describing rotations in $T^4$. In the $K_3=T^4/Z_2$ case, the $Z_2$ reflection acts on fields with $k$ indices of the type $A$ by including an overall factor of $(-1)^k$. Thus the operator~\eqref{eq:O1ads3} is $Z_2$-even and is part of the spectrum in both ${\cal M}=T^4,K_3$ theories.} to rotations in $T^4$, the superscript $\pm$ are other $SU(2)$ indices related to the R-charges $j,\bar{j}$ (in the example above we have $h=\bar{h}=j=\bar{j}=1/2$ and the subscript $r$ runs over the $N$ copies of the seed CFT with central charge $c=6$ used in the symmetric orbifold construction.\footnote{We follow the conventions of~\cite{Guo:2022ifr} for the orbifold description of the D1D5 CFT and refer to that paper for more details.} One can easily write explicit operators describing other three flavors by taking the symmetric combinations of the indices $A,B$ in the fermion bilinear appearing in~\eqref{eq:O1ads3}, for instance 
\begin{equation}
  \label{eq:O1ads3f2}
  O' = \sum_{r=1}^N \psi_{(r)}^{+ 1} \tilde\psi_{(r)}^{+ 1} \;.
\end{equation}
The operators~\eqref{eq:O1ads3} and~\eqref{eq:O1ads3f2} are in the untwisted sector of the symmetric orbifold and their invariance under the permutations of any pairs of the seed CFT is a simple consequence of the sum over the copies. In this paper we will not consider states in the twisted sectors of the symmetric orbifold.

Let us now describe the $1/2$-BPS multi-particle states in the free orbifold language introduced above. The most natural approach from the CFT point of view is to take the leading OPE term in the product of single-particle chiral primaries. For instance the double particle state obtained by squaring~\eqref{eq:O1ads3} reads
\begin{equation}
  \label{eq:O1ads3O2}
  O^2 = 2 \left[\sum_{r<s} \psi_{(r)}^{+ A} \tilde\psi_{(r)}^{+ B}\epsilon_{AB}\, \psi_{(s)}^{+ C} \tilde\psi_{(s)}^{+ D}\epsilon_{CD} - \sum_{r} \psi_{(r)}^{+1} \psi_{(r)}^{+2} \tilde{\psi}_{(r)}^{+1} \tilde{\psi}_{(r)}^{+2}\right] \;.
\end{equation}
In the explicit free-theory expression we separated the contributions where the all fields act on the same copy of the CFT from the rest because they are independently invariant under the symmetric orbifold permutations. Indeed the two combinations represent physical states separately, so one can define a different composite operator as follows
\begin{equation}
  \label{eq:O1ads3O2:}
  :\! O^2 \!:\, = \sum_{r<s} \psi_{(r)}^{+ A} \tilde\psi_{(r)}^{+ B}\epsilon_{AB}\, \psi_{(s)}^{+ C} \tilde\psi_{(s)}^{+ D}\epsilon_{CD}  \;.
\end{equation}
By following what is done in ${\mathcal N}=4$ SYM~\cite{Aprile:2020uxk}, we will refer to~\eqref{eq:O1ads3O2} as a true double-particle state of dimension $(1,1)$ and define the single-particle operators with $h=\bar{h}=1$ as the states orthogonal to it (and to all other double-particles). With this definition, the operator in~\eqref{eq:O1ads3O2:} represents a linear combination of single and double-particle states. The pattern can be extended to higher orders: of course the number of possible definitions of the $n^{\rm th}$ power of $O$ increases, but they all differ by single-particle and lower orders multi-particle operators. For the purposes of this paper, we will need just $O^n$, defined by subsequent OPE as done in~\eqref{eq:O1ads3O2}, and $:O^n$, which is easily defined only in the orbifold theory in terms of a product where the elementary fields always act on different copies of the seed CFT as in~\eqref{eq:O1ads3O2:}. Since these composite operators define two different basis of the same $1/2$-BPS states, they should exist at a generic point in the superconformal manifold.

We are particularly interested in the supergravity regime, which is of course far from the orbifold CFT limit. The spectrum of the relevant ${\mathcal N} = (2, 0)$ supergravity on AdS$_3 \times$~S$^3$ corresponds to the single-particle operators, in the sense described above, which are in $1/2$-BPS multiplets~\cite{Deger:1998nm}. The multi-particles operators constructed from these protected single-particle states are still described within supergravity by imposing the appropriate boundary conditions~\cite{Witten:2001ua} on the same fields used to describe the supergravity spectrum. When the number of single-particle constituents is large, of the order of the central charge $n\sim N$, we are dealing with a heavy state which deforms the background to an asymptotically AdS solution encoding the details of the dual state. The known smooth geometries are dual to coherent superpositions\footnote{Sometime the superposition is somewhat degenerate and a single term in the coherent sum survives: in these cases the dual solution is locally AdS$_3 \times$~S$^3$ and the dual state is a superdescendant of the vacuum. These geometries become non-trivial when extended to the asymptotically flat region, but here we will focus entirely on the AdS/CFT paradigm and so will not consider these degenerate cases.} of multi-particle states~\cite{Skenderis:2006ah}. It turns out that the precise definition of the composite operators entering in the coherent state description is relevant at the level of the corresponding supergravity solution. For instance, as already discussed in~\cite{Ceplak:2021wzz}, the $1/2$-BPS solutions constructed in~\cite{Lunin:2001jy,Kanitscheider:2007wq} should be interpreted as dual to coherent states built with the composite operators defined according to the product introduced in~\eqref{eq:O1ads3O2:}. A well-studied case is the state\footnote{With the conventions~\eqref{eq:O1ads3} we have $\langle O| O \rangle = 2 N$.}
\begin{equation}
  \label{eq:skst}
  :\! O_{H} \! : = \sum_{n=0}^N \sqrt{\binom{N}{n}} \;\left(\frac{\alpha}{2}\right)^n \left(1-\frac{\alpha^2}{4}\right)^{\frac{N-n}{2}}  [:\! O^{n}\! :] = \sum_{n=0}^N \left(\frac{\alpha}{2\sqrt{2}}\right)^n \left(1-\frac{\alpha^2}{4}\right)^{\frac{N-n}{2}}  :\! O^{n}\! : \;,
\end{equation}
where the square parenthesis indicates that the composite operator has been normalised to one
\begin{equation}
  \label{eq:nn}
  [:O^n:] = \left(2^n \binom{N}{n}\right)^{-\frac{1}{2}} :O^n: 
\end{equation}
and $0\leq \alpha\leq 2$ is the parameter that appears in the dual geometry, see the discussion below~\eqref{met3d}. This geometry was first constructed in~\cite{Kanitscheider:2007wq} and it was noticed in~\cite{Mayerson:2020tcl,Houppe:2020oqp} that this solution sits in a 3D gauged supergravity truncation of the full theory. It was further pointed out in~\cite{Ganchev:2021iwy} that this solution is part of larger family of solutions with the same supersymmetries and quantum numbers. In~\cite{Ganchev:2023sth} the dual CFT interpretation of this class of configurations was clarified: the new solutions interpolate between the coherent state~\eqref{eq:skst} and one that is built out of the true multi-particle operators defined trough the standard OPE, as in~\eqref{eq:O1ads3O2}. This latter solution was dubbed as the ``Special Locus" in the original literature. Though the precise form of the state dual to the Special Locus has never been written before, we will show below that the observation of \cite{Ganchev:2023sth} and the matching with some known expectation values of CPOs in this state leads uniquely to the following coherent state
\begin{equation}
  \label{eq:splocst}
O_H = \sum_{n=0}^{2N} b_n \left(\frac{\alpha}{2}\right)^n N^{\frac{n}{2}} [O^n]  \;,
\end{equation}
with
\begin{equation}
b_n =\frac{1}{\sqrt{n!}} \left(\frac{(2N)!}{(2N-n)! (2N)^n }\right)^{\frac{1}{2}} = \frac{1}{\sqrt{n!}} + O(N^{-1})\;.
\end{equation}
As before, $[O^n]$ is the normalised $n$-particle operator, but now with the product defined in terms of the standard OPE. The relation between the normalised and the non-normalised operators reads
\begin{equation}\label{eq:normSL}
[O^n] = \left(n! \frac{(2N)!}{(2N-n)!}\right)^{-\frac{1}{2}} O^n\;,
\end{equation}
so we have
\begin{equation}\label{eq:OHsl}
O_H   = \sum_{n=0}^{2N} \frac{1}{n!} \left(\frac{\alpha}{2\sqrt{2}}\right)^n O^n\;.
\end{equation}

Notice that both sums in~\eqref{eq:VEVgrav} and~\eqref{eq:splocst} truncate and this is a peculiarity of the AdS$_3$/CFT$_2$ case when the CFT duals have an infinite dimensional superalgebra which is at least $\mathcal{N}=2$. One can consider the super-charge of level $r=-3/2$ and, by requiring that its action on a quasi-primary of dimension $h$ and charge $j$ under a holomorphic $U(1)$ R-charge $J$ has positive norm, it is easy to find a constraint fixing the maximum value of $j$ given $h$ and $c$~\cite{Lerche:1989uy}. Applying this to a $1/2$-BPS multi-particle state made out of CPOs with charge $j_{CPO}$, one sees that $h=j$ is possible only if the number $p$ of constituents satisfies $p\leq N/j_{CPO}$. This is the stringy exclusion principle pointed out in~\cite{Maldacena:1998bw}: trying to extend the sum in~\eqref{eq:splocst} beyond $p=2N$ (corresponding to $j_{CPO}=1/2$) would make the state non-BPS. In the case~\eqref{eq:VEVgrav} the sum truncates before because of the way  composite operators are defined: when $p>N$ it is not possible to avoid placing two excitations $O_{(r)}$ on the same copy of the symmetric orbifold, thus going beyond the class of composite operators defined in Eq.~\eqref{eq:O1ads3O2:}. It is remarkable that the definition of the  heavy state in~\eqref{eq:VEVgrav}, which is natural just at the free orbifold point, has a simple geometric description in the supergravity regime.

We conclude this introductory part by recalling the 3D metric, in the gauged supergravity description, of the solutions dual to the CFT states introduced above. By following~\cite{Ganchev:2023sth}, we parametrize this metric as
\begin{equation}
ds_3^2  =  R^2_{\rm AdS} \left[-\Omega_1^{2} \, \bigg(d t + \frac{k}{(1- \xi^{2})} \, d\psi \bigg)^{\!\!2} + \,\frac{\Omega_0^{2}}{(1-\xi^{2} )^{2}} \, \big( d \xi^2 ~+~ \xi^2 \, d \psi^2 \big) \right] \,,
\label{met3d}
\end{equation}
where $0\leq \xi<1$ is the holographic coordinate (with $\xi\to 1$ representing 
the conformal boundary) and $t$ is the time direction of the CFT$_2$ in units of 
$R_y$, the radius of the space-like direction which is parametrized by $\theta=\psi-t$. For the state in~\eqref{eq:skst}, we have
\begin{equation}
  \label{eq:vanilsup}
  \Omega_0 = \sqrt{1 - \frac{\alpha^2}{4} (1-\xi^2) } \,, \quad k = \frac{\xi^2}{\Omega_1}\,, \quad \Omega_1=1-\frac{\alpha^2}{4}\;.
\end{equation}
We refer to Section~2.5.3 of~\cite{Ganchev:2023sth} for the full solution in the gauged supergravity language (one should set to zero the parameters $\alpha_2$ and $n_{1,2}$ appearing in that more general discussion and identify $\alpha_1$ there with $\alpha$ here). Instead the geometry dual to the state~\eqref{eq:OHsl} involves the following 3D metric
\begin{equation}
  \label{eq:sploc}
  \Omega_0 =1-\frac{8\,\alpha^2\,(1-\xi^2)}{64 -\alpha^4\,\xi^{2}}\;,\quad k=\frac{\xi^2}{\Omega_1}\,\bigg(1-\frac{\alpha^4\,(1-\xi^2)}{64-\alpha^4\,\xi^{2}}\bigg)\;,\quad \Omega_1=\frac{8-\alpha^2}{8+\alpha^2}\;.
\end{equation}
We refer again to~\cite{Ganchev:2023sth} for the full solution, see Section~2.6.1 there (where now one should set to zero the parameters $\alpha_2$ and $n_{1,2}$ and again identify $\alpha_1$ and $\alpha$.). 

The evidence for the holographic identifications between the geometries \eqref{eq:vanilsup} and \eqref{eq:sploc} with the states \eqref{eq:skst} and \eqref{eq:OHsl} comes from the study of the protected HHL 3-point correlators capturing the expectation values of a light single-particle operator in the presence of the same heavy operator inserted at plus or minus infinite time. On the bulk side they are captured by the decay at the boundary of the microstate geometry of the supergravity fields dual to the light operators, while on the CFT side they can be computed in the free orbifold theory. This precision holography analysis was initiated in~\cite{Kanitscheider:2006zf,Skenderis:2007yb} and then extended in~\cite{Giusto:2015dfa,Giusto:2019qig,Rawash:2021pik}. For our two states, from their CFT descriptions~\eqref{eq:skst} and~\eqref{eq:OHsl} one can derive
\begin{equation}
  \label{eq:Ovev}
  \begin{aligned}
     \frac{\langle :\! O_H\!\!:|\, J\, | :\! O_H\!\!:\rangle}{\langle :\! O_H\!\!:| :\! O_H\!\!:\rangle} & = \frac{N}{2} \frac{\alpha^2}{4}\;, \qquad & \frac{\langle O_H| J | O_H\rangle}{\langle O_H| O_H\rangle} & = \frac{N \alpha^2}{8+\alpha^2}\;, \\
    \frac{\langle :\! O_H\!\!: |\, [O]\, | :\! O_H\!\!: \rangle}{\langle :\! O_H\!\!: | :\! O_H\!\!:\rangle} & = \frac{\sqrt{N}}{\sqrt{2}}\frac{\alpha}{\sqrt{2}} \sqrt{1-\frac{\alpha^2}{4}}\;,\qquad &  \frac{\langle O_H| [O] | O_H\rangle}{\langle O_H| O_H\rangle} & = \frac{\sqrt{N}}{\sqrt{2}} \frac{4\sqrt{2}\alpha}{8+\alpha^2}\;.
   \end{aligned}
\end{equation}
These results match what is found by extracting the decay of the dual bulk scalar $s_1^{(7)}$, see\footnote{Eq.~(4.44) of~\cite{Ganchev:2023sth} has a typo: a factor of $\alpha_1$ in the numerator was forgotten, as clear by setting $n_1=0$ in Eq.~(4.58) of the same paper.} Eq.~(4.36) and Eq.~(4.44) of~\cite{Ganchev:2023sth}, with the dictionary $O \leftrightarrow \frac{\sqrt{N}}{\sqrt{2}} s_1^{(7)}$. This provides a non-trivial check of the identification between heavy states and geometries. A similar precision holography check was discussed in~\cite{Ganchev:2023sth} at the next order. In particular it was showed that the expectation values of all single-particle operators of dimension 2 vanish in the solution with the geometry~\eqref{eq:sploc}, as expected from the orthogonality between single-particle and multi-particle states~\cite{Aprile:2020uxk}. This is not the case for the other solution corresponding to~\eqref{eq:vanilsup}, which is consistent with the use of the different definition of composite operators~\eqref{eq:skst}, instead of~\eqref{eq:splocst}. As highlighted in~\eqref{eq:O1ads3O2}, this definition yields a mixture a single and double particle states already at the second order and this is reflected in non-trivial expectation values of some single-particle operator of dimension 2~\cite{Rawash:2021pik}.

\subsection{Holographic HHLL correlators from gravity}
\label{ssec:ads3HHLL}

By following the standard AdS/CFT approach, the quadratic fluctuations around a microstate geometry capture holographic four-point correlators with two heavy and two light operators which are respectively dual to the particular geometry and the particular fluctuation used. Of course, if the calculations on the bulk side are performed in the supergravity approximation, the results obtained will be valid in the corresponding CFT regime, {\em i.e.} at strong coupling and at large values of the central charge. In the context for the AdS$_3$/CFT$_2$ duality the study of such correlators for D1D5 microstate geometries was initiated\footnote{There is a special class of heavy states dual to locally AdS$_3 \times$~S$^3$ geometries which yield particularly simple HHLL correlators~\cite{Galliani:2016cai} that are protected. We will not considered this case here.} in~\cite{Galliani:2017jlg}. Instead of considering bulk fluctuations which are dual to CPOs as done in that reference, it is easier to consider the field dual to the superdescendant obtained by acting on the CPO~\eqref{eq:O1ads3f2} with a holomorphic and an anti-holomorphic supercharge~\cite{Bombini:2017sge}. In this way one obtains a neutral operator under the R-symmetry group, which is described by a massless 6D Klein-Gordon field $\Phi$ in the theory obtained after compactifying on the 4D space ${\cal M}_4$. By following the same logic discussed in Appendix~\ref{app:bpsbubble}, we decompose the 6D field along the S$^3$ scalar harmonics, $\Phi= \sum\Phi_I Y^I$, and then focus on the lowest KK mode $\Phi_0$. This field satisfies the massless 3D Klein-Gordon equation
\begin{equation}
  \label{eq:PhiAds3}
  \Box_3 \Phi_0=0\;,
\end{equation}
where $\Box_3$ is calculated in the metric~\eqref{met3d}. In all examples we analyse in this paper, one can further reduce the problem to an Ordinary Differential Equation (ODE), by going to Fourier space
\begin{equation}
  \label{eq:Phifs}
\Phi_0= \sum_{\myell=-\infty}^\infty \int_{-\infty}^\infty \!\frac{d\omega}{2\pi}\,e^{i\omega t}\, e^{i l\theta}\,{\phi}(\xi) \,.
\end{equation}
We are looking for the solution which is regular everywhere and is a linear combination of a non-normalizable delta-function source and a normalizable mode at the AdS boundary
\begin{equation}
  \label{eq:adsxp}
  \Phi_0 \approx \delta(\tau)\delta(\theta) + w^2 \Phi_B(\tau,\theta)\;,
\end{equation}
where we changed coordinates from $\xi$ to $w$ so as to have asymptotically a canonical AdS$_3$ metric
\begin{equation}\label{eq:canads3}
ds^2_3 = \frac{d w^2}{w^2} + \frac{1}{w^2} \Big(-d\tau^2+ d\theta^2+O(w^2)\Big)\,.
\end{equation}
The HHLL correlator involving the superdescendant (scalar) perturbation is captured by the normalizable solution $\Phi_B$ and we can reconstruct the correlator with the CPO, $\Psi_B$, by integrating the following Ward Identity~\cite{Bombini:2017sge}
\begin{equation}
  \label{eq:PsiB}
  \Phi_B(\tau,\theta) = \frac{1}{4} \left(\frac{\partial^2}{\partial t^2}- \frac{\partial^2}{\partial \theta^2}\right) \Psi_B(\tau,\theta)\,.
\end{equation}

We first consider the metric~\eqref{eq:vanilsup}. In this case~\eqref{eq:PhiAds3} takes the following particularly simple form
\begin{equation}
  \label{eq:ads3vs}
  {\phi}''(\xi )+\frac{\phi'(\xi )}{\xi }- \frac{ \left(l^2 \left(\alpha ^2 \left(\xi ^2-1\right)+4\right)-4 \xi ^2 \omega ^2\right)}{\left(\alpha ^2-4\right) \xi ^2 \left(\xi ^2-1\right)}  {\phi(\xi)}= 0\;,
\end{equation}
which can be mapped to the Hypergeometric differential equation for generic values of $\alpha$. We are interested in the regular solution at $\xi=0$, thus we obtain
\begin{equation}
  \label{eq:ads3vs1}
  \phi(\xi) \sim \xi^{|l|} \, {}_2F_1\left(\frac{|l|}{2}-\frac{\gamma}{2},\frac{|l|}{2} + \frac{\gamma}{2},|l|+1;\xi^2 \right)\;,
\end{equation}
where
\begin{equation}
  \label{eq:gammavs}
  \gamma=  \sqrt{\frac{\omega^2 -\frac{\alpha^2 l^2}{4}}{1-\frac{\alpha^2}{4}}}\;.
\end{equation}
Close to the AdS$_3$ boundary,
\begin{equation}
  \label{eq:z2wVS}
  \xi^2 \approx 1-\left(1-\frac{\alpha^2}{4}\right) w^2 \to 1\;,
\end{equation}
the solution~\eqref{eq:ads3vs1} is a linear combination of the normalizable and non-normalizable local solutions. By using~(15.3.11) of~\cite{abramowitz+stegun}
\begin{align}
  \label{eq:2f1d}
   \xi^{|l|} {}_2F_1\left(\hat{a},\hat{b},\hat{a}+\hat{b}+1;\xi^2 \right) \approx  ~\frac{\Gamma (\hat{a}+\hat{b}+1)}{\Gamma (\hat{a}+1) \Gamma (\hat{b}+1)} \Bigg[ & \left(1 + \hat{a}\,\hat{b} (1-\xi^2) \ln(1-\xi^2) + \ldots \right) \\ \nonumber & + \hat{a}\,\hat{b} (1-\xi^2 )\,\left(H_{\hat{a}} + H_{\hat{b}} - 1 - \frac{|l|}{2\,\hat{a}\,\hat{b}}+\ldots  \right) \Bigg]\;,
\end{align}
where  $H_{\hat{a}}= \psi^{(0)}(\hat{a}+1)+\gamma_E$ are the Harmonic numbers and can be written as
\begin{equation}
  \label{eq:harmH}
  H_{\hat{a}} = 
  \sum_{k=1}^\infty \left(\frac{1}{k}-\frac{1}{k+\hat{a}}\right)\,.
\end{equation}
For the solution \eqref{eq:ads3vs1}, the values of $\hat{a}$, $\hat{b}$ are
\begin{equation}\label{eq:ahatbhatvs}
\hat{a}=\frac{|l|+\gamma}{2}\quad,\quad \hat{b}=\frac{|l|-\gamma}{2}\,.
\end{equation}
It is now straightforward to extract the correlator: the leading term in the first line of~\eqref{eq:2f1d} reconstructs the delta function in~\eqref{eq:adsxp}, so the second line can be identified with the result of the momentum-space correlator $\Phi_B(\omega,l)$. Then by using~\eqref{eq:adsxp},~\eqref{eq:PsiB},~\eqref{eq:z2wVS} and~\eqref{eq:2f1d}, we can write the HHLL correlator $\Psi_B$ as follows
\begin{align}
  \label{eq:Psivs}
  \Psi_B  & = \sum_{\myell=-\infty}^\infty \int_{-\infty}^\infty \!\frac{d\omega}{2\pi}\,e^{i\omega t}\, e^{i l\theta} \left(\frac{1-\frac{\alpha^2}{4}}{\frac{l^2-\omega^2}{4}}\, \hat{a}\, \hat{b} \right) \left[\sum_{k=1}^\infty \left(\frac{1}{k+\hat{a}}+\frac{1}{k+\hat{b}}\right)\right]
  \\ \nonumber & = \sum_{\myell=-\infty}^\infty \int_{-\infty}^\infty \!\frac{d\omega}{2\pi}\,e^{i\omega t}\, e^{i l\theta} \sum_{k=1}^\infty \left[\frac{2}{\gamma+ 2k + |l|} - \frac{2}{\gamma - 2k - |l|} \right]
\end{align}
In the first line we focused just on the factors involving the Harmonic number in~\eqref{eq:2f1d}, as the others yield only delta function supported contributions when written in configuration space. In the second line we have noticed that the explicit values of $\hat{a}$, $\hat{b}$ \eqref{eq:ahatbhatvs} satisfy 
\begin{equation}
  \label{eq:abexpl}
  \hat{a}\,\hat{b} = \frac{l^2}{4}-\frac{\gamma^2}{4} = \frac{l^2-\omega^2}{4\left(1-\frac{\alpha^2}{4}\right)}\;.
\end{equation}
Thus the product $\hat{a}\,\hat{b}$ is proportional to the differential operator appearing in the Ward Identity~\eqref{eq:PsiB} and the whole round parenthesis in the first line of~\eqref{eq:Psivs} is just $1$. In summary the second line of~\eqref{eq:Psivs} is the HHLL correlator for the chiral primary up to rational terms annihilated by the differential operator itself, which can be fixed by matching the contribution of protected operators in the OPEs. In summary the identificatioin between the perturbation $\Psi_B$ and the HHLL correlator with the CPOs reads as follows
\begin{equation}
    \label{eq:HHLLads3R1}
    \Psi_B(z,{\bar z}) = \langle [:\! O_H\!\,:] (0) \, [:\!\overline{O}_H\!:\,] (\infty) \,[\overline{O}] (1) \, [O] (z,{\bar z})\rangle\;,
  \end{equation}
  which holds of course  in the supergravity limit. 
  
We can perform the $\omega$ integral in~\eqref{eq:Psivs} as usual: by taking the Feynman contour in order to obtain the time ordered correlator, we can close the contour in the upper half plane for $t>0$ and use Cauchy theorem to pick the poles on the negative real axis. Then the correlator can be written as a double sum
\begin{equation}
  \label{eq:frVS}
  \Psi_B = \left(1-\frac{\alpha^2}{4}\right)\sum_{\myell=-\infty}^\infty   e^{i l\theta} \sum_{k=1}^\infty  \frac{\exp\left[-i\sqrt{\left(1-\frac{\alpha^2}{4}\right) (|l| + 2 k)^2+ \frac{\alpha^2  l^2}{4}} t \right]}{\sqrt{1+\frac{\frac{\alpha^2}{4}}{1-\frac{\alpha^2}{4}}\frac{l^2}{(|l|+2k)^2}} }\;,
\end{equation}
in agreement with~\cite{Bombini:2017sge}.

It is now straightforward to expand the result for small values of $\alpha$ and perform the sums by using the approach discussed in Appendix~\ref{app:heun} obtaining explicit expressions for $\Psi^{(n)}_B$
\begin{equation}
  \label{eq:PsialphaVS}
  \Psi_B = \sum_{n=0}^\infty \frac{1}{n!} \left(\frac{\alpha}{2}\right)^{2n} \Psi^{(n)}_B \;.
\end{equation}
For each term in the perturbative expansion it is possible to evaluate explicitly the sums over $k$ and $l$ in terms of polylogarithms as discussed in~\cite{Ceplak:2021wzz}. We summarize the formulas needed at the end of Appendix~\ref{app:heun}. The results for the correlators up to $n=4$ were already obtained in~\cite{Ceplak:2021wzz} where they are written in terms of the generalised Bloch-Wigner functions. Here we rewrite the first three cases ($n=1,2,3$) in terms of the ladder integrals ${\mathcal P}_n$ introduced in~\eqref{ladderdef}, see the ancillary file included with the {\tt arXiv} submission.

Let us now consider exactly the same calculation in the metric~\eqref{eq:sploc}. Again we can obtain an ODE after going to Fourier space~\eqref{eq:Phifs}. In terms of the variables
\begin{equation}
  \label{eq:xix}
  x=\xi^2\;,\quad {\phi}(\xi) = x^{-\frac{1}{2}} \tilde{\psi}(x)
\end{equation}
one obtains
\begin{equation}
  \label{eq:phisl}
  \tilde{\psi} ''(x) -\Bigg[ \frac{\left(l^2-1\right) }{4 x^2}+ \frac{\left(\alpha ^2+8\right)^2 (l^2-\omega^2) }{4 (x-1) x \left(\alpha ^4 x-64\right)}\Bigg]\tilde{\psi}(x)=0\;.
\end{equation}
This goes beyond the hypergeometric case which was enough to solve the radial equation~\eqref{eq:ads3vs} relevant for the geometry~\eqref{eq:sploc}. Instead~\eqref{eq:phisl} is of the Heun type with four regular singular points at $x=0,1,\frac{64}{\alpha^4}, \infty$. Thanks to recent progress exploiting the connections with ${\cal N}=2$ supersymmetric gauge theories and their AGT dual~\cite{Nekrasov:2009rc,Alday:2009aq}, it has been possible to solve in perturbation theory several interesting problems where the Heun equation appears both in the context of black hole physics~\cite{Aminov:2020yma,Bonelli:2021uvf,Bianchi:2021xpr,Bianchi:2021mft,Bonelli:2022ten,Consoli:2022eey,Bianchi:2023sfs,Bautista:2023sdf} and holography~\cite{Dodelson:2022yvn,Bianchi:2022qph,Giusto:2023awo}. This approach applies also to the case of Eq.~\eqref{eq:phisl} and we collect in Appendix~\ref{app:heun} a summary of the formulas needed. Since we are interested in the small $\alpha$ expansion, it is convenient to make the change of variables\footnote{This makes it manifest that the problem is symmetric under the exchange $\alpha \leftrightarrow 8/\alpha$, since the potential in~\eqref{eq:phislz} is related to the one in~\eqref{eq:phisl} by this map. This feature is related to the fact that the states with $n<N$ in~\eqref{eq:OHsl} are related to those with $N-n$ by a spectral flow. The case $n=N$ is the peak of the coherent state with $\alpha=\sqrt{8}$, which is the fixed point of the map above. Interestingly for this value the radial equation becomes hypergeometric.} $x=\frac{64}{\alpha^4} \hat{z} $ and consider $\psi(\hat{z})=\tilde{\psi}(\hat{z})$. We get
\begin{equation}
  \label{eq:phislz}
  \psi''(\hat{z}) +\left[(l^2-\omega^2) \left(\frac{\left(\alpha^2+8\right)}{4 \left(\alpha^2-8\right) (\hat{z}-1)}-\frac{\left(\alpha^2+8\right)^2 }{4 \alpha^4 \hat{z}}+\frac{1024 \left(\alpha^2+8\right)}{\alpha^4 \left(\alpha^2-8\right) \left(\alpha^4-64 \hat{z}\right)}\right) -\frac{l^2-1}{4 \hat{z}^2}\right]\psi(\hat{z})=0\;,
\end{equation}
with singular points at $\hat{z}=0,1,\frac{\alpha^4}{64}, \infty$. Even if this equation is more challenging than the one encountered in the previous example, it turns out that the behaviour at the AdS boundary of the regular solution  is still given by a formula similar to~\eqref{eq:2f1d} but with with a more complicated relation between the parameters $\hat{a}$, $\hat{b}$ and $\omega$, $l$. We provide in Appendix~\ref{app:heun} a sketch of this derivation, while here it is sufficient to say that the relation involves the Nekrasov-Shatashvili function $F$~\eqref{eq:Nek-Shat} of the ${\cal N}=2$ $SU(2)$ theory relevant for the Heun equation~\eqref{eq:phislz}
\begin{equation}
  \label{eq:hatahatb}
  \hat{a}+\hat{b} = |l|\;,\quad \hat{a}\, \hat{b} = \frac{l^2}{4} - a^2\quad \Leftrightarrow\quad \hat{a} = -a +\frac{|l|}{2}\;,\quad \hat{b} = a +\frac{|l|}{2}\;.
\end{equation}
Notice that $a$ is a function of $\omega$ and $l$~\eqref{eq:aSW} that can be straightforwardly calculated as an expansion in $\alpha$ (and it is sufficient to include the $1$-instanton contribution~\eqref{eq:1inst} to arrive up to order $\alpha^6$). For the geometry~\eqref{eq:sploc} under study the asymptotic relation between $\hat{z}$ and the canonical AdS radial coordinate~\eqref{eq:canads3} reads
\begin{equation}
  \label{eq:zhat2w}
  \xi^2 \approx 1-\frac{8-\alpha^2}{8+\alpha^2} w^2 \to 1\quad \Rightarrow \quad \hat{z} \approx \frac{\alpha^4}{64} \left(1-\frac{8-\alpha^2}{8+\alpha^2} w^2\right)\;.
\end{equation}
Thus we need to connect the regular solution at $\hat{z}=0$ with the local solutions at $\hat{z}=\frac{\alpha^4}{64}$. This step is summarised in Appendix~\ref{app:heun} and it turns out that the HHLL correlator takes the same functional form as in the previous case, but with a more complicated relation between ${a}$ and $\omega$, $l$, $\alpha$
\begin{equation}
  \label{eq:Psisl}
  \begin{aligned}
    \Psi_B  & = \sum_{\myell=-\infty}^\infty \int_{-\infty}^\infty \!\frac{d\omega}{2\pi}\,e^{i\omega t}\, e^{i l\theta}  \left(\frac{e^{-\partial_{a_t} F}\frac{8-\alpha^2}{8+\alpha^2}}{\frac{l^2-\omega^2}{4}}\, \hat{a} \,\hat{b} \right)  \left[\sum_{k=1}^\infty \left(\frac{1}{k+\hat{a}}+\frac{1}{k+\hat{b}}\right)\right] \\ & = \sum_{\myell=-\infty}^\infty \int_{-\infty}^\infty \!\frac{d\omega}{2\pi}\,e^{i\omega t}\, e^{i l\theta} \left[\sum_{k=1}^\infty \left(\frac{1}{k+\hat{a}}+\frac{1}{k+\hat{b}}\right)\right]\;,
    \end{aligned}
\end{equation}
where in the second step we used~\eqref{eq:simpmir} to simplify the round parenthesis. As before, we perform the $\omega$ integral by picking the contribution of the poles on the negative real axis and, to do so, we can use~\eqref{eq:asbs} to write $\hat{b}$ in terms of $\omega$ up to order $\alpha^6$. Since we used~\eqref{eq:PsiB} to reconstruct $\Psi_B$ from $\Phi_B$, there is a freedom to add terms that are in the kernel of the Ward Identity which again can be fixed by looking at protected contributions to the OPEs. The perturbation $\Psi_B$ around this geometry corresponds to the HHLL correlator
\begin{equation}
   \label{eq:HHLLads3R2}
    \Psi_B(z,{\bar z}) = \langle [O_H] (0) \, [\overline{O}_H] (\infty) \, [\overline{O}] (1) \, [O] (z,{\bar z})\rangle\;,
\end{equation}
which involves a different heavy operators with respect to~\eqref{eq:HHLLads3R1}. By taking into account the normalization in~\eqref{eq:OHsl}, it is natural to define the same small $\alpha$ expansion as in \eqref{eq:PsialphaVS}. For each term $\Psi^{(n)}_B$, the integral over $\omega$ can be done by using Cauchy's theorem. One obtains series over $k$ and $l$ which can be again performed by using the formulas needed at the end of Appendix~\ref{app:heun}. 

We summarise in the ancillary file included with the {\tt arXiv} submission, the results for $\Psi^{(n)}_B$ with $n=1,2,3$ following from the expansion of~\eqref{eq:HHLLads3R2}.

\subsection{The interpretation of the gravity result as a LLLL correlator}
\label{ssec:ads3LLLL}

As already discussed in the introduction, our working hypothesis is that the re-interpretation of the HHLL results in terms of light correlators with multi-particle operators is straightforward: as suggested by~\eqref{eq:corrperteta}, each term $\Psi^{(n)}_B$ should capture the supergravity contributions of order $1/N^n$ to a light correlator with two single-particle and two $n$-particle operators. Since we are working in the supergravity approximation, these results take into account only the contributions of tree level diagrams in the bulk, as depicted in Fig.~\ref{fig1}. Thus only {\em connected} diagrams contribute to each $\Psi^{(n)}_B$ as they are the only ones with the required scaling $1/N^n$ among all tree level contributions. Thus we indicate the contribution to the light correlators obtained from $\Psi_B$ as 
\begin{equation}
  \label{eq:diad3}
  \Psi_B^{(n)}(z,{\bar z}) = N^n \langle [O^n](0) \,[\overline{O}^n](\infty) \,[\overline{O}](1) \,[O](z,{\bar z})\rangle^c\,,
\end{equation}
where the subscript $c$ highlights that it captures only the connected tree level terms. Of course the nature of the multi-particle operators depends on the starting point and~\eqref{eq:diad3} is appropriate for the results obtained from~\eqref{eq:Psisl}. An analogue of Eq.~\eqref{eq:diad3} holds also for the perturbative expansion of~\eqref{eq:frVS}, but now with the multi-particle operators $:O^n:$ defined as in~\eqref{eq:O1ads3O2:}.

As usual in QFT, it is possible to reconstruct the full correlator once the connected contributions are known. However the precise form of the relation depends on the definition of the multi-particle operators adopted. The case of operators $:O^n:$ was discussed in~\cite{Ceplak:2021wzz} and, for instance, we have
\begin{align}
  \label{eq:con2cor1}
  \langle [:\!{O}^2\!:](1) [:\!\bar{O}^2\!:](2) & [\bar{O}](3) [{O}](4)\rangle^c   = \langle [:\!{O}\!:]^2(1) [:\!\bar{O}^2\!:](2) [\bar{O}](3) [{O}](4)\rangle \\ &  - 2 \langle [{O}](1) [\bar{O}](2)\rangle  \langle [{O}](1) [\bar{O}](2) [\bar{O}](3) [{O}](4)\rangle + (\langle [{O}](1) [\bar{O}](2)]\rangle)^2  \langle [\bar{O}](3) [{O}](4)\rangle\;, \nonumber
\end{align}
where the label in the round parenthesis keeps track schematically of the location of each operator. Instead when dealing with the multi-particle operators obtained by using the standard OPE~\eqref{eq:O1ads3O2}, we have\footnote{In our AdS$_3$ analysis the perturbation described by the wave $\Psi_B$ sits in a different tensor multiplet with respect to the fields used to construct the multi particle operators. For this reason we do not have correlators such as $\langle O(1) \bar{O}(3)\rangle$, which instead are present in the AdS$_5$ case where all supergravity fields are in the same multiplet, see for instance~\eqref{eq:doublecon}.}
\begin{align}\label{eq:con2cor2}
  \langle {O}^2(1) \bar{O}^2(2) \bar{O}(3) {O}(4)\rangle^c & = \langle O^2(1) \bar{O}^2(2) \bar{O}(3) {O}(4)\rangle \\ & - 4 \langle {O}(1) \bar{O}(2)\rangle  \langle {O}(1) \bar{O}(2) \bar{O}(3) {O}(4)\rangle + 4 (\langle {O}(1) \bar{O}(2)\rangle)^2  \langle \bar{O}(3) {O}(4)\rangle\;, \nonumber
\end{align}
which is now more easily written in terms of the unnormalized operators. From the equation above and the operator normalisations in \eqref{eq:normSL}, it follows that the $1/N^2$ contribution to the {\em full} correlator with two double-particle states receives contributions also from the disconnected term.

A first check of the approach discussed above is of course provided by the case $n=1$, i.e. the correlator with four single-particle operators. At this order, the two geometries~\eqref{eq:vanilsup} and~\eqref{eq:sploc} yield the same result which has all the features expected expected from such correlator~\cite{Giusto:2018ovt} and can also agrees with the independent result of~\cite{Rastelli:2019gtj}). As usual with tree-level holographic correlators, their space-time expression can be simplified by introducing the $\bar{D}_{\Delta_1 \Delta_2 \Delta_3 \Delta_4}$ functions which are directly related to the AdS integral of a $4$-point diagram with a single vertex connecting the external states 
(see Appendix D of \cite{Dolan:2001tt}).
In terms of these function the order $\alpha^2$ result takes the following simple form
\begin{equation}\label{eq:psi1gravads3}
{\Psi}_B^{(1)}(z,{\bar z}) = - \frac{1}{U} + \frac{V}{U}\, {\bar D}_{1122} \,, \quad \mbox{with} \quad U= (1-z)(1-\bar{z})\;, ~~ V = z \bar{z}\;.
\end{equation}

The functions mentioned above are directly related to the 4-point, one loop conformal integral. It is maybe not surprising that its multiloop generalization appears in the expressions for $\Psi_B^{(n)}$ with $n>1$. It is then convenient to introduce the following generalization of the standard $\bar{D}$ functions
\begin{equation}
D^{(n)}\equiv \frac{\mathcal{P}_n}{z-\bar{z}} \,,
\end{equation}
where $\mathcal{P}_n$ is the ladder integral~\eqref{ladderdef}. For $n=1$ we have $D^{(1)}=\bar{D}_{1111}$, which can be used as a seed to construct the general $\bar{D}_{\Delta_1 \Delta_2 \Delta_3 \Delta_4}$ (where the weights sum to an even number, which covers all the cases we need). For instance, we have ${\bar D}_{1122} = - U \partial_U D^{(1)}$. Then, at order $\alpha^4$, the result from the expansion~\eqref{eq:frVS}, already obtained in~\cite{Ceplak:2021wzz}, can be written compactly as follows
\begin{equation}\label{eq:psi2VS}
{\Psi}_B^{(2)}(z,{\bar z}) = -2 \,V\,\Bigl[\partial_U D^{(2)}-(1+V-U )\,\partial^2_U D^{(2)}  - \partial_U D^{(1)}\Bigr]\,.
\end{equation}
At this order the two geometries we studied explicitly start to differ and this is reflected in the 4-point correlators. However, given the CFT interpretation of the dual states~\eqref{eq:O1ads3f2} and~\eqref{eq:O1ads3O2}, the difference between the two operators is accounted for by a {\em single}-particle state of weight $h=\bar{h}=1$. As pointed out in~\cite{Ceplak:2021wzz}, this should imply that the terms of trascendentality higher than two are the same in the two results. We can test this picture by comparing explicitly the two results. One finds that
\begin{equation}
  \label{eq:SLmVS}
  \delta {\Psi}_B^{(2)} \equiv {\Psi}_B^{(2)}\eqref{eq:vanilsup} - {\Psi}_B^{(2)}\eqref{eq:sploc} = \left[\frac{V}{U} \bar{D}_{1133} + \frac{1}{V U}\right] \;,
\end{equation}
where we are indicating in brackets the geometry used to derive the correlators ${\Psi}_B^{(2)}$. As expected, the difference matches exactly the supergravity result for the correlator $\langle O_2 \bar{O}_2 \bar{O} O \rangle$ in the notation of~\cite{Rastelli:2019gtj,Giusto:2019pxc}.

The same pattern persists at order ${\cal O}(\alpha^6)$: the result for the geometry~\eqref{eq:vanilsup} was already discussed in~\cite{Ceplak:2021wzz}, but it can be rewritten compactly in terms of ladder integral
\begin{equation}\label{eq:psi3VS}
  \begin{aligned}
    {\Psi}_B^{(3)} = -6 \,V\, \Bigl[ & \partial_U D^{(3)} -3 (1+V-U )\,\partial^2_U D^{(3)}  + ((1+V-U )^2+2V)\,\partial^3_U D^{(3)} \\ &~ -\partial_U D^{(2)} + (1+V-U)\,\partial^2_U D^{(2)} \Bigr]\,.
  \end{aligned} 
\end{equation}
By following the same argument discussed for the $n=2$ case, we expect that the result obtained from the geometry~\eqref{eq:sploc} reproduces the first line, while the terms of trascendentality four or lower in the second line are different. It is straightforward to check that this is the case by using the explicit expression in the ancillary file included with the {\tt arXiv} submission.

By using the results summarized in Appendix~\ref{appLadder} it is straightforward to rewite \eqref{eq:psi2VS} and \eqref{eq:psi3VS} in Mellin space. It is convenient to factorize the R-symmetry dependence by introducing $\mathcal{H}^{(n)} = \frac{U}{V}\,{\Psi}_B^{(n)}$ (for $n=1$ one should take away the shift by $-1/U$ which is in the kernel of the Ward Identity). Our conventions for the Mellin transform in AdS$_3$ are given in~\eqref{conventionsMellinads3} and the Mellin amplitude ${\cal M}_{[1^n][1^n]11}$ can be written in terms of rational functions times the kernels introduced {in Appendix~\ref{appLadder}, see also~\eqref{tripleboxmellinamp}}. For $n=2$~\eqref{eq:psi2VS} yields
\begin{equation}
  {\cal M}_{[1^2][1^2]11}(s,t) =  \Big[ {K}_2(u,t) A(u,t) + \big(
\psi^{(0)}(-u)-\psi^{(0)}(-t) \big) B(u,t) + C(u,t) \Big]\;,
\end{equation}
with
\begin{align}
A(u,t)= \frac{1-\frac{2 u t}{s+2}}{s+1}\;,\quad
B(u,t)=  \frac{2 (u-t)}{(s+1) (s+2)}\;,\quad 
C(u,t)=  \frac{2 s}{(s+1) (s+2)}\;. 
\end{align}
A similar rewriting is possible also for $n=3$, for which~\eqref{eq:psi3VS} yields
\begin{equation}
  \label{eq:n=3m}
  \begin{aligned}
  {\cal M}_{[1^3][1^3]11}(s,t) &=  \Big[ {K}_4(u,t) a(u,t) + \,{K}_3(u,t)
b(u,t) \,+ \tilde{K}_2(u,t) c(u,t) +  \\
&\rule{3.3cm}{0pt} +
  {K}_2(u,t) A(u,t) + K_1(u,t) B(u,t) + C(u,t)\ \Big] \,,
  \end{aligned}
\end{equation}
with
\begin{equation}
  \label{eq:abcn3}
  \begin{aligned}
    a(u,t) & = \frac{4 (s - t) (-1 + s + t + 3 s t)}{(-1 + s + t) (s + t)
(1 + s + t)}\;, \\
    b(u,t) & =  \frac{4 (s-t) (3 s t+s+t-1)}{(s+t-1) (s+t) (s+t+1)}\;,\quad
    c(u,t) = \frac{6 \left(s^2-4 s t-s+t^2-t\right)}{(s+t-1) (s+t)
(s+t+1)}\;,\\
    A(u,t)& = \frac{3 \left(2 s^2 t-3 s^2+2 s t^2+16 s t+s-3
t^2+t-4\right)}{(s+t-1) (s+t) (s+t+1)} \;, \\
    B(u,t)& = -\frac{6 (s-t) (s+t+5)}{(s+t-1) (s+t) (s+t+1)} \;,\quad
    C(u,t) =  \frac{12 (s+t+2)}{(s+t-1) (s+t) (s+t+1)}  \,.
    \end{aligned}
\end{equation}
For convenience we summarized these Mellin space results in the ancillary file included with the {\tt arXiv} submission.

\section{AdS$_5$ /CFT$_4$ multi-particle correlators}
\label{sec:ads5}

The results for the holographic multi-particle correllators of the D1D5 CFT have a close analogue in the context of the AdS$_5$/$\mathcal{N}=4$ SYM duality. In the next subsection, we review the half-BPS spectrum of the CFT$_4$ and its dual supergravity description, emphasising similarities and differences with the 2D case. As in the previous section, the study of the linearised perturbations of heavy half-BPS states yields the generating function of correlators containing two single-particle and two $p$-particle operators in the supergravity limit. We derive the explicit form of these correlators up to $p=3$.
\subsection{The half-BPS sector of the spectrum}
We start by recalling some known facts about the half-BPS spectrum of $\mathcal{N}=4$ SYM with the aim of introducing the ingredients that make up the heavy operators admitting a semiclassical supergravity description. The HHLL correlators containing such operators will be computed in the supergravity approximation in the next subsection. Our discussion will parallel that presented in the previous section for the D1D5 CFT.

The spectrum of half-BPS states of $\mathcal{N}=4$ SYM consists of single and multi-particle operators.  Their description in the free theory starts by introducing the single-trace operators, $T_p$:
 \begin{equation}
 T_p(x,y) = \mathrm{Tr}\! \left(y_I\, \phi^I(x) \right)^p\;,
 \end{equation}
where $\phi^I$ ($I=1,\ldots,6$) are the scalars of the $\mathcal{N}=4$ multiplet, $x\in \mathbb{R}^4$ is a space-time point and $y\in \mathbb{C}^6$ is null, $y\cdot y=0$. These have protected dimension $\Delta=p$ and belong to the $[0,p,0]$ representation of the R-symmetry group $SU(4)_R$. In the $SU(N)$ theory the lightest single-trace is $T_2$ of dimension $\Delta=2$. Multi-trace operators formed by products of single-traces with the same $x$ {\it and} $y$, $T_{p_1}(x,y) \ldots T_{p_n}(x,y)$, also preserve half of the supersymmetries and are not affected by short distance singularities. 

At strong coupling, the half-BPS spectrum is naturally described by the linear perturbations of AdS$_5\times$S$^5$ in type IIB supergravity \cite{Kim:1985ez}. The operators dual to these supergravity modes are single-particle operators, $\mathcal{O}_p$, of the form
\begin{equation}
{\cal O}_p= T_p + \sum_{q_1+\ldots+ q_n=p} C_p^{q_1\ldots q_n} \,T_{q_1}\ldots T_{q_n} \,.
\end{equation}  
The coefficients $C_p^{q_1\ldots q_n}$ are suppressed by powers of $1/N$ and, thus, at leading order in the large $N$ expansion the single-particle operators, ${\cal O}_p$, coincide with the corresponding single-trace operators, $T_p$, but more generally the ${\cal O}_p$'s contain an admixture of multi-trace operators, as shown above, which becomes relevant for precision holography in the large $N$ expansion. Non-perturbatively, the  ${\cal O}_p$'s are uniquely defined as the operators orthogonal to all multi-trace operators.
At the lowest levels, $\Delta=2,3$, there is a single half-BPS state of dimension 
$\Delta$ and, thus, $\mathcal{O}_2=T_2$ and $\mathcal{O}_3=T_3$. At dimension $\Delta=4$ there 
are two degenerate states, $T_4$ and $T_2^2$, and the single-particle operator is
$$
\mathcal{O}_4=T_4  - \frac{2N^2-3}{N(N^2+1)}T_2^2\,.
$$ 
A general formula for any ${\cal O}_p$ is given in \cite{Aprile:2020uxk}. Note that the single-particle operator ${\cal O}_p$ exists only for $p\leq N$ and it vanishes when $p$ exceeds $N$. 
This can be checked explicitly with the formulae of \cite{Aprile:2020uxk} and it follows directly from the trace relations and the required orthogonality between ${\cal O}_p$ and all the multi-traces. Thus, the space of half-BPS operators with charge $p>N$ only contains multi-trace operators.  
Moreover  \cite{Aprile:2020uxk} showed that for $p\sim N$ with both $p,N\gg1$,  he operator ${\cal O}_p$ becomes precisely a giant graviton, as expected from the gravity picture. 

Note that in $\mathcal{N}=4$ SYM, the half-BPS single-particle state of dimension $\Delta=p$ is unique and thus the flavor group is trivial. This is an important qualitative difference with respect to the D1D5 CFT, which originates from the fact that the 10D supergravity theory on AdS$_5\times$ S$^5$ has maximal supersymmetry and thus only one supermultiplet. 

As for the AdS$_3$/CFT$_2$ duality, it is interesting to consider the gravitational description of multi-particle states where the number of single-particle constituents is of order of the central charge, which for $\mathcal{N}=4$ SYM is proportional to $N^2$. For half-BPS states this problem has been beautifully solved in \cite{Lin:2004nb}: The generic half-BPS geometry is uniquely specified by a configuration of droplets in a two-dimensional plane. The simplest deformation of pure AdS$_5\times$S$^5$ is the geometry associated with the ripple on a circle, whose profile is described in polar coordinates, $(r,\phi)$, by the curve $r^2(\phi) = 1+ \alpha \cos(p \phi)$; this solution was studied in \cite{Skenderis:2007yb} and more recently in \cite{Turton:2024afd}. 
At linear order in $\alpha$, this solution reduces to the perturbation dual to the single-particle state $\mathcal{O}_p$ \cite{Kim:1985ez} and it is natural to conjecture that higher orders in $\alpha$ are related with multi-particle operators, but determining the precise form of these operators is non-trivial. 

The problem become more tractable for the lowest harmonic, $p=2$. It was shown in \cite{Turton:2025svk} that for $p=2$ and at order $\alpha^2$ the dual state has the form $\mathcal{O}_2^2 + \# \,\mathcal{O}_4$. The mixing with $\mathcal{O}_4$ also has a natural supergravity explanation. It was indeed found in \cite{Cvetic:2000nc} that there exists a consistent truncation of type IIB supergravity on S$^5$ that projects out all the scalar fields but the ones in the $[0,2,0]$ rep; it can be checked that the ripple geometry does not fit in this truncation and this is consistent with the presence of a $[0,4,0]$ component in the dual state. By the same logic, if a geometry lies in the truncation of \cite{Cvetic:2000nc} its dual state can only involve multi-traces of the type $\mathcal{O}_2^n$. In this truncation, there is a unique, regular and normalisable solution with the  same linearised limit as the ripple geometry, which was constructed in \cite{Giusto:2024trt}. This geometry turned out to be equivalent to the ``BPS AdS bubble" found long ago in \cite{Chong:2004ce,Liu:2007xj} and which was shown to be equivalent to the particular LLM geometry corresponding to an ellipsoidal bubble \cite{Chen:2007du}. We use here the form of the solution of \cite{Liu:2007xj} and we only quote the asymptotically AdS$_5$ part of the metric, which is all we need for the computation of our HHLL correlator (the full 10D solution is given in Appendix~\ref{app:bpsbubble}): 
\begin{equation}\label{eq:5Dmetric}
ds^2_5=-H_1^{-2/3} f \,dt^2 + H_1^{1/3}(f^{-1} dr^2+ r^2 d\Omega_3^2)\,,
\end{equation}
with
\begin{equation}\label{eq:defH1}
f =  1+ r^2\, H_1\quad,\quad H_1=\sqrt{1+\frac{2(1+q_1)}{r^2}+\frac{1}{r^4}}-\frac{1}{r^2} \,.
\end{equation}
The geometry depends on the parameter $q_1$ which controls the ratio between the dimension of the state $\Delta_H$ and $N^2$. The simplest limit is $q_1=0$ (or equivalently $\Delta_H/N^2 = 0$), in which the metric reduces to global AdS$_5$. We show in Appendix~\ref{app:bpsbubble} the equivalence of this solution with the one of \cite{Giusto:2024trt}. This requires relating the parameter $q_1$ with the parameter $\epsilon$ used in \cite{Giusto:2024trt}, via
\begin{equation}\label{eq:q12epsilon}
q_1=\cosh\epsilon-1\,,
\end{equation}
and replacing the radial coordinate $r$ with the coordinate $\xi$ through a quite complicated transformation given in \eqref{eq:r2xi}. 

The form of the full 10D metric depends on the choice of the R-symmetry orientation, $y$, of the constituent state, $\mathcal{O}_2(x,y)$. 
We can choose $y=\frac{1}{\sqrt{2}}(1,i,0,0,0,0)$ and denote by $O_2$ the operator $\mathcal{O}_2$ with this particular R-symmetry orientation:
\begin{equation}
O_2 \equiv \frac{1}{2} \mathrm{Tr} (\phi_1+i\,\phi_2)^2 = \mathrm{Tr}(X^2)\;.
\end{equation}
Note that this implies that our supergravity calculation provides the HHLL correlator and its light limit only for this particular R-symmetry configuration, but the superconformal Ward identity recalled in Section~\ref{sec:CFTgen} will allow us to reconstruct the most general correlators. The explicit form of the 10D metric is not needed for the main purpose of this article and it is recalled in the Appendix~\ref{app:bpsbubble}. It is worth mentioning that this supergravity solution had appeared long ago in \cite{Liu:2007xj}, in a different, and simpler, parametrization. We give the relation between the two parametrizations in Appendix~\ref{app:bpsbubble}. 

The reader might have noticed a difference with the AdS$_3$ analysis, where we have introduced two different heavy states, denoted as $O_H$ and $:\!O_H\!:$ in \eqref{eq:skst} and \eqref{eq:OHsl}; both these states are constructed out of the lightest single-particle operator, $O$, but are based on different definitions of the multi-particle operators, $O^n$ and $:\!O^n\!:$. These two states are actually part of a one-parameter family of solutions that fit inside the consistent truncation containing the lowest KK mode. As we have reviewed above, the AdS$_5$ truncation of \cite{Cvetic:2000nc} contains a unique solution with the appropriate linearised limit, whose dual state should be though as the analogue of \eqref{eq:OHsl}. We will focus on this solution in the following. 

Our first task is to find the precise expression of this state. The general form implied by the above considerations is a superposition of states of the form $O_2^n$, without any mixing with single-particles other than $O_2$. For later convenience we use a basis of multi-particle states with unit norm, which we denote by $[O_2^n]$. One finds that 
\begin{equation}\label{eq:normo2}
[\mathcal{O}_2^n] = \mathcal{N}_n^{-1} \left(\frac{O_2}{\sqrt{2a}}\right)^n\quad,\quad \mathcal{N}_n = \sqrt{n!} \left[ \left(\frac{a}{2}\right)_n\,\left(\frac{a}{2}\right)^{-n}\right]^{1/2}\,,
\end{equation} 
where 
\begin{equation}
a\equiv N^2-1\,,
\end{equation}
is a number proportional to the central charge of the $SU(N)$ gauge theory. Note that, unlike in AdS$_3$, one can sum multi-particle states $[O_2^n]$ with arbitrarily large values of $n$, since for a 4D CFT there is no analogue of the stringy exclusion principle that applies to the D1D5 CFT \cite{Maldacena:1998bw}. Then one can write for the heavy state $O_H$ the general ansatz
\begin{equation}\label{eq:OHeta}
O_H = \sum_{n=0}^\infty b_n\, \alpha^n\, a^{n/2}\,[\mathcal{O}_2^n]\,,
\end{equation}
where $b_n$ are coefficients to be determined, $\alpha$ is a reparametrization of the supergravity parameter\footnote{We find it convenient to use here the parameter $\epsilon$, rather than $q_1$, because we will need to use the results \eqref{eq:VEVgrav}, which are expressed in that parametrization.} $\epsilon$, whose explicit form also needs to be found, and we have inserted appropriate powers of $a$ for later convenience. 

Our goal is to determine the coefficients $b_n$ and the relation between $\alpha$ and $\epsilon$, which we encode in a series of unknown coefficients $c_k$ ($k\ge 1$):
\begin{equation}\label{eq:eta2espsilon}
\alpha = \sum_{k=1}^\infty c_k \,\epsilon^k\,.
\end{equation}
Since $O_H$ reduces to the vacuum for $\epsilon=\alpha=0$, one has $b_0=1$ and one can set $b_1=1$ by appropriately rescaling $\alpha$. The remaining coefficients can be inferred by 
comparing the expectation values of protected single-trace operators in the state $O_H$, extracted from the geometry through the holographic dictionary \cite{Skenderis:2007yb}, with the values computed in the (free) field theory. On the gravity side, we can use the results of \cite{Giusto:2024trt}, which computed the expectation values of the R-current, $J$, and of $O_2$:
\begin{equation}\label{eq:VEVgrav}
\langle J \rangle_{sugra} = a\,\sinh^2\left(\frac{\epsilon}{2}\right)\quad ,\quad \langle [O_2] \rangle_{sugra} = c_{\mathcal{O}_2} \sinh \epsilon\,,
\end{equation}
where $c_{\mathcal{O}_2}$ is a normalisation constant that was left undertermined in \cite{Giusto:2024trt} and that we shall fix in \eqref{eq:holodictionary}; the $\langle J \rangle_{sugra}$ above is the exact result in $a$ obtained from
\cite{Giusto:2024trt} by replacing $N^2\to N^2-1$. The same expectation values can be computed in the free CFT by using the representation \eqref{eq:OHeta} for $O_H$:
 \begin{equation}\label{eq:VEVcft}
 \begin{aligned}
\langle J \rangle &\equiv \frac{\langle O_H | J | O_H \rangle}{\langle O_H| O_H \rangle} = \alpha\,\partial_\alpha \log\left(\sum_{n=0}^\infty b_n^2\,a^n\,\alpha^{2n}\right)\,,\\
\langle [O_2] \rangle  &\equiv \frac{\langle O_H | [O_2] | O_H \rangle}{\langle O_H| O_H \rangle} =\frac{\sum_{n=0}^\infty b_{n+1} b_n \,\mathcal{N}_{n+1}\,\mathcal{N}_n^{-1}\,a^{n+\frac{1}{2}}\,\alpha^{2n+1}}{\sum_{n=0}^\infty b_n^2\,a^n\,\alpha^{2n}}\,,
\end{aligned}
\end{equation}
with $\mathcal{N}_n$ given in \eqref{eq:normo2}. The identity for $\langle J \rangle$ follows simply from the fact that $[O_2^n]$ is an eigenvector of $J$ with eigenvalue $2n$, while to compute $\langle [O_2] \rangle$ we have used that when $[O_2]$ acts on the $[O_2^n]$ contained in $|O_H\rangle$ it generates a state proportional to $[O_2^{n+1}]$. 

Matching \eqref{eq:VEVgrav} and \eqref{eq:VEVcft} yields
\begin{equation}
\label{eq:holodictionary}
b_n =\frac{\mathcal{N}_n}{n!} = \frac{1}{\sqrt{n!}}\, \left[ \left(\frac{a}{2}\right)_n\,\left(\frac{a}{2}\right)^{-n}\right]^{1/2}\quad,\quad  \alpha = \frac{1}{\sqrt{2}} \tanh\left( \frac{\epsilon}{2}\right)\quad,\quad c_{\mathcal{O}_2} = \frac{\sqrt{a}}{2\sqrt{2}}\\.
\end{equation}
Hence the heavy state~\eqref{eq:OHeta} can be compactly written as
\begin{equation}\label{eq:OHAdS5}
O_H = \sum_{n=0}^\infty \frac{\mathcal{N}_n}{n!} \alpha^n \,a^{n/2} \,[O_2^n] = e^{\frac{\alpha}{\sqrt{2}}\,O_2}= \sum_{n=0}^\infty \frac{1}{\sqrt{n!}} \left(1+O(a^{-1})\right) \alpha^n \,a^{n/2} \,[O_2^n]\,.
\end{equation}
By using~\eqref{eq:holodictionary}, one can calculate the norm of $O_H$ obtaining $\langle O_H| O_H\rangle = (1-2\alpha^2)^{-a/2}$ which is finite in the allowed range $0\leq \alpha <1/\sqrt{2}$.

\subsection{Holographic HHLL correlators from gravity}
\label{sec:WaveEq}

It is known that HHLL correlators can be computed by studying the quadratic 
fluctuations of a light operator $O_L$ around the background geometry 
dual to a heavy operator $O_H$. We consider here the $\mathcal{N}=4$ SYM state 
$O_H$ in \eqref{eq:OHAdS5}, whose 10D geometry is given in \eqref{eq:10Dmetric}.  
The choice of $O_L$ will be dictated by simplicity: the lightest supergravity 
operator is ${O}_2$, the CPO in the graviton multiplet, whose linearised equations around the geometry \eqref{eq:10Dmetric} are, 
however, fairly involved; the field with the simplest equations of motion is the lowest S$^5$ harmonic 
of the axion-dilaton field, which we denote by $\Phi$. As we explain in more detail in Appendix~\ref{app:bpsbubble}, 
the wave equation of $\Phi$ in the metric dual to $O_H$ is simply the Klein-Gordon equation,
\begin{equation}\label{eq:5Dwe}
\Box_{5} \Phi=0\,,
\end{equation} 
where the 5D box, $\Box_5$, is defined by the metric reduced on S$^5$ \eqref{eq:5Dmetric}. The dual of $\Phi_0$ is an R-charge neutral operator of dimension $\Delta_L=4$ given by a superdescendant of $\mathcal{O}_2$:  
\begin{equation}
O_L= \bar{Q}^4 O_2
\end{equation}
where $Q^4$ is the product of the four supercharges acting non-trivially on $O_2$.

With this choice of $O_L$, it is thus enough to restrict ourselves to the metric $ds^2_5$ \eqref{eq:5Dmetric} of the AdS$_5$ consistent truncation. The HHLL correlator is extracted from the boundary limit of the regular solution of the Klein-Gordon equation for $\Phi_0$ \eqref{eq:5Dwe}. In the coordinates of \eqref{eq:5Dmetric}, the AdS boundary is reached for $r\to \infty$, where the 5D metric can be brought to the canonical Fefferman-Graham form
\begin{equation}\label{eq:wads5}
ds^2_5 = \frac{d w^2}{w^2} + \frac{1}{w^2} (-dt^2+ d\Omega_3^2+O(w^2))\,,
\end{equation}
by a change of coordinates connecting $r$ and $w$. For our purposes it is sufficient the work at the first order close to the AdS boundary where
$  w= \frac{1}{r} \left(1 + O(r^{-2})\right)$.
It will also be convenient in the following to isolate an $S^2$ within the spatial $S^3$ by writing
\begin{equation}
d\Omega_3^2 = d\theta^2 + \sin^2\theta \,d\Omega_2^2\,,
\end{equation}
with $\theta\in[0,\pi]$.

We look for a solution $\Phi_0$ that is regular everywhere in the bulk and has the asymptotic ($w\to 0$) behaviour
\begin{equation}\label{eq:Phi0as}
\Phi \approx \delta_N + w^4\, \Phi_B(t,\theta)\,,
\end{equation}
with $\delta_N$ the delta-function supported at $t=0$ and at the north-pole ($\theta=0$) of the spatial S$^3$ at the boundary. Our goal is to determine the function $\Phi_B(t,\theta)$, which, according to the holographic dictionary, represents the 4-point function
\begin{equation}\label{eq:PhiBcor}
\Phi_B(t,\theta) = \langle O_H(t=-\infty) \,\overline{O}_H(t=\infty) \,O_L(t=0,\theta=0) \,O_L(t,\theta)\rangle\,.
\end{equation}
Given the invariance of the problem with respect to the $S^2$ angular coordinates, $\Omega_2$, we can expand the wave function in an orthonormal basis of $\Omega_2$-invariant scalar harmonics on $S^3$
\begin{equation}
Y_\myell(\theta) = \frac{1}{\sqrt{2}\pi} \frac{\sin[(\myell+1)\theta]}{\sin\theta}\quad (\myell=0,1,\dots,\infty)\,.
\end{equation}
with $\int d\Omega_3 \,Y_\myell Y_\myell' = \delta_{\myell,\myell'}$. Hence we can write
\begin{equation}\label{eq:phi0fourier}
\Phi= \sum_{\myell=0}^\infty \int_{-\infty}^\infty \!\frac{d\omega}{2\pi}\,e^{i\omega t}\, Y_\myell(\theta)\,{\phi}(r) \,.
\end{equation}
In the same basis, the delta-function supported at the north pole can be written as
\begin{equation}\label{eq:deltaN}
\delta_N = \sum_{\myell=0}^\infty \frac{\myell+1}{\sqrt{2}\pi} \,Y_\myell(\theta)\,.
\end{equation}
Then, the wave equation \eqref{eq:5Dwe} reduces to the ODE
\begin{equation}\label{eq:ODE1}
r^{-3}\,\partial_r \!\left(r^3 f\, \partial_r \phi(r) \right) + \left(\frac{H_1}{f}\,\omega^2- \frac{\myell(\myell+2)}{r^2} \right)\phi(r)=0\,.
\end{equation}
Since we have found that the state $O_H$ rewrites in a particular simple way in terms of the parameter $\alpha$, see \eqref{eq:OHAdS5}, 
we will switch to this parametrization and replace
\begin{equation}\label{eq:q1alpha}
q_1=\frac{4\alpha^2}{1-2\alpha^2}
\end{equation}
in the functions $f$ and $H_1$.

As it happened in the AdS$_3$ case, it turns out that \eqref{eq:ODE1} is a Heun equation.
This is not immediately obvious because some change of variables are needed. First,  
it is convenient to rationalize  the square root in the definition of $H_1$ in~\eqref{eq:defH1}.
This can be done by introducing a new variable $x$ and, similarly to what was done in~\eqref{eq:xix}, rescale the wave function
\begin{equation}
  \label{eq:ratH1}
  r^2 = \frac{\sqrt{2} \alpha  \left(x+\frac{1}{x}\right)-\left(2 \alpha^2+1\right)}{1-2 \alpha^2}\;,\qquad \phi(r) = x^{\frac{1}{2}} \tilde{\psi}(x)\;.
\end{equation}
Then, in order to bring the problem exactly as in AdS$_3$ form, 
see~\eqref{eq:phisl} and~\eqref{eq:phislz}, we perform a second change of variables
\begin{equation}
  \label{eq:x2zads5}
  x=\frac{2 \alpha^2+\left(1-2 \alpha^2\right) \hat{z}}{\sqrt{2} \alpha }\;, \qquad \tilde{\psi}(x) = \hat{z}^{-1} (1-\hat{z})^{-1} \psi(\hat{z})\;.
\end{equation}
In terms of these variables~\eqref{eq:ODE1} reads
  \begin{equation}
  \label{heunAdS5}
\left(\partial_{\hat{z}}^2+\frac{\frac{1}{4}-a_1^2}{(\hat{z}-1)^2}
-\frac{\frac{1}{2}-a_0^2-a_1^2-a_t^2+a_{\infty}^2+u}{\hat{z}(\hat{z}-1)} 
+\frac{\frac{1}{4}-a_t^2}{(\hat{z}+\frac{q_1}{2})^2}+\frac{u}{\hat{z}(\hat{z} +\frac{q_1}{2})}+\frac{\frac{1}{4}-a_0^2}{\hat{z}^2}\right) \psi(\hat{z})=0 \,,
\end{equation}
where $q_1$ is \eqref{eq:q1alpha} and
\begin{equation}
  \label{valuesdictads5}
  \begin{gathered}
    a_0^2= \frac{1}{4} (l+1)^2\;,\quad a_1^2 = \frac{1}{4} \left(-l (l+2)+2 \omega^2+1\right)\;,\quad a_t^2 = 1\;,\quad a_\infty^2= 1\;,\\ 
    u= \frac{1}{4} \left(2 \alpha^2 \omega^2+8 \alpha^2-\left(2 \alpha^2-1\right) l (l+2)-\omega^2+4\right)\,.
  \end{gathered}    
\end{equation}
In the new coordinate the center of space, $r=0$, corresponds to $\hat{z}=0$, while the AdS boundary, $r\to\infty$, is mapped to $\hat{z}=-\frac{2\alpha^2}{1-2\alpha^2}=-\frac{q_1}{2}$.\footnote{However, the change of variables is a double covering: the interval $r\in[0,\infty)$ is covered by
$\hat{z}\in(-\frac{q_1}{2},0]$ and $\hat{z}\in[1,\infty)$.}
The calculation now follows the same steps as for the AdS$_3$ case~\eqref{eq:phislz}, where the key ingredient is the connection formula in~\eqref{eq:phiSL_dec}, that links the bulk center, $\hat{z}=0$, with the AdS boundary, $\hat{z}=-\frac{q_1}{2}$. The result is determined by the Nekrasov-Shatashvili function $F$ in ~\eqref{eq:Nek-Shat}, 
to be evaluated in~\eqref{valuesdictads5}, and by the parameters $\hat{a}$ and $\hat{b}$, which satisfy
\begin{equation}
   \hat{a}+\hat{b} = l\;,\quad  \hat{a} \,\hat{b} = \frac{l^2}{4} - a^2\quad \Leftrightarrow\quad \hat{a} = -a +\frac{l}{2}\;,\quad \hat{b} = a +\frac{l}{2}\;.
\end{equation}
(Note that $a$ can be obtained by inverting $F$ in \eqref{eq:ua}.) 
Hence, we find
\begin{equation}\label{eq:PhiBads5}
\Phi_B(t,\theta) = \frac{\pi}{12\,\sqrt{2}} \sum_{\myell=0}^\infty\sum_{k=1}^\infty \int_{-\infty}^\infty \!\frac{d\omega}{2\pi i}\,e^{i\omega t}\, Y_\myell(\theta)
(\myell+1) \left[\frac{e^{-\partial_{a_t} F}}{\left(1-2\alpha^2\right)^2} \,\hat{a}\,\hat{b}\,(\hat{a}+1)\,(\hat{b}+1)\right] \left[\frac{1}{k+\hat{a}}+\frac{1}{k+\hat{b}} \right]\,,
\end{equation}
where we have chosen in hindsight the overall normalisation to have a canonically normalised correlator. Again, it turns out that the first square parenthsis above all $\alpha$-dependent corrections cancel.\footnote{We checked this up to order $t^3\sim \alpha^6$ which is the order needed for the triple particle correlators discussed in this work.} In particular we have
\begin{equation}
  \label{eq:F2difop}
   \left[\frac{e^{-\partial_{a_t} F}}{\left(1-2\alpha^2\right)^2} \,\hat{a}\,\hat{b}\,(\hat{a}+1)\,(\hat{b}+1)\right] = \frac{1}{16} \left(l^2 - \omega^2\right) \left((l+2)^2-\omega^2\right)\;.
\end{equation}
The combination on the r.h.s.~can be interpreted as the eigenvalue of a particular differential operator. In order to see this we introduce the coordinates
\begin{equation}
z=e^{i(t+\theta)}\quad,\quad {\bar z} = e^{i(t-\theta)}\quad,\quad U = (1-z)(1-{\bar z})\quad,\quad V= z \,{\bar z}
\end{equation}
and the operator 
\begin{equation}\label{eq:defcalD}
\mathcal{D}\equiv U\partial_U^2 + V \partial_V^2 +(U+V-1)\partial_U \partial_V +2 \partial_U + 2 \partial_V\,.
\end{equation}
Note that $\mathcal{D}=\frac{1}{z-\bar{z}}\partial_{z}\partial_{\bar{z}}(z-\bar{z})$. Then the r.h.s. of~\eqref{eq:F2difop} is directly related to~\eqref{eq:defcalD} thanks to the identity
\begin{equation}\label{eq:scftidentity2}
\mathcal{D}^2 (e^{i\omega t} Y_\myell(\theta)) = \frac{e^{-4\,i t}}{16} (\myell^2-\omega^2) ((\myell+2)^2-\omega^2) \,e^{i\omega t} Y_\myell(\theta)\,.
\end{equation}
This makes it possible to write the correlator~\eqref{eq:PhiBads5} as follows
\begin{equation}\label{eq:scftWI}
\Phi_B(z,{\bar z}) \equiv e^{-4 i t}\,\Phi_B(t,\theta) = \frac{1}{12} \mathcal{D}^2 \Psi_B(z,{\bar z})\,,
\end{equation}
with
\begin{align}
\label{eq:psiB}
\Psi_B(z,{\bar z}) 
&= \frac{\pi}{\sqrt{2}} \sum_{\myell=0}^\infty \sum_{k=1}^\infty \int_{-\infty}^\infty \!\frac{d\omega}{2\pi i}\,e^{i\omega t }\,Y_\myell(\theta) (\myell+1)\,\Bigg[\frac{1}{k+\hat{a}}+\frac{1}{k+\hat{b}} \Bigg] \\
&= \sum_{\myell=0}^\infty \sum_{k=1}^\infty \int_{-\infty}^\infty \!\frac{d\omega}{2\pi i}\,\frac{l+1}{2}\, (z {\bar z})^{\frac{\omega-l}{2}}\, \frac{z^{l+1}-{\bar z}^{l+1}}{z-{\bar z}}\,\Bigg[\frac{1}{k+\frac{l}{2}+a(\omega,l;\alpha)}+\frac{1}{k+\frac{l}{2}-a(\omega,l;\alpha)}\Bigg]\,.
\label{eq:psiBlastline}
\end{align}
The identities in \eqref{eq:scftWI} have a neat CFT meaning: the factor $e^{-4 i t}$ relating $\Phi_B(t,\theta)$ and $\Phi_B(z,{\bar z})$ is the Jacobian that arises from the transformation of dimension 4 primary operator $O_L$  from the $(t,\theta)$ cylinder to the $(z,{\bar z})$ plane; the differential operator $\mathcal{D}^2$ that links $\Phi_B(z,{\bar z})$ and $\Psi_B(z,{\bar z})$ implements the superconformal Ward identity \cite{Drummond:2006by,Goncalves:2014ffa} that connects\footnote{The factor $1/12$ in \eqref{eq:scftWI} is needed to connect canonically normalised correlators, see e.g.~the identity $\mathcal{D}^2 U^{-2} = 12\,U^{-4}$.} the correlator (on the plane) with the insertion of $O_L = \overline{Q}^4 O_2$ with the correlator containing the chiral primary $O_2$ and its conjugate $\overline{O}_2$ (note that the superconformal descendant $O_L$ is instead real $O_L=\overline{O}_L$):
\begin{equation}\label{eq:WIcorr}
\langle O_H(z=0) \,\overline{O}_H(z\to \infty) \,O_L(z=1) \,O_L(z,{\bar z})\rangle=\frac{1}{12}\mathcal{D}^2 \langle O_H(z=0) \,\overline{O}_H(z\to \infty) \,\overline{O}_2(z=1) \,O_2(z,{\bar z})\rangle\,,
\end{equation}
and thus $\Psi_B(z,{\bar z})$ will be related with the correlator on the r.h.s. of the equation above, the precise connection being established in Section~\ref{sec:gravcft}. In the light of this interpretation, the identities \eqref{eq:F2difop}, \eqref{eq:scftidentity2} constitute a non-trivial consistency check between the supergravity calculation and the superconformal invariance of the CFT. Note also that  the Ward identity \eqref{eq:scftWI} determines $\Psi_B(z,{\bar z})$, given $\Phi_B(z,{\bar z})$, only up to terms proportional to the kernel of the operator $\mathcal{D}^2$. Our derivation leads naturally to the ansatz \eqref{eq:psiB}, where this ambiguity has been fixed: we will show that this choice is also natural from the CFT point of view.

The function $\Psi_B$ is thus the HHLL correlator for the light operator $O_2$. It can be 
expanded in $\alpha$ after solving perturbatively the equation for the poles: $k+\frac{l}{2}+a(\omega,l;\alpha) =0$. 
The first few terms of the expansion of $a(\omega,l)$ are 
\begin{equation}
a(\omega,l;\alpha)= \frac{\omega}{2} +\gamma_1(\omega,l) \alpha^2  + \gamma_2(\omega,l) \alpha^4+ \ldots
\end{equation}
where
\begin{align}
\gamma_1(\omega,l) &= \frac{ (\myell^2-\omega^2) ((\myell+2)^2-\omega^2)}{4(\omega^2-1)\omega} \\[.2cm]
\gamma_2(\omega,l)&=\frac{4}{3}\gamma_1(\omega,l) \Bigg[ \frac{3l(l+2)(2\omega^2-l(l+2))(15\omega^4-35\omega^2+8)}{32(\omega^2-4)(\omega^2-1)^2\omega^2} -\frac{29\omega^4-73\omega^2+8}{32(\omega^2-1)^2}+1\Bigg]
\end{align}
From the $\alpha$ expansion of $\Psi_B(z,\bar{z};\alpha)$ we define the functions $\Psi_B^{(n)}(z,{\bar z})$:
\begin{equation}\label{eq:psiexpads5}
\Psi_B(z,{\bar z};\alpha)  =  \sum_{n=0}^\infty \frac{\alpha^{2n}}{n!} \,\Psi_B^{(n)}(z,{\bar z})\,;
\end{equation}
as we will show in the next subsection, $\Psi_B^{(n)}(z,{\bar z})$ captures the connected part of the correlator with $n$-particle correlators, 
and it is thus the interesting object to study. It is instructive to look at the structure of the integrand of $\Psi_B^{(n)}(z,{\bar z})$, which is obtained from \eqref{eq:psiBlastline}. For this purpose it is enough to expand the energy factor, 
\begin{equation}
E_{\rm gravity}(\omega,l;\alpha)=\frac{1}{k+\frac{l}{2}+a(\omega,l;\alpha)}+\frac{1}{k+\frac{l}{2}-a(\omega,l;\alpha)}\,.
\end{equation}
At each order in $\alpha^{2n}$ the result is a sum of terms indexed by the degree of the zero-th order poles $\omega=\pm(l+2k)$. For fixed $n$, 
the highest degree of the pole will be $n+1$ and the coefficient of the highest pole is given by the $n$-th power of $\gamma_1$. Thus, at order $\alpha^{2n}$, one can write an expansion of the form 
\begin{align}
\label{expagravityCFT3}
E_{\rm gravity}(\omega,l;\alpha)\Big|_{\alpha^{2n}} &= \frac{(+\gamma_1(\omega,l))^n}{(k+\frac{l-\omega}{2})^{n+1}} +\frac{(-\gamma_1(\omega,l) )^n}{(k+\frac{l+\omega}{2})^{n+1}}  +\ldots\,,
\end{align}
where the dots include all possible terms containing poles of degree strictly less than $n+1$; note that these lower degree poles are absent at order $\alpha^0$ and $\alpha^2$ and thus, in these cases, the expression above is complete without the dots.
In  formula \eqref{expagravityCFT3}, the function $\gamma_1$ is the ``anomalous dimension" of the bound states between $O_2^n$ and $O_2$ in the BMN picture that we mentioned in the introduction. 
As such, it determines the leading $\log(z\bar{z})$ term, which by explicit computation we can express as the double sum
\begin{equation}\label{maxloggravity}
 \,\Psi_B^{(n)}(z,{\bar z})\Big|_{\alpha^{2n}} = \log^n(z\bar{z}) \sum_{l\ge 0}\sum_{k\ge1}  (l+1)  \Bigg[ \frac{ -4(k-1)k(k+l)(k+l+1)}{(-1+2k+l)(2k+l)(1+2k+l)} \Bigg]^n\!\!(z\bar{z})^k\, \frac{z^{l+1}-\bar{z}^{l+1}}{z-\bar{z}} +\ldots\,.
\end{equation}
where all the terms omitted give subleading powers in the $\log(z\bar{z})$ stratification. Note also that the numerator of $\gamma_1(\omega,l)$ is the same as the eigenvalue of 
the differential operator ${\cal D}^2$ in \eqref{eq:scftidentity2}.\footnote{A similar relation between the numerator of the anomalous dimensions and the eigenvalue of a Casimir was noticed for the tree level single-particle correlators  \cite{Aprile:2018efk,Caron-Huot:2018kta,Drummond:2006by}. The operator there was $\Delta^{(8)}_{[0,0,0]}={\cal D}^2 U^2V^2{\cal D}^2$. }

The term with $n=0$ in \eqref{maxloggravity} corresponds to the trivial state $O_H(\alpha=0)=1$ and thus to the free propagator $\langle \bar{O}_2O_2\rangle$. The series over $k$ and $l$ are geometric and can be readily resummed to 
\begin{equation}
\Psi_B^{(0)}(z,{\bar z}) = \frac{V}{U^2}\,.
\end{equation}
However, this result is ambiguous due to the kernel of the differential operator ${\cal D}^2$ that we had to integrate to go from $\Phi_B$ to $\Psi_B$. In fact, up to an element in the kernel of $\mathcal{D}^2$, we can replace $\Psi_B^{(0)}(z,{\bar z}) \to  \frac{1}{U^2}$ which agrees with the free propagator for the dimension 2 operator $O_2$. 

To compute the full expression for $\Psi_B^{(n)}(z,{\bar z})$ systematically, including the subleading powers in $\log(z\bar{z})$, one can use the following rewriting of \eqref{eq:psiB}
\begin{equation}
\Psi_B(z,{\bar z}) = {-\frac{1}{4\sin\theta}} \partial_\theta \sum_{\myell\in\mathbb{Z}}^\infty \sum_{k=1}^\infty \int_{-\infty}^\infty \!\frac{d\omega}{2\pi i}\,e^{i\omega t+i l \theta}\,\Bigg[\frac{1}{k+\frac{|l|-1}{2}+a(\omega,|l|-1;\alpha)}+\frac{1}{k+\frac{|l|-1}{2}-a(\omega,|l|-1;\alpha)}\Bigg]\,.
\end{equation}
After performing the $\omega$ integral and expanding in $\alpha$, one can rewrite the full correlator $\Psi_B^{(n)}(z,{\bar z})$ in terms of double series of the form 
\begin{equation}\label{eq:polysums}
\sum_{\myell\in \mathbb{Z}}^\infty  \sum_{k=1}^\infty e^{-i (2k+|\myell|) t+ i \,\myell\, \theta} \frac{\myell^{2p}}{(2k+|\myell|\pm q)^n}
\end{equation}
for some integers $p$, $q$ and $n$. Such series can be resummed in terms of $\mathrm{Li}_n(e^{-i(t\pm\theta)})$ and elementary functions. The order of the polylogarithms grows with the perturbative order: the correlator at order $\alpha^{2n}$ contains $\mathrm{Li}_{n'}(e^{-i(t\pm\theta)})$ with $n'\le 2n$.  

We summarise in the ancillary file included with the {\tt arXiv} submission the results for $\Psi^{(n)}_B$ with $n=1,2,3$ following from the expansion of~\eqref{eq:psiB}, 
and we list at the end of Appendix~\ref{app:heun} the series that are needed for resummation, up to $n=3$.

Following our procedure, one can show that the first non-trivial correction, of order $\alpha^2$, can be expressed in terms of the standard D-functions 
\begin{equation}\label{eq:psi1grav}
{\Psi}_B^{(1)}(z,{\bar z}) =-4 \, V^2 {\bar D}_{2422} \,.
\end{equation}
This result was also obtained starting from a different asymptotically AdS$_5$ solution~\cite{Turton:2024afd}. The precise form of the CFT state dual to the geometry considered in~\cite{Turton:2024afd} is not known, but it agrees with~\eqref{eq:OHeta} up to order $\alpha$, while starting at order $\alpha^2$ the two states start to differ~\cite{Turton:2025svk}. Thanks to the all order dictionary~\eqref{eq:OHAdS5}, the geometry considered here makes it possible to extract the higher order corrections, $\Psi_B^{(n)}(z,{\bar z})$ with $n>1$. As in the AdS$_3$ case discussed in the previous section, also the AdS$_5$ results can be written in terms of the ladder integrals ${\mathcal P}_n$~\eqref{ladderdef} and have a dual CFT interpretation both as a ``light" limit of a HHLL correlator, which will be further explored in Appendix~\ref{app:ReggeLimit}, and as a series of LLLL correlators, as we will show in Section~\ref{sec:gravcft}.

\subsection{The interpretation of the gravity result as a LLLL correlator}
\label{sec:gravcft}

Section~\ref{sec:WaveEq} describes the computation of the functions ${\Psi}_B^{(n)}(z,{\bar z})$, derived from the perturbative (in $\alpha$) expansion of the wave function $\Phi_0$ in the geometry \eqref{eq:5Dmetric}. Using the holographic dictionary \eqref{eq:holodictionary} for the state $O_H$ dual to this geometry, we provide here a precise identification of ${\Psi}_B^{(n)}(z,{\bar z})$ with a CFT correlator containing two single-trace and two $n$-trace operators.

Considering first finite $\alpha$, the Ward identity \eqref{eq:WIcorr} suggests to identity $\Psi_B(z,{\bar z})$ with the HHLL  correlator $\langle O_H(0) \,\overline{O}_H(\infty) \,\overline{O}_2(1) \,O_2(z,{\bar z})\rangle$, computed at tree level in the $1/N$ expansion and at the supergravity point corresponding to string 't Hooft coupling. This is, however, imprecise for two reasons: First, one could add to $\Psi_B(z,{\bar z})$ terms that are annihilated by the Ward identity operator $\mathcal{D}^2$; with the choice for $\Psi_B(z,{\bar z})$ that we have made in \eqref{eq:psiB}, one captures only the ``dynamical'' part of the correlator, that corresponds to the term denoted as $\mathcal{H}_{\vec{p}}$ in \eqref{partial_nreno}. Second, the wave function only computes the connected part of the correlator. Focusing only on the dynamical part of the correlator, we are thus led to the identification 
\begin{equation}\label{eq:psiBconn}
\Psi_B(z,{\bar z}) = \langle [O_H](0) \,[\overline{O}_H](\infty) \,[\overline{O}_2] (1) \,[O_2] (z,{\bar z})\rangle - \langle [O_H](0) \,[\overline{O}_H](\infty) \,[\overline{O}_2] (1) \rangle \langle [O_H](0) \,[\overline{O}_H](\infty) \,[O_2](z,{\bar z})\rangle \,,
\end{equation}
where we recall the the square parenthesis denote the canonically normalised operators. We can now use the form \eqref{eq:OHAdS5} for the state $O_H$ to expand the r.h.s. of the equation above for $\alpha\to 0$ and $N\to \infty$ and assume that one can identify terms of the same order in $\alpha$ at leading order in $1/N$ on the two sides of the equation. Note that this assumption is non-trivial, because it entails an exchange of the $\alpha\to 0$ and the $N\to \infty$ limits: while in the supergravity computation leading to $\Psi_B(z,{\bar z})$ one first takes $N\to \infty$ keeping $\alpha$  finite and then sends $\alpha\to 0$, the interpretation of the r.h.s. as the generator of correlators with light operators is obtained by first expanding for $\alpha\to 0$ and then sending $N\to \infty$. The ultimate justification for this assumption is that it leads, as we will see, to consistent results, and we suspect that this happens thanks to the supersymmetric nature of the operators contained in the correlators. 

Then, equating powers of $\alpha$ on the two sides of Eq. \eqref{eq:psiBconn} implies:
\begin{equation}\label{eq:corrdictionary}
\begin{aligned}
&\Psi_B^{(0)}(z,{\bar z}) =\frac{1}{U^2}\quad ,\quad \Psi_B^{(n)}(z,{\bar z}) =a^n \langle [O_2^n](0) \,[\overline{O}_2^n](\infty) \,[\overline{O}_2](1) \,[O_2](z,{\bar z})\rangle^c\,,
\end{aligned}
\end{equation}
where $\langle [O_2^n](0) \,[\overline{O}_2^n](\infty) \,[\overline{O}_2](1) \,[O_2](z,{\bar z})\rangle^c$ denotes the connected part of the correlator thought as the limit of a $2(n+1)$-point amplitude where the first $n$ and the second $n$ points are made to coincide. The relation between the connected and the full correlators are given by the usual QFT formulas; these become quite lengthy as $n$ grows, but for our purposes it is sufficient to keep the terms with four-point functions, which are the only ones to have non-trivial dynamical parts. Denoting by dots the remaining terms, one has, for $n\le 3$:
\begin{equation}\label{eq:doublecon}
\begin{aligned}
\langle \mathcal{O}_2(1) \mathcal{O}_2(2) \mathcal{O}_2(3) \mathcal{O}_2(4)\rangle^c&=\langle \mathcal{O}_2(1) \mathcal{O}_2(2) \mathcal{O}_2(3) \mathcal{O}_2(4)\rangle+\ldots\\
 \langle \mathcal{O}_2^2(1) \mathcal{O}_2^2(2) \mathcal{O}_2(3) \mathcal{O}_2(4)\rangle^c &= \langle \mathcal{O}_2^2(1) \mathcal{O}_2^2(2) \mathcal{O}_2(3) \mathcal{O}_2(4)\rangle - 4 \langle \mathcal{O}_2(1) \mathcal{O}_2(2) \rangle \,\langle \mathcal{O}_2(1) \mathcal{O}_2(2) \mathcal{O}_2(3) \mathcal{O}_2(4)\rangle^c+\ldots\,\\
 \langle \mathcal{O}_2^3(1) \mathcal{O}_2^3(2) \mathcal{O}_2(3) \mathcal{O}_2(4)\rangle^c & = \langle \mathcal{O}_2^3(1) \mathcal{O}_2^3(2) \mathcal{O}_2(3) \mathcal{O}_2(4)\rangle- 9 \langle \mathcal{O}_2(1) \mathcal{O}_2(2) \rangle \,\langle \mathcal{O}_2^2(1) \mathcal{O}_2^2(2) \mathcal{O}_2(3) \mathcal{O}_2(4)\rangle^c\\
 &- 9 \langle \mathcal{O}_2^2(1) \mathcal{O}_2^2(2) \rangle \,\langle \mathcal{O}_2(1) \mathcal{O}_2(2) \mathcal{O}_2(3) \mathcal{O}_2(4)\rangle^c+\ldots\,,
\end{aligned}
\end{equation}
where, for brevity, we have denoted $\mathcal{O}_2(i)\equiv \mathcal{O}_2(x_i,y_i)$. Note also that the relations \eqref{eq:doublecon} are expressed in terms of the non-normalised operators $\mathcal{O}_2^n$ and, thus, to make contact with the normalised correlators appearing in \eqref{eq:corrdictionary} one has to use the normalisations in \eqref{eq:normo2}.

To summarise, Eq. \eqref{eq:corrdictionary} establishes the dictionary between the outputs of the gravity computation, $\Psi_B^{(n)}(z,{\bar z})$, and the dynamical part of the connected CFT correlators with two single-particle and two $n$-particle operators. The relations \eqref{eq:doublecon} allow us, in some cases, to reconstruct the full correlators from their connected parts. We conclude this section with some comments on this dictionary. The tree-level gravity computation leading to $\Psi_B(z,{\bar z})$ only encodes the leading large $N$ limit of the correlators appearing in \eqref{eq:corrdictionary}. Since the connected part of the normalised correlator $\langle [O_2^n](0) \,[\overline{O}_2^n](\infty) \,[\overline{O}_2](1) \,[O_2](z,{\bar z})\rangle$ goes at large $N$ like $a^{-n}$, $\Psi_B^{(k)}(z,{\bar z})$ has a finite large $N$ limit. The first non-trivial correlator, $\Psi_B^{(1)}(z,{\bar z})$, provides a check of our computation:\footnote{This observation has already appeared in \cite{Turton:2024afd}, which also extends it to correlators with higher-dimensional single-trace 1/2-BPS operators.} from the gravity result \eqref{eq:psi1grav} and \eqref{eq:corrdictionary} for $n=1$, we obtain
\begin{equation}
\langle [O_2](0) \,[\overline{O}_2](\infty) \,[\overline{O}_2](1) \,[O_2](z,{\bar z})\rangle^c = -\frac{4}{a} \,V^2 {\bar D}_{2422} \,,
\end{equation}
which agrees with the tree-level result for this 4-point function \cite{Dolan:2004iy}, up to a free-field term which is not included in the dynamical part. Moving to the next-order term, $n=2$, the leading contribution to the connected correlator is of order $1/a^2$. One should note that the full correlator  $\langle [O_2^2] \,[\mathcal{O}_2^2] \,[\overline{O}_2]\,[O_2]\rangle$ receives, at that order, both tree level and one-loop contributions: the latter come from disconnected Witten diagrams (see Figure 1.c in \cite{Ceplak:2021wzz}) and correspond to the second term in the r.h.s. of the second line of \eqref{eq:doublecon}. These type of contributions are subtracted in the definition of the connected correlator and, as expected, they do not appear in the gravity result, which only captures tree-level diagrams. However, since the 1-loop corrections to the 4-point single-trace correlator $\langle [O_2]\, [\bar{O}_2] \,[\bar{O}_2] [O_2]\rangle$ are known \cite{Alday:2017xua,Aprile:2017bgs,Alday:2017vkk,Alday:2018kkw,Aprile:2019rep}, one can use our result for $\Psi_B^{(2)}(z,{\bar z})$ and the relation \eqref{eq:doublecon} to reconstruct the full correlator $\langle [O_2^2] \,[\mathcal{O}_2^2] \,[\overline{O}_2]\,[O_2]\rangle$  up to order $1/a^2$. A similar argument applies to the correlator with triple-particles, $\langle [O_2^3] \,[\mathcal{O}_2^3] \,[\overline{O}_2]\,[O_2]\rangle$, whose connected part starts at order $1/a^3$; to reconstruct the full correlator at this order one would need, from \eqref{eq:doublecon}, both the two-loop correction to $\langle [O_2]\, [\bar{O}_2] \,[\bar{O}_2] [O_2]\rangle$ \cite{Aprile:2018efk,Drummond:2022dxw} and the one-loop correction to $\langle [O_2^2]\, [\bar{O}_2]^2 \,[\bar{O}_2] [O_2]\rangle$, which is, however, not yet available.

\section{The general CFT correlator}\label{sec:CFTgen}

We will now show how the full CFT correlators can be reconstructed from the gravity results 
presented in previous sections. We begin by recalling some basics about four-point functions of 
half-BPS operators in ${\cal N}=4$ SYM. Very generally, these can be written as function of cross ratios as follows, 
\begin{equation}\label{fa_4ptfunc}
\langle {\cal O}_{p_1}(\vec{x}_1,\vec{y}_1) {\cal O}_{p_2}(\vec{x}_2,\vec{y}_2) {\cal O}_{p_3}(\vec{x}_3,\vec{y}_3) {\cal O}_{p_4}(\vec{x}_4,\vec{y}_4) \rangle =  
{\cal P}_{\vec{p}} \ \hat{\cal C}_{\vec{p} }( U,V,\sigma,\tau)
\end{equation}
where $\vec{p}=\{p_1,p_2,p_3,p_4\}$ is a vector of BPS charges, the cross ratios are defined as follows,
\begin{equation}\label{def_cross_ratiosFA}
U=\frac{\vec{x}^{\,2}_{12}\vec{x}^{\,2}_{34}}{  \vec{x}^{\,2}_{13} \vec{x}^{\,2}_{24} } \qquad;\qquad  
V=\frac{\vec{x}^{\,2}_{14} \vec{x}^{\,2}_{23} }{  \vec{x}^{\,2}_{13}\vec{x}^{\,2}_{24} }  \qquad;\qquad  
\sigma=\frac{ \vec{y}^{\,2}_{13} \vec{y}^{\,2}_{24} }{ \vec{y}^{\,2}_{12} \vec{y}^{\,2}_{34} }\qquad;\qquad 
\tau=\frac{ \vec{y}^{\,2}_{14} \vec{y}^{\,2}_{23} }{ \vec{y}^{\,2}_{12} \vec{y}^{\,2}_{34} }
\end{equation}
and ${\cal P}_{\vec{p}}$ is a prefactor  built out of propagators and normalizations.  
We will specify ${\cal P}_{\vec{p}}$ case by case.  

The function $\hat{\cal C}_{\vec{p} }( U,V,\sigma,\tau)$ is further
constrained by the ${\cal N}=4$ superconformal Ward identity \cite{Nirschl:2004pa} 
and a ``partial non-renormalization" theorem \cite{Eden:2000bk}.
As a result, correlators of half-BPS operators admit a unique decomposition
 into a free theory contribution, and a coupling dependent contribution ${\cal H}_{\vec{p}}$
 with reduced kinematics: 
\begin{equation}\label{partial_nreno}
\hat{\cal C}_{ \vec{p} }= {\cal G}^{\rm free}_{ \vec{p} }(U,V,\sigma,\tau) + 
{\cal I}(U,V,\sigma,\tau) {\cal H}_{ \vec{p} }(U,V,\sigma,\tau) \;,
\end{equation}
where  
\begin{equation}\label{intriligator_factor}
{\cal I}= V+\sigma V(V-1-U)+\tau(1-U-V) +\sigma\tau U(U-1-V)+\sigma^2 U V + \tau^2 U\;.
\end{equation}
We shall refer to ${\cal H}_{\vec{p}}$ as the dynamical function. The free theory part is the same for all values of the couplings and 
can be computed at $g_{YM}=0$. This is natural in perturbation theory, however, it is important to emphasise that
at strong coupling, the connected part of ${\cal G}^{\rm free}_{ \vec{p} }$ and ${\cal H}_{\vec{p}}$ come 
from a sum of  Witten diagrams and the split \eqref{partial_nreno} is not manifest on the gravity side. 
We will discuss the implication of this statement in the next section.

The dynamical function spans a number of $su(4)$ reps equals to $\frac{1}{8}\kappa(\kappa+2)$,
where $\kappa$ is called the extremality parameter:
\begin{equation}
\kappa=   \min(p_1+p_2,p_3+p_4) - \max( p_{43},p_{12} )-2 \,.
\end{equation}
The reduction by the factor of $-2$ comes from the ${\cal I}$ polynomial. 
When $\kappa=0,1$ the correlator is extremal, next-to-extremal,  respectively. These have no dynamics a
nd coincide with their free value \cite{Erdmenger:1999pz,Eden:2000gg}. The first case with dynamics is
the case $\kappa=2$, which we call next-to-next-to-extremal, for short N$^2$E. 
The dynamical function ${\cal H}_{\vec{p}}$ of  N$^2$E correlators is the simplest, since it does not depend on $\sigma, \tau$.

The gravity correlators $\Psi_B^{(n)}$  that we studied in section  \ref{sec:gravcft} correspond to 
a particular choice of polarizations $y_{i}$ in the correlators 
\begin{equation}
\langle {\cal O}_2^n {\cal O}_2^n {\cal O}_2 {\cal O}_2 \rangle\;,
\end{equation}
specifically,  $\sigma=1$, $\tau=0$. 
Since these correlators are N$^2$E correlators the knowledge of $\Psi_B^{(n)}$ would be enough to reconstruct  
the dynamical function of the correlator. 
However, as discussed already in section \ref{sec:gravcft}, the expansion of the HHLL correlator on the gravity 
side computes the leading term of the \emph{connected} correlator, 
thus it computes the leading term of the connected contribution to the dynamical function. 
The relation between the connected dynamical function, called ${\cal H}^c$ in the following, 
and the dynamical function ${\cal H}$, follows from basic definitions in QFTs. 
We quote below the first cases of interest:
\begin{align}\label{connFA_general1}
{\cal H}_{2222}= &\ {\cal H}^c_{2222}  \\[.2cm]
{\cal H}_{[2^2][2^2]22}= &\ {\cal H}^c_{[2^2][2^2]22}  +2^2 \frac{  \langle {\cal O}_2 {\cal O}_2  \rangle \langle {\cal O}_2 {\cal O}_2 \rangle}{ \langle {\cal O}^2_2 {\cal O}^2_2  \rangle }  {\cal H}^c_{2222}\\[.2cm]
{\cal H}_{[2^3][2^3]22}= &\ {\cal H}^c_{[2^3][2^3]22}  + 3^2 \frac{  \langle {\cal O}_2 {\cal O}_2  \rangle \langle {\cal O}^2_2 {\cal O}^2_2 \rangle}{ \langle {\cal O}^3_2 {\cal O}^3_2  \rangle }{\cal H}^c_{[2^2][2^2]22} +  3^2 \frac{  \langle {\cal O}^2_2 {\cal O}^2_2 \rangle \langle {\cal O}_2 {\cal O}_2  \rangle  }{ \langle {\cal O}^3_2 {\cal O}^3_2  \rangle }{\cal H}^c_{2222} \;.
\label{connFA_general3}
\end{align}
With the definition of $\Psi_B$ in \eqref{eq:psiB}, the gravity computation gives 
\begin{equation}\label{CFTfromgravity}
{\cal H}^c_{[2^n][2^n]22}\Big|_{\frac{1}{a^n}}= \frac{U^2}{V^2} \Psi_B^{(n)}\,,
\end{equation}
where the prefactors correspond to, $V^2={\cal I}(U,V,1,0)$, and the $U^2$ comes 
from \eqref{def_cross_ratiosFA} by recalling that $\Psi_B$ is computed 
in the frame where $\vec{x}_1=0$, $\vec{x}_2=\infty$, $\vec{x}_3=1$ and $\vec{x}_4=(z,\bar{z})$. 

We see from \eqref{connFA_general1}-\eqref{connFA_general3} that in order to reconstruct a given ${\cal H}_{[2^n][2^n]22}$ up to order $\frac{1}{a^n}$ 
one has to consider the contribution from both $\Psi_B$ and disconnected contributions from ${\cal H}_{[2^m][2^m]22}$ with $m<n$.
The interesting question then is to understand how all these contributions fit together at a given order.
To demonstrate this interplay it will be advantageous to write the dynamical function in Mellin space, 
and in particular, extract from the position space the amplitude $\mathcal{M}_{[2^n][2^n]22}(s,t)$ defined as follows,
 \begin{equation}\label{conventionsMellin}
 \mathcal{H}_{[2^n][2^n]22}(U,V)=\oint \frac{ds dt}{(2\pi i)^2} \,U^{s+2} V^t \,\Gamma^2(-s) \Gamma^2(-t) \Gamma^2(s+t+4) \,\mathcal{M}_{[2^n][2^n]22}(s,t)\;.
 \end{equation}
In the next sections we give the explicit results for $n=1,2,3$, and discuss the aforementioned interplay between connected and disconnected contributions, 
first at the level of transcendental functions and then from the point of view of the OPE.

\subsection{$\langle {\cal O}_2 {\cal O}_2 {\cal O}_2 {\cal O}_2 \rangle$ }\label{sec2222}

The first correlator to consider is the four-point function of the scalar CPO in the ${\cal N}=4$ stress tensor multiplet: ${\cal O}_2$.
The tree level correlator, corresponding in \eqref{CFTfromgravity} to $n=1$,  has been the first full fledge holographic 
correlator to be computed with Witten diagrams techniques, and also the first one to be bootstrapped 
enterily within CFT methods \cite{Dolan:2006ec}. The CFT methods can be extended beyond 
tree level, and it was shown in \cite{Aprile:2017bgs} that
 the one-loop correlator is determined in this way up to a single contact interactions, 
that is fixed by 
the study of the integrated correlator computed via supersymmetric 
localization techniques \cite{Chester:2019pvm}. Since the $\langle {\cal O}_2 {\cal O}_2 
{\cal O}_2 {\cal O}_2 \rangle$ will contribute to multi-particle 
correlators with $n\ge 2$, as a disconnected contribution, 
here below, we present a summary of results.

The $\langle {\cal O}_2 {\cal O}_2 {\cal O}_2 {\cal O}_2 \rangle$ correlator is fully crossing symmetric. 
The prefactor and the free theory are
\begin{equation}\label{free2222}
{\cal P}_{2222}= 4 a^2 g_{12}^2 g_{34}^2\qquad;\qquad
{\cal G}_{2222}=1+U^2\sigma^2 +\frac{U^2\tau^2}{V^2}+ \frac{4}{a}\Bigg[ U\sigma +\frac{U\tau}{V} + \frac{U^2\sigma\tau}{V}\Bigg]\,.
\end{equation}
For the dynamical function, crossing symmetry implies
\begin{equation}
{\cal H}_{2222}(U,V)=\frac{1}{V^2} {\cal H}_{2222}\left(\frac{U}{V},\frac{1}{V}\right)=  \frac{U^2}{V^2} {\cal H}_{2222}(V,U)\,,
\end{equation}
which in Mellin space translates into
\begin{equation}
{\cal M}(s,t)={\cal M}(s,u)={\cal M}(t,s)\;,
\end{equation}
where recall that $u=-s-t-4$.

The tree-level Mellin amplitude is 
\begin{equation}\label{tree2222}
{\cal M}_{2222}\Big|_{\frac{1}{a}}= +\frac{4}{a}\frac{1}{(s+1)(t+1)(u+1)}\,.
\end{equation}
The one-loop Mellin amplitude is
\begin{equation}\label{tree22222}
{\cal M}_{2222}\Big|_{\frac{1}{a^2}}=\frac{16}{a^2} \Big[ {\cal M}^{1\myell}(s,t)+{\cal M}^{1\myell}(s,u) +{\cal M}^{1\myell}(u,t) \Big]\;,
\end{equation}
where 
\begin{equation}\label{bbMoneloop}
{\cal M}_{}^{1\myell}(s,t)= \Big[ {K}(s,t) A(s,t) + \big( \psi^{(0)}(-s)-\psi^{(0)}(-t) \big) B(s,t) + C(s,t) \Big]
\end{equation}
and we used the definitions
\begin{equation}\label{doubleboxwith2}
K(s,t)=  - \big(\psi^{(0)}(-s)-\psi^{(0)}(-t)\big)^{\!2} +\big(\psi^{(1)}(-s)+\psi^{(1)}(-t)\big)- \pi^2
\end{equation}
and
\begin{align}\label{oneloop2222}
\begin{array}{rl}
\displaystyle
A(s,t)=\,\,\,\,\,\,\,\frac{1}{20}  
&
\displaystyle
\!\!\left[ \frac{ s^2 t^2 (-15-\frac{12}{4+u} +\frac{72}{5+u} )}{ (1+u)(2+u)(3+u) } 
+ \frac{ st\,(25 -\frac{6}{3+u} -\frac{60}{4+u} )}{(1+u)(2+u)} -\frac{4\,(3-\frac{2}{2+u}-\frac{1}{3+u} )}{1+u}\right]\\[.5cm]
\displaystyle
B(s,t)=\, \frac{ (s-t) }{20} 
&\displaystyle
\!\!\left[\frac{st\,(-15-\frac{12}{4+u} +\frac{72}{5+u}) }{(1+u)(2+u)}  +\frac{10}{4+u} +\frac{8}{2+u}+\frac{2}{1+u}   -10 \right] \\[.5cm]
\displaystyle
C(s,t)=\,\,\,\,\,\, \,\frac{1}{20} &
\displaystyle
\!\!\left[ \frac{ st\, (-15-\frac{12}{4+u} +\frac{72}{5+u} ) }{ 1+u } +\frac{6}{5+u}-\frac{20}{4+u} +\frac{2}{1+u} \right] -1+ B_0 \,,
\end{array}
\end{align}
where $B_0=\frac{5}{4}$. As pointed out in \cite{Aprile:2017bgs}, 
the bootstrap approach does not fix the one-loop result completely, but leaves one ambiguity, parametrised above by $B_0$. 
The value $B_0=\frac{5}{4}$ was later fixed by using results for the integrated correlator \cite{Chester:2019pvm}.

Let us describe first the structure of \eqref{tree2222}. At tree level the Mellin 
amplitude is just a rational function,  $\frac{4}{a}\frac{1}{(s+1)(t+1)(u+1)}$. 
Instead the one-loop  amplitude \eqref{tree22222} is stratified in 
transcendental function of $s,t$,  
\begin{equation}\label{Kpsi1}
K(s,t)\;,\qquad  \psi^{(0)}(-s)-\psi^{(0)}(-t) \;,\qquad 1\;,
\end{equation}
and their crossing orientations. 
The first one, 
$K(s,t)$, is the Mellin amplitude of the double box function, 
see appendix \ref{appLadder}. Note that  $K(s,t)$ has trascendentality two, while the  double box function is a weight four function. 
The additional weight two is accounted by the Gamma functions in \eqref{conventionsMellin}. 
To match this counting one can simply look at the degree of the poles in $s$ and $t$ in the total integrand, and note that $K(s,t)$ indeed contributes 
to the infinite tower of simultaneous triple poles in $s$ and $t$.  
In the small $U,V$ expansion 
the oder of the poles determines the stratification in powers of $\log(U)$ and $\log(V)$. 
In particular, $K(s,t)$ is the only one that contributes to $\log^2(U)\log^2(V)$ and lower logs. 
The second function in \eqref{Kpsi1}, $\psi^{(0)}(-s)-\psi^{(0)}(-t)$, is a weight one contributions, and 
contributes to triple poles in $s$ or $t$.\footnote{In the flat space limit, the leading asymptotics of $K(s,t)\rightarrow -\log^2(\frac{-s}{-t})-\pi^2$, 
and contains the double logs $\log(-s)\log(-t)$, while that of  $\psi^{(0)}(-s)-\psi^{(0)}(-t)$ is $\log(-s)-\log(-t)$.} 
Repeating the same logic, $K(s,u)$, $\psi(-s)$, and $K(u,t)$, $\psi(-t)$  contribute to $\log^2(U)$ or $\log^2(V)$ and lower logs.

The tree level amplitude, $\frac{4}{a}\frac{1}{(s+1)(t+1)(u+1)}$, was first obtained on the gravity side, by
rearranging a certain combination of contact and exchange Witten diagrams. Since in Mellin space these diagrams have simple 
poles in correspondence of the exchanged bulk fields, one expects to find only simple poles in the tree level ${\cal M}$, which is indeed the case.
The residue at the poles $s,t,u=-1$ has a physical meaning in the CFT: it is related to the twist two sector
that we will review in section \ref{OPEsect1}.
On the other hand, one needs to be careful when interpreting 
the simple poles in the one loop result, i.e.~$A(s,t),B(s,t),C(s,t)$ and their crossing orientations.
The poles $s,t,u=-2,\ldots,-5$ are spurious poles and their total residue must vanish. 
This can be checked from the explicit one-loop amplitude provided above. 
The residue at $s,t,u=-1$ has the same meaning as in the tree level amplitude, but it vanishes in the one-loop case. 
The reason will be that in the $\frac{1}{a}$ expansion it 
can be set to zero consistently as a consequence of the fact that the 
twist two sector truncates at tree level \cite{Aprile:2017bgs}.\footnote{In the bootstrap approach of \cite{Aprile:2017bgs}, $A(s,t)$ and $B(s,t)$ are obtained from the double logarithmic discontinuity, then,
vanishing of the residue at $s,t,u=-1,\ldots, 5$, together with crossing symmetry and the flat space limit, is enough to bootstrap $C(s,t)$ up to a contact term.}

Finally, a feature of the one-loop Mellin amplitude worth emphasising is the fact that it admits 
a representation as a sum over (shifted and dressed) double boxes, first proposed in \cite{Huang:2024dck,Huang:2024rxr}, namely
\begin{equation}\label{repofEllis}
{\cal M}^{1\myell}(s,t)= \sum_{\substack{ ij\ge 0,  |i|+|j|\leq 2} } \frac{ p_{ij}(s,t) }{ (s+i)+(t+j)+1} K(s+ i, t+j)\;,
\end{equation}
where $p_{ji}(s,t)=p_{ij}(t,s)$ are some polynomials and the sum goes over the lattice points drawn below,
\begin{equation}\label{gridEllis}
\begin{array}{c}
\begin{tikzpicture}
\draw[->] (-3,0)--(3,0); \draw  (3.1,-.1) node[scale=.8] {$i$};
\draw[->] (0,-3)--(0,3); \draw  (-.1,3.1) node[scale=.8] {$j$};

\foreach \x/\y in { 0/2, 1/1, 2/0} \draw (\x,\y) circle (2pt) node[anchor=north,scale=.5] {$(\x,\y)$};
\draw[dashed,gray] (-1,3) -- (3,-1); \draw  (3.2,-1.1) node[scale=.7,gray] {$u+1$};

\foreach \x/\y in { 0/1,1/0} \draw (\x,\y) circle (2pt) node[anchor=north,scale=.5] {$(\x,\y)$};
\draw[dashed,gray] (-1.5,2.5) -- (2,-1); \draw  (2.2,-1.1) node[scale=.7,gray] {$u+2$};

\foreach \x/\y in { 0/0} \draw (\x,\y) circle (2pt) node[anchor=north,scale=.5] {$(\x,\y)$};
\draw[dashed,gray] (-2,2) -- (1,-1); \draw  (1.2,-1.1) node[scale=.7,gray] {$u+3$};

\foreach \x/\y in { -1/0,0/-1} \draw (\x,\y) circle (2pt) node[anchor=north,scale=.5] {$(\x,\y)$};
\draw[dashed,gray] (-2.5,1.5) -- (1,-2); \draw  (1.1,-2) node[scale=.7,gray] {$u+4$};

\foreach \x/\y in { -2/0,-1/-1, 0/-2} \draw (\x,\y) circle (2pt) node[anchor=north,scale=.5] {$(\x,\y)$};
\draw[dashed,gray] (-3,1) -- (1,-3); \draw  (1.1,-3) node[scale=.7,gray] {$u+5$};
\end{tikzpicture}
\end{array}
\end{equation}
A simple reason why this representation is conceivable is the fact that the set of $K(s+i,t+j)$ span the 
same basis \eqref{Kpsi1} of the amplitude, up to some rational prefactors 
that read
\begin{equation}\label{rationalprefactorEllis}
\begin{array}{c}
\begin{tikzpicture}
\draw (0,0) node[scale=.85] {$
\begin{array}{c|c|c|c|c|c|c|c}
 		 & K(s+2,t) & K(s+1,t) & K(s-1,t) & K(s-2,t) & K(s,t) & K(s-1,t-1) & K(s+1,t+1) \\[2ex]
\hline 
\rule{0pt}{.8cm}K(s,t) & 1 & 1 & 1 & 1 & 1 & 1 & 1 \\[2ex]
\hline 
\rule{0pt}{.8cm}\Psi_1(s,t)& -\frac{2(3+2s)}{(s+1)(s+2)} & -\frac{2}{1+s} & \frac{2}{s} & \frac{2(-1+2s)}{(-1+s)s} & 0 & -\frac{2(s-t)}{st} & \frac{2(s-t)}{(s+1)(t+1)} \\[2ex]
\hline
\rule{0pt}{.8cm}  1 & -\frac{2}{(s+1)(s+2)} & 0 & -\frac{2}{s^2} & -\frac{2(1-3s+3s^2)}{(-1+s)^2s^2} & 0 & -\frac{2(s^2-st+t^2)}{s^2t^2} & \frac{2}{(s+1)(t+1)}
\end{array}$};
\end{tikzpicture}
\end{array}
\end{equation}
(We defined $\Psi_1(s,t)=\psi^{(0)}(-s)-\psi^{(0)}(-t)$, and the cases of $K(s,t+j)$ are obtained by crossing $s\leftrightarrow t$.)
Therefore an ansatz of the form $\sum_{i,j}r_{ij}(s,t)K(s+i,t+j)$ where $r_{ij}(s,t)$ are some rational functions would be 
general enough to describe the one-loop amplitude. 

However, the particular form in \eqref{repofEllis}, with the factors $(s+i)+(t+j)+1$ in the denominators, is more restricting, 
and it is important to understand  why this restriction is compatible with the amplitude \eqref{oneloop2222} in the first place.
To start with consider $i=j=0$, that in \eqref{repofEllis} corresponds to $K(s,t)$ and is associated with the denominator $(u+3)$. 
This association works becasue the factor $(u+3)$ only appears in $A(s,t)$ and does not appear 
in $B(s,t)$ and $C(s,t)$, thus $i=j=0$ is the central node in the grid \eqref{gridEllis}. Similarly, consider $i=1,j=0$, that  in \eqref{repofEllis} 
corresponds to $K(s+1,t)$ and is associated with the denominator $(u+2)$. This association works  because 
the factor $(u+2)$ does not appear in $C(s,t)$, therefore the last zero in the column of $K(s+1,t)$ 
in \eqref{rationalprefactorEllis}  can be justified. The rest of the grid  \eqref{gridEllis} is fixed by the  
$s\leftrightarrow t$ symmetry.

At this point, note that the rational 
prefactors in \eqref{rationalprefactorEllis} have spurious poles not present in the amplitude, $A(s,t)$,$B(s,t)$,$C(s,t)$, 
therefore they have to cancel in the final result, giving constraints on the $p_{ij}$.\footnote{For example by the 
imposing a zero in $p_{ij}$ for each spurios poles in the column 
corresponding to $K(s+i,t+j)$.}  The precise form of the polynomials $p_{ij}$ is not important.\footnote{A representative 
can be found in  \cite{Huang:2024rxr,Huang:2024dck}. We have added a small discussion about this in the ancillary file included with the {\tt arXiv} submission.}
We are mentioning this result here because we will discuss the analogous 
representation for ${\cal H}_{[2^2][2^2]22}$ in the following section.

\subsection{$\langle {\cal O}_2^2 {\cal O}_2^2 {\cal O}_2 {\cal O}_2 \rangle$}\label{sec2d2d22}

We now study the mixed correlator with two single and two double particle insertions.  
The results up to order $\frac{1}{a^2}$ were given in \cite{Aprile:2024lwy} and here we extend that analysis by
giving more details about the properties of the full correlator and the dynamical function. In the order
\begin{align}\label{prefa_dd22}
{\cal P}_{ [2^2][2^2]22 } & =16 a^2 (a+2) g_{12}^4 g_{34}^2 \;, \\[.2cm]
{\cal G}_{[2^2][2^2]22} & =1+2 U^2\sigma^2 +\frac{2U^2\tau^2}{V^2}+ \frac{8}{a}\Bigg[ U\sigma +\frac{U\tau}{V} \Bigg] + \frac{24}{a} \frac{U^2\sigma\tau}{V} + 
 \frac{4}{a}\Bigg[ U^2\sigma^2 +\frac{U^2\tau^2}{V^2} \Bigg] \;, \\
 {\cal H}_{[2^2][2^2]22}&={\cal H}^c_{[2^2][2^2]22} +\frac{2a}{a+2}  {\cal H}_{2222} \;,
\end{align}
where the $16 a^2 (a+2)=  8 a(a+2)\times 2 a $ comes from the two-point functions $\langle {\cal O}^2_2 {\cal O}^2_2\rangle\times \langle {\cal O}_2 {\cal O}_2\rangle$.

Note that ${\cal H}_{[2^2][2^2]22}$ has an all order $\frac{1}{a}$ 
expansion inherited from that of ${\cal H}^c_{[2^2][2^2]22}$, ${\cal H}_{2222}$, 
and $\frac{2a}{a+2}$. 
The first two orders of the $\frac{1}{a}$ expansion are
\begin{align}
&
{\cal H}_{ [2^2][2^2]22 }\Big|_{\frac{1}{a} \ }=\,+2{\cal H}_{2222}\Big|_{\frac{1}{a}}\;,  \label{tree2d2d22}\\
&
{\cal H}_{ [2^2][2^2]22 }\Big|_{\frac{1}{a^2} }=  {\cal H}^c_{ [2^2][2^2]22 }\Big|_\frac{1}{a^2}- 4 {\cal H}_{2222}\Big|_{\frac{1}{a}} + 2 {\cal H}_{ 2222 }\Big|_{\frac{1}{a^2} }\,. \label{oneL2d2d22}
\end{align}
The new term at order $\frac{1}{a^2}$ is ${\cal H}^c_{ [2^2][2^2]22 }$, and is given by
the gravity result through \eqref{CFTfromgravity}, namely
\begin{equation}
{\cal H}^c_{[2^n][2^n]22}\Big|_{\frac{1}{a^2}}= \frac{U^2}{V^2} \Psi_B^{(2)}\,.
\end{equation}
It reads
\begin{align}
 {\cal H}^c_{ [2^2][2^2]22 }\Big|_\frac{1}{a^2}  =\ &  \frac{U^2}{V^2}\Bigg[ R^{(1)}_{11} \mathcal{P}_2 + 
 			R^{(2)}_{10} \left[z \partial_z - \bar{z} \partial_{\bar{z}} \right] \mathcal{P}_2+ 
			R^{(3)}_{11} \mathcal{P}_1 \log V
		      + R^{(4)}_{10} \log^2 V+R^{(5)}_{10} \log V\log U   \notag\\
 		     & \rule{.8cm}{0pt}  + R^{(6)}_{11} \mathcal{P}_1 +R^{(7)}_{10} \log V+R^{(8)}_{10} \log U +R^{(9)}_{8} \Bigg]\,, \label{Hc2d2d22fromletter}
\end{align}
where ${\cal P}_{L}(z,\bar{z})$ are the ladder integrals, see appendix \ref{appLadder}, 
and $R^{(a)}_k=R^{(a)}_k(z,\bar{z})$ are rational functions of $z,\bar{z}$ with denominators proportional to $(z-\bar{z})^k$ given in the ancillary file included with the {\tt arXiv} submission.

A formula analogous to \eqref{oneL2d2d22} holds in Mellin space. 
As shown in  \cite{Aprile:2024lwy}, the Mellin amplitude corresponding to \eqref{Hc2d2d22fromletter} is, 
\begin{equation}\label{Mellindoublepsection4}
{\cal M}_{[2^2][2^2]22}^c(s,t)=  \Big[ {K}(u,t) A(u,t) + \big( \psi^{(0)}(-u)-\psi^{(0)}(-t) \big) B(u,t) + C(u,t) \Big]\;,
\end{equation}
where 
\begin{align}
A(u,t)=& \frac{32}{(1+s)(2+s)(3+s)} \left[ -16 + 3 ut -3s+\frac{18 ut}{4+s}-\frac{18 u^2 t^2}{(4+s)(5+s)} \right]\;, \label{Aof2d2d22}\\[.2cm]
B(u,t)= & \frac{32(u-t)}{(1+s)(2+s)} \left[ -\frac{1}{(1+u)(1+t)} - \frac{ 18 u t-18 (5+s)}{(4+s)(5+s)} \right] \;, \\[.2cm]
C(u,t)= & \frac{ 32}{(1+s)} \left[ \frac{3}{2}- \frac{ \frac{1}{2} }{(1+u)(1+t)} + \frac{6 - 18 ut + 12(5+s)}{(4+s)(5+s)} \right]\;. \label{Aof2d2d223}
\end{align}
The total amplitude ${\cal M}_{[2^2][2^2]22}$ is the sum of ${\cal M}_{[2^2][2^2]22}^c$ and 
the disconnected contributions proportional to ${\cal H}_{2222}$ that can be read from the previous section, see \eqref{tree2222}.
We note that the same basis \eqref{Kpsi1} that shows up for the ${\cal M}^{1\myell}_{2222}$ is showing up also here for ${\cal M}_{[2^2][2^2]22}^c$. 

The $\langle {\cal O}_2^2 {\cal O}_2^2 {\cal O}_2 {\cal O}_2 \rangle$ correlator satisfies invariance under (residual) crossing symmetry, $\vec{x}_1\leftrightarrow \vec{x}_2$ (or $\vec{x}_3\leftrightarrow \vec{x}_4$). This means that
\begin{equation}\label{symm4422}
{\cal H}_{[2^2][2^2]22}(U,V)=\frac{1}{V^2} {\cal H}_{[2^2][2^2]22}\left(\frac{U}{V},\frac{1}{V}\right)\;,
\end{equation}
which in Mellin space corresponds to
\begin{equation}\label{symm4422c}
{\cal M}_{[2^2][2^2]22}(s,t)= {\cal M}_{[2^2][2^2]22}(s,u)\,.
\end{equation}
This invariance is obvious for ${\cal H}_{2222}$, however it is less obvious for ${\cal H}^c_{[2^2][2^2]22}$.
From the explicit expression of
${\cal M}_{[2^2][2^2]22}^c$ we see that the amplitude is symmetric under \eqref{symm4422c} but out of the two possible combinations,  $K(u,t)$ and $K(s,t)+K(s,u)$, we only find
the first one, and similarly, we only find $\psi^{(0)}(-t)-\psi^{(0)}(-u)$ and no $\psi^{(0)}(-s)$ contribution.

The presence of $K(u,t)$ only,  implies that ${\cal M}_{[2^2][2^2]22}^c$  
contributes to triple poles in $t$ but does not contribute to triple poles in $s$. 
In particular, the tower of the triples poles in $t$ originates from the simple poles 
of ${\cal M}_{[2^2][2^2]22}^c$ and the double poles of the Gamma functions in  \eqref{conventionsMellin}.
Thus we see that ${\cal H}^c_{ [2^2][2^2]22 }$ has a $\log^2(V)$ discontinuity, 
but does not have a $\log^2(U)$ discontinuity, consequently no $\log^2(U)\log^2(V)$ discontinuity. 
This can be checked explicitly from the position space result by properly expanding 
the transcendental functions in the small $U$, small $V$ regime. The subleading $\log(U)$ discontinuity instead is non trivial. 

In section \ref{OPEsect12} we shall give a CFT explanation about the absence of 
the $\log^2(U)$ discontinuity in ${\cal H}^c_{[2^2][2^2]22}$, using considerations about the  ${\cal O}_2^2{\cal O}^2_2$ OPE. 
Since we will also study the ${\cal O}_2{\cal O}^2_2$ OPE, it is 
worth mentioning that these two OPEs are different and therefore we expect to find different physics in the 
two orientation of the correlators. To better understand the latter, we briefly describe  ${\cal H}_{2[2^2]2[2^2]}$.

The orientation $\langle {\cal O}_2 {\cal O}_2^2 {\cal O}_2 {\cal O}_2^2 \rangle$ is obtained from 
$\langle {\cal O}_2^2 {\cal O}_2^2 {\cal O}_2 {\cal O}_2 \rangle$
 upon sending $\vec{x}_1\leftrightarrow \vec{x}_4$, ie.~$U\rightarrow \frac{1}{U}$, $V\rightarrow \frac{V}{U}$. 
The correlator in this case is given by 
\begin{equation}
{\cal P}_{ 2[2^2]2[2^2]2 }  = 16 a^2 (a+2)  g_{12}^2 g_{24}^2 g_{34}^2\,
\end{equation}
with  
\begin{equation}\label{crossing22d22d}
{\cal H}_{2[2^2]2[2^2]2 }=  {\cal H}_{ [2^2][2^2]22}\left(\frac{1}{U},\frac{V}{U}\right)\qquad;\qquad {\cal M}_{2[2^2]2[2^2]}(s,t)={\cal M}_{[2^2][2^2]22}(u,t)\,.
\end{equation}
The Mellin amplitude ${\cal M}_{2[2^2]2[2^2]}^c$ is thus 
\begin{equation}
{\cal M}_{2[2^2]2[2^2]}^c(s,t)=  \Big[ {K}(s,t) A(s,t) + \big( \psi^{(0)}(-s)-\psi^{(0)}(-t) \big) B(s,t) + C(s,t) \Big]\,.
\end{equation}
From \eqref{Aof2d2d22}-\eqref{Aof2d2d223} we find trivially that
\begin{align}
A(s,t)=& \frac{32}{(1+u)(2+u)(3+u)} \left[ -16 + 3 st -3u+\frac{18 st}{4+u}-\frac{18 s^2 t^2}{(4+u)(5+u)} \right]\;, \label{Aof22d22d}\\[.2cm]
B(s,t)= & \frac{32(s-t)}{(1+u)(2+u)} \left[ -\frac{1}{(1+s)(1+t)} - \frac{ 18 s t- 18 (5+u)}{(4+u)(5+u)} \right]\;, \\[.2cm]
C(s,t)= & \frac{ 32}{(1+u)} \left[ \frac{3}{2}- \frac{ \frac{1}{2} }{(1+s)(1+t)} + \frac{6 - 18 st + 12(5+u)}{(4+u)(5+u)} \right]\;.
\end{align}
The presence of ${K}(s,t)$ now shows that ${\cal H}^{c}_{2[2^2]2[2^2]}$ has a $\log^2(U)$ 
discontinuities and in particular a $\log^2(U)\log^2(V)$ contribution. In the CFT analysis of the 
correlator we have to consider together both ${\cal H}^{c}_{2[2^2]2[2^2]}$ and the disconnected 
contribution from ${\cal H}^{}_{2222}$, thus we have to add up their $\log^2(U)\log^2(V)$ 
contributions. Then,  one thing that we can look at is the fixed $s$ large $t$ limit of 
these amplitudes, since this will have to do with the large spin limit in the block expansion. 
From \eqref{Aof22d22d} and \eqref{tree2222} we find
\begin{equation}\label{largespin22d22d}
\lim_{t\gg 1}{A}_{2[2^2]2[2^2]}^c(s,t)\sim \frac{(1+s)}{t^2}\qquad;\qquad \lim_{t\gg 1} {A}^{1\myell}_{2222}(s,t)\sim \frac{(12+25s+15s^2)}{t},
\end{equation}
thus the behaviour of ${\cal M}_{2[2^2]2[2^2]}^c$ is subleading compared 
to the one loop amplitude, therefore we expect to find different 
contributions in spin when we perform the CFT analysis. We will comment on this in the 
discussion section. Interestingly, each one of the rational coefficient functions of ${\cal M}_{2[2^2]2[2^2]}^c$ 
has the same scaling as the tree level amplitude ${\cal M}_{2222}$.

\paragraph{Finite difference representation.}
At this point, as for ${\cal H}_{2222}$, it is interesting to ask if the one-loop contributions to ${\cal H}_{[2^2][2^2]22}$ 
or  ${\cal H}_{2[2^2]2[2^2]2 }$ can be written solely in terms of (dressed and shifted) double box functions. 
In order to make a cleaner comparison with the discussion around \eqref{repofEllis} for 
${\cal H}_{[2^2][2^2]22}$,  we shall focus on the orientation  ${\cal H}_{2[2^2]2[2^2]2 }$, considering the possibility of rewriting the correlator as
\begin{equation}
\sum_{\substack{ ij\ge 0,  |i|+|j|\leq 2} } \frac{ P_{ij}(s,t) }{ (s+i)+(t+j)+1} K(s+ i, t+j)
\end{equation}
for some other polynomials $P_{ij}(s,t)$ with $P_{ji}(s,t)=P_{ji}(t,s)$. 
It is clear that results for this orientation apply to 
${\cal H}_{[2^2][2^2]22}$ by crossing.

Recalling that ${\cal H}_{2[2^2]2[2^2]2 }$ contains the one-loop term \eqref{repofEllis} of ${\cal H}^{}_{2222}$ as a disconnected contribution, 
we first define the correlator ${\cal H}^{t}_{2[2^2]2[2^2]2 }$ where we subtract this contributions, namely
\begin{align}
{\cal H}^{t}_{2[2^2]2[2^2]2 }&\equiv {\cal H}_{2[2^2]2[2^2]2 }\Big|_{\frac{1}{a^2}}-2{\cal H}^{}_{2222}\Big|_{\frac{1}{a^2}}
\end{align}
Then, from \eqref{oneL2d2d22} we find 
\begin{align}\label{first_time_tree2d2d22}
\ \ \ \ \ \ \ \ \ {\cal H}^{t}_{2[2^2]2[2^2]2 }&=  {\cal H}^c_{ [2^2][2^2]22 }\Big|_{\frac{1}{a^2}} - 4 {\cal H}_{2222}\Big|_{\frac{1}{a}}  \\
&=  {\cal H}^c_{ 2[2^2]2[2^2] } \Big|_{\frac{1}{a^2}}+ 16 U^2 \overline{D}_{2422}(U,V)\,.
\end{align}
The Mellin amplitude ${\cal M}^t_{2[2^2]2[2^2]}$ is (obviously) ${\cal M}^t_{2[2^2]2[2^2]}={\cal M}^{c}_{2[2^2]2[2^2]}- \frac{16}{(s+1)(t+1)(u+1)}$.

The superscript ``$t$" in \eqref{first_time_tree2d2d22} stands for \emph{tree}, since both 
contributions can be understood as tree-like contributions: ${\cal M}_{2222}|_{\frac{1}{a}}$ 
is a connected tree level contribution by definition, and  we can think of 
${\cal M}^c_{ [2^2][2^2]22 }|_{\frac{1}{a^2}}$ as coming from the coincident limit of 
tree level diagrams in a six point function, as pointed out in \cite{Ceplak:2021wzz}. 
The interesting result we want to present is that ${\cal M}^{t}_{2[2^2]2[2^2]2 }$ can be put in the form
\begin{equation}\label{repofEllisfor22d22d}
{\cal M}^{t}_{2[2^2]2[2^2]2 }=
\sum_{\substack{ ij\ge 0,  |i|+|j|\leq 2} } \frac{ p_{ij}(s,t) }{ (s+i)+(t+j)+1} K(s+ i, t+j)
\end{equation}
for some polynomials $p_{ij}(s,t)$, but ${\cal H}^{c}_{2[2^2]2[2^2]2 }$ on its own cannot! Specifically, 
if one allows for an arbitrary coefficient in front of the $\overline{D}_{2422}$, compatibility 
with \eqref{repofEllisfor22d22d} fix it to unity.  This simply means that ${\cal M}^{c}_{2[2^2]2[2^2]}$ 
can be written as in \eqref{repofEllisfor22d22d} up to a tree level term. Let us understand this result by 
repeating the discussion about the compatibility of the expression \eqref{repofEllisfor22d22d} 
with the structure of the functions $A(s,t),B(s,t),C(s,t)$, as we did for 
${\cal M}^{1\myell}_{2222}$ around \eqref{rationalprefactorEllis}. In the order,
\begin{itemize}
\item[1.]
Similarly to ${\cal M}^{1\myell}_{2222}$, we see from \eqref{Aof22d22d} that
the factor $(u+3)$ is present in $A(s,t)$ but absent from $B(s,t)$ and $C(s,t)$, and that
the factor $(u+2)$ is absent in $C(s,t)$. 
\item[2.]
Differently from  ${\cal M}^{1\myell}_{2222}$, the denominators of $B(s,t)$ and $C(s,t)$ 
contain also the factors $(s+1)$ and $(t+1)$. These are not present in \eqref{repofEllisfor22d22d}, 
but they can be generated from shifts of $K(s,t)$, for example, see \eqref{rationalprefactorEllis} and 
the columns corresponding to $K(s+2,t),K(s+1,t),K(s,t+2),K(s,t+1)$ and $K(s+1,t+1)$.
Even though this is an additional  restriction on the $p_{ij}$, in principle, \eqref{repofEllisfor22d22d} 
is still compatible with the amplitude.
\end{itemize}
It is remarkable that the $p_{ij}$ exist such that \eqref{repofEllisfor22d22d} actually works.
We give them in the ancillary file included with the {\tt arXiv} submission.
Considering the results for certain N$^2$E single particle correlators obtained 
in \cite{Huang:2024rxr}, our result here 
suggests that the representation \eqref{repofEllisfor22d22d} 
might provide the natural generalization of $\overline{D}$ function for a general class of one-loop amplitudes.

\subsection{$\langle {\cal O}_2^3 {\cal O}_2^3 {\cal O}_2 {\cal O}_2 \rangle$}\label{sec_tripleparticle}

We will conclude this section by giving explicit results for
$\langle {\cal O}_2^3 {\cal O}_2^3 {\cal O}_2 {\cal O}_2 \rangle$. 
This is the first computation of a multi-particle correlator of this kind in ${\cal N}=4$ SYM.
Our main purpose here will be to show how 
knowledge of both $\langle {\cal O}_2 {\cal O}_2 {\cal O}_2 {\cal O}_2 \rangle$
and (the new one) $\langle {\cal O}^2_2 {\cal O}^2_2 {\cal O}_2 {\cal O}_2 \rangle$ 
is crucial to understand new features in $\langle {\cal O}_2^3 {\cal O}_2^3 {\cal O}_2 {\cal O}_2 \rangle$. 
The discussion here will also illustrate what is the general situation for $\langle {\cal O}_2^n {\cal O}_2^n {\cal O}_2 {\cal O}_2 \rangle$.

For the $\langle {\cal O}_2^3 {\cal O}_2^3 {\cal O}_2 {\cal O}_2 \rangle$ we have in the order, 
\begin{align}
{\cal P}_{[2^3][2^3]22}&=96 a^2 (a+2)(a+4) g_{12}^6 g^2_{34}\\[.2cm]
{\cal G}_{[2^3][2^3]22}& = 1+ 3 \Bigg[ U^2\sigma^2 +\frac{U^2\tau^2}{V^2} \Bigg] 
+ \frac{12}{a} \Bigg[ U\sigma+ \frac{U\tau}{V}\Bigg] + \frac{60}{a} \frac{U^2\sigma\tau}{V} +  \frac{12}{a} \Bigg[ U^2\sigma^2 +\frac{U^2\tau^2}{V^2} \Bigg] \label{free3p3p22}\\
{\cal H}_{[2^3][2^3]22}&= {\cal H}^{c}_{[2^3][2^3]22} + \frac{3a}{a+4} {\cal H}_{2222}+\frac{3a}{a+4} {\cal H}^c_{[2^2][2^2]22}\,.
\end{align}
Expanding in $\frac{1}{a}$, the first two orders are determined solely by disconnected 
contributions, thus they are determined by ${\cal H}_{2222}$ and ${\cal H}^c_{[2^2][2^2]22}$ at the relevant orders,
\begin{align}
&
{\cal H}_{ [2^3][2^3]22 }\Big|_{\frac{1}{a}\  }= 3 {\cal H}_{2222}\Big|_{\frac{1}{a} }  \label{3d3d22order1}\\
&
{\cal H}_{ [2^3][2^3]22 }\Big|_{\frac{1}{a^2} }= 3 {\cal H}^c_{[2^2][2^2]22}\Big|_{\frac{1}{a^2}} - 12 {\cal H}_{2222}\Big|_{\frac{1}{a}}  + 3 {\cal H}_{2222}\Big|_{\frac{1}{a}^2}\,.\label{3d3d22order2}
\end{align}
In particular, to get ${\cal H}_{ [2^3][2^3]22 }|_{\frac{1}{a^2} }$ we also need ${\cal H}^c_{[2^2][2^2]22}|_{\frac{1}{a^2}}$ from the previous section.

The new connected contribution ${\cal H}^c_{[2^3][2^3] 22}$ comes at the next order,
\begin{equation}\label{twoloop3d3d22}
{\cal H}_{ [2^3][2^3]22 }\Big|_{\frac{1}{a^3} }=  
{\cal H}^c_{ [2^3][2^3]22 }\Big|_{\frac{1}{a^3}}  
-12 {\cal H}^c_{[2^2][2^2] 22}\Big|_{ \frac{1}{a^2} } + 3  {\cal H}^c_{[2^2][2^2] 22} \Big|_{\frac{1}{a^3}} 
+48 {\cal H}_{2222}\Big|_{\frac{1}{a}} - 12 {\cal H}_{2222}\Big|_{\frac{1}{a^2}} +3 {\cal H}_{2222}\Big|_{\frac{1}{a^3}}
\end{equation}
Differently from the $\langle {\cal O}^2_2 {\cal O}^2_2 {\cal O}_2 {\cal O}_2 \rangle$  correlator,  
the gravity result for ${\cal H}^c_{ [2^3][2^3]22 }$ now  is not enough to fix the correlator 
at this order because ${\cal H}^c_{[2^2][2^2] 22}|_{\frac{1}{a^3}}$ is an additional \emph{loopy} 
contribution \emph{not} determined by our gravity computation.

The position space expression for ${\cal H}^c_{ [2^3][2^3]22 }$ is obtained from 
\begin{equation}
{\cal H}^c_{[2^n][2^n]22}\Big|_{\frac{1}{a^3}}= \frac{U^2}{V^2} \Psi_B^{(3)}\,.
\end{equation}
and reads
\begin{align}\label{positionspace3p}
{\cal H}^c_{ [2^3][2^3]22 }\Big|_{\frac{1}{a^3}}=\  &\frac{U^2}{V^2}\Bigg[ R^{(1)}_{15} {\cal P}_3(z,\bar{z})+ R^{(2)}_{14} \partial_{-} {\cal P}_3(z,\bar{z}) + R^{(3)}_{15} \partial_{+} {\cal P}_3 (z,\bar{z})+ R^{(4)}_{14} \partial_{+}\partial_{-} {\cal P}_3(z,\bar{z}) \\
& \rule{.8cm}{0pt}  +R^{(5)}_{15} {\cal P}_2(z,\bar{z}) + R^{(6)}_{15}  {\cal P}_1(z,\bar{z}) \log^2 V +  \notag \\[.27cm] 
& \rule{.8cm}{0pt}  +R^{(7)}_{14} \partial_-{\cal P}_2(z,\bar{z})  + R^{(8)}_{14} \log U\log^2 V  + R^{(9)}_{14} \log^3V + R^{(10)}_{15} {\cal P}_1(z,\bar{z}) \log V \notag \\
& \rule{.8cm}{0pt}  +R^{(11)}_{14} \log^2 V + R^{(12)}_{14} \log V \log U + R^{(13)}_{15} {\cal P}_1(z,\bar{z})  +  R^{(14)}_{14} \log V +  R^{(15)}_{14} \log U + R^{(16)}_{12}\Bigg]\;, \notag
\end{align}
where ${\cal P}_{L}(z,\bar{z})$ are the ladder integrals, {$\partial_{\pm}\equiv z \partial_{z}\pm \bar{z} \partial_{\bar{z}}$}, 
and $R^{(a)}_k=R^{(a)}_k(z,\bar{z})$ 
are rational functions of $z,\bar{z}$ with denominators proportional to $(z-\bar{z})^k$ given in the ancillary file included with the {\tt arXiv} submission.

A visible feature of the position space result is the fact that only the triple box integral ${\cal P}_3$ appears at top weight six.
This is particularly nice because at weight six more functions could have appeared. Indeed
we can contrast our result for ${\cal H}^c_{ [2^3][2^3]22 }$ in \eqref{positionspace3p} with the result 
for ${\cal H}_{2222}$ at two loops \cite{Drummond:2022dxw}, since both contribute to the full correlator, as can be seen from \eqref{twoloop3d3d22}.
Notably,  ${\cal P}_3$ does not appear in ${\cal H}_{2222}$ at two loops, 
and one finds instead the weight six zig-zag integrals first introduced in \cite{Drummond:2012bg}. 
Moreover the function ${\cal H}_{2222}$ at two loops is fairly more complicated.

At this point it is important to emphasise that both ${\cal H}^c_{ [2^3][2^3]22 }$ in \eqref{positionspace3p} and ${\cal H}_{2222}$ 
at two loops enter as quantum corrections to the full dynamical function ${\cal H}_{ [2^3][2^3]22 }$, 
but we find that ${\cal H}^c_{ [2^3][2^3]22 }$ is simple! An argument to justify 
this simplicity is to think about the HHLL correlator on the gravity side, 
considering the idea that ${\cal H}^c_{ [2^3][2^3]22 }$ retains features
 inherited from the dual geometry, which therefore are simpler. For example, the Regge 
 limit computed from a geodesic in the heavy-heavy background. We investigate this in appendix \ref{app:ReggeLimit} 
 and show perfect match with the result coming from the correlator. 
In this sense, it would be very interesting to pin down what features of the quantum 
expansion of multi-particle correlators are `geometrical'. This might give an alternative physical 
explanation to the fact that the maximal logs of ${\cal H}_{2222}$ in the loop expansion do not generate a ladder series.

The Mellin amplitude corresponding to ${\cal H}^c_{[2^3][2^3]22}$ is
\begin{align}\label{Mellin3d3d22}
{\cal M}_{[2^3][2^3]22}^c(s,t) &=  \Big[ {K}_4(u,t) a(u,t) + \,{K}_3(u,t)   b(u,t) \,+ \tilde{K}_2(u,t) c(u,t) +  \\
&\rule{3.3cm}{0pt} + 
{K}_2(u,t) A(u,t) + K_1(u,t) B(u,t) + C(u,t)\ \Big]\;, \notag
\end{align}
where we have defined\footnote{The function $K_2(s,t)$ is the same function as in \eqref{Kpsi1} and \eqref{Mellindoublepsection4}, we only added a label for the transcendental weight.}  
\begin{align}\label{tripleboxmellinamp}
{K}_4(s,t)&=
\Psi_1^4 + 3 \Psi_2^2-2\Psi_1 \Psi_3 -2 \pi^2 \Psi_2 - 2 \pi^2 K_{2} -\pi^4 
\\[.2cm]
\label{tripleboxmellinamp2}
{K}_3(s,t)&= \Psi^3_1(u,t) -\frac{1}{2} \Psi_3(u,t)  + \pi^2 \Psi_1 \\[.2cm]
\tilde{K}_2(s,t)&=\Psi_2-\frac{2}{3} \pi^2\qquad;\qquad
{K}_2(s,t)=
-\Psi_1^2 + \Psi_2 -\pi^2  
\qquad;\qquad
{K}_1(s,t)=\Psi_1
\end{align}
with $\Psi_i(s,t)\equiv \psi^{(i-1)}(-s)+(-1)^{i} \psi^{(i-1)}(-t)$. Note that $\Psi_i$ carries transcendental weight $i$. 
The coefficient functions $a(s,t),$\ldots,$C(s,t)$ are given in the ancillary file included with the {\tt arXiv} submission. 
For reference we quote the first two
\begin{align}
&
a(s,t)=\frac{256}{(u+1)(u+2)(u+3)}\Bigg[  \\
&
\rule{1.7cm}{0pt}
14 -15 st +\frac{9}{2}u + \frac{15 st(4+5 st)}{4(u+4)_1} -\frac{3s^2 t^2(58+15 st)}{4(u+4)_2}\notag
+\frac{45s^2 t^2 (1+2st)}{2(u+4)_3}-\frac{135s^2t^2(-2+st)^2}{2(u+4)_4}\Bigg]
\end{align}
and
\begin{align}
&
b(s,t)=\frac{384(s-t)}{(u+1)(u+2)(u+3)}\Bigg[  \\
&
\rule{1.7cm}{0pt}
46 + \frac{2(10-77 st)}{(u+4)_1} -\frac{st(116-195 st)}{(u+4)_2}\notag
+\frac{15st(4+21st-9 s^2t^2))}{(u+4)_3}-\frac{360st(2-st)(1-st)}{(u+4)_4}\Bigg]\;.
\end{align}
The other functions $c(s,t), A(s,t),B(s,t),C(s,t)$ 
have a complexity similar to that of $a(s,t)$ and $b(s,t)$, and in addition $c(s,t), A(s,t)$ and $C(s,t)$ have a contribution proportional to  $\frac{1}{(s+1)(t+1)(u+1)}$.

The function $K_2(s,t)$ in \eqref{tripleboxmellinamp2} is the Mellin amplitude of the double-box that we encountered already in ${\cal H}^c_{[2^2][2^2]22}$. 
It is the same function as in \eqref{Kpsi1} and \eqref{Mellindoublepsection4}, we only added a label for the transcendental weight, since now there are other functions as well.
The function $K_4(s,t)$ is the Mellin amplitude corresponding to the triple-box, which we derive in appendix \ref{appLadder}, see section \ref{appladderexplicit}. 
As we mentioned around \eqref{Kpsi1}, in Mellin space the trascendental weight goes down by two because of the universal factor of Gamma functions. 
To recover the same counting as in position space it is simple to look at the degree of the poles in $s$ and $t$, and therefore at the stratification in powers of $\log(U)$ and $\log(V)$, 
which obviously matches the transcendental degree in position space. 
The amplitude $K_3(s,t)$ is inferred from $\partial_+{\cal P}_3$, and finally ${\tilde K}_2(s,t)$ is inferred from ${\cal P}_1\log^2(U)$, see the discussion at the end of section \ref{appF3}.

One thing to note about the coefficients functions $a(s,t),$\ldots,$C(s,t)$, visible already in $a(s,t)$ and $b(s,t)$ displayed above, is the common denominator which consists of 
the seven factors $(u+k)$ with $k=1,\ldots 7$. In the case of ${\cal M}_{[2^2][2^2]22}^c(s,t)$ in \eqref{Aof22d22d} 
we found the five factors $(u+k)$ with $k=1,\ldots 5$, thus now we also have $(u+6)$ and $(u+7)$. However, there are also structural differences compared to 
the ${\cal M}_{[2^2][2^2]22}^c(s,t)$ in \eqref{Aof22d22d} and in fact one can check that a simple extension of the pattern given in \eqref{gridEllis} is now insufficient
to describe the function in Mellin space as a difference operator acting on $K_4(s,t)$. It would be nice to find the correct generalization.

\section{Operator product expansion}\label{sec:OPE}

In the regime where ${\cal N}=4$ SYM is dual to AdS$_5\times$S$^5$ supergravity, 
the spectrum of the theory consists of particles and multi-particle operators built out 
of the supergravity fields. Stringy states are decoupled.
This simple statement puts several constraints on the spectral properties of correlators, and in particular 
there are two type of consistency checks that can be considered:
\begin{itemize}
\item[(S)] Consistency with respect to the protected sector and multiplet recombination.
\item[(L)] Consistency with respect to the anomalous dimensions of known double-particle operators. 
\end{itemize}
Both types of checks are non elementary and non trivial.  

Type (S) requires a small introduction. The spectrum of ${\cal N}=4$ SYM is organised  
into short and long multiplets. Long multiplets acquire anomalous dimension in the interacting theory, and  
stringy states are those long multiplets whose anomalous dimension goes off to infinity in the supergravity limit. 
As recognized in a series of seminal works \cite{Arutyunov:2000ku,Arutyunov:2000im,Bianchi:2000hn,Arutyunov:2002rs,Dolan:2006ec}, 
the decoupling of stringy states gives valuable information 
when considered in combination with the ``partial non-renormalization" theorem \cite{Eden:2000bk}.
To see this recall the split of the correlator
into a free and a dynamical part that follows from this theorem. See  \eqref{partial_nreno}. Now, a free theory diagram
exchanges all the allowed operators of the theory, including the stringy states, 
therefore in order to decouple the latter from the supergravity OPE  
it must happen that the free theory and the dynamical function are related to each other.  
This relation is particularly relevant for stringy states that in the free theory have 
quantum numbers at the unitarity bound, i.e.~free twist $\tau=2a+b+2$ in the $su(4)$ rep $[a,b,a]$, because
these states are short in the free theory. In the interacting theory all stringy states acquire anomalous 
dimension, both at weak and strong coupling, and therefore in the interacting theory they recombine to form long multiplets.
However, many genuine short multi-particle operators exists at the unitary bound that instead are protected and remain in the supergravity OPE.
Therefore, when we look at the block decomposition of a free theory 
diagram projected onto the unitarity bound, generically we will find a 
combination of both protected multi-particle operators and stringy states. 
If this is the case, in order to decouple the latter from the supergravity OPE 
we have to unmix the two contributions. This is a non trivial problem because 
explicit three-point couplings of the protected operators are not readily available. 
Nevertheless the problem can be tackled, and it was solved 
for double-particle protected operators up to order $\frac{1}{a^2}$ in \cite{Doobary:2015gia,Aprile:2019rep,Aprile:2025nta}.
We shall see that the method developed in \cite{Doobary:2015gia,Aprile:2019rep,Aprile:2025nta} applies to both
\begin{equation}\label{sec12introcorr}
\langle  {\cal O}_2^2 {\cal O}_2^2 {\cal O}_2  {\cal O}_2  \rangle\qquad;\qquad  \langle  {\cal O}_2 {\cal O}_2^2 {\cal O}_2  {\cal O}^2_2  \rangle\,.
\end{equation}
Instead, the discussion for $\langle  {\cal O}_2 {\cal O}_2^n {\cal O}_2  {\cal O}_2^n  \rangle$ with $n\ge 3$ goes beyond the double particle spectrum.

We have organized the OPE analysis as follows. In sections \ref{OPEsect1} and \ref{OPEsect2}, 
we discuss consistency checks of type (S), giving predictions for
\begin{equation}\label{pred_intro1}
{\cal H}^c\Big|_{\log^0(U)}\,.
\end{equation}
In section \ref{OPEsect12} we discuss OPE consistency checks of type (L), giving predictions for
\begin{equation}\label{pred_intro2}
{\cal H}^c\Big|_{\log^n(U)}\,.
\end{equation}
In section \ref{triplecorrelatorsection} we discuss OPE checks on $\langle  {\cal O}_2^3 {\cal O}_2^3 {\cal O}_2  {\cal O}_2 \rangle$, 
and we comment on the general case $\langle  {\cal O}_2^n {\cal O}_2^n {\cal O}_2  {\cal O}_2 \rangle$ with $n\ge 3$.
We shall see in particular that the gravity computation shows that three-point functions of the form $\langle{\cal O}_2^n {\cal O}_2^n{\cal L}_{\tau,l}\rangle$, 
where ${\cal L}_{\tau,l}$ are certain long operators,
satisfy large $N$ factorization formulae that are not obvious in the CFT.

\subsection{$\langle {\cal O}_2^2 {\cal O}_2^2 {\cal O}_2 {\cal O}_2 \rangle$ and the twist two sector} \label{OPEsect1}

We consider here the OPE  of ${\cal O}_2^2{\cal O}_2^2$
and the OPE of  ${\cal O}_2{\cal O}_2$.  In the limit $\vec{x}_1\rightarrow \vec{x}_2$ 
and $\vec{x}_3\rightarrow \vec{x}_4$, their common OPE
determines the block decomposition,
\begin{equation}\label{first_time_block_expa}
\langle {\cal O}_2^2(\vec{x}_1,\vec{y}_1) {\cal O}_2^2(\vec{x}_2,\vec{y}_2)  {\cal O}_2(\vec{x}_3,\vec{y}_3) {\cal O}_2(\vec{x}_4,\vec{y}_4) \rangle
={\cal P}_{[2^2][2^2]22}\sum_{ \underline{R} } A_{\underline{R}}\,B_{\underline{R}}\,.
\end{equation}
The $B_{\underline{R}}$ are the superblocks for the operator exchanged, whose quantum numbers we are denoting collectively by 
$\underline{R}$. The $A_{\underline{R}}$ are the four-point OPE coefficients, defined by
$$
A_{\underline{R}} =\sum_{O,O' \in\underline{R}} C_{[2^2][2^2] O}\ \mathbb{G}^{O,O'}\ C_{22 O'}\;,
$$
where $C_{p_ip_jO}$ are three-point couplings and $\mathbb{G}_{O,O'}$ is the two-point function metric. 
The sum over ${\underline{R}}$ splits into four sectors, depending on the type of primary multiplet exchanged:
\begin{equation}\label{operatorlabel}
\begin{array}{c|c|c|c|c}
\underline{R}&\frac{1}{2}\textrm{-BPS} &\frac{1}{4}\textrm{-BPS}&\textrm{semishort} & \textrm{long} \\[.3ex]
\hline
\rule{0pt}{.6cm}\textrm{operator\ label} & {\cal B} & {\cal Q} & {\cal S} & {\cal L}\\[.3ex]
\end{array}
\end{equation}
For each multiplet exchanged there is a corresponding four-point superblock. See 
appendix \ref{N4blocks}. 

The ${\cal O}_2 {\cal O}_2$ OPE is well studied, see eg.~\cite{Arutyunov:2000ku,Arutyunov:2000im,Bianchi:2000hn,Dolan:2004iy,Doobary:2015gia}, 
and is contained in the ${\cal O}_2^2 {\cal O}_2^2$ OPE. The common OPE is thus the one of  ${\cal O}_2 {\cal O}_2$. 
It will be important to recall a couple of features of this OPE starting from the \emph{free theory}. In the free theory we find schematically,
\begin{equation}\label{O2O2OPE}
 {\cal O}_2 {\cal O}_2\   \sim \ 
1\, \oplus\,  \underbrace{ {\cal O}_2  + {\cal K}_2 + \ldots  }_{\rule{0pt}{.22cm}\tau=2} \, \oplus\,  \underbrace{ {\cal O}_2^2\ +\ \ldots}_{\tau\ge4}\,.
\end{equation}
where:
\begin{itemize}
\item[(1)]
In the $\tau=2$ sectors we find operators at the unitary bound for the $[0,0,0]$ rep. The stringy states are the twist two single-
trace operators  ${\cal K}_{2,l}$. The spin zero operator is the well known Konishi multiplet. From large $N$ counting we expect
\begin{equation}
C_{{\cal O}_2{\cal O}_2;{\cal K}_2}\sim \frac{1}{N}\;.
\end{equation}
In the $SU(N)$ theory there are \emph{no} double particle operators at twist two.
\item[(2)] In the $\tau\ge4$ sector we find protected double-particle operators at twist four in the $[0,2,0]$ and $[1,0,1]$ reps (and long operators in the $[0,0,0]$ rep starting from twist four). 
We point out that the protected semi-short operators at twist four are schematically of the form 
${\cal O}_2\partial^l{\cal O}_2$, with leading three-point couplings  
\begin{equation}\label{3ptleading}
C_{{\cal O}_2{\cal O}_2; {\cal O}_2\partial^l{\cal O}_2} \sim 1\,.\ 
\end{equation}
(We will need this observation in the next section.) 
\end{itemize} 
The free theory  of the $\langle {\cal O}_2^2 {\cal O}_2^2 {\cal O}_2 {\cal O}_2\rangle$ correlator was given already in section \ref{sec2d2d22}, but we recall it here below for convenience:
\begin{align}
{\cal P}_{[2^2][2^2]22} &=16 a^2 (a+2) g_{12}^4 g_{34}^2 \label{prefa_dd22m} \\[.2cm]
{\cal G}_{[2^2][2^2]22} & =1+2 U^2\sigma^2 +\frac{2U^2\tau^2}{V^2}+ \frac{8}{a}\Bigg[ U\sigma +\frac{U\tau}{V} \Bigg] + \frac{24}{a} \frac{U^2\sigma\tau}{V} + 
 \frac{4}{a}\Bigg[ U^2\sigma^2 +\frac{U^2\tau^2}{V^2} \Bigg]\;.
\end{align}
Note that the normalized free theory ${\cal G}_{[2^2][2^2]22}$ is exact in $a$.

We are interested in extracting the free theory block coefficient corresponding to the semishort operators of twist $2$ in the $[0,0,0]$ rep, 
and from the considerations above we know that
these come exclusively from the stringy states ${\cal K}_{2,l}$.
The  relevant diagrams are
 \begin{equation}
\langle {\cal O}_2^2 {\cal O}_2^2 {\cal O}_2 {\cal O}_2 \rangle\Big|_{\substack{ \tau=2\\[.08cm] {\rm OPE}}}=  {\cal P}_{[2^2][2^2]22} \times \frac{8}{a}\Bigg[ U\sigma +\frac{U\tau}{V} \Bigg] 
 \end{equation}
 We refer to appendix \ref{N4blocks} for details on 
 how to extract the corresponding OPE coefficient.  
The results is 
\begin{equation}\label{konishi}
A^{free}_{{\cal K},{2,\myell,[000]}}= \frac{8}{a} \times \frac{2 (l+2)!^2}{(2l+4)!} \frac{1+(-1)^l}{2}\,. 
\end{equation}
and it is well known in the literature.

In the \emph{interacting theory} the stringy states ${\cal K}_{2,l}$ acquire an anomalous dimensions, thus they
do not belong to the semishort sector anymore, rather they recombine to form a \emph{long} rep. 
In the block expansion this recombination is 
done via a crucial relation between superblocks,
\begin{equation}\label{recombinationblocks}
B_{}^{ {\cal S}}\Big|_{ \substack{ \tau=2\\ [000] } }\ =
B^{ {\cal L}}_{}\Big|_{ \substack{ \tau=2\\ [000]} }  
{-}\ 
B_{}^{ {\cal S} }\Big|_{ \substack{\tau=4\\ [101]} }\,.
\end{equation} 
The precise relation is given in appendix \eqref{recombination_fullformula}, as well as the 
spin dependence.\footnote{The reason is that this formula 
requires an understanding of $\underline{R}$ in Young diagram 
notation, which we give in app.~\ref{N4blocks}.} 
In practice, given the block expansion of the free theory one uses the identity $A_{{\cal K},{2,l,[000]}} B^{ {\cal S}} = A_{{\cal K},{2,l,[000]}} B^{ {\cal L}}_{} + \ldots$ 
to obtain the block expansion of the interacting theory. In the latter, the 
stringy contributions appear in the long sector ${\cal L}$.\footnote{
Not only that, recombination 
also shifts the free value of certain twist $4$ protected contributions.}

Now that the stringy states have been recombined into the long 
sector it is possible to decouple them. There are two aspects of this decoupling. First, the 
dynamical function by construction contributes only to the long sector. Second,  
in order for the decoupling to occur, the total long twist two contributions $A_{{\cal L},{2,l,[000]}}$ 
must vanish, namely
\begin{equation}\label{recoKonishi}
A_{{\cal L},{2,l,[000]}}=A^{free}_{{\cal K},{2,l,[000]}} + {\cal H}_{[2^2][2^2]22}\Big|_{\tau=2,l,[000]}=0\,.
\end{equation}
This equation gives a non trivial prediction for ${\cal H}_{[2^2][2^2]22}$ at twist two, 
at all spins. In the space of conformal blocks this is
\begin{align}\label{2d2d2check1}
&
{\cal H}_{ [2^2][2^2]22}\Big|_{\frac{1}{a},\tau=2} \ = - \frac{8}{a} \times \frac{2 (l+2)!^2}{(2l+4)!} \frac{1+(-1)^l}{2} \\[.2cm]
&
{\cal H}_{[2^2][2^2]22}\Big|_{\frac{1}{a^n},\tau=2} \ = 0  \rule{5cm}{0pt} n\ge 2 \label{2d2d2check2}
\end{align}
Equivalently we can give directly the result after resummation in spin.
\begin{equation}
{\cal H}_{ [2^2][2^2]22}\Big|_{\frac{1}{a}} =  -8U\Bigg[\frac{(1+V)}{(-1+V)^2 V}-\frac{2\log(V)}{(-1+V)^3}\Bigg] + U^2(\ldots) \;. 
\end{equation}
Both predictions are matched by our \eqref{tree2d2d22} and \eqref{oneL2d2d22}  as we explain below.
\begin{itemize}
\item[$\bullet$] Eq.~\eqref{2d2d2check1}. This follows directly from ${\cal H}_{2222}$, in particular from
\begin{equation}
{\cal H}_{2222}\Big|_{\frac{1}{a}}= -4 U^2\overline{D}_{2422} = -4U\Bigg[\frac{(1+V)}{(-1+V)^2 V}-\frac{2\log(V)}{(-1+V)^3}\Bigg] + U^2(\ldots)
\end{equation}
and the relation \eqref{tree2d2d22}, namely ${\cal H}_{[2^2][2^2]22}\Big|_{\frac{1}{a}}= 2 {\cal H}_{2222}\Big|_{\frac{1}{a}}$.
\item[$\bullet$]
Eq.~\eqref{2d2d2check2}. Interestingly, this is a prediction only for ${\cal H}^t_{ [2^2][2^2]22 }$ because 
the one-loop function ${\cal H}_{2222}$ by construction does not contribute at twist 
two \cite{Aprile:2017bgs,Aprile:2019rep}.\footnote{\label{footkonishi}The reason is simply 
that the free theory ${\cal G}_{2222}$ truncates at order $\frac{1}{a}$. See \eqref{free2222}. 
Therefore at order $\frac{1}{a^2}$ one can impose the absence of twist 
two contributions simply by letting the function start at twist four. As pointed out in section \ref{sec2222} the one loop
 Mellin amplitude has vanishing residue at $s=-1$, this is the simple pole corresponding to the twist two sector.}
The result is non trivial and follows from a non trivial cancellation of twist two contributions 
$$
{\cal H}^t_{[2^2][2^2]22}\Big|_{\log^0(U)}={\cal H}^c_{ [2^2][2^2]22 } +16 U^2\overline{D}_{2422}\Big|_{\log^0(U)}= U^2( \ldots )\;,
$$
i.e.~the one from ${\cal H}^c_{ [2^2][2^2]22 }$ cancel the one from 
$U^2\overline{D}_{2422}$,  that as we saw above is in fact non zero.
\end{itemize}

\subsection{$\langle {\cal O}_2 {\cal O}_2^2 {\cal O}_2 {\cal O}_2^2 \rangle$ and the twist four protected operators} \label{OPEsect2}

In this section we consider the  ${\cal O}_2 {\cal O}_2^2$ OPE.
It will be convenient to apply a crossing transformation 
to $\langle {\cal O}_2^2 {\cal O}_2^2 {\cal O}_2 {\cal O}_2\rangle$, following \eqref{crossing22d22d},  and write the OPE in the conventional limit $x_1\rightarrow x_2$ and $x_3\rightarrow x_4$. 
Therefore we will consider the block decomposition
\begin{equation}\label{first_time_block_expa2}
\langle {\cal O}_2(\vec{x}_1,\vec{y}_1) {\cal O}_2^2(\vec{x}_2,\vec{y}_2)  {\cal O}_2(\vec{x}_3,\vec{y}_3) {\cal O}^2_2(\vec{x}_4,\vec{y}_4) \rangle
={\cal P}_{2[2^2]2[2^2]} \sum_{ \underline{R} } A_{\underline{R}}\,B_{\underline{R}}\,.
\end{equation}
where now
$$
A_{\underline{R}} =\sum_{O,O' \in\underline{R}} C_{2[2^2]O}\ \mathbb{G}^{O,O'}\ C_{2[2^2] O'}
$$
and again $\underline{R}={\cal B},{\cal S},{\cal L}$ depending on the type of multiplet exchanged.

As for the ${\cal O}_2{\cal O}_2$ OPE, we first focus on the \emph{free} theory. 
Again we want to highlight a couple of important features 
of the ${\cal O}_2 {\cal O}_2^2$ OPE, schematically summarized as follows,
\begin{equation}
{\cal O}_2 {\cal O}_2^2\ \sim\ 
{\cal O}_2\, \oplus\,  \underbrace{  {\cal O}_2^2  + {\cal K}_{4} + \ldots  }_{\tau=4} \, \oplus\,  \underbrace{{\cal O}_2^3+ \ldots}_{\tau\ge6}\,.
\end{equation}
where
\begin{itemize}
\item[(1)] In the $\tau=4$ sector we find operators at the unitarity bound for the $[0,2,0]$ rep. We find both protected
double-particle operators ${\cal O}_2\partial^l {\cal O}_2$, already encountered in the ${\cal O}_2{\cal O}_2$ OPE, 
and a number stringy states, denoted collectively by ${\cal K}_{4,l}$. 
The precise nature of these stringy states will not be important, however, it is simple to 
see that they exist. For example, we can find  the double trace ${\cal K}_2{\cal O}_2$ and the twist four single-traces.
From large $N$ counting we find that 
\begin{equation}\label{scaleNdd}
C_{{\cal O}_2{\cal O}_2^2;\,{\cal O}_2\partial^l {\cal O}_2} \sim \frac{1}{N} 
\end{equation}
and similarly for ${\cal K}_2{\cal O}_2$, 
thus we expect non trivial mixing between these operators.
\item[(2)] In the $\tau\ge6$ sector we find both protected and long operators. The long operators belong 
to the $[0,2,0]$ rep and importantly, the one with leading order three-point coupling are
 triple-particle operators starting 
from twist six. Consequently we expect to find new CFT data that cannot be extracted from the single-particle correlators.
\end{itemize}
It follows from the first consideration above that when we extract 
the block coefficient $A^{free}_{\tau=4,l,[020]}$ from   
the $\langle {\cal O}_2{\cal O}^2_2{\cal O}_2{\cal O}^2_2\rangle$  correlator
we find one coefficient for both the sum over stringy states and the protected contributions.
Instead we want to isolate the contribution due to  ${\cal O}_2\partial^\myell{\cal O}_2$, 
which is the only operator present in the supergravity OPE, and recombine
the stringy states into the long sector. 
Disentangling the two contributions is non trivial  
because the protected three-point couplings are not readily available.

Following \cite{Doobary:2015gia}, we can resolve the mixing between the ${\cal O}_2\partial^\myell{\cal O}_2$ 
and the ${\cal K}^{}_{4,l}$ stringy states
by considering that the protected double-particle operators
${\cal O}_2\partial^\myell{\cal O}_2$ are also exchanged in the 
${\cal O}_2{\cal O}_2$ OPE, and crucially, in the ${\cal O}_2{\cal O}_2$ OPE they are the only 
twist four semishort operators exchanged both in the free and the interacting theory.
In this OPE we actually find
\begin{equation}
{\cal O}_2\partial^l{\cal O}_2\Big|_{[0,2,0],l=0,2,4,\ldots}\qquad;\qquad {\cal O}_2\partial^l{\cal O}_2\Big|_{[1,0,1],l=1,3,5,\ldots}
\end{equation}
For the ${\cal O}_2{\cal O}_2^2$ OPE we are interested only in the rep $[0,2,0]$. 
The above considerations imply that when we assemble the Gram matrix,
\begin{equation}\label{gramM}
G=\left(\begin{array}{cc} \langle {\cal O}_2{\cal O}_2{\cal O}_2{\cal O}_2\rangle & \langle {\cal O}_2{\cal O}_2{\cal O}_2{\cal O}^2_2\rangle  \\[.2cm]
\langle {\cal O}_2{\cal O}^2_2{\cal O}_2{\cal O}_2\rangle  & \langle {\cal O}_2{\cal O}^2_2{\cal O}_2{\cal O}^2_2\rangle 
\end{array}\right)\,,
\end{equation}
and we project this matrix onto the $[0,2,0]$ semishort reps at twist $\tau=4$ ,
we must find $\det G=0$ in the \emph{interacting} theory because there is only one protected operator, ${\cal O}_2\partial^\myell{\cal O}_2$. Thus, $\det G=0$ gives the prediction for the 
${\cal O}_2\partial^\myell{\cal O}_2$ semishort contributions to the ${\cal O}_2{\cal O}^2_2$ OPE. 

Let us denote ${\cal S}^{\vec{p}}_{4,[020]}\equiv A^{\vec{p}}_{{\cal S},\tau=4,l,[aba]}$ the block coefficient for the various semishort contributions entering \eqref{gramM}, 
then from $\det G=0$ we find 
\begin{equation}\label{pred020}
{\cal S}^{2[2^2]2[2^2]}_{4,[020]}=\frac{ \left( {\cal S}^{[2^2] 2 ; 22}_{4,[020]} \right)^2 }{  {\cal S}^{2 2 ; 22}_{4,[020]} }\;,
\end{equation}
where we used ${\cal S}^{2[2^2];22}_{4,[020]}={\cal S}^{22;2[2^2]}_{4,[020]}$.
Note that this prediction is unrelated to $\langle {\cal O}_2{\cal O}^2_2{\cal O}_2{\cal O}^2_2\rangle$ and instead
follows from the knowledge of  $\langle {\cal O}_2{\cal O}_2{\cal O}_2{\cal O}_2\rangle$
and $\langle {\cal O}_2{\cal O}_2{\cal O}^2_2{\cal O}_2\rangle$ in free theory.
The first one is given already in \eqref{free2222}, while the latter
is a near-extremal correlator which is non-renormalized and so can be
computed in free theory, see~\eqref{free22d22}. This
non-renormalization theorem is consistent with our strong coupling
results obtained from the analysis of the HHLL correlator: a non-trivial dynamical contribution to $\langle {\cal
O}_2{\cal O}_2{\cal O}^2_2{\cal O}_2\rangle$ would appear as a
${\mathcal O}(\alpha^3)$ term in~\eqref{eq:psiexpads5}, which is
indeed absent.

As expected from the large $N$ counting of three-point functions in \eqref{3ptleading} and \eqref{scaleNdd},  at leading order
the denominator of \eqref{pred020} is $O(1)$, while the numerator $\frac{1}{a}$ suppressed, thus 
the overall prediction will start at order $\frac{1}{a}$. 
Putting together the various ${\cal S}^{\vec{p}}_{4,[020]}$, the prediction is
\begin{align}\label{prediction020}
{\cal S}^{[2^2]2[2^2]2}_{4,[020]}
=\frac{32}{a} \frac{ (1+\frac{2}{a})}{(l+3)(l+4) + \frac{4}{a} }\frac{(l+4)!^2}{(2l+6)!} \frac{1+(-1)^l}{2}\;.
\end{align}
We refer to appendix \ref{N4blocks} for details.
We emphasise that
 \eqref{prediction020} is an exact result  in $a$ because 
there are no other protected operators at twist four than the ${\cal O}_2\partial^l{\cal O}_2$ operators.

At this point we go back to  the free theory $\langle {\cal O}_2{\cal O}^2_2{\cal O}_2{\cal O}^2_2\rangle$
and similarly to the case of twist two in [0,0,0], \eqref{recombinationblocks}, we proceed with multiplet recombination. 
Thus, from the knowledge of the protected sector we find that the stringy contribution is,
\begin{equation}\label{disentangleK4}
A^{free}_{{\cal K},{4,l,[020]}}=- {\cal S}^{}_{4,[020]}+A^{free}_{4,l,[020]}\,,
\end{equation}
where $A^{free}_{4,l,[020]}$ in \eqref{disentangleK4} is the block coefficient of,\footnote{For 
reader's convenience,
$\langle {\cal O}_2 {\cal O}_2^2 {\cal O}_2 {\cal O}_2^2 \rangle|_{\tau=4}
={\cal P}_{ 2[2^2]2[2^2]2 }  \times\Big[ \frac{8}{a}U\sigma +\frac{24}{a}\frac{U\tau}{V} \Big]$ where  ${\cal P}_{ 2[2^2]2[2^2]2 }=16 a^2 (a+2)  g_{12}^2 g_{24}^2 g_{34}^2$.}
\begin{equation}
A^{free}_{4,l,[020]}= \frac{1}{a}\left\{\begin{array}{cl}
+4(l+1)(l+6) \frac{(l+2)!(l+4)!}{(2l+6)!} &\ \textrm{spin\ odd}\\[.5cm]
 +4 (18+7l+l^2) \frac{(l+2)!(l+4)!}{(2l+6)!} &\ \textrm{spin\ even} 
 \end{array}\right.
 \end{equation}
Then, the constraint equation that follows from multiplet recombination and the absence of stringy states in the supergravity OPE is
that the total twist four contribution $A_{{\cal L},{4,l,[020]}}$ must vanish, namely
\begin{equation}\label{recombinetwist4}
A_{{\cal L},{4,l,[020]}}=A^{free}_{{\cal K},{4,l,[020]}} + {\cal H}_{2[2^2]2[2^2]}\Big|_{\tau=4,l,[020]}=0\,.
\end{equation}
Explicitly,
\begin{equation}
A_{{\cal L},{4,l,[020]}}=
\left\{\begin{array}{ll}
+\frac{4}{a}(l+1)(l+6) \frac{(l+2)!(l+4)!}{(2l+6)!} &\ \textrm{spin\ odd}\\[.5cm]
+\frac{4(l+2)(l+5)}{a} \frac{(l+3)(l+4)-\frac{12}{a}}{(l+3)(l+4)+\frac{4}{a} }\frac{ (l+2)!(l+4)!}{(2l+6)!}&\ \textrm{spin\ even}
 \end{array}\right.
 \end{equation}
As for \eqref{prediction020}, these results are exact in $a$. Further details on the computation are given in appendix \ref{N4blocks}.

Solving the predictions for ${\cal H}_{ 2[2^2]2[2^2] }$ in the $\frac{1}{a}$ expansion, we find that at order $\frac{1}{a}$,
\begin{equation}\label{22d22dcheck1}
{\cal H}_{ 2[2^2]2[2^2]}\Big|_{\frac{1}{a},\tau=4}=
\left\{\begin{array}{c}
-4(l+1)(l+6) \frac{(l+2)!(l+4)!}{(2l+6)!} \frac{1-(-1)^l}{2} \\[.5cm] 
 -4(l+2)(l+5) \frac{(l+2)!(l+4)!}{(2l+6)!} \frac{1+(-1)^l}{2}
\end{array}\right.\,,
\end{equation}
and at order $\frac{1}{a^2}$,
\begin{equation}\label{22d22dcheck2}
{\cal H}_{ 2[2^2]2[2^2] }\Big|_{\frac{1}{a^2},\tau=4} =
\left\{\begin{array}{c}
0 \\[.5cm] 
 +64(l+2)(l+5)\frac{ (l+2)!^2}{(2l+6)!} \frac{1+(-1)^l}{2}
\end{array}\right. \,.\rule{.6cm}{0pt}
\end{equation}
The results \eqref{22d22dcheck1} and \eqref{22d22dcheck2} after resummation are
\begin{align}\label{resummation1}
 {\cal H}_{22^222^2}=& \,\  -\frac{8}{a} U\Bigg[\frac{(1+V)}{(1-V)^2 V}+\frac{2\log(V)}{(1-V)^3}\Bigg] + U^2(\ldots) \\
 &+ \frac{32}{a^2} U \Bigg[ \frac{1}{(1-V)^2V} - \frac{ \log(V)+\log^2(V)+2{\rm Li}_2(1-V) }{(1-V)^3}\Bigg]+  U^2(\ldots)
 \label{resummation2}
\end{align}
and we find a perfect match with the functions given in section \ref{sec2d2d22}. There are several comments to highlight here:
\begin{itemize}
\item[1.]
The dynamical function ${\cal H}_{2[2^2]2[2^2]}$ at order $\frac{1}{a}$ is $-8U^2\overline{D}_{2422}$, 
ie.~the same $\overline{D}_{2422}$ as in ${\cal H}_{2222}$! But in \eqref{22d22dcheck1} we have 
to decompose it in blocks w.r.t.~to the external charges $p_{43}=2$ and $p_{12}=-2$ and twist $\tau\ge 4$.
This decomposition is non standard compared to the literature on single-particle correlators. 
\item[2.]
The spin dependence changes if we compare \eqref{22d22dcheck1} at order $\frac{1}{a}$ 
and \eqref{22d22dcheck2} at order $\frac{1}{a^2}$. This is an indication that 
${\cal H}^c$ must have new features, compared to $\overline{D}_{2422}$. In fact we see 
from the resummation \eqref{resummation1} the presence of $\log^2(V)$, and this cannot be obtained from a tree level function.
\item[3.]
The one-loop function ${\cal H}_{2222}$ -- interpreted as a contribution to ${\cal H}_{22^222^2}$ --  
does not contribute to the twist four in $[0,2,0]$.\footnote{The fact that ${\cal H}_{2222}$ has an 
empty twist two sector in [0,0,0], see footnote \ref{footkonishi}, now implies that ${\cal H}_{2222}$ does not contribute to 
the twist four sector in $[0,2,0]$.} The prediction \eqref{22d22dcheck2} is directly a prediction 
for ${\cal H}^t$, and thus for ${\cal H}^c$.
\end{itemize}
Worth emphasising that all the non trivial details about the ${\cal O}_2{\cal O}_2^2$ OPE,
could not possibly work without the inclusion of the ${\cal H}^c_{2[2^2]2[2^2]}$ contribution. 
This provides a new consistency check of AdS/CFT in the context of multi-particle correlators.

\subsection{$\langle {\cal O}_2^2 {\cal O}_2^2 {\cal O}_2 {\cal O}_2 \rangle$ and the long sector}\label{OPEsect12}

Continuing the discussion about the ${\cal O}_2{\cal O}_2$ OPE and its implications 
for ${\cal H}_{[2^2][2^2]22}$, in this section we study what information can be used 
from the long double particle sector \cite{Aprile:2017bgs,Aprile:2017xsp,Aprile:2018efk}.  

The relevant double particle operators will be denoted with ${\cal D}_{\tau,l}$ and belong 
to the $[0,0,0]$ rep, studied in \cite{Aprile:2018efk}.  They exist for twist $\tau\ge 4$ at any spin, and they are 
degenerate in twist.\footnote{In $[0,0,0]$ these are  
$\{ {\cal O}_2\Box^{\tau-4}\partial^l{\cal O}_2, {\cal O}_3\Box^{\tau-6}\partial^l{\cal O}_3,\ldots\}$.}
These double particle operators have three-point couplings with ${\cal O}_2^n{\cal O}_2^n$ with the following large $N$ expansion,   
\begin{equation}
C_{[2^n][2^n];{\cal D}}= C^{(0)}_{[2^n][2^n];{\cal D}}+ \frac{1}{a} C^{(1)}_{[2^n][2^n];{\cal D}} +\ldots
\end{equation}
and their anomalous dimension, defined by
\begin{equation}
 \Delta_{\cal D}=\tau+l+ \frac{2}{a} \eta^{(1)}_{\cal D} + \ldots
\end{equation}
are known exactly from previous work in \cite{Aprile:2017bgs,Aprile:2017xsp,Aprile:2018efk}. 

Our starting point is the observation that large $N$ factorization relates
$C^{(0)}_{[2^2][2^2]; {\cal D}_{\tau,l} }$  and $C^{(0)}_{[2][2]; {\cal D}_{\tau,l}}$, 
through the obvious relation
\begin{equation}\label{new3pt}
C^{(0)}_{[2^2][2^2]; {\cal D}_{\tau,l} }= 2\, C^{(0)}_{22; {\cal D}_{\tau,l} }\,.
\end{equation} 
Then, from the OPE we deduce that 
\begin{equation}\label{leadingCimplicationdp}
\hat{\cal C}_{[2^2][2^2]22}\Big|_{\frac{\log^n(U)}{a^n}}=\frac{1}{n!}
\sum_{\tau,l } \left( \sum_{{\cal D}} C^{(0)}_{[2^2][2^2] ;{\cal D}}\ \eta^n_{{\cal D}}\ C^{(0)}_{22 ;{\cal D}} \right)\!B^{\cal{L}}_{\tau,l,[000]}=2\, \hat{\cal C}_{2222}\Big|_{\frac{\log^n(U)}{a^n}}\,.
\end{equation}
The case $n=0$ corresponds to the disconnected free theory diagrams contributions in \eqref{prefa_dd22m}. 

From \eqref{leadingCimplicationdp} follows that at each order in the large $N$ expansion the function $ {\cal H}_{[2^2][2^2]22}-{\cal H}_{2222}$ does not have a maximal logarithmic discontinuity. For example, at order $\frac{1}{a^2}$ we can define 
\begin{equation}
 {\cal H}^{t}_{[2^2][2^2]22} \Big|_{\frac{1}{a^2}}\equiv {\cal H}_{[2^2][2^2]22}\Big|_{\frac{1}{a^2}}-  2 {\cal H}_{2222}\Big|_{\frac{1}{a^2}}\,.
\end{equation}
This ${\cal H}^{t}_{[2^2][2^2]22}$ 
\emph{does not} have a $\log^2(U)$ discontinuity. A posteriori ${\cal H}^{c}_{[2^2][2^2]22}$ 
will be contained in ${\cal H}^{t}_{[2^2][2^2]22}$, and the claim that ${\cal H}^{t}_{[2^2][2^2]22}$ 
\emph{does not} have a $\log^2(U)$ discontinuity is indeed what we have found 
from the gravity computation. See section \ref{sec2d2d22}.
At the next order we can define
\begin{equation}\label{2d2d221oneloop}
{\cal H}^{1\myell}_{[2^2][2^2]22} \Big|_{\frac{1}{a^3}}\equiv {\cal H}_{[2^2][2^2]22}\Big|_{\frac{1}{a^3}}-  2 {\cal H}_{2222}\Big|_{\frac{1}{a^3}}\,.
\end{equation}
This ${\cal H}^{1\myell}_{[2^2][2^2]22}$ \emph{does not} have a $\log^3(U)$ discontinuity, and so on so forth.

We will now show that the CFT data of the non degenerate ${\cal O}_2\partial^l{\cal O}_2$ 
long operators \cite{Aprile:2017xsp},  gives a prediction for the $\log(U)$ discontinuity of ${\cal H}^t$. 
This follows from the two OPE equations
\begin{align}\label{newprediction1}
 \hat{\cal C}_{[2^2][2^2]22}\Big|_{\frac{1}{a}}  
-2\, \hat{\cal C}_{2222} \Big|_{\frac{1}{a}}  & = \sum_{l} \Big[ C^{(1)}_{[2^2][2^2]; {\cal D}_4} -2 C^{(1)}_{22 ;{\cal D}_4}\Big] C^{(0)}_{22;{\cal D}_4} {B}^{\cal L}_{4,l,[000]} + \big( \textrm{twist}\ge 6 \big)  \\[.2cm]
\hat{\cal C}_{[2^2][2^2]22}\Big|_{\frac{\log(U)}{a^2}}  
-2\, \hat{\cal C}_{2222} \Big|_{\frac{\log(U)}{a^2}} & = \sum_{l}  
\Big[ C^{(1)}_{[2^2][2^2] ;{\cal D}_4} -2 C^{(1)}_{22; {\cal D}_4}\Big]\eta_{{\cal D}_4 } C^{(0)}_{22;{\cal D}_4}
 {B}^{\cal L}_{4,l,[000]} + \big( \textrm{twist}\ge 6 \big)\label{newprediction2}
\end{align}
which can be derived from appendix \ref{expaOPElong}. 

Let us begin by making some important comments about \eqref{newprediction1}.
Firstly,  $\hat{\cal C}_{[2^2][2^2]22}$ and $\hat{\cal C}_{2222}$ have a block expansion with various terms 
and derivatives of the blocks. See \eqref{expaOPElong_dp1} and \eqref{expaOPElong_dp2}. However 
their difference at twist four simplifies dramatically because of the relation between leading three-point couplings in \eqref{new3pt}. 
Secondly, on the lhs of \eqref{newprediction1} we find  
\begin{equation}\label{twist4prediction2d2d22free1}
\hat{\cal C}_{[2^2][2^2]22}\Big|_{\frac{1}{a}}  
-2\, \hat{\cal C}_{2222} \Big|_{\frac{1}{a}}= \frac{4}{a} U^2\sigma^2 + \frac{4}{a} \frac{U^2\tau^2}{V^2} + \frac{16}{a} \frac{U^2\sigma\tau}{V} 
\end{equation}
because the $\overline{D}_{2422}$ functions cancel each other. 
Therefore, from  \eqref{newprediction1} we obtain the result
\begin{equation}
 \frac{4}{a} U^2\sigma^2 + \frac{4}{a} \frac{U^2\tau^2}{V^2} + \frac{16}{a} \frac{U^2\sigma\tau}{V} = 
\sum_{l} \Big[ C^{(1)}_{[2^2][2^2] ;{\cal D}_4} -2 C^{(1)}_{22 ;{\cal D}_4}\Big] C^{(0)}_{22;{\cal D}_4}\, {B}^{\cal L}_{4,l,[000]} + \big( \textrm{twist}\ge 6 \big)
\end{equation}
which is an equation that determines $C^{(1)}_{[2^2][2^2] {\cal D}_4}$ from free theory data. 
The result for the combination under the sum is
 \begin{equation}
\Big[ C^{(1)}_{[2^2][2^2] ;{\cal D}_4} -2 C^{(1)}_{22; {\cal D}_4}\Big]C^{(0)}_{22;{\cal D}_4} 
= \frac{4}{3} (l+2)(l+5)\frac{(l+3)!^2}{(2l+6)!} \frac{1+(-1)^l}{2}\,.
 \end{equation}
  The explicit result for $C^{(1)}_{[2^2][2^2] ;{\cal D}_4}$ is given in \eqref{C1twist4explicit}. 
  
Then, consider the  $\log(U)$ contribution at order $\frac{1}{a^2}$ given in \eqref{newprediction2}. 
Since we are looking at the $\log(U)$ discontinuity, there cannot be a free theory 
contribution and the equation \eqref{newprediction2} is directly an equation for the dynamical 
functions ${\cal H}_{[2^2][2^2]22}$ and ${\cal H}_{2222}$, namely
\begin{align}
\hat{\cal C}_{[2^2][2^2]22}\Big|_{\frac{\log(U)}{a^2}}  
-2\, \hat{\cal C}_{2222} \Big|_{\frac{\log(U)}{a^2}} & =  {\cal I}\times\Bigg[ {\cal H}_{[2^2][2^2]22}-2{\cal H}_{2222}\Bigg]_{\frac{\log(U)}{a^2}}\;.
\end{align}
The combination in parenthesis is precisely ${\cal H}^t_{[2^2][2^2]22}$!
Again, the block expansion of each term contains separately many 
contributions but given the relation between leading three-point couplings in \eqref{new3pt}, 
there is only one contribution left at twist four.
Thus, we obtain the prediction 
\begin{equation}
{\cal H}^t_{[2^2][2^2]22}\Big|_{\frac{\log(U)}{a^2}}  = \sum_{l} 
\Big[ C^{(1)}_{[2^2][2^2] ;{\cal D}_4} -2 C^{(1)}_{22; {\cal D}_4}\Big]\eta_{{\cal D}_4 } C^{(0)}_{22;{\cal D}_4}
 {H}_{4,l,[000]} + \big( \textrm{twist}\ge 6 \big)
\end{equation}
where the reduced long block $H_{\tau,l,[aba]}$ is defined in \eqref{allLongs}.

At this point, we know from \cite{Aprile:2017bgs,Aprile:2017xsp} that
\begin{equation}
\eta_{{\cal D}_4}=-\frac{48}{(l+1)(l+6)} \frac{1+(-1)^l}{2}\,,
\end{equation}
therefore we can predict the twist four sector 
\begin{equation}\label{etapred5.2sec}
{\cal H}^t_{[2^2][2^2]22}\Big|_{\frac{\log(U)}{a^2}} =\sum_{l} \Bigg[ - 64 \frac{(l+2)(l+5)}{(l+1)(l+6)}\frac{(l+3)!^2}{(2l+6)!} \frac{1+(-1)^l}{2}\Bigg] {H}_{4,l,[000]}+ \big( \textrm{twist}\ge 6 \big)\;.
\end{equation}
Equivalently, we give its resummation
\begin{align}
&
{\cal H}^t_{[2^2][2^2]22}\Big|_{\frac{\log(U)}{a^2}} =  \label{predictionlogU}\\
&\rule{1cm}{0pt}
-32 U^2\Bigg[\frac{(1+10V+V^2)}{(1-V)^4V} +\frac{12(1+V)\log(V)}{(1-V)^5} +\frac{2(1+4V+V^2)\log^2(V)}{(1-V)^6} +  U^3(\ldots) \Bigg]\;. \notag
\end{align}
As for the remarks on the presence of $\log^2(V)$ in 
\eqref{resummation2}, we see above the presence of $\log(U)\log^2(V)$ which is a weight three contribution beyond those of a weight two tree-level function. 
Remarkably, this is precisely the same result that we obtain when we include ${\cal H}^{c}_{[2^2][2^2]22}$, see  \eqref{Hc2d2d22fromletter}.

\subsection{$\langle {\cal O}^3_2 {\cal O}_2^3 {\cal O}_2 {\cal O}_2\rangle$ and higher multi-particle correlators}\label{triplecorrelatorsection}

In this section we study the correlator $\langle {\cal O}^3_2 {\cal O}_2^3 {\cal O}_2 {\cal O}_2\rangle$ and the 
consistency checks that can be performed on ${\cal H}_{[2^3][2^3]22}$ from the known OPE data. These are
the absence of twist two stringy states and the predictions for the twist four long 
operators in the $[0,0,0]$ rep.\footnote{We cannot perform type (S) checks in 
$\langle  {\cal O}_2  {\cal O}^3_2  {\cal O}_2  {\cal O}_2^3 \rangle$ without knowledge about the twist six protected
triple-particle operators in $[0,4,0]$. These determine the $\frac{1}{a}$ term and then mix with 
the protected double particles at higher orders.}  Both checks will require knowledge from  
$\langle {\cal O}_2 {\cal O}_2 {\cal O}_2 {\cal O}_2\rangle$ and $\langle {\cal O}^2_2 {\cal O}^2_2 {\cal O}_2 {\cal O}_2\rangle$, 
and therefore will be non trivial in various ways. Specifically we will need  three-point couplings relations already obtained in \ref{OPEsect1} 
and \ref{OPEsect12}, and new three-point coupling relations that we will derive. These checks generalize to higher multi-particle correlators 
$\langle {\cal O}^n_2 {\cal O}^n_2 {\cal O}_2 {\cal O}_2\rangle$. However, already for $n=3$,  we shall see that
the analytic structure of the higher multi-particle correlators implies relations between three-point couplings that are not obvious from 
the CFT.

To study $\langle {\cal O}^3_2 {\cal O}^3_2 {\cal O}_2 {\cal O}_2\rangle$ we will generalize the discussion  
in section \ref{OPEsect12}.  
There we undestood that the OPE 
naturally splits ${\cal H}_{[2^2][2^2]22}$ into a disconnected contribution with 
maximal logarithmic discontinuities, that is $2{\cal H}_{2222}$,  and the rest.  At order $\frac{1}{a^2}$ 
we defined the rest to be ${\cal H}^t_{[2^2][2^2]22}$. Then, the CFT analysis revealed 
new predictions for ${\cal H}^t_{[2^2][2^2]22}$. Since we knew from gravity that 
$$
{\cal H}^t_{[2^2][2^2]22}={\cal H}^c_{[2^2][2^2]22}\Big|_{\frac{1}{a^2}}-4 {\cal H}_{2222}\Big|_{\frac{1}{a}}
$$ 
the aforementioned predictions for ${\cal H}^t_{[2^2][2^2]22}$ gave constraints on ${\cal H}^c_{[2^2][2^2]22}$ that we tested successfully.

Let us now consider ${\cal H}_{ [2^3][2^3]22 }$ at order $\frac{1}{a^3}$, and generalize the reasoning above. This function was given 
in section \ref{sec_tripleparticle}, see \eqref{twoloop3d3d22},
where we showed that
\begin{equation}\label{H3p3p22OPEsec}
{\cal H}_{ [2^3][2^3]22 }\Big|_{\frac{1}{a^3} }=  
{\cal H}^c_{ [2^3][2^3]22 }\Big|_{\frac{1}{a^3}}  
-12 {\cal H}^c_{[2^2][2^2] 22}\Big|_{ \frac{1}{a^2} } + 3  {\cal H}^c_{[2^2][2^2] 22} \Big|_{\frac{1}{a^3}} 
+48 {\cal H}_{2222}\Big|_{\frac{1}{a}} - 12 {\cal H}_{2222}\Big|_{\frac{1}{a^2}} +3 {\cal H}_{2222}\Big|_{\frac{1}{a^3}}
\end{equation}
The claim is that the OPE analysis singles out the following combination
\begin{equation}\label{specialcombotrp}
{\cal H}^t_{[2^3][2^3]22}\Big|_{\frac{1}{a^3}}={\cal H}^c_{ [2^3][2^3]22 }\Big|_{\frac{1}{a^3}} - 12 {\cal H}^{c}_{[2^2][2^2]22}\Big|_{\frac{1}{a^2}}+ 24 {\cal H}_{2222}\Big|_{\frac{1}{a}}
\end{equation}
and the constraint on \eqref{specialcombotrp} are
\begin{align}
&\label{obvioustriplep}
\Bigg[ {\cal H}^t_{[2^3][2^3]22}\Bigg]_{\frac{\log^3(U)}{a^3}}=0\\
\label{simpleclaimtriplep}
 &\Bigg[ 
{\cal H}^t_{[2^3][2^3]22} \Bigg]_{\frac{\log^0(U)}{a^3},\, {\rm twist\, 2}}=0 \\
&\label{complicatedclaimtriplep}
 \Bigg[ 
{\cal H}^t_{[2^3][2^3]22} \Bigg]_{\frac{\log^n(U)}{a^3},\, {\rm twist\, 4}}=0\qquad;\qquad n=1,2
\end{align}
Very remarkably, all these predictions are satisfied when we include our gravity result for ${\cal H}^c_{[2^3][2^3]22}$. 

Let us first understand why the combination \eqref{specialcombotrp} is singled out by the OPE analysis. 
Consider arranging the expression in \eqref{H3p3p22OPEsec} as
\begin{equation}\label{rewfull3d3d22}
{\cal H}_{ [2^3][2^3]22 }\Big|_{\frac{1}{a^3}} - 3 {\cal H}_{2222}\Big|_{\frac{1}{a^3}}- 
3  {\cal H}^{1\myell}_{[2^2][2^2] 22} \Big|_{\frac{1}{a^3}}  = 
\Bigg[ {\cal H}^c_{ [2^3][2^3]22 }\Big|_{\frac{1}{a^3}} - 12 {\cal H}^{c}_{[2^2][2^2]22}\Big|_{\frac{1}{a^2}}+ 24 {\cal H}_{2222}\Big|_{\frac{1}{a}}\Bigg]\;,
\end{equation}
where the rhs in $[\ldots]$ is precisely \eqref{specialcombotrp} and 
on the lhs we subtracted two disconnected contributions. The first one is ${\cal H}_{2222}|_{\frac{1}{a^3}}$, and is obvious.  
The second one, ${\cal H}^{1\myell}_{[2^2][2^2] 22}|_{\frac{1}{a^3}}$  is defined by \eqref{2d2d221oneloop}.
It can be computed explicitly from \eqref{prefa_dd22}, and it is equal to
\begin{equation}\label{1loop2d2d22}
{\cal H}^{1\myell}_{[2^2][2^2]22}\Big|_{\frac{1}{a^3}}= {\cal H}^c_{[2^2][2^2]22}\Big|_{\frac{1}{a^3} } + 8 {\cal H}_{2222}\Big|_{\frac{1}{a}} - 4 {\cal H}_{2222}\Big|_{\frac{1}{a^2}}\,.
\end{equation}
The crucial point about the two disconnected subtractions is that  both  contributions 
have a twist four sector that is predicted by the OPE analysis. It follows that \eqref{specialcombotrp} 
is the combination that is subject to the new predictions at twist four, given those for 
${\cal H}_{ [2^3][2^3]22 }|_{\frac{1}{a^3}} - 3 {\cal H}_{2222}|_{\frac{1}{a^3}}$ and those for 
${\cal H}^{1\myell}_{[2^2][2^2]22}|_{\frac{1}{a^3}}$.

\paragraph{Details about the OPE predictions.} Starting from \eqref{obvioustriplep},  the fact that ${\cal H}^t_{[2^3][2^3]22}$ is defined by 
subtracting ${\cal H}_{ [2^3][2^3]22 }|_{\frac{1}{a^3}}-3 {\cal H}_{2222}|_{\frac{1}{a^3}}$ 
demonstrates that it does not have a $\log^3(U)$ discontinuity. In fact, from large $N$ factorization 
we readily find that
\begin{equation}\label{disconnected3pt3p3p}
C^{(0)}_{[2^3][2^3];{\cal D}_{\tau,l}} = 3 C^{(0)}_{22;{\cal D}_{\tau,l}} 
\end{equation}
for all double particle operators $ {\cal D}_{\tau,l}$ in $[0,0,0]$ (see section \ref{OPEsect12} and \cite{Aprile:2017bgs,Aprile:2017xsp,Aprile:2018efk}).
Note that \eqref{disconnected3pt3p3p} is the generalization of the relation
$C^{(0)}_{[2^2][2^2]; {\cal D}_{\tau,l} }= 2\, C^{(0)}_{22; {\cal D}_{\tau,l} }$ 
that we encountered in \eqref{new3pt}.

Now let us comment on the derivation of the constraints \eqref{simpleclaimtriplep}-\eqref{complicatedclaimtriplep}. 
The simplest one is \eqref{simpleclaimtriplep}, namely the absence of twist two stringy states. 
This constraint follows from the fact 
that the free theory, see \eqref{free3p3p22}, truncates at order $\frac{1}{a}$, and therefore 
there are no twist two stringy states  at the next orders,  $\frac{1}{a^n}$ with $n\ge 2$. In particular, 
we must find the result quoted in \eqref{simpleclaimtriplep}.\footnote{The case $n=2$ of 
${\cal H}_{[2^3][2^3]22}$ is automatically  satisfied by the same for property for ${\cal H}^t_{[2^2][2^2]22}$, 
noting that (see \eqref{3d3d22order2})  $${\cal H}_{ [2^3][2^3]22 }|_{\frac{1}{a^2} }= 
3 {\cal H}^t_{[2^2][2^2]22}|_{\frac{1}{a^2}} + 3 {\cal H}_{2222}|_{\frac{1}{a^2}}.$$}

The derivation of the constraints \eqref{complicatedclaimtriplep} instead requires 
a statement about subleading three-point couplings with the twist four double 
particle operator ${\cal D}_{4,l}$,
\begin{equation}\label{subleadingCpredtriplep}
C^{(n)}_{[2^3][2^3];{\cal D}_4} = 3 C^{(n)}_{[2^2][2^2]:{\cal D}_4} - 3 C^{(n)}_{22;{\cal D}_4}\qquad;\qquad n=1,2\,.
\end{equation}
These relations can be understood as follows.
For the case $n=1$ we observe that
\begin{align}\label{betterprediction3d3d22}
\hat{\cal C}_{[2^3][2^3]22}\Big|_{\frac{1}{a}}  
-3\, \hat{\cal C}_{2222} \Big|_{\frac{1}{a}}& =
 \frac{48}{a} \frac{U^2\sigma\tau}{V} +  \frac{12}{a} \Bigg[ U^2\sigma^2 +\frac{U^2\tau^2}{V^2} \Bigg]
= 3\Bigg[ \hat{\cal C}_{[2^2][2^2]22}\Big|_{\frac{1}{a}}  
-2\, \hat{\cal C}_{2222} \Big|_{\frac{1}{a}} \Bigg] \,.
\end{align}
where we used that the difference $\hat{\cal C}_{[2^n][2^n]22}-n\hat{\cal C}_{2222}$ does not have $\frac{1}{a}$ 
dynamical funtions, i.e.~the $\overline{D}$ functions cancel out, and thus free theory is the only contribution.
At this point, using \eqref{expaOPElong_dp1}, the result follows since there is only one ${\cal D}_{4}$ exchanged operator for each spin.
For the case $n=2$, we observe that
\begin{align}\label{betterprediction3d3d22bis}
\hat{\cal C}_{[2^3][2^3]22}\Big|_{\frac{1}{a^2}}  
-3\, \hat{\cal C}_{2222} \Big|_{\frac{1}{a^2}}& ={\cal I}\Bigg[ 3 {\cal H}^c_{[2^2][2^2]22}\Big|_{\frac{1}{a^2}} - 12 {\cal H}_{2222}\Big|_{\frac{1}{a}} \Bigg] 
= 3\Bigg[ \hat{\cal C}_{[2^2][2^2]22}\Big|_{\frac{1}{a^2}}  
-2\, \hat{\cal C}_{2222} \Big|_{\frac{1}{a^2}} \Bigg] \,.
\end{align}
where we used \eqref{oneL2d2d22} and the fact that free theory truncates at order $\frac{1}{a}$.
Then, using \eqref{expaOPElong_dp2} the result follows, again because there is only one ${\cal D}_{4}$ exchanged operator for each spin.  

The last step to demonstrate the constraints in \eqref{complicatedclaimtriplep} is to look at the OPE 
predictions at twist four for the $\log^2(U)$ and the $\log(U)$ of ${\cal H}_{ [2^3][2^3]22 } - 3 {\cal H}_{2222}-3{\cal H}^{1\myell}_{[2^2][2^2]22}$.
Using the expansions quoted in \eqref{expaOPElong_dp3} and the result for the 
leading three-point couplings, the block expansion simplifies to
\begin{align}
&
\label{log2U3p3ptwist4}
\Bigg[{\cal H}^{t}_{[2^3][2^3]22}
\Bigg]_{\frac{\log^2(U)}{a^3}}\!\!\!\!\!\!  =\ \tfrac{1}{2}  
\sum_{l} \Bigg[ ( C^{(1)}_{[2^3][2^3];{\cal D}_4}-3C^{(1)}_{[2^2][2^2];{\cal D}_4}+3C^{(1)}_{22;{\cal D}_4})\big(\eta^{(1)}_{{\cal D}_4} \big)^{\!2}C^{(0)}_{22 {\cal D}_4}\Bigg]{H}_{4,l,[000]} + \big( \textrm{twist}\ge 6 \big) 
\end{align}
At this point, the relation 
\eqref{subleadingCpredtriplep} for $n=1$, namely 
$C^{(1)}_{[2^3][2^3];{\cal D}_4} = 3 C^{(1)}_{[2^2][2^2]:{\cal D}_4} - 3 C^{(1)}_{22;{\cal D}_4}$
gives the result quoted in \eqref{complicatedclaimtriplep}.
For the $\log(U)$ predictions one finds a longer expression,
\begin{align}
&
\!\!\!\!\!\Bigg[
{\cal H}^{t}_{[2^3][2^3]22}\Bigg]_{\frac{\log(U)}{a^3}}\!\!\!\!\!\! = \ 
\sum_{l} \Bigg[ ( C^{(1)}_{[2^3][2^3];{\cal D}_4}-3 C^{(1)}_{[2^2][2^2];{\cal D}_4}+3 C^{(1)}_{22;{\cal D}_4} )( \eta^{(1)}_{{\cal D}_4} C^{(1)}_{22;{\cal D}_4}+\eta^{(2)}_{{\cal D}_4}C^{(0)}_{22 {\cal D}_4}) \notag\\
&
\rule{2.1cm}{0pt}+ ( C^{(2)}_{[2^3][2^3];{\cal D}_4}-3 C^{(2)}_{[2^2][2^2];{\cal D}_4} + 3 C^{(2)}_{22;{\cal D}_4})\eta^{(1)}_{{\cal D}_4} C^{(0)}_{22 {\cal D}_4}\Bigg]{H}_{4,l,[000]} + \ldots + \big( \textrm{twist}\ge 6 \big)
\end{align}
where $\ldots$ contain omitted terms involving derivatives of the twist four blocks. 
Again, by using \eqref{subleadingCpredtriplep} for $n=2$, we find the result quoted in \eqref{complicatedclaimtriplep}

\paragraph{Absence of $\log^2(U)$ discontinuity.}  
Not only the twist four sector of the $\log^2(U)$ discontinuity of ${\cal H}^t_{[2^3][2^3]22}$ 
vanishes, but the whole $\log^2(U)$ discontinuity.  As a result,  the gravity result for ${\cal H}^t_{[2^3][2^3]22}$, 
has more information than the one extracted about the twist four sector.
This is non trivial information because beyond twist four not only double particle operators
 are exchanged but also long triple particle operators. The triple particle operators ${\cal T}_{\tau,l}$ 
 exist starting from twist six. These operators are degenerate both in twist and spin. Their three-point couplings with ${\cal O}_2^n{\cal O}_2^n$ 
and their anomalous dimension have the following  large $N$ expansion 
\begin{equation}
C_{[2^n][2^n];{\cal T}}= \frac{1}{\sqrt{a}}\Big[ C^{(\frac{1}{2})}_{[2^n][2^n];{\cal T}} + \ldots\Big] \qquad;\qquad \Delta=\tau+l +\frac{2}{a} \eta^{(1)}_{\cal T} +\ldots
\end{equation}
Extending the analysis done in \eqref{log2U3p3ptwist4} we find
 \begin{align}\label{absencelog2U3p}
 \Bigg[{\cal H}^{t}_{[2^3][2^3]22}
\Bigg]_{\frac{\log^2(U)}{a^3}}\!\!\!\!\!\!  = \big( \textrm{twist}\, 4 \big)  
+\tfrac{1}{2}  
\sum_{\tau,l} \Bigg[ & \sum_{{\cal D}} ( C^{(1)}_{[2^3][2^3];{\cal D} }-3C^{(1)}_{[2^2][2^2];{\cal D}}+3C^{(1)}_{22;{\cal D}})\big(\eta^{(1)}_{{\cal D}} \big)^{\!2}C^{(0)}_{22 {\cal D}} + \notag\\
&
\sum_{{\cal T}} ( C^{(\frac{1}{2})}_{[2^3][2^3];{\cal T} }-3C^{(\frac{1}{2})}_{[2^2][2^2];{\cal T}}+3C^{(\frac{1}{2})}_{22;{\cal T}})\big(\eta^{(1)}_{{\cal T}} \big)^{\!2}C^{(\frac{1}{2})}_{22 {\cal T}}
\Bigg]{H}_{\tau,l,[000]} \;.
 \end{align}
Setting the above expression to zero is not obvious because the operators exchanged are degenerate.
However, considering that at large spin we would find different types of contributions for the double particles and the triple particles, because of the fact that the triple particle operators are degenerate in twist and spin,
we are led to conclude that each individual contribution must vanish, 
therefore
\begin{equation}\label{prediction_nolog2U_3p}
C^{(1)}_{[2^3][2^3];{\cal D} }=3C^{(1)}_{[2^2][2^2];{\cal D}}-3C^{(1)}_{22;{\cal D}}\qquad; \qquad
C^{(\frac{1}{2})}_{[2^3][2^3];{\cal T} }=3C^{(\frac{1}{2})}_{[2^2][2^2];{\cal T}}-3C^{(\frac{1}{2})}_{22;{\cal T}}\,.
\end{equation}
Note that \eqref{prediction_nolog2U_3p} are non trivial relations for \emph{subleading} three-point 
couplings that involve long operators, and that cannot be proven from four-point functions alone.

\paragraph{Large $N$ factorization for three-point functions.} 
It is interesting at this point to collect the results obtained for three-point couplings, and rephrase them
 as a statement about large $N$ factorization for three-point functions, introducing a notion of connected versus disconnect contributions. 
By this mean the following:

\begin{itemize}
\item[$\bullet$] For the ${\cal O}^2_2{\cal O}_2^2$ and the double particle operators we find that
\begin{equation}\label{largeNfacto2pt1}
\langle {\cal O}_2^2{\cal O}_2^2 {\cal D}_{\tau,l} \rangle= \langle {\cal O}_2^2{\cal O}_2^2  {\cal D}_{\tau,l} \rangle_c +  4 \langle {\cal O}_2{\cal O}_2\rangle \langle {\cal O}_2{\cal O}_2 {\cal D}_{\tau,l} \rangle
\end{equation}
where the connected component is subleading and such that
\begin{equation}
\frac{ \langle {\cal O}_2^2{\cal O}_2^2  {\cal D}_{\tau,l} \rangle_c}{ \langle {\cal O}^2_2{\cal O}^2_2\rangle }\sim \frac{1}{a}\,.
\end{equation}
\item[$\bullet$]
For the ${\cal O}^3_2{\cal O}_3^2$ and the double particle operators we find that
\begin{equation}\label{largeNfacto3pt1}
\langle {\cal O}_2^3{\cal O}_2^3  {\cal D}_{\tau,l} \rangle = \langle {\cal O}_2^3{\cal O}_2^3 {\cal D}_{\tau,l} \rangle_c
+ 9 \langle {\cal O}_2 {\cal O}_2 \rangle \langle {\cal O}_2^2 {\cal O}_2^2  {\cal D}_{\tau,l} \rangle 
+9\big( \langle {\cal O}^2_2 {\cal O}^2_2 \rangle - 4  \langle {\cal O}_2 {\cal O}_2 \rangle^{\!2}\big) \langle {\cal O}_2 {\cal O}_2  {\cal D}_{\tau,l} \rangle 
\end{equation}
where the connected component is subleading and such that
\begin{equation}
 \frac{ \langle {\cal O}_2^3{\cal O}_2^3 {\cal D}_{\tau,l} \rangle_c}{\langle {\cal O}_2^3 {\cal O}_2^3\rangle}\sim \frac{1}{a^2} 
\end{equation}
\item[$\bullet$]
For the ${\cal O}^3_2{\cal O}_3^2$ and the triple particle operators we find that
\begin{equation}
\label{largeNfacto3pt2}
\langle {\cal O}_2^3{\cal O}_2^3  {\cal T}_{\tau,l} \rangle = \langle {\cal O}_2^3{\cal O}_2^3 {\cal T}_{\tau,l} \rangle_c
+ 9 \langle {\cal O}_2 {\cal O}_2 \rangle \langle {\cal O}_2^2 {\cal O}_2^2  {\cal T}_{\tau,l} \rangle 
+9\big( \langle {\cal O}^2_2 {\cal O}^2_2 \rangle - 4  \langle {\cal O}_2 {\cal O}_2 \rangle^{\!2}\big) \langle {\cal O}_2 {\cal O}_2  {\cal T}_{\tau,l} \rangle
\end{equation}
where the connected component is subleading and such that
\begin{equation}
\frac{ \langle {\cal O}_2^3{\cal O}_2^3 {\cal T}_{\tau,l} \rangle_c}{ \langle {\cal O}_2^3{\cal O}_2^3\rangle} \sim \frac{1}{a^{\frac{3}{2}}}\,.
\end{equation}
\end{itemize}
Given the large $N$ expansion of each term, and from the knowledge of the two-point functions $\langle{\cal O}_2^n{\cal O}_2^n\rangle$, 
we match the results obtained from the considerations about the the four-point functions, namely
\begin{align}
C^{(0)}_{[2^n][2^n];{\cal D}}&=n C^{(0)}_{22;{\cal D}}\rule{1.65cm}{0pt} n=2,3,\\
C^{(1)}_{[2^3][2^3];{\cal D} }&=3C^{(1)}_{[2^2][2^2];{\cal D}}-3C^{(1)}_{22;{\cal D}}\qquad;\qquad C^{(\frac{1}{2})}_{[2^3][2^3];{\cal T} }=3C^{(\frac{1}{2})}_{[2^2][2^2];{\cal T}}-3C^{(\frac{1}{2})}_{22;{\cal T}}\,.
\label{lastline_facto3p}
\end{align}
Let us note that factorization can only be manifested on the external $\frac{1}{2}$-BPS operators, 
and in fact the same factors are there for both \eqref{largeNfacto3pt1} and \eqref{largeNfacto3pt2}.
This is consistent with \eqref{lastline_facto3p}.

From the analytic structure of the function ${\cal H}^c_{[2^n][2^n]22}$, in particular from 
the fact that only a particular orientation of the ladder ${\cal P}_n$ appears, 
we know that out of all allowed logarithmic discontinuities, $\frac{1}{a^n}\log^{1}(U),\ldots \frac{1}{a^n}\log^{n}(U)$, 
the gravity result only gives a non vanishing single logarithmic discontinuity, thus all the higher logarithmic discontinuities must 
come from disconnected contributions. For $n=2,3$, this agrees with the factorization of three-point functions given above. 
But this factorization pattern necessarily has to generalise to higher values of $n$ in order to be consistent with the gravity result. 
We find this pattern quite remarkable, since it is not entirely obvious how to prove it in the CFT. 

Finally, let us note that for given value of $n$, 
we only make a statement about three-point functions with long operators whose particle number $\leq n$. 
For ${\cal O}_2^2{\cal O}_2^2$  this means long double 
particles (i.e.~no statement about triple particles), and  for 
${\cal O}_2^3{\cal O}_2^3$ this means long double and triple particles. Considering for example  ${\cal O}_2^4{\cal O}_2^4$ 
we would then stop at the quadrupole particle operators. In each case the OPE contains higher multi-particle operators 
but for these it seems that there is no meaningful separation into a disconnected contribution (as it is the case for 
the ${\cal O}_2^2{\cal O}_2^2$ OPE  and the triple particle operators, which appear 
at the same order as for the ${\cal O}_2{\cal O}_2$ OPE).

\subsection{A first look at long triple-trace operators}

The mixed correlator with two double and two single particle operators 
encode average CFT data for long triple trace operators, which is new, and available from 
from analysing 
the block expansion of the correlator in the long sector. 
Here we would like to highlight two important results that follow from that analysis.

(1) In the $[0,0,0]$ rep long operators are multi-particle operators with twist greater equal than four. The double particle operators are known: there is a single twist four double particle operators, and degenerate 
operators starting from twist six (as we recalled in section \ref{OPEsect12}) \cite{Aprile:2017xsp}. Then, starting from twist six we are interested in the triple trace operators.
Say ${\cal T}$ is a triple trace operators, it is expected that its leading three-point couplings are
\begin{equation}\label{discussiontriple0001}
C^{}_{22;{\cal T}}\sim \frac{1}{N}\qquad;\qquad  C^{}_{[2^2][2^2];{\cal T}}\sim \frac{1}{N}\,.
\end{equation}
From the study of ${\cal H}_{2^22^222}$ we can indeed show that 
\begin{equation}\label{discussiontriple0002}
\rule{1.1cm}{0pt}C^{}_{22;{\cal T}}\neq 0\rule{.5cm}{0pt}\qquad;\qquad C^{}_{[2^2][2^2]; {\cal T}} -2 C^{}_{22 ;{\cal T}}\neq 0 \,.
\end{equation}
We prove this by contradiction: if \eqref{discussiontriple0001} is false 
then the double particle operators are the only operators exchanged in the
$\log(U)$ discontinuity of ${\cal H}_{2^22^222}$ at order ${\frac{1}{N^4}}$. Since their CFT data is known from \cite{Aprile:2017xsp} 
it is possible to predict this $\log(U)$ discontinuity. Note that at this order ${\cal H}^c_{2^22^222}$ contributes, and 
special case of the twist four predictions was discussed already in section \ref{OPEsect12}. However, 
the higher twist predictions are not consistent with the correlator. This proves the claim in \eqref{discussiontriple0002}.  
See appendix \ref{apptriplepsec1} for more details. 

(2) In the $[0,2,0]$ rep long operators are multi-particle operators with twist greater equal than six. 
The double particle operators are known \cite{Aprile:2018efk}, and we are interested in triple particle operators.
In the correlator $\langle {\cal O}_2 {\cal O}_2^2 {\cal O}_2 {\cal O}_2^2\rangle$, we expect
\begin{equation}
C_{2[2^2], {\cal T}}\sim 1
\end{equation}
therefore we expect triple particle operators in the disconnected free theory and in the leading logarithmic discontinuities. It will be interesting to compare with the case of 
the single particle correlator $\langle {\cal O}_2 {\cal O}_4 {\cal O}_2 {\cal O}_4\rangle$ , where we know that double particle operators are exchanged at leading order, since
\begin{equation}
C_{24, {\cal D}}\sim 1\,.
\end{equation}
Since we cannot resolve the degeneracy of triple particle operators directly, i.e~just from the knowledge of a single correlator, 
it is instructive to compare the behavior at large spin between 
$\langle {\cal O}_2 {\cal O}_4 {\cal O}_2 {\cal O}_4\rangle$ and $\langle {\cal O}_2 {\cal O}_2^2 {\cal O}_2 {\cal O}_2^2\rangle$.  This comparison can be read from Table \ref{table1}.

\begin{figure*}[!htbp]
\begin{center}
\begin{tabular}{|c|c|c|}
\hline
\rule{0pt}{.5cm} {\rm fixed twist}& $\langle {\cal O}_2 {\cal O}_4 {\cal O}_2 {\cal O}_4\rangle$  &  $\langle {\cal O}_2 {\cal O}_2^2 {\cal O}_2 {\cal O}_2^2\rangle$    \\[0.5ex]
\hline\hline
\rule{0pt}{.8cm}$\frac{\log^0(U)}{N^0}$\,$\sum C^2$  & $O(l^6) $ &    $O(l^6) $  \\[2ex] 
\hline
\rule{0pt}{.8cm} $\frac{\log^1(U)}{N^2}$\,$\sum C\eta C$   &  $O(l^4)$   &   $O(l^4)$  \\[2ex] 
 \hline
 \rule{0pt}{.8cm} $\frac{\log^2(U)}{N^4}$\,$\sum C\eta^2 C$  &  $O(l^2)$   & $O(l^2)$+$O(l^2)\log(l)$   \\[2ex] 
 \hline
\end{tabular} 
\end{center}
\renewcommand{\figurename}{Table}
\caption{Large spin behaviour of max log CFT data, normalized by ${(l+\frac{\tau}{2})!^2}/{(2l+2+\tau)!}$. The first row 
corresponds to disconnected contributions, then, in the order, tree level and one-loop data. Complete results are given in appendix \ref{apptriplep2}.
In the last row last column we splitted the result into that of ${\cal H}^t_{2 2^2 2 2^2}$, which goes like $O(l^2)$, and that of 
${\cal H}_{2222}$, which interpreted as a contribution to ${\cal H}_{2 2^2 2 2^2}$, contains also a
harmonic sum $(l+2)(l+7) {\rm HarmonicNumber}(l+4)$, whose asymptotics is a $O(l^2)\log(l)$.
\label{table1}}
\end{figure*}

The the second column of Table \ref{table1} refers to the single particle correlators $\langle {\cal O}_2 {\cal O}_4 {\cal O}_2 {\cal O}_4\rangle$ 
and the result there are well understood. 
At fixed twist the number of degenerate double particle operators does 
not depend on spin, and their anomalous dimension behaves as $O(1/l^2)$ at large spin, 
see explicit results in \cite{Aprile:2018efk}. Therefore, at given twist, the maximal log 
contribution is computed from CFT data of exchanged double particle operators and this 
sums up into a rational function of 
spin that multiplies the known factorials, ${(l+\frac{\tau}{2})!^2}/{(2l+2+\tau)!}$. The same is true for any other single particle correlator.

The third column of Table \ref{table1} refers to $\langle {\cal O}_2 {\cal O}_2^2 {\cal O}_2 {\cal O}_2^2\rangle$ 
and in particular to the dynamical function ${\cal H}_{22^222^2}$. 
The first two orders in the $1/N$ expansion (thus the first two rows of the Table)
are determined by disconnected contributions whose functional form 
is that of the single particle operators. This is also the result given in \cite{Bissi:2024tqf}.  
From our new results it is possible to go beyond that and study the total one loop contribution, 
which is shown in the last row. There we find an expression that contains (among other terms) a harmonic sum. 
The latter gives an additional $\log(l)$ term at large spin, and thus a different large spin behaviour compared to the single particle case. 
We interpret this as a neat signal of exchanged triple particle operators, since the latter are degenerate also in spin. Interestingly, 
this $\log(l)$ actually comes from the one-loop function ${\cal H}_{2222}$ interpreted as a contribution to ${\cal H}_{2[2^2]2[2^2]}$.\footnote{The 
same ${\cal H}_{2222}$, interpreted as one-loop contribution to 
 $\langle {\cal O}_2 {\cal O}_2 {\cal O}_2 {\cal O}_2\rangle$ behaves as in the second column of Table \ref{table1}.}
It would be nice 
to understand the emergence of the extra $\log(l)$ behaviour from
semiclassics, eg. along the lines of \cite{Kravchuk:2024wmv}.

\section{Discussion and outlook}\label{sec:discussion}

In this paper we studied a particular class of multi-particle operators in ${\cal N}=4$ SYM and the D1D5 SCFT. The key idea is to start from the lightest operator $O$ in the theory and define a heavy state $O_H$ as a coherent superposition of multi-particle operators $O^n$, see Eq.~\eqref{introOperator5} and~\eqref{introOperator3}. When the parameter $\alpha$ defining the state is ${\mathcal O}(1)$, the coherent sum is picked at values $n\sim \alpha^2 c$ and so describes a state with conformal dimension $\Delta \sim c$. On the bulk side, the heavy operators $O_H$ deform the geometry and, since they are semiclassical coherent states, they are described by regular, asymptotically AdS solutions. As a first step we studied the supergravity perturbations around these geometries: they compute the 2-point correlators of a light state in these non-trivial background and thus the holographic 4-point correlators $\langle O_H \bar{O}_H O O \rangle$ at strong coupling in the CFT language. As a first result of this paper we showed that these HHLL correlators are described by a Heun equation at all orders in $\alpha$.

A second key observation was that one can formally extend the results of the HHLL correlators to the regime where $\alpha \sim {\mathcal O}(c^{-1/2})$. While this is outside the regime of supergravity, the results satisfy all the CFT constraints necessary for being reinterpreted as {\em light} 4-point correlators. In particular, we argued that this limit captures the connected tree-level supergravity contribution to the LLLL correlators. We checked that, at the leading order, ${\mathcal O}(\alpha^0)$, we reproduce the 2-point function in the vacuum and at the next order, ${\mathcal O}(\alpha^2)$, we reproduce the known light 4-point correlators with single particle states. By using this idea at subleading order in $\alpha$, we obtained explicit expressions both in position and Mellin space for the leading connected part of the correlators $\langle {\mathcal O}^n \bar{\mathcal O}^n \mathcal{O} \mathcal{O} \rangle$ for $n=2$ and $n=3$. The space of functions appearing in this result is particularly simple: the ladder integrals up to $n$-loop for the configuration space case and polygamma up to order $n-1$ in Mellin space.

We then focused on the AdS$_5$ case. For $n=2$ we used known results for the 1-loop contribution to the 4-point single-particle correlator to reconstruct the full double-particle correlator. A similar step is possible also in the $n = 3$ case, but requires the knowledge of both the full two-loop four-point correlator among single-particle operators~\cite{Huang:2021xws,Drummond:2022dxw} and the connected double particle correlator beyond tree level. Finally we studied in detail various OPE limits, providing non-trivial consistency checks on our results and their interpretation as light multi-particle four-point correlators. In particular, we showed consistency of the correlator with respect to the protected sector, the absence of stringy states through multiplet recombination, and the long double particle spectrum. The CFT analysis of our correlators also led to new data for the long triple particle spectrum and novel results for the large $N$ factorization of three-point couplings.

The approach discussed in this paper is amenable to various generalizations and applications. We highlight below those the we see as most interesting and/or doable.

\paragraph{New correlators from new geometries.} Of course a natural avenue is to consider other coherent states describing more complicated multi-particle states, derive the corresponding supergravity solutions and use them to calculate correlators with new multi-particle operators. The programme of finding less supersymmetric geometries describing condensates of supergravity modes is well developed in the AdS$_3$ context, see~\cite{Shigemori:2020yuo} for a review, and it may be possible to follow similar steps also in AdS$_5$. With a richer dataset of holographic correlators one could attempt to unmix the contributions from different superprimaries and extract precise information about the triple-particle superprimaries, as was done in~\cite{Aprile:2017xsp} for the double-particle sector.

\paragraph{New correlators from the ``AdS bubble''.} It is possible to extract new multi-particle correlators even using as a starting point the same geometry considered in this work. Here we focused on the simplest scalar perturbation that is trivial in the S$^5$ part. By considering perturbations which are proportional to $Y_{p-2}$, a non-trivial S$^5$ scalar harmonics for $p>2$, one can derive correlators with single-particle CPOs of the type $\mathrm{Tr}(X^p)$. Again such results would provide valuable information in the analysis of strongly coupled CFT data in $\mathcal{N}=4$ SYM. As discussed below, they will also be useful in the comparison between higher point correlators and our results with multi-particle operators.

\paragraph{Integrated correlators and extensions beyond supergravity.} A way to drastically simplify the kinematics of correlators is to perform an integral over the spacetime dependence and extract a function of the couplings and the quantum number of the external states~\cite{Binder:2019jwn}. These integrated correlators have been studied in various regimes, including the one relevant for this paper~\cite{Paul:2023rka,Brown:2023why,Brown:2024yvt}. Thus it should be possible to integrate our results for the HHLL correlation functions and compare with the results that have been obtained from localization. The latter depend both on $\alpha$, $N$ and the 't Hooft coupling and therefore interpolate between the free and the classical supergravity regimes. While getting a similar result for an unintegrated correlator is currently out of reach, it may be possible to include in our approach the first higher derivative corrections both to the background and to the equations of the perturbation and to derive their effect on the multi-particle correlators.

\paragraph{Different scalings of $\alpha$ and $c$.} In this paper we focused on two regimes, the one with $\alpha \sim {\mathcal O}(1)$, corresponding to states that deform the bulk geometry, and the one with with $\alpha \sim {\mathcal O}(c^{-1/2})$, corresponding to light states. We find that our results smoothly interpolates between the two regimes. Thus it would be interesting to study the {\em same} results in other scaling limits. For instance, when $\alpha \sim {\mathcal O}(c^{-1/4})$, it may be possible to rewrite our results as a linear combination of correlators involving giant gravitons. In this regime, the expansion parameter is $\alpha^2 \sim c^{-1/2}$, which is what is expected for diagrams in the bulk in presence of D-branes.
Note that the four-point correlator of two external giant gravitons and two single particle operators has recently been computed in \cite{Chen:2025yxg}, and the integrated correlator was discussed earlier in  \cite{Brown:2024tru}.
 Of course the challenge from our geometric approach is to disentangle from the full result the contribution coming from a single giant graviton. It is likely that this can be more easily done by starting from a different basis of supergravity solutions. It is also possible to study whether the limit $\alpha \gg 1$, which captures multi-particle states with $n\gg c$, smoothly connects with the large charge limit. This large charge regime has been under intense investigation in several SCFT, see e.g. \cite{Bourget:2018obm}, and also in ${\cal N}=4$ SYM, see e.g.~\cite{Caetano:2023zwe}: HHLL correlators are being studied in this context~\cite{Brown:2024yvt,Wentoappear} and it would be interesting to understand if they have any relation with the $\alpha \gg 1$ limit of our results.

\paragraph{Relation to higher point correlators.} Of course, another way to derive the class of correlators discussed in this paper is to start from a higher point function with all single particle states and then take the OPE limits needed to construct the multi-particle operators. Strongly coupled correlators with more than four states are under intense investigation in several holographic theories and, for ${\cal N}=4$ SYM, some explicit correlation functions have been derived with both 5 and 6 single-particle operators~\cite{Goncalves:2019znr,Goncalves:2023oyx,Goncalves:2025jcg}. It is possible to specialise these results to the case where one or two pairs of states are mutually $1/2$-BPS. Then, by taking the appropriate OPEs, one should obtain the correlators discussed in the paper. In configuration space, these OPEs correspond to smooth limits which simply reduce the complicated kinematics of a higher point function to that of a 4-point correlator. We checked that the result of~\cite{Goncalves:2019znr} is consistent with the absence of odd terms in $\alpha$ in~\eqref{eq:psiexpads5} and work is in progress to extend this analysis to the more general cases of~\cite{Goncalves:2023oyx,Goncalves:2025jcg}.

\section*{Acknowledgements}

It is a pleasure to acknowledge fruitful scientific exchange with A.~Bissi, S.~Chester, J.~Drummond, R.~Emparan, G.~Fardelli, F.~Galvagno, A.~Georgoudis, C.~Heissenberg, P.~Heslop, C.~Iossa, J.~F.~Morales, H.~Paul, M.~Santagata, D.~Turton, A.~Tyukov, J.~Vilas Boas, C.~Wen.
F.~Aprile is supported by the Ramon y Cajal program through the fellowship RYC2021-031627-I funded by MCIN/AEI/10.13039/501100011033 and by the European Union NextGeneration EU/PRTR and by the
MSCA programme - HeI - \href{https://cordis.europa.eu/project/id/101182937}{DOI 10.3030/101182937}. 
R.~Russo. is partially supported by the UK EPSRC grant ``CFT and Gravity: Heavy States and Black Holes'' EP/W019663/1 and the STFC grants ``Amplitudes, Strings and Duality'', grant numbers ST/T000686/1 and ST/X00063X/1. 
No new data were generated or analysed during this study.

\appendix

\section{The gauge theory approach to the Heun equation}
\label{app:heun}

The Heun equation is a second order ODE with four regular singular points; particular cases include confluent or reduced equations where some of the singularities collide or some of the parameters are set to zero. As seen in Section~\eqref{ssec:ads3HHLL}, the fluctuations around the geometry~\eqref{eq:sploc} are captured by~\eqref{eq:phislz} which is of the general Heun form,
\begin{equation}
  \label{eq:gen_heun}
\left(\partial_{\hat{z}}^2+\frac{\frac{1}{4}-a_1^2}{(\hat{z}-1)^2}-\frac{\frac{1}{2}-a_0^2-a_1^2-a_t^2+a_{\infty}^2+u}{\hat{z}(\hat{z}-1)} +\frac{\frac{1}{4}-a_t^2}{(\hat{z}-t)^2}+\frac{u}{\hat{z}(\hat{z}-t)}+\frac{\frac{1}{4}-a_0^2}{\hat{z}^2}\right) \psi(\hat{z})=0 \,.
\end{equation}
By comparing the equation above with~\eqref{eq:phislz} one obtains the following dictionary
\begin{equation}
  \label{eq:apara}
  a_0^2 = \frac{l^2}{4}\;,~~ a_1^2 = a_t^2 =  \frac{1}{4}\;,~~ u= -\frac{\left(\alpha^2+8\right) (\omega^2-l^2)}{4 \left(8-\alpha^2\right)}\;,~~ a_\infty^2 = \frac{l^2}{4}\;,~~ t = \frac{\alpha^4}{64}\;.
\end{equation}
As usual we need to start from the regular solution in the centre of the space ($\hat{z}=0$) and expand it at the AdS boundary ($\hat{z}=t$ in the case of~\eqref{eq:phislz}). One of the key ingredients to solve this problem is the so-called Nekrasov-Shatashvili function $F$,
\begin{equation}
  \label{eq:Nek-Shat}
  F = c_1\left(a, a_0, a_t, a_1, a_{\infty}\right) t + {\mathcal O}(t^2)\,
\end{equation}
where
\begin{equation}\label{eq:1inst}
c_1\left(a, a_0, a_t, a_1, a_{\infty}\right)=\frac{\left(4 a^2-4 a_0^2+4 a_t^2-1\right)\left(4 a^2+4 a_1^2-4 a_{\infty}^2-1\right)}{8-32 a^2}\;,
\end{equation}
see for instance~(C.1.9) of~\cite{Bonelli:2022ten}. In the gauge-theory language the relation between $a$ and $u$ is obtained by inverting the so-called Matone relation~\cite{Matone:1995rx}
\begin{equation}
  \label{eq:ua}
  u = -a^2 + a_t^2 -\frac{1}{4} + a_0^2 + t \partial_t F\;,
\end{equation}
which can be written in terms of $\omega$ and $l$ by using~\eqref{eq:apara}
\begin{equation}
  \label{eq:aSW}
  \begin{aligned}
    a^2= \frac{\omega^2}{4}+\frac{\omega^2-l^2}{16} \left(\alpha^2  + \frac{\alpha^4 \left(l^2+3 \omega^2-4\right)}{32 \left(\omega^2-1\right)} +\frac{\alpha^6 \left(l^4-2 l^2+\omega^4-2 \omega^2+2\right)}{128 \left(\omega^2-1\right)^2}\right)+{\cal O}(\alpha^8)\;.
  \end{aligned}
\end{equation}

The CFT correlator is obtained by taking the ratio of the leading normalizable decay at the AdS boundary of wavefunction and the leading non-normalizable decay. For our purposes we can use Eq.~(4.1.16) of~\cite{Bonelli:2022ten} to write the regular solution at the centre of space\footnote{We indicate with a hat the parameters appearing in~(4.1.1) of~\cite{Bonelli:2022ten} so as to avoid confusion with the notation in the main part of this paper. They are related to the parameters used in~\eqref{eq:gen_heun} by Eq.~(4.1.4) of~\cite{Bonelli:2022ten}.}
\begin{equation}
  \label{eq:w0-}
  w^{(0)}_- = {\rm HeunG}(t,\hat{q},\hat{\alpha},\hat{\beta},\hat{\gamma},\hat{\delta},\hat{z})
\end{equation}
in terms of the local solutions around $t$, $w^{(t)}_+$ and $w^{(t)}_-$, which capture the normalizable and non-normalizable decay respectively. By taking the ratio between the two we obtain the following asymptotic behaviour close to the AdS boundary
\begin{equation}
  \label{eq:phiSL_dec}
  \psi(\hat{z}) \approx \left[1+ \frac{\Gamma\left(\hat{\epsilon}-1\right) \Gamma\left(\frac{1+\hat{\gamma}-\hat\epsilon}{2} -a \right) \Gamma\left(\frac{1+\hat\gamma-\hat\epsilon}{2} + a \right)}{\Gamma\left(1-\hat\epsilon\right) \Gamma\left(\frac{-1+\hat\gamma+\hat\epsilon}{2} -a \right) \Gamma\left(\frac{-1+\hat\gamma+\hat\epsilon}{2} + a \right)} e^{-\partial_{a_t} F} t^{\hat\epsilon-1}(t-\hat{z})^{1-\hat\epsilon} +\ldots \right]\;.
\end{equation}
The parameter $\hat\epsilon$ in our case vanishes since
\begin{equation}
  \label{eq:hateps}
  \hat\epsilon=1-2 a_t\;,\quad \mbox{and}\quad \hat\gamma =1-2a_0=1+|l|.
\end{equation}
where we chose $a_t=\frac{1}{2}$ and $a_0=-\frac{|l|}{2}$.
Then strictly speaking~\eqref{eq:phiSL_dec} above does not apply since the first Gamma function at the numerator becomes singular. We follow the approach of~\cite{Dodelson:2022yvn,Giusto:2023awo} and keep $\hat\epsilon$ generic in the intermediate steps: we will be able to set it to zero after writing the Green function in terms of physical observables. The position of the poles in $\omega$ is determined by the $a$-dependent Gamma functions at the numerator and are located at
\begin{equation}
  \label{eq:polpos}
  a=\pm\left(k+\frac{|l|-\hat\epsilon}{2}\right)\;,\quad\mbox{with}\quad k=1,2,\ldots\;.
\end{equation}
Let us focus on the poles on the negative real axis (lower sign) as done in~\eqref{eq:Psivs}. These poles follow from the last Gamma in the numerator which can be also written as $\Gamma(\hat{b}+1)$, in terms of the parameters introduced in~\eqref{eq:hatahatb}. When calculating the residues of these poles, the last Gamma of the denominator cancels the singularity of the first one in the numerator and the $\hat\epsilon\to 0$ becomes smooth. In summary the behaviour of the second term of~\eqref{eq:phiSL_dec} around each negative pole is
\begin{equation}
  \label{eq:epsto0}
  -k(k+|l|) \frac{e^{-\partial_{a_t} F}}{\hat{b}+k} \left(1-\frac{\hat{z}}{t}\right) \sim \left(\frac{l^2}{4} -a^2\right) \frac{e^{-\partial_{a_t} F} }{\hat{b}+k} \frac{8-\alpha^2}{8+\alpha^2} w^2 =\frac{1}{\hat{b}+k} \left(\hat{a}\,\hat{b}\, e^{-\partial_{a_t} F} \frac{8-\alpha^2}{8+\alpha^2}\right) w^2\,.
\end{equation}
For the poles on the positive real axis one obtains the same structure as in~\eqref{eq:epsto0} but with $\hat{a}$ and $\hat{b}$ exchanged. In the first step we used the relation between $k$ and $a$ valid on the pole to isolate a factor of $\hat{a} \,\hat{b}$ in order to reproduce the same pattern seen in the hypergeometric case~\eqref{eq:2f1d}, which makes it easier to integrate the Ward Identity~\eqref{eq:PsiB}. In the last step we put together a group of terms which simplify when expressed explicitly in terms of $\omega$ and $l$:
\begin{equation}
  \label{eq:simpmir}
  \hat{a}\,\hat{b}\, e^{-\partial_{a_t} F} \frac{8-\alpha^2}{8+\alpha^2} = \frac{l^2-\omega^2}{4}\;. 
\end{equation}
We checked this relation up to three instanton, i.e. order $\alpha^{14}$ in the case at hand. It would be interesting to prove it exactly by using the properties of the Nekrasov-Shatashvili function appearing in~\eqref{eq:simpmir}.

For completeness, let us write $\hat{b}$ in terms of $\omega$
\begin{equation}
  \label{eq:asbs}
  \begin{aligned}
    \hat{b} & = \frac{l+\omega }{2} + \frac{\alpha^2 \left(\omega^2-l^2\right)}{16 \omega }+\frac{\alpha^4 \left(-3 l^4 \omega^2+2 l^4+2 \left(l^2-1\right) \omega^4+\omega^6\right)}{512 \omega^3 \left(\omega^2-1\right)}\\
    & \quad + \frac{\alpha^6 \left((\omega^2-l^2) \left(l^4 \left(5 \omega^4-5 \omega^2+2\right)-2 l^2 \left(\omega^6+\omega^4\right)+\left(\omega^4-\omega^2+2\right) \omega^4\right)\right)}{4096 \left(\omega^5 \left(\omega^2-1\right)^2\right)} +{\cal O}(\alpha^8) \;,
  \end{aligned}
\end{equation}
which follows from~\eqref{eq:hatahatb} and~\eqref{eq:aSW} (then by using the first of~\eqref{eq:hatahatb} one can obtain also $\hat{a}$). In order to perform the $\omega$ integral in~\eqref{eq:Psisl} via Cauchy's theorem, we have to calculate $\partial_\omega \hat{b}$ to extract the contribution to the residue from the harmonic sum. Then we need to evaluate both this result and the factor of $e^{i\omega\tau}$ on the poles by writing $\omega$ in terms of $k$ and $l$, which can be done by equating~\eqref{eq:aSW} and~\eqref{eq:polpos}. All these steps are straightforward in perturbation theory and at this stage we can set $\hat\epsilon=0$ as the singularities have cancelled as described before~\eqref{eq:epsto0}.

It turns out that the same Heun equation is relevant also in the AdS$_5$ geometry~\eqref{eq:5Dmetric}, for the wave equation in~\eqref{eq:ODE1}. 
In terms of the variable \eqref{eq:ratH1}-\eqref{eq:x2zads5} the Heun equation is
\begin{equation}
  \label{eq:heunads5}
  \begin{aligned}
  \psi''(\hat z) + & \left[\frac{(l+2) (-l)}{4 \hat{z}^2}+\frac{l (l+2)-2 \omega^2}{4 (\hat{z}-1)^2} +\frac{\left(3-2 \alpha^2\right) \omega^2+\left(2 \alpha^2-1\right) (l (l+2)-4)-8}{4 (\hat{z}-1)}\right. \\ & ~-\frac{3}{4 \left(\frac{2 \alpha^2}{1-2 \alpha^2}+\hat{z}\right)^2} -\frac{\left(2 \alpha^2+1\right) \omega^2-8 \alpha^2-\left(1-2 \alpha^2\right) l (l+2)-4}{8 \alpha^2 \hat{z}}\\ &~ \left. +\frac{\left(1-2 \alpha^2\right) \left(-2 \alpha^2 \omega^2-8 \alpha^2+\left(2 \alpha^2-1\right) (l+2) l+\omega^2-4\right)}{8 \alpha^2 \left(\frac{2 \alpha^2}{1-2 \alpha^2}+\hat{z}\right)}\right]\psi(\hat{z})=0
  \end{aligned}
\end{equation}
Notice that at infinity the singularity of the order $1/\hat{z}$ cancels, so this equation is of the form of~\eqref{eq:gen_heun} with the following relation
\begin{equation}
  \label{eq:dictads5}
  \begin{gathered}
    a_0^2= \frac{1}{4} (l+1)^2\;,\quad a_1^2 = \frac{1}{4} \left(-l (l+2)+2 \omega^2+1\right)\;,\quad a_t^2 = 1\;,\quad a_\infty^2= 1\;,\\ u= \frac{1}{4} \left(2 \alpha^2 \omega^2+8 \alpha^2-\left(2 \alpha^2-1\right) l (l+2)-\omega^2+4\right)\;,\quad t= -\frac{2 \alpha^2}{1-2 \alpha^2}\;.
  \end{gathered}    
\end{equation}
The calculation follows the same steps as for the AdS$_3$ case~\eqref{eq:phislz}, where the key ingredient is~\eqref{eq:phiSL_dec}. Let us just summarize briefly the few steps where there are some novelties. First, of course, we now have to use the dictionary~\eqref{eq:dictads5} and, choosing $a_t=1$ and $a_0=-\frac{|l|+1}{2}$, we have
\begin{equation}
  \label{eq:hatepsads5}
  \hat\epsilon=-1\;, \qquad\hat\gamma=2+|l|\;
\end{equation}
and a new relation between the radial coordinate $w$~\eqref{eq:wads5} and $\hat{z}$, which reads
\begin{equation}
  \label{eq:w2hatz}
  1-\frac{\hat{z}}{t} = \frac{w^2}{1-2\alpha^2} \left(1+O(w^2)\right)\;.
\end{equation}
Also the delta function for the non-normalizable term takes a more complicated form in AdS$_5$, see~\eqref{eq:deltaN}. Thus we need to multiply~\eqref{eq:phiSL_dec} by $\frac{Y_l(\theta)}{\sqrt{2}\pi}(l+1)$. With these ingredients one obtains~\eqref{eq:PhiBads5} with 
\begin{equation}
  \label{eq:bhatads5}
  \hat{b} = \frac{l+\omega}{2} +\frac{\alpha^2 \left((l+2)^2-\omega^2\right) \left(l^2-\omega^2\right)}{4 \omega  \left(\omega^2-1\right)}+ \ldots \;.
\end{equation}
The subleading term, up to order $\alpha^6$, are included in the ancillary file of the {\tt arXiv} version of our paper.

In all cases, the holographic correlator expanded in $\alpha$ reduces to a sum of terms of the form \eqref{eq:polysums}. The double series over $k$ and $l$ can be resummed in terms of polylogs, and elementary functions. We list below the series that appear in the computation up to $O(\alpha^6)$ and we include the resummed expressions in the ancillary file of the {\tt arXiv} version of our paper:
\begin{equation}
I_0[p,q]\equiv  \sum_{\myell\in \mathbb{Z}}^\infty  \sum_{k=1}^\infty \frac{\myell^{2p}}{(|\myell| + 2k)^q} \,e^{-i ( |\myell| + 2k) \tau+ i \myell \sigma} \,;
\end{equation}

\begin{equation}
I_1[p,q]\equiv  \sum_{\myell\in \mathbb{Z}}^\infty  \sum_{k=1}^\infty \frac{\myell^{2p}}{(|\myell| + 2k-1)^q} \,e^{-i ( |\myell| + 2k) \tau+ i \myell \sigma}\,;
\end{equation}

\begin{equation}
I_2[p,q]\equiv  \sum_{\myell\in \mathbb{Z}}^\infty  \sum_{k=1}^\infty \frac{\myell^{2p}}{(|\myell| + 2k-2)^q} \,e^{-i ( |\myell| + 2k) \tau+ i \myell \sigma} \quad (2p-q\ge 0)\,;
\end{equation}

\begin{equation}
I_3[p,q]\equiv  \sum_{\myell\in \mathbb{Z}}^\infty  \sum_{k=1}^\infty \frac{(\myell^2-1)^2 \, \myell^{2p}}{(|\myell| + 2k-3)^q} \,e^{-i ( |\myell| + 2k) \tau+ i \myell \sigma} \quad (q\le 2)\,;
\end{equation}

\begin{equation}
I_4[p,q]\equiv  \sum_{\myell\in \mathbb{Z}}^\infty  \sum_{k=1}^\infty \frac{(\myell^2-4)^2\, \myell^{2p}}{(|\myell| + 2k-4)^q} \,e^{-i ( |\myell| + 2k) \tau+ i \myell \sigma} \quad (q\le 2)\,;
\end{equation}

\begin{equation}
{\tilde I}_1[p,q]\equiv  \sum_{\myell\in \mathbb{Z}}^\infty  \sum_{k=1}^\infty \frac{\myell^{2p}}{(|\myell| + 2k+1)^q} \,e^{-i ( |\myell| + 2k) \tau+ i \myell \sigma}\,;
\end{equation}

\begin{equation}
{\tilde I}_2[p,q]\equiv  \sum_{\myell\in \mathbb{Z}}^\infty  \sum_{k=1}^\infty \frac{\myell^{2p}}{(|\myell| + 2k+2)^q} \,e^{-i ( |\myell| + 2k) \tau+ i \myell \sigma}\,.
\end{equation}

\section{The BPS AdS$_5$ bubble}
\label{app:bpsbubble}

We describe here the 10D geometry dual to the $\mathcal{N}=4$ SYM state $O_H$ in \eqref{eq:OHAdS5} and provide the relation between the various forms of this solution found in the literature.

The goal of finding the geometry dual to a coherent state $O_H$ made by a superposition of the multi-particle operators $O_2^n$ was pursued in \cite{Giusto:2024trt}. Given the 1/2-BPS nature of the state $O_H$, its dual description must fit in the class of the LLM geometries, which are parametrised by a droplet configuration in a 2D plane. Singling out the droplet configuration leading precisely to a state of the form \eqref{eq:OHAdS5} -- without any mixing with operators other than $O_2$ at all orders in $\alpha$ -- was, however, non-obvious. The alternative strategy used in \cite{Giusto:2024trt} was to look for a solution within the consistent truncation of \cite{Cvetic:2000nc}, which projects out all the scalar degrees of freedom other than those in the $[0,2,0]$ $SU(4)$ representation containing the chiral primary $O_2$. It was shown that, within this truncation, there is a unique regular and normalisable solution that reduces to the linearised solution dual to $O_2$ \cite{Kim:1985ez} at first order in $\alpha$, and the holographic duality between this geometry and $O_H$ passes, in fact, some non-trivial check. Only later, it has been realised that the same solution had already been found in \cite{Liu:2007xj} (see the ``BPS AdS bubble", sec. 2.2). Since the two geometries are related by a non-trivial change of coordinates, we show here their equivalence.

The solution of \cite{Liu:2007xj} is described by the asymptotically AdS$_5$ metric, $ds^2_5$, given in \eqref{eq:5Dmetric}, together with a gauge field, $A^1$, and two scalars, $X_1$, $\varphi_1$:
\begin{equation}
A^{1} = - H_1^{-1} dt \quad,\quad X_1 = H_1^{-2/3} \quad,\quad \cosh \varphi_1 = \frac{d}{d r^2}(r^2 \,H_1)= \frac{1+\frac{1+q_1}{r^2}}{\sqrt{1+\frac{2(1+q_1)}{r^2}+\frac{1}{r^4}}} \,.
\end{equation}
Note that we have set $g=1$ with respect to the conventions of  \cite{Liu:2007xj}. All these fields become trivial when $q_1=0$ and the solution reduces to the AdS vacuum. 

The ten-dimensional uplift of this solution is \cite{Giusto:2024trt}
\begin{equation}\label{eq:10Dmetric}
ds^2_{10} = \Delta^{1/2}\, ds^2_5 + \Delta^{-1/2} \,G_{\alpha\beta}\, (d y^\alpha + A^\alpha)(d y^\beta+A^\beta)\,,
\end{equation}
where $G_{\alpha\beta} dy^\alpha dy^\beta$ is the following asymptotically S$_5$ metric (parametrised by $\tilde\theta$, $\phi$ and the S$^3$, $\tilde\Omega_3$)
\begin{equation}
\begin{aligned}
G_{\alpha\beta} dy^\alpha dy^\beta &= \left(e^{2\mu}(e^\lambda\sin^2\phi+e^{-\lambda}\cos^2\phi)\sin^2\tilde\theta+e^{-2\mu}\cos^2\tilde\theta\right)d\tilde\theta^2 \\
&+e^{2\mu}(e^\lambda \cos^2\phi+e^{-\lambda}\sin^2\phi)\cos^2\,\tilde\theta d\phi^2-e^{2\mu}\sinh\lambda\sin(2\phi) \sin (2\tilde\theta) \,d\tilde\theta d\phi+e^{-\mu}\sin^2\tilde\theta \,d\tilde\Omega_3^2 \,,
\end{aligned}
\end{equation}
 the only non-trivial component of the S$^5$ KK gauge fields, $A^\alpha$, is
 \begin{equation}
A^\phi \equiv A^{1}= -e^{-3\mu} \,dt\,,
\end{equation}
and the wrap factor $\Delta$ is given by
\begin{equation}
\Delta = \left(\frac{\mathrm{det} \,G}{\mathrm{det} \,G_0}\right)^{-2/3}=e^{-2\mu} (e^\lambda\cos^2\phi+e^{-\lambda}\sin^2\phi)\cos^2\tilde\theta+e^\mu\sin^2\tilde\theta\,,
\end{equation}
with $G_0$ the unit round S$^5$ metric. The two scalars $\lambda$ and $\mu$ are functions of only the radial coordinate $r$ and are related to the functions $H_1$ and $\varphi_1$ of \cite{Liu:2007xj} by
\begin{equation}
\lambda=\varphi_1\quad,\quad e^{3\mu}=H_1\,.
\end{equation}
Apart from the 10D metric, the solution also has a non-trivial RR 5-form, which is not relevant for our purposes.

On the other hand, the solution of \cite{Giusto:2024trt} uses a different parametrisation for the the asymptotically AdS$_5$ metric:
\begin{equation}\label{eq:5Dmetricxi}
ds^2_5 = \Omega_0^2\left[\frac{d\xi^2}{(1-\xi^2)^2}+\frac{\xi^2}{1-\xi^2} \,d\Omega_3^2\right]-\frac{\Omega_1^2}{1-\xi^2} \,dt^2 \,,
\end{equation}
where the radial variable $r$ is replaced by $\xi\in[0,1]$; the parameter $q_1$ maps to the parameter $\epsilon$ used in \cite{Giusto:2024trt} by the relation \eqref{eq:q12epsilon} already quoted in the text.  The whole solution can be expressed algebraically -- through Eqs. (5.15) in \cite{Giusto:2024trt}) --  in terms of the function $\lambda(\xi)$, defined by Eq. (5.14) in \cite{Giusto:2024trt}. The metrics in \eqref{eq:5Dmetricxi} and \eqref{eq:5Dmetric} are related by
\begin{equation}\label{eq:matchbubble}
\frac{\Omega_1^2}{1-\xi^2} = H_1^{-2/3} f \quad,\quad \Omega_0^2 \frac{\xi^2}{1-\xi^2}= H_1^{1/3} r^2\quad,\quad \frac{\Omega_0^2}{(1-\xi^2)^2}= H_1^{1/3} f^{-1} (\partial_\xi r)^2\,.
\end{equation}
One can verify that these matching conditions are verified if one relates the parameters $q_1$ and $\epsilon$ as in \eqref{eq:q12epsilon} and the radial coordinates $r$ and $\xi$ via
\begin{equation}\label{eq:r2xi}
(\partial_s r)^2 = r^2 (1+r^2 H_1)\quad\mathrm{with}\quad \tanh s= \sqrt{1-\xi^2}\,.
\end{equation}
This leads to a fairly complicated function $r(\xi)$, which can be expressed implicitly through the Appel function $F_1$. Hence, the parametrisation of \cite{Liu:2007xj} is simpler and more convenient to study the geometry at finite $\epsilon$. 

With the help of the 10D metric \eqref{eq:10Dmetric}, we can also discuss how the 5D equation for the lowest harmonic of the dilaton arises from 10D. The linearised equation for the axion-dilaton field, $\Phi$, in the background \eqref{eq:10Dmetric} is simply
\begin{equation}\label{eq:10Dwe}
\Box_{10} \Phi=0\,,
\end{equation}
with $\Box_{10}$ the box operator associated with $ds^2_{10}$. One can verify, using the form of $G_{\alpha\beta}$, that this equation is not separable, i.e. if one decomposes $\Phi$ in S$^5$ spherical harmonics:\footnote{Note that the coordinate describing the AdS$_5$ vacuum is $\tilde \phi\equiv \phi-t$ and, thus, the spherical harmonics should be expressed in terms of this $\tilde\phi$ in order to identity the proper holographic fields $\Phi_I$.} $\Phi=\sum_I \Phi_I Y^I$, the equations for the different $\Phi_I$'s are generically coupled. The situation is simpler for the trivial harmonic $\Phi_0$, which, given the factorised form of $ds^2_{10}$, can be immediately seen to satisfy the massless 5D Klein-Gordon equation 
\begin{equation}\label{eq:5Dwbis}
\Box_{5} \Phi_0=0\,.
\end{equation} 
This is the equation discussed in the main text.

\section{Regge limit}
\label{app:ReggeLimit}

The results~of the previous section were derived by studying the quadratic fluctuations around a non-trivial asymptotically AdS geometry produced by a state of mass much smaller than the AdS scale. Thus they are naturally interpreted as the first terms of the perturbative expansion of the Euclidean HHLL correlator~\eqref{eq:PhiBcor} in the regime where $\Delta_H/N^2$ is finite but small. It is interesting to analytically continue these expressions to the Lorentzian region. This can be done by treating $z$ and$\bar{z}$ as independent variables and then by continuing $z$ through the logarithmic branch cut starting at zero
\begin{equation}
  \label{eq:z-ac}
  z \to e^{-2\pi i} z\;, \quad \bar{z}~\mbox{ fixed}\;.
\end{equation}
In particular the regime where both $z$ and $\bar{z}$ approach one simultaneously in the Lorentzian patch corresponds to the Regge limit. In the context of the light 4-point correlators in ${\mathcal N}=4$ SYM, this was studied long time ago~\cite{Cornalba:2006xk,Cornalba:2006xm} deriving the AdS/CFT analogue of the leading eikonal for the $2\to 2$ scattering in flat space~\cite{Amati:1987wq,tHooft:1987vrq}. In order to study this regime it is convenient to parametrize the crossratios as
\begin{equation}
  \label{eq:zsiet}
  z=1-\sigma_R\;,\quad \bar{z}=1-\sigma_R \,\eta_R\;,
\end{equation}
then the Regge limit is captured by sending $\sigma_R \to 0$ at fixed values of $\eta_R$. In this regime the correlator is expected to exponentiate when written in the appropriate variables~\cite{Cornalba:2007zb}. In the following we will interpret the CFT results as HHLL correlators where $\Delta_H$ is proportional to the CFT central charge. As discussed in~\cite{Kulaxizi:2018dxo,Karlsson:2019qfi,Kulaxizi:2019tkd}, the Regge limit can be studied along the same lines in this case as well: the analytic continuation~\eqref{eq:z-ac} provides the Lorentzian correlator and in the limit $\sigma_R \to 0$ one enters the regime where the light operator has a large kinetic energy. However the energy is not large enough for the light state to produce a sizeable gravitational backreaction nor to puff up into an extended object such as an AdS giant graviton. Instead the semiclassical interpretation of the Regge limit is in terms of a geodesic describing the motion of the light perturbation in the background produced by the heavy state. Again this is very much in parallel to what happens in flat space, see~\cite{DAppollonio:2010krb}. In this section, we use the semiclassical limit as derived from the geodesic motion as a consistency check on the (more complicated) derivation of the full correlator and its expansion for $\Delta_H/N^2\ll 1$. We do so by following closely~\cite{Kulaxizi:2018dxo,Karlsson:2019qfi}, but using the the $1/2$-BPS background considered in this paper at the place of the AdS-Schwarzschild solution. A similar check was performed in~\cite{Ceplak:2021wak} for a class of $1/2$-BPS microstate solutions in the context of the AdS$_3$/CFT$_2$ duality and it turns out that the AdS$_5$ analysis is very similar.

As a first step, we define the eikonal phase $\delta$ in terms of a null geodesic that propagates along the equator of the S$^3$ inside AdS$_5$ in the geometry~\eqref{eq:5Dmetric}
\begin{equation}
  \label{eq:phas-shift}
\delta + \delta_{\rm AdS} \equiv -p_t \Delta t - p_\varphi \Delta\varphi = 2 |p_t| \int\limits_{r_0}^\infty \frac{dr}{f} \sqrt{H_1-\beta^2 \frac{f}{r^2}}\,.
\end{equation}
Here $p_t$ and $p_\varphi$ are the conserved momenta related to the shift symmetries of the background along time and the angle along the equator, $\beta$ indicates their ratio $\beta=p_\varphi/p_t$ and $r_0$ is the turning point of the geodesic, i.e. the largest solution of $H_1-\beta^2 \frac{f}{r^2}=0$
\begin{equation}
  \label{eq:turnpoint}
  r_0 = \sqrt{\frac{1}{\left(1-\beta^2\right)^2}+(q_1+2) q_1}- q_1-1 \;.
\end{equation}
The intervals $\Delta t$ and $\Delta\varphi$ indicate the variation of the corresponding coordinates between the initial and the final point of the geodesic
\begin{equation}
  \label{eq:Deltatth}
  \Delta t = 2 \int\limits_{r_0}^\infty dr\; \frac{\dot{t}}{\dot{r}}\;,\quad
  \Delta \varphi  = 2 \int\limits_{r_0}^\infty dr\; \frac{\dot{\varphi}}{\dot{r}}\;,
\end{equation}
where the dots represent the derivatives with respect to the affine parameter. In order to evaluate $\delta$ perturbatively in the small $q_1$ expansion it is convenient to introduce $y=r_0/r$
\begin{equation}
  \label{eq:delexpl}
  \delta + \delta_{\rm AdS}  =  2 |p_t| \int\limits_0^1 dy \left[\frac{\left(1-\beta^2\right) \sqrt{1+2 (1+q_1) \frac{y^2}{r^2_0}+ \frac{y^4}{r_0^4}}- \frac{y^2}{r_0^2}}{1+2(1+q_1) \frac{y^2}{r_0^2}+\frac{y^4}{r_0^4}}\right]^{1/2}\;.
\end{equation}
After expanding the integrand for $q_1 \ll 1$, the integral can be easily performed at any order in perturbation theory by using
\begin{equation}
  \label{eq:yint}
  \int\limits_0^1 \frac{(y^2)^{p_1}\sqrt{1-y^2}}{\left(y^2 + \frac{\beta^2}{1-\beta^2}\right)^{p_2}}\, dy = \frac{\sqrt{\pi}}{4} \left(\frac{\beta^2}{1-\beta^2}\right)^{-p_2} \frac{\Gamma\left(p_1+\frac{1}{2}\right)}{\Gamma(p_1+2)}\,{}_2F_1\left(p_2,p_1+\frac{1}{2};p_1+2,-\frac{1-\beta^2}{\beta^2}\right)\,.
\end{equation}
By construction, $\delta$ starts at order $q_1$, since the pure AdS contribution is written separately in $\delta_{\rm AdS}$, and the first three terms read
\begin{align} \nonumber
  \delta & = \pi |p_t| \left(1-\beta^2\right)^2\left\{\frac{\alpha^2}{2} + \alpha^4 \left[1-\frac{15}{16} \left(\beta^2-1\right)^2\right] + \alpha^6 \left[2-5 \left(\beta^2-1\right)^2+\frac{105}{32} \left(\beta^2-1\right)^4\right]+\mathcal{O}(\alpha^8)\right\}
\\ & = 8\pi \frac{h^2 \bar{h}^2}{(h+\bar{h})^3} \alpha^2 \Big[1+2\alpha^2 \left(1-15\nu^2\right)+4\alpha^4 \left(1 -40 \nu^2 + 420 \nu^4\right)\Big]+O\left(\alpha^8\right)
\label{eq:delexpl2}
\end{align}
where in the last step we introduced the variables $h\geq\bar{h}$ as follows~\cite{Kulaxizi:2018dxo}
\begin{equation}
|p_t|=h+\bar h\;,\quad \beta=\frac{h-\bar h}{h+\bar h}\;,\quad \nu = \left(\frac{h \bar{h}}{(h+\bar{h})^2}\right)\,.
\end{equation}
The parametrization on terms of $h$, $\bar{h}$ has two advantages: first they are directly related to the (average) anomalous dimension of the operators exchanged in the $z,\bar{z}\to 1$ OPE, second in terms of these variable the Regge limit of the correlator $\Psi_b$~\eqref{eq:psiexpads5} exponentiates
\begin{equation}
  \label{eq:corr-Regge}
  \Psi_B^R = \int_0^\infty\!\! dh \int_0^h \!\! d\bar{h} (h \bar{h})^{\Delta-2} \left[ \frac{z^{h+1} \bar{z}^{\bar{h}} - z^{\bar{h}} \bar{z}^{h+1}}{z-\bar{z}}\right] e^{i \delta(h,\bar{h})}\;,
\end{equation}
where the square parenthesis is determined by the heavy limit of the 4D conformal blocks in the channel $z,\bar{z}\to 1$. Notice that while in our case the eikonal phase is symmetric in the exchange $h\leftrightarrow \bar{h}$ the integrand~\eqref{eq:corr-Regge} is not. However in the Regge limit it is possible to exploit the symmetry of $\delta$ to write the correlator in the following simpler form~\cite{Karlsson:2019qfi}
\begin{equation}
  \label{eq:corr-Regge2}
  \Psi_B^R = \frac{z}{z-\bar{z}} \int_0^\infty\!\! dh \int_0^\infty \!\!d\bar{h} \;(h \bar{h})^{\Delta-2} (h-\bar{h}) z^{h} \bar{z}^{\bar{h}} e^{i \delta(h,\bar{h})}\;.
\end{equation}

By focusing on the case $\Delta=2$, which is the one relevant for our result, we obtain
\begin{equation}
  \label{eq:corr-Regge3}
  \Psi_B^R = \left[\frac{z}{z-\bar{z}}  \left(z \partial_z-\bar{z}\partial_{\bar{z}}\right)\right] \int_0^\infty\!\! dh \int_0^\infty \!\!d\bar{h} \; z^{h} \bar{z}^{\bar{h}} e^{i \delta(h,\bar{h})}\;.
\end{equation}
Perturbatively in $q_1$ all integrals can be performed by using
\begin{align}
  \label{eq:intRegge}
   I(a,c)  & \equiv \int\limits_0^\infty\! dh \int\limits_0^\infty \!d\bar{h} \;\frac{h^{a+1} \bar{h}^{a+1}}{(h+\bar{h})^c}  \; z^{h} \bar{z}^{\bar{h}} \\  & = \frac{\Gamma^2(a+2)\,\Gamma(2a+4-c)}{\Gamma(2a+4)} \,\eta_R^{c-a-2}\sigma_R^{c-2a-4}\, {}_2F_1(a+2,c;2a+4;1-\eta_R)\,.\nonumber
\end{align}
In order to compare with the explicit results for the HHLL correlators above, we switch to $\alpha$~\eqref{eq:q1alpha} as expansion parameter. At order $\alpha^0$, we can set $\delta$ to zero and Eq.~\eqref{eq:corr-Regge3} reproduces the leading term as $\sigma_R\to 0$ of the $2$-point function in AdS. By using the term ${\mathcal O}(\alpha^2)$ in~\eqref{eq:delexpl2}, one readily derives the corresponding leading contribution to the Regge correlator
\begin{equation}
  \label{eq:CRe2}
  \begin{aligned}
  \Psi_B^{R} & \equiv \Psi_B^{R(0)} + \alpha^2 \Psi_B^{R(2)} + {\mathcal O}(\alpha^4) = {\Delta} I(-1,0) + \alpha^2 8 \pi i {\Delta} I(1,3) + {\mathcal O}(\alpha^4) \\ & = \frac{1}{\eta_R^2 \sigma_R^4} + \alpha^2 16 \pi i \frac{\left(1-\eta_R^2\right)\left(1+28 \eta_R  +\eta_R^2\right)+12 \left(1+3 \eta_R + \eta_R^2\right) \eta_R  \log (\eta_R )}{(1- \eta_R)^7 \,\eta_R\,  \sigma_R^5}+{\mathcal O}(\alpha^4)\;,
  \end{aligned}
\end{equation}
where ${\Delta} \equiv \frac{1}{\sigma_R (1-\eta_R)}  \left(\partial_{\sigma_R} - \frac{1+\eta_R}{\sigma_R} \partial_{\eta_R}\right)$ follows from rewriting the differential operator in the square parenthesis of~\eqref{eq:corr-Regge3}. One can check that this same result is obtained by performing the analytic continuation~\eqref{eq:z-ac} on the correlator~\eqref{eq:psiexpads5} and then taking the leading contribution as $\sigma_R\to 0$ at arbitrary values of $\eta_R$. At the next order (${\mathcal O}(\alpha^4)$) there are two contributions. The first comes from expanding at the quadratic order the eikonal phase in~\eqref{eq:corr-Regge3} and using the $\alpha^2$ term for $\delta$. This superleading Regge term is real and scales as $1/\sigma^6$, while the contribution scaling as $1/\sigma^5$ is imaginary and is obtained by using the $\alpha^4$ term for $\delta$. Explicitly we have
\begin{align}
  \label{eq:CRe4}
  \Psi_B^{R(4)} = &  - 32\pi^2  {\Delta} I(3,6) + 16 \pi i {\Delta} \Big[I(1,3) -15  I(3,7) \Big]\;,
\end{align}
where the first term in the first line matches the expansion at quadratic order of $e^{i\delta}$ by keeping only the leading term in~\eqref{eq:delexpl2} (and scales as $1/\sigma_R^6$), while the second term reproducs the subleading contribution in~\eqref{eq:delexpl2} in the linear expansion of $e^{i\delta}$ (and scales as $1/\sigma_R^5$). Again there is a perfect match between the analytically continued CFT result~\eqref{eq:psiexpads5} and what is obtained from~\eqref{eq:corr-Regge3}, both for the real part at order $1/\sigma_R^6$, and for the imaginary leading term at order $1/\sigma_R^5$. As expected, the most singular term as $\sigma_R\to 0$ at order $\alpha^4$ is determined by the result at order $\alpha^2$, thus checking the first step of the eikonal exponentiation. Similarly at order $\alpha^6$, we have 
\begin{equation}
  \label{eq:CRe6}
  \Psi_B^{R(6)} = {\Delta} \left\{ -i \frac{(8\pi)^3}{3!}  I(5,9) - 128 \pi^2 \Big[I(3,6) -15  I(5,10) \Big] + 32\pi i \Big[I(1,3) - 40 I(3,7) + 420 I(5,11) \Big]\right\} \;,
\end{equation}
where the first two terms reconstruct the exponentiation of the leading and subleading contributions of~\eqref{eq:delexpl2}, while the third matches the contribution from the term of order $\alpha^6$ in~\eqref{eq:delexpl2}.

\section{ ${\cal N}=4$ OPE and four-point functions }\label{N4blocks}

For completeness, we present a short/pragmatic review of the ${\cal N}=4$ superblock technology \cite{Doobary:2015gia}
for studying four-point functions of $\frac{1}{2}$-BPS operators, as in our case. 

Consider a four-point correlator of $\frac{1}{2}$-BPS operators in ${\cal N}=4$ SYM, 
\begin{equation}
\langle O_{p_1}(\vec{x}_1,\vec{y}_1) O_{p_2}(\vec{x}_2,\vec{y}_2) O_{p_3}(\vec{x}_3,\vec{y}_3) O_{p_4}(\vec{x}_4,\vec{y}_4) \rangle
\end{equation}
Maintaining the OPE channel, $x_1\rightarrow x_2$ and $x_3\rightarrow x_4$, 
we can always achieve a configuration such that ${\rm max}(p_{12},p_{43})\ge 0$. Define
 \begin{equation}
 \gamma_{\min}=\max(p_{12},p_{43})\qquad;\qquad \gamma_{\max}=\min(p_1+p_2,p_3+p_4)\,.
 \end{equation}
In the setup specified above, say $B_{\underline{R} }$ is a superblock corresponding 
to a superprimary with quantum numbers ${\underline{R} }$, 
we can write the superblock decomposition as
\begin{equation}\label{scblockdeco}
\langle O_{p_1}O_{p_2} O_{p_3} O_{p_4} \rangle =   {\cal P}_{\vec{p}} \sum_{\underline{R} } A_{\underline{R} }\, B_{\underline{R} }\,,
 \end{equation}
 where 
\begin{equation}
{\cal P}_{\vec{p}}=g_{12}^{ \frac{p_1+p_2}{2} } g_{34}^{ \frac{p_3+p_4}{2} } g_{13}^{\frac{p_3-p_4}{2} } g_{14}^{\frac{p_1+p_4-p_2-p_3}{2}} g_{24}^{\frac{p_2-p_1}{2}} 
\left( \frac{g_{13}g_{24} }{g_{12}g_{34}}\right)^{\!\!\!\frac{\max(p_{12},p_{43} ) }{2} }\,.
 \end{equation}
 with $g_{ij}=\frac{ \vec{y}_i.\vec{y}_j }{(\vec{x}_{ij})^2}$ free propagators.
Under our assumptions, $\gamma_{\min}\ge 0$, the propagator structure ${\cal P}_{\vec{p}}$ is 
an allowed propagator structure in the free theory, corresponding to one of the two diagrams below
 \begin{equation}
 \begin{array}{ccc}
 \begin{tikzpicture}[scale=1.5]
 
\def\latoxuno{0}
\def\latoyuno{0}
\def\latoxdue{1}
\def\latoydue{-1}

\draw (\latoxuno,\latoyuno) --  (\latoxuno, \latoydue);
\draw (\latoxuno, \latoyuno) --  (\latoxdue, \latoyuno);
\draw (\latoxdue, \latoyuno) --  (\latoxdue, \latoydue);
\draw (\latoxdue, \latoyuno) --  (\latoxuno, \latoydue);

    \draw[fill=white!60,draw=white] (\latoxuno, \latoyuno) circle (2.5pt);
    \draw[fill=white!60,draw=white] (\latoxuno, \latoydue) circle (2.5pt);
    \draw[fill=white!60,draw=white] (\latoxdue, \latoyuno) circle (2.5pt);
    \draw[fill=white!60,draw=white] (\latoxdue, \latoydue) circle (2.5pt);

   \draw(\latoxuno-.1, \latoyuno+.1) node[scale=.8] {${\cal O}_1$};
   \draw(\latoxuno-.1, \latoydue-.1) node[scale=.8] {${\cal O}_2$};
   \draw(\latoxdue+.1, \latoyuno+.1) node[scale=.8] {${\cal O}_4$};
   \draw(\latoxdue+.1, \latoydue-.1) node[scale=.8] {${\cal O}_3$};
   
   \draw(\latoxuno+0.6, \latoydue-0.5) node[scale=.8] {$\max=p_{43}$};
   
\end{tikzpicture}   
&
\rule{2cm}{0pt}
&
 \begin{tikzpicture}[scale=1.5]
 
\def\latoxuno{0}
\def\latoyuno{0}
\def\latoxdue{1}
\def\latoydue{-1}

\draw (\latoxuno, \latoyuno) --  (\latoxuno, \latoydue);
\draw (\latoxuno, \latoyuno) --  (\latoxdue, \latoyuno);
\draw (\latoxdue, \latoyuno) --  (\latoxdue, \latoydue);
\draw (\latoxuno, \latoyuno) --  (\latoxdue, \latoydue);

    \draw[fill=white!60,draw=white] (\latoxuno, \latoyuno) circle (2.5pt);
    \draw[fill=white!60,draw=white] (\latoxuno, \latoydue) circle (2.5pt);
    \draw[fill=white!60,draw=white] (\latoxdue, \latoyuno) circle (2.5pt);
    \draw[fill=white!60,draw=white] (\latoxdue, \latoydue) circle (2.5pt);

   \draw(\latoxuno-.1, \latoyuno+.1) node[scale=.8] {${\cal O}_1$};
   \draw(\latoxuno-.1, \latoydue-.1) node[scale=.8] {${\cal O}_2$};
   \draw(\latoxdue+.1, \latoyuno+.1) node[scale=.8] {${\cal O}_4$};
   \draw(\latoxdue+.1, \latoydue-.1) node[scale=.8] {${\cal O}_3$};
   
      \draw(\latoxuno+0.6, \latoydue-0.5) node[scale=.8] {$\max=p_{12}$};

\end{tikzpicture} 
\end{array}
\end{equation}
The sum over reps $\underline{R}$ in \eqref{scblockdeco} is the sum over exchanged 
superprimaries. In the OPE between $\frac{1}{2}$-BPS operators, the reps $\underline{R}$ 
are symmetric traceless reps in spin $l$ and have $su(4)$ Dynkin labels $[a,b,a]$. 
All ${\cal N}=4$ multiplet with this specifics can be exchanged in the OPE. 
The $\frac{1}{2}$-BPS, $\frac{1}{4}$-BPS, semishort multiplets of this type
 in the notation of \cite{Dolan:2002zh} are
\begin{equation}\label{DOapp1}
{\cal B}^{\frac{1}{2},\frac{1}{2}}_{[0,p,0](0,0)}\quad;\quad {\cal B}^{\frac{1}{4},\frac{1}{4}}_{[q,p,q](0,0)}
\end{equation}
(obtained from the shortening conditions \cite[(5.6),(5.26)]{Dolan:2002zh}) and the diagonal semishorts rep
\begin{equation}\label{DOapp2}
{\cal C}^{\frac{1}{4},\frac{1}{4}}_{[q,p,q](\frac{l}{2},\frac{l}{2})}
\end{equation}
see \cite[Table 4]{Dolan:2002zh} (obtained from the shortening conditions 
\cite[(5.15),(5.29)]{Dolan:2002zh} which include the more general reps $[k,p,q](j,\bar{j})$). 
Finally there are long operators.
 
Rather than using the above notation, 
it is more convenient to understand all these reps by using a unified formalism, where 
$\underline{R}=\{\gamma,\underline{\lambda}\}$ is parametrized by $\gamma$ and a 
Young diagram $\underline{\lambda}$. The origin of this identification is nicely explained 
in \cite[section 3.2]{Aprile:2021pwd} and summarized in the following table,\footnote{The interested 
reader can match  the number of components of these multiplets versus 
\eqref{DOapp1}-\eqref{DOapp2} by looking at \cite[appendix B]{Chicherin:2015edu}.}
\begin{equation}\label{table}
\begin{array}{|c||c|c|c|c|}
\multicolumn{5}{c}{ 
\begin{array}{l}
\rule{0pt}{0cm}
\end{array}
}\\
\hline
\underline{\lambda}     					&     \tau=\Delta{-}l			& l     			&  \mathfrak{R} 			  	& \text{multiplet} \\\hline
[\varnothing]                                     					&    \gamma        			&0    				& [0,\gamma,0]                            	& \frac{1}{2}\text{-BPS}       \\\hline
\left[1^\mu\right]                                 		 	& \gamma             			&0				&   [\mu,\gamma{-}2\mu,\mu]		&\frac{1}{4}\text{-BPS} \\ 
\left[\lambda, 1^\mu\right]\ (\lambda\geq 2)   	& \gamma             			&\lambda{-}2      	&  [\mu,\gamma{-}2\mu{-}2,\mu] 	&  \text{semi-short}  \\\hline
{ [\lambda_1,\lambda_2,2^{\mu_2},1^{\mu_1}]\  (\lambda_2\geq 2)} 
									& \gamma{+}2\lambda_2{-}4	&\lambda_1{-}\lambda_2&[\mu_1,\gamma{-}2\mu_1{-}2\mu_2{-}4,\mu_1]&\text{long} \\ \hline 
\end{array}
\end{equation}

In a given four point function, the possible values of $\gamma$ are
 \begin{equation}
 \gamma=\gamma_{\min},\gamma_{\min}+2,\ldots,\gamma_{\max}\,.
 \end{equation}
 Then,   the max height of the Young diagram for given $\gamma$ is $ \beta=\frac{\gamma-\gamma_{\min}}{2}\in\mathbb{N}$.

The blocks are
\begin{equation}
 B_{\underline{R} }= 
\left[ \left( \frac{g_{12} g_{34}}{g_{13} g_{24} }  \right)^{\!\!\!\frac{ \gamma-\gamma_{\min}}{2} }\!\!\! F_{\gamma,\underline{\lambda} }  \right] 
\end{equation}
and the function $F_{\gamma,\underline{\lambda} }$ is a non trivial function of the cross ratios:
 \begin{equation}
\frac{g_{12} g_{34}}{g_{13} g_{24} }= \frac{y_1y_2}{x_1x_2}\qquad;\qquad \frac{g_{14} g_{23}}{g_{13} g_{24} }=\frac{(1-y_1)(1-y_2)}{(1-x_1)(1-x_2)}\;.
 \end{equation}
{Note: compared to the main text $z=1-x_1$, $\bar{z}=1-x_2$ and referring to \eqref{def_cross_ratiosFA}, we have
 \begin{equation}
 U=x_1x_2\qquad;\qquad V=(1-x_1)(1-x_2)\qquad;\qquad \sigma=\frac{1}{y_1y_2}\qquad;\qquad \tau=\frac{(1-y_1)(1-y_2)}{y_1y_2}\;.
\end{equation}

While there are various possible
choices of cross ratios for the correlator, it will become soon clear that $x_1,x_2,y_1,y_2$ is 
the natural choice to perform the OPE analysis.

 \subsection{Superblocks}

A way to construct $F_{\gamma,\underline{\lambda} }$ is to 
solve the second order Casimir eigenvalue equation. This can always be done 
by considering an hypergeometric expansion of the form
 \begin{equation}\label{convblocks}
 F_{\gamma,\underline{\lambda}}=\Bigg[ P_{\underline{\lambda}} + P_{\underline{\lambda}+\underline{\Box}}
 + \ldots \Bigg]  = \sum_{\underline{\mu}: \underline{\lambda}\subseteq\underline{\mu}} (T_{\gamma})^{\underline{\mu}}_{\underline{\lambda}}\, P_{\underline{\mu}}
 \end{equation}
 where  the $P_{\underline{\mu}}(x_1,x_2,y_1,y_2)$ are superSchur polynomials that 
 can be shown to be in triangular correspondence with the descendants of the primary.  
 In particular, the primary is $P_{\underline{\lambda}}(x_1,x_2,y_1,y_2)$.
The $F_{\gamma,\underline{\lambda}}$ can be resummed and so the whole block, 
which then takes the form 
\begin{mdframed}
\begin{equation}
{B}_{\gamma,\underline{\lambda}}
={\cal P}_{\vec{p}}\Bigg[ k + {S}(x_1,x_2,y_1,y_2) + \frac{ \prod_{ij}(x_i-y_j) }{(y_1 y_2)^2} 
{H}(x_1,x_2,y_1,y_2) 
\Bigg]
\end{equation}
\end{mdframed}
where $k$ is a constant, and 
\begin{equation}
{S}= \frac{ \prod_{ij} (x_i-y_j) }{ (x_1-x_2)(y_1-y_2) }\Bigg[  
\frac{ f(x_2,y_1) }{x_2 y_1(x_1-y_2)} + \frac{ f(x_1,y_2) }{x_1 y_2(x_2-y_1)} - \frac{ f(x_1,y_1) }{x_1 y_1(x_2-y_2)} -\frac{ f(x_2,y_2) }{x_2 y_2(x_1-y_1)} 
\Bigg]\ 
\end{equation}
This particular combination only depends on a single-variable function $f(x,y)$.  Note also that
\begin{equation}
 \frac{ \prod_{ij}(x_i-y_j) }{(y_1 y_2)^2}= {\cal I}(U,V,\sigma,\tau)
\end{equation}
with the latter given in \eqref{intriligator_factor}.

\subsubsection*{Summary: Long Block}

 This is the simplest case: $k=0$, $f(x_1,y_1)=0$, and
 \begin{equation}
 {H}_{\gamma,\underline{\lambda}}= 
(-1)^{\lambda_1-\lambda_2} \left(\frac{x_1 x_2}{y_1 y_2} \right)^{\!\!\beta } \frac{   \Big( F^{}_{\lambda_1}(x_1)F^{}_{\lambda_2-1}(x_2)- x_1\leftrightarrow x_2\Big)
 }{x_1-x_2} \frac{ 
  \Big( G^{}_{\lambda'_1}(y_1)G^{}_{\lambda'_2-1}(y_2)- y_1\leftrightarrow y_2\Big)
  }{y_1-y_2}
 \end{equation}
 where $\lambda_i$ denotes the row length and $\lambda'_j$ denotes the column length of the Young diagram, and we introduced the notation
 \begin{align}
 F^{}_{\lambda}(z) &= {z^{\lambda-1}}\,_2F_1\Big[ \,^{\lambda+\frac{\gamma-p_{12}}{2},\ \lambda+\frac{\gamma-p_{43}}{2}}_{\ \ \ \ \ \ \ 2\lambda+\gamma} ;z\Big],\\
 G^{}_{\lambda}(z) &= {z^{\lambda+1}}\,_2F_1\Big[ \,^{\lambda-\frac{\gamma-p_{12}}{2},\ \lambda-\frac{\gamma-p_{43}}{2}}_{\ \ \ \ \ \ \ 2\lambda-\gamma} ;z\Big]
 \end{align}
The Young diagram of a long block contain always the $2\times 2$ square of boxes, thus $\lambda_i\ge 2$ and $\lambda'_j\ge 2$.  
For long blocks the parameter $\gamma$ is redundant  and cancels out when we relabel the
block with $\tau,l, [aba]$ defined through \eqref{table}, namely
 \begin{equation}
 \lambda_1=2+l+\frac{\tau-\gamma}{2}\quad;\quad \lambda_2=2+\frac{\tau-\gamma}{2}\quad;\quad \lambda'_1=-\frac{b-\gamma}{2}\quad;\quad \lambda'_2=-a-\frac{b-\gamma}{2}
 \end{equation}
   It is simple to show that 
 \begin{align}
 &
\!\!\!\!\!{H}_{\tau,l,[aba]}=
(-1)^l(x_1 x_2)^{ \frac{ \tau-\gamma_{\min} }{2} }
\frac{ 
x_1^{l+1}\,_2F_1\Big[\,^{2+\frac{\tau}{2}+l-\frac{p_{12}}{2},\, 2+\frac{\tau}{2}+l-\frac{p_{43}}{2} }_{\ \ \ \ \ \ \ \ \ 4+\tau+2l};x_1 \Big] 
\,_2F_1\Big[\,^{1+\frac{\tau}{2}-\frac{p_{12}}{2},\, 1+\frac{\tau}{2}-\frac{p_{43}}{2}}_{\ \ \ \ \ \ \ \ 2+\tau}  ;x_2 \Big] - x_1\leftrightarrow x_2}{ (x_1-x_2) }  \times \notag\\[.2cm]
& \rule{1cm}{0pt} \frac{1}{ (y_1 y_2)^{\frac{b-\gamma_{\min} }{2}  }} 
\frac{ 
y_1y_2^{-a}\,_2F_1\Big[\,^{-1-\frac{b}{2}-a+\frac{p_{12}}{2},\, -1-\frac{b}{2}-a+\frac{p_{43}}{2} }_{\ \ \ \ \ \ \ \ \ -2-b-2a};y_2 \Big] 
\,_2F_1\Big[\,^{-\frac{b}{2}+\frac{p_{12}}{2},\, -\frac{b}{2}+\frac{p_{43}}{2}}_{\ \ \ \ \ \ \ \ -b}  ;y_1 \Big] - y_1\leftrightarrow y_2}{ (y_1-y_2) }
\label{allLongs}
 \end{align}
 While the Young diagrams define a long block for $\tau\ge 4+2a+b$, which in fact is  the correct value for the \emph{free} theory,
 the expression in \eqref{allLongs} is valid below this threshold.

\subsubsection*{Summary: $\frac{1}{4}$-BPS and semishort}

The $\frac{1}{4}$-BPS and  semishort blocks can be considered together since they both 
correspond to a Young diagram with at most one row and one column, namely $[\lambda,1^{\lambda'-1}]$ 
with $\lambda'\ge 1$. We shall call Young diagrams of this type atypical. The typical Young diagrams are those that contain the $2\times 2$ square of boxes. 
The latter label the long blocks, as shown in the previous section. The parameter $\gamma$ for the short blocks is now a part of the quantum numbers, 
and cannot be gauged away, as we did above for the long blocks. Analytic properties are 
manifest w.r.t.~the Young diagram labels. This is not the case for $l,[aba]$. For example, note in 
\eqref{table} that $l$ varies discontinuously when going from $\frac{1}{4}$-BPS to semishorts. 

These blocks have a single-variable seed function,
\begin{equation}
 f_{\gamma,[\lambda,1^{\lambda'-1}]}(x,y)= (-1)^{\lambda}\left( \frac{x}{y} \right)^{\!\!\beta }\Bigg[ 
x^{\lambda} \,_2F_1\Big[\,^{\lambda+\frac{\gamma-p_{12}}{2},\ \lambda+\frac{\gamma-p_{43} }{2} }_{ \rule{1cm}{0pt}  2\lambda+\gamma};x \Big]\
y^{\lambda'} \,_2F_1\Big[ \,^{ \lambda'-\frac{\gamma-p_{12}}{2} ,\ \lambda-\frac{\gamma-p_{43}}{2} }_{\rule{1cm}{0pt}2\lambda'-\gamma};y\Big]
 \Bigg]
\end{equation}
and \emph{also} a contribution to the two-variable function
\begin{equation}
 {H}_{\gamma,[\lambda,1^{\lambda'-1}]}=    (-1)^{\lambda-1}\sum_{h=1}^{\beta} \left(\frac{x_1 x_2}{y_1 y_2} \right)^{\!\!\beta }
 \frac{ \Big( F_{1-h}(x_1)F_{\lambda}(x_2)  - x_1\leftrightarrow x_2 \Big) }{x_1-x_2 } 
 \frac{ \Big( G_{h}(y_1) G_{\lambda'}(y_2) - y_1\leftrightarrow y_2 \Big) }{ y_1-y_2}
\end{equation}
Considering the chiral twist, say $x_2=y_2$, the two-variable function is set to zero by $\prod_{i,j}(x_i-y_j)$,
 and the block reduces to (a $GL(1|1)$ block),
\begin{equation}
{B}_{\gamma,[\lambda,1^{\lambda'-1}]}\Big|_{x_2=y_2}=
\frac{(x_1-y_1)}{x_1y_1} f_{\gamma,[\lambda,1^{\lambda'-1}]}
\end{equation}

\subsubsection*{Recombination}

The key relation, which follows from a similar relation for the superSchur polynomials, is
\begin{equation}
f_{\gamma,[\lambda,1^{\lambda'-1}] }(x,y) + f_{\gamma+2,[\lambda-1,1^{\lambda'}]}(x,y)=0\,.
\end{equation}
Now, a long block at the unitarity bound $\tau=2a+b+2$ can be written as the sum of two ``atypical" blocks, 
\begin{equation}
{B}^{\cal L}_{\gamma,[\lambda,1^{\lambda'-1}]}={B}^{}_{\gamma,[\lambda,1^{\lambda'-1}]} + {B}^{}_{\gamma+2,[\lambda-1,1^{\lambda'}]}  =  
 \frac{ \prod_{ij}(x_i-y_j) }{(y_1 y_2)^2} {\cal H}_{\tau=\gamma, l=\lambda-2,a=\lambda'-1,b=\gamma-2\lambda'}
\end{equation}
where the rhs is given by \eqref{allLongs}. Let us note that when we recombine the stringy states we use
\begin{equation}\label{recombination_fullformula}
{B}^{}_{\gamma,[\lambda,1^{\lambda'-1}]} ={B}^{\cal L}_{\gamma,[\lambda,1^{\lambda'-1}]} - {B}^{}_{\gamma+2,[\lambda-1,1^{\lambda'}]}\,.
\end{equation}
Semishorts blocks are defined for $\lambda\ge 2$, as given in the table \eqref{table}, and the 
Young diagram notation makes recombination obvious. Otherwise, one has to be careful:  
a $\lambda=2$  semishort on the lhs of \eqref{recombination_fullformula},  thus spin $0$,
recombines into a spin $0$ long block and a $\frac{1}{4}$-BPS contribution. 
The odd spins semishorts on the lhs  of \eqref{recombination_fullformula} with spin 
greater than $1$ recombine into odd spins long blocks and even 
spin semishorts for the reps $[a+1,b,a+1]$ with spin greater than $0$.
The even spins semishorts on the lhs  of \eqref{recombination_fullformula} 
with spin greater than $2$ recombine into even spins long blocks 
and odd spin semishorts for the reps $[a+1,b,a+1]$ with spin greater than $1$.

\subsubsection*{Summary: $\frac{1}{2}$-BPS}

In this case the block has 
$$k=1$$ 
as a seed function. Then there is a contribution from a single-variable function ${\cal S}$, determined by
\begin{equation}
f_{\gamma,[\varnothing]}(x,y)= \left( \frac{x_1}{y_1} \right)^{\!\! \beta } \sum_{h=1}^{\beta}
x^{1-h} \,_2F_1\Big[\,^{1-h+\frac{\gamma-p_{12}}{2},\ 1-h+\frac{\gamma-p_{43} }{2} }_{ \rule{1cm}{0pt}  2(1-h)+\gamma};x \Big]\
y^{h} \,_2F_1\Big[ \,^{ h-\frac{\gamma-p_{12}}{2} ,\ h-\frac{\gamma-p_{43}}{2} }_{\rule{1cm}{0pt}2h-\gamma} ,y\Big]
\end{equation}
and a two-variable function ${\cal H}$ determined by
\begin{equation}
{H}_{\gamma,[\varnothing]}=\left(\frac{x_1 x_2}{y_1 y_2} \right)^{\!\!\beta } \sum_{1\leq i<j\leq \beta }
\frac{ \Big[ F_{1-i}(x_1)F_{1-j}(x_2) - x_1\leftrightarrow x_2 \Big] }{x_1-x_2}
\frac{ \Big[ G_{i}(y_1)G_{j}(y_2) - y_1\leftrightarrow y_2 \Big] }{y_1-y_2}
\end{equation}

\subsection{OPE equations in the long sector}\label{expaOPElong}

Here we collect some useful formulae for the large $N$ expansion of the  correlators 
$$\langle{\cal O}_2^n{\cal O}_2^n{\cal O}_2{\cal O}_2\rangle$$ 
in the long sector.
Recall that the full correlator is decomposed in blocks $B_{\underline{R}}$, but 
since we are interested in the long sector, and in particular the 
logarithmic discontinuities,  the relevant part of the correlator is ${\cal H}$ and the 
relevant long block is ${H}_{\tau,l,[aba]}$ given in \eqref{allLongs}. It will be convenient to define 
\begin{equation}
\partial {H}_{\tau,l,[aba]} \equiv U^{\tau/2}\partial_{\tau} U^{-\tau/2} {H}_{\tau,l,[aba]}.
\end{equation}
as the quantity that sometimes we refer to as the ``derivative of the blocks". For higher derivatives, e.g.~$\partial^n {H}_{\tau,l,[aba]}$, these are obtained by the same formula with $\partial_{\tau}^n$.

Generically the OPE will include multi-particle exchange. Here we focus on double 
and triple particle exchange, which are relevant to the analysis of $n=2,3$ correlators.
It is understood that one has to add all contributions, whenever they contribute at the 
same order in the large $N$ expansion. For example, at twist four there is only 
a double particle operator, but starting from twist six there are double and triple particle operators. 

In order to avoid cluttering the notation, 
we will present separately the contribution from double and triple particle operators.\\ 

{\bf Double particles exchange: conventions for the CFT data}.
Let ${\cal D}_{\tau,l}$ be a double particle operator of spin $l$ and free twist $\tau=\Delta-l$. 
The double particle CFT data has the following expansion,
\begin{align}
C_{[2^n][2^n];{\cal D}}&=C^{(0)}_{[2^n][2^n];{\cal D}} + \frac{1}{a} C^{(1)}_{[2^n][2^n];{\cal D}} +\frac{1}{a^2} C^{(2)}_{[2^n][2^n];{\cal D}} +\ldots  \ \\
\Delta_{{\cal D}_{\tau,l}}&= \tau+l +\frac{2}{a} \eta_{(1);\cal{D}}+ \frac{2}{a^2}\eta_{(2);\cal{D}} +\ldots \;.
\end{align}
In the following the label ${\cal D}$ is implicit all the times, as well as the sum over operators.

{\bf Double particle exchange at order $\frac{1}{a}$}
\begin{align}\label{expaOPElong_dp1}
&
\log^1(U)\times \Bigg[C^{(0)}_{[2^n][2^n]}\,\eta_{(1)} C^{(0)}_{22}  {H}_{\tau,l,[aba]}  \Bigg] \\
&
\log^0(U) \times \Bigg[  2 C^{(0)}_{[2^n][2^n] }\,\eta_{(1)}C^{(0)}_{22 } \partial {H}_{\tau,l,[aba]} + \bigg[ C^{(1)}_{[2^n][2^n]}C^{(0)}_{22}+C^{(0)}_{[2^n][2^n]}C^{(1)}_{22} \bigg] {H}_{\tau,l,[aba]}  \Bigg] \notag
\end{align}

{\bf Double particle exchange at order $\frac{1}{a^2}$}
\begin{align}\label{expaOPElong_dp2}
&
\!\!\!\!\!\!\!\!\!\log^2(U)\times\Bigg[ \frac{1}{2}C^{(0)}_{[2^n][2^n]}\,\eta_{(1)}^2 C^{(0)}_{22}  {H}_{\tau,l,[aba]}  \Bigg] \\
&
\!\!\!\!\!\!\!\!\!\log^1(U) \times \Bigg[ 2 C^{(0)}_{[2^n][2^n]}\,\eta_{(1)}^2 C^{(0)}_{22 } \partial{H}_{\tau,l,[aba]} \ 
+\bigg[ C^{(0)}_{[2^n][2^n]}\eta_{(2)} C^{(0)}_{22}+C^{(1)}_{[2^n][2^n]}\eta_{(1)}C^{(0)}_{22}+C^{(0)}_{[2^n][2^n]}\eta_{(1)}C^{(1)}_{22} \bigg] {H}_{\tau,l,[aba]}  \Bigg] \notag \\
&
\!\!\!\!\!\!\!\!\!\log^0(U) \times \Bigg[  2 C^{(0)}_{[2^n][2^n]}\,\eta_{(1)}^2 C^{(0)}_{22} \partial^2 {H}_{\tau,l,[aba]} 
+\ \,2\!\! \sum_{i+a+j=2} C^{(i)}_{[2^n][2^n]}\eta_{(a)} C^{(j)}_{22} \partial{H}_{\tau,l,[aba]} +\notag\\
&
\rule{6.2cm}{0pt}\,+\  \bigg[ C^{(1)}_{[2^n][2^n]2}C^{(1)}_{22}+C^{(2)}_{[2^n][2^n]}C^{(0)}_{22}+C^{(0)}_{[2^n][2^n]}C^{(2)}_{22} \bigg]{H}_{\tau,l,[aba]} \rule{1.4cm}{0pt} \Bigg]  \notag
\end{align}
where $\eta_{(0)}=0$.

{\bf Double particle exchange at order $\frac{1}{a^3}$}

\begin{align}\label{expaOPElong_dp3}
&
\!\!\!\!\!\!\!\!\!\log^3(U)\times\Bigg[ \frac{1}{6}C^{(0)}_{[2^n][2^n]}\,\eta_{(1)}^3 C^{(0)}_{22}  {H}_{\tau,l,[aba]}  \Bigg] \\[.2cm]
&
\!\!\!\!\!\!\!\!\!\log^2(U) \times \Bigg[ \ C^{(0)}_{[2^n][2^n]}\,\eta_{(1)}^3 C^{(0)}_{22 } \partial{H}_{\tau,l,[aba]} \,\,\,+\  \,\frac{1}{2}\sum_{\substack{i+a+b+j=3\\a,b\leq 2}}\!\!\!\!C^{(i)}_{[2^n][2^n]}\eta_{(a)}\eta_{(b)} C^{(j)}_{22}  {H}_{\tau,l,[aba]} \Bigg]\notag\\
&
\!\!\!\!\!\!\!\!\!\log^1(U)\times\Bigg[ 2 C_{[2^n][2^n]}^{(0)} \eta_{(1)}^3 C_{22}^{(0)} \partial^2{H}_{\tau,l,[aba]}\,
+\  \,2 \sum_{\substack{i+a+b+j=3\\a,b\leq 2}}\!\!\!\!C^{(i)}_{[2^n][2^n]}\eta_{(a)}\eta_{(b)} C^{(j)}_{22} \partial{H}_{\tau,l,[aba]} 
+ \sum_{\substack{i+a+j=3\\a\leq 3}} C^{(i)}_{[2^n][2^n]} \eta_{(a)}C_{22}^{(j)} {H}_{\tau,l,[aba]} \Bigg] \notag
\end{align} 
and finally
\begin{align}
&
\log^0(U)\times\Bigg[ \frac{4}{3} C_{[2^n][2^n]}^{(0)} \eta_{(1)}^3 C_{22}^{(0)} \partial^3{H}_{\tau,l,[aba]}\,
			+\  \,2 \sum_{\substack{i+a+b+j=3\\a,b\leq 2}}\!\!\!\!C^{(i)}_{[2^n][2^n]}\eta_{(a)}\eta_{(b)} C^{(j)}_{22} \partial{H}^2_{\tau,l,[aba]} \notag\\[.2cm]
&
\rule{2.8cm}{0pt}\,+ \sum_{\substack{i+a+j=3\\a\leq 3}} C^{(i)}_{[2^n][2^n]} \eta_{(a)}C_{22}^{(j)} \partial{H}_{\tau,l,[aba]}+
\sum_{\substack{i+j=3}} C^{(i)}_{[2^n][2^n]}C_{22}^{(j)} \partial{H}_{\tau,l,[aba]} \Bigg]
\end{align}
where $\eta_{(0)}=0$.
~\\

{\bf Triple particles exchange: conventions for the CFT data}.

Let ${\cal T}_{\tau,l}$ be a triple particle operator of spin $l$ and free twist $\tau=\Delta-l$. 
The triple particle CFT data has the following expansion,
\begin{align}
C_{[2^n][2^n];{\cal T}}&=\frac{1}{\sqrt{a}} \Big[ C^{(\frac{1}{2})}_{[2^n][2^n];{\cal T}} + \frac{1}{a} C^{(\frac{3}{2})}_{[2^n][2^n];{\cal T}} +  +\ldots \Big] \ \\
\Delta_{{\cal T}_{\tau,l}}&= \tau+l +\frac{2}{a} \eta_{(1);{\cal T}}+ \frac{2}{a^2}\eta_{(2);{\cal T}} +\ldots 
\end{align}
In the following the label ${\cal T}$ is implicit all the times, as well as the sum over operators.

{\bf Triple particle exchange at order $\frac{1}{a}$}
\begin{align}\label{expaOPElong_tp1}
&
\!\!\!\!\!\!\!\!\!\!\!\!\!\!\!\!\!\!\!\!\!\!\!\!\!\!\!\!\!\!\!\!\!\!\!\!\!\!\!\!\!\!\!\!\!\!\!\!\!\!\!\!\!\!\!\!\!\!\!\!\!\!\!\!\!\!\!\!\!\!\!\!\!\!\!\!\!\!\!\!\log^1(U)\times 0 + 
\log^0(U) \times \bigg[ C^{(\frac{1}{2})}_{[2^n][2^n] }C^{(\frac{1}{2})}_{22} \bigg] {H}_{\tau,l,[aba]}
\end{align}

{\bf Triple particle exchange at order $\frac{1}{a^2}$}
\begin{align}\label{expaOPElong_tp2}
&
\log^2(U)\times 0+ \log^1(U)\times \Bigg[C^{(\frac{1}{2})}_{[2^n][2^n]}\,\eta_{(1)} C^{(\frac{1}{2})}_{22}  {H}_{\tau,l,[aba]}  \Bigg] \\
&
\log^0(U) \times \Bigg[  2 C^{(\frac{1}{2})}_{[2^n][2^n] }\,\eta_{(1)}C^{(\frac{1}{2})}_{22 } \partial {H}_{\tau,l,[aba]} + \bigg[ C^{(\frac{1}{2})}_{[2^n][2^n]}C^{(\frac{3}{2})}_{22}+C^{(\frac{3}{2})}_{[2^n][2^n]}C^{(\frac{1}{2})}_{22} \bigg] {H}_{\tau,l,[aba]}  \Bigg] \notag
\end{align}

{\bf Triple particle exchange at order $\frac{1}{a^3}$}
\begin{align}\label{expaOPElong_dp4}
&
\!\!\!\!\!\!\!\!\!\log^3(U)\times 0+\log^2(U)\times\Bigg[ \frac{1}{2}C^{(\frac{1}{2})}_{[2^n][2^n]}\,\eta_{(1)}^2 C^{(\frac{1}{2})}_{22}  {H}_{\tau,l,[aba]}  \Bigg] \\[.2cm]
&
\!\!\!\!\!\!\!\!\!\log^1(U) \times \Bigg[ 2 C^{(\frac{1}{2})}_{[2^n][2^n]}\,\eta_{(1)}^2 C^{(\frac{1}{2})}_{22 } \partial{H}_{\tau,l,[aba]} \ + 
\bigg[ C^{(\frac{1}{2})}_{[2^n][2^n]}\eta_{(2)} C^{(\frac{1}{2})}_{22}+
C^{(\frac{3}{2})}_{[2^n][2^n]}\eta_{(1)}C^{(\frac{1}{2})}_{22}+C^{(\frac{1}{2})}_{[2^n][2^n]}\eta_{(1)}C^{(\frac{3}{2})}_{22} \bigg] {H}_{\tau,l,[aba]}  \Bigg] \notag \\[.2cm]
&
\!\!\!\!\!\!\!\!\!\log^0(U) \times \Bigg[  2 C^{(\frac{1}{2})}_{[2^n][2^n]}\,\eta_{(1)}^2 C^{(\frac{1}{2})}_{22} \partial^2 {H}_{\tau,l,[aba]} 
+ 2 \bigg[ C^{(\frac{1}{2})}_{[2^n][2^n]}\eta_{(2)} C^{(\frac{1}{2})}_{22}+C^{(\frac{3}{2})}_{[2^n][2^n]}\eta_{(1)}C^{(\frac{1}{2})}_{22}+C^{(\frac{1}{2})}_{[2^n][2^n]}\eta_{(1)}C^{(\frac{3}{2})}_{22} \bigg] \partial{H}_{\tau,l,[aba]} +  \notag\\
&
\rule{6.3cm}{0pt}\,+\  \bigg[ C^{(\frac{3}{2})}_{[2^n][2^n]2}C^{(\frac{3}{2})}_{22}+C^{(\frac{5}{2})}_{[2^n][2^n]}C^{(\frac{1}{2})}_{22}+C^{(\frac{1}{2})}_{[2^n][2^n]}C^{(\frac{5}{2})}_{22} \bigg]{H}_{\tau,l,[aba]} \rule{1.4cm}{0pt} \Bigg]  \notag
\end{align}
This concludes the list of formulae.

\subsection{Predicting the protected sector of $\langle {\cal O}_2 {\cal O}_2^2 {\cal O}_2 {\cal O}_2^2\rangle$}

The traditional approach to the ${\cal N}=4$ superconformal partial wave decomposition 
of free theory is to split the process in various simple pieces, by doing the following.
First of all we observe that any free theory diagram $G$ becomes a monomial of the form
\begin{equation}
G= {\cal P}_{\vec{p}}\times \frac{U^a}{V^b} \sigma^i \tau^j \qquad;\qquad a=\frac{\gamma-\gamma_{\min}}{2}\quad;\quad b,i,j\ge 0
\end{equation}
The power of $U$ gives the value of $\gamma$. Then, we decompose $G$ as
\begin{align}
\label{decok}
G\Big|_{\substack{x_1=y_1\\ x_2=y_2}}&= {\cal P}_{\vec{p}}\times k \\
\label{decofG}
G\Big|_{x_2=y_2}&={\cal P}_{\vec{p}}\times\left[ k+\frac{(x_1-y_1)}{x_1y_1} f_G(x_1,y_1)\right] \\
G&={\cal P}_{\vec{p}}\times\Bigg[ k + {S}(x_1,x_2,y_1,y_2) + \frac{ \prod_{ij}(x_i-y_j) }{(y_1 y_2)^2} 
{H}_G(x_1,x_2,y_1,y_2)  \Bigg]
\end{align}
The constant $k$ gives the contribution to the $\frac{1}{2}$-BPS block. 
The contribution of the semishorts is found in two steps. Decompose $f_G$ 
into $[aba]$ reps,  
\begin{equation}
 f_G(x,y)=
  \sum_{h=1}^{ \beta } f_{G,h}(x)  \Bigg[ y^{h- \beta}\,_2F_1\Big[ \,^{h-\frac{\gamma-p_{12}}{2} ,\ h-\frac{\gamma-p_{43}}{2} }_{\rule{1cm}{0pt}2h-\gamma} ,y\Big] \Bigg] 
\end{equation}
This defines $f_{G,h}(x)$, which is then decomposed into Young diagram spin,
\begin{equation}\label{sl2deco}
f_{G,h}(x)= \sum_{\lambda\ge 1-h} c_{h,\lambda}\Bigg[ x^{\beta+\lambda} \,_2F_1\Big[\,^{\lambda+\frac{\gamma-p_{12}}{2},\ \lambda+\frac{\gamma-p_{43} }{2} }_{ \rule{1cm}{0pt}  2\lambda+\gamma};x_1 \Big]\bigg]
\end{equation}
The coefficients $c_{h,\lambda}$ are put in correspondence with the single variable functions of the superblocks: 
$\lambda=1$ gives $\frac{1}{4}$-BPS contribution, $\lambda\ge 2$ gives the semishort contributions, and
the values $1-\lambda' \leq \lambda\leq 0$ correspond the $\frac{1}{2}$-BPS contribution. 

The procedure explained in this section is superseded by the use of the newly introduced master formula presented in \cite{Aprile:2025nta}. 
However, for a pedagogical purpose, it can be useful to exemplify the traditional approach in some cases, 
as those relevant to sections \ref{OPEsect1} and \ref{OPEsect2}. This is what we shall present next.

\subsubsection*{Twist 4 semishorts in 2222}
These semishorts comes from the diagrams 
${\cal G}_{2222}=U^2\sigma^2 +\frac{U^2\tau^2}{V^2} + \frac{4}{a}  \frac{U^2\sigma\tau}{V}$ where ${\cal P}_{2222}=4a^2 g_{12}^2g_{34}^2$. See \eqref{free2222}. 
Here $\gamma_{\min}=0$, thus $\gamma=4$ and $\beta=2$ which means a Young diagram with at most two rows.
Following \eqref{decok}-\eqref{decofG}, we find
\begin{align}
f_{\lambda'=1}(x_1)&=  \frac{ x_1^2(2-2x_1+x_1^2)}{(x_1-1)^2}  - \frac{4}{a}\ \frac{x_1^2}{x_1-1}\\[.2cm]
f_{\lambda'=2}(x_1)&= \frac{1}{2}\frac{(x_1-2)x_1(-2+2x_1+x_1^2)}{ (x_1-1)^2 } + \frac{2}{a}\ \frac{x_1(x_1-2)}{x_1-1}
\end{align}
The term $\lambda'=1$ (would give $\frac{1}{4}$-BPS in $[121]$ according to the table which however is not there and) gives  semishorts in $[020]$. The term $\lambda'=2$ gives $\frac{1}{4}$-BPS in $[202]$ and semishorts in $[101]$.
From \eqref{sl2deco} we find
\begin{equation}
A^{2222}_{{\cal S},4,\myell,[020]}= \Bigg[   (l+3)(l+4)  + \frac{4}{a}   \Bigg]\frac{ 2(l+3)!^2}{(2l+6)}\frac{1+(-1)^l}{2}
\end{equation}
Note that $p_{12}=p_{43}=0$.

\subsubsection*{Twist 4 semishorts in 2222$^2$}
The  2222$^2$ correlator is NE, given by 
\begin{equation} \label{free22d22} 
{\cal P}_{2222^2}=  8 a^2 \sqrt{a+2}\, g_{12} g_{14}g_{24}g_{34}^2  \qquad;\qquad 
{\cal G}_{2222^2}= \frac{4\sqrt{a+2}}{a} \Bigg[ 1+U\sigma+\frac{U\tau}{V} \Bigg]
\end{equation}
where the normalization counts $| {\cal O}_2 {\cal O}_2|\sqrt{|{\cal O}_2 {\cal O}_2|\,|{\cal O}_2^2{\cal O}_2^2 |}$.
Here $\gamma_{\min}=2$, 
and the actual values of $\gamma$ are restricted to be just $\gamma=2,4$, since the correlator is protected. 
For the twist four contributions $\beta=1$, which means aYoung diagram with at most one row.
Following \eqref{decok}-\eqref{decofG}, we find
\begin{align}\label{singlev22d22}
f_{\lambda'=1}(x_1)&= \frac{4\sqrt{a+2}}{a} \Bigg[ x_1+\frac{x_1}{1-x_1}\Bigg]
\end{align}
From \eqref{sl2deco} we find
\begin{equation}
A^{22^2 22}_{{\cal S},4,[020]}= \frac{4\sqrt{a+2}}{a} (l+4) \frac{2(l+3)!^2}{(2l+6)!} \frac{1+(-1)^l}{2}
\end{equation}
Note that $p_{12}=-2$ and $p_{43}=0$.

\subsubsection*{Twist 4 sector in 22$^2$22$^2$}
The twist four sector comes from the diagrams $G={\cal P}_{22^222^2}\left[ \frac{8}{a}U\sigma +\frac{24}{a}\frac{U\tau}{V} \right]$ where ${\cal P}_{22^222^2}=16a(a+2)$. 
Here $\gamma_{\min}=2$, thus $\gamma=4$. As above $\beta=1$, which means aYoung diagram with at most one row.
Following \eqref{decok}-\eqref{decofG}, we find
\begin{align}
f_{\lambda'=1}(x_1)&= \frac{8}{a} x_1+ \frac{24}{a}\frac{x_1}{1-x_1}
\end{align}
From \eqref{sl2deco} we find
\begin{equation}
A^{}_{4,\myell,[020]}= \frac{4}{a}\Bigg[  (l+3)(l+4)+(-1)^l 6 \Bigg]\frac{(l+2)!(l+4)!}{(2l+6)!} 
\end{equation}
Note that $p_{12}=-2$ and $p_{43}=2$.

\subsubsection*{Twist 2 semishorts in 2$^2$2$^2$22}
This is a classic computation that (up to the color factor) is done in the same way for all $ppqq$ correlators. 
The free theory diagrams are ${\cal G}_{[2^2][2^2]22}=\left[ \frac{8}{a} U\sigma +\frac{8}{a} \frac{U\tau}{V} \right]$ with ${\cal P}_{2^22^222}=$.
Here $\gamma_{\min}=0$, thus $\gamma=2$. Then
\begin{align}
f_{\lambda'=1}(x_1)&= \frac{8}{a}\Bigg[ x_1+\frac{x_1}{1-x_1}\Bigg]
\end{align}
From \eqref{sl2deco} we find \eqref{konishi}. Note that $p_{12}=0$ and $p_{43}=0$.

\subsection{Further CFT data in the long twist four sector}

In this section we give additional details about the OPE analysis of the long twist four double particle operators ${\cal O}_2\partial^l{\cal O}_2$ in the $[0,0,0]$ rep, 
that we introduced in section \ref{OPEsect12}.
These  operators are non degenerate and therefore we can obtain their three-point couplings with the external operators by performing the  OPE analysis of the four-point functions. 
Specifically, in section \ref{OPEsect12} we obtain the relation
\begin{equation}\label{appFC1}
\Big[ C^{(1)}_{[2^2][2^2] ;{\cal D}_4} -2 C^{(1)}_{22; {\cal D}_4}\Big]C^{(0)}_{22;{\cal D}_4} = \frac{4}{3} (l+2)(l+5)\frac{(l+3)!^2}{(2l+6)!} \frac{1+(-1)^l}{2}
\end{equation}
which can be solved for $C^{(1)}_{[2^2][2^2] ;{\cal D}_4}$, namely
\begin{equation}
C^{(1)}_{[2^2][2^2] ;{\cal D}_4}  = \frac{ \frac{4}{3} (l+2)(l+5)\frac{(l+3)!^2}{(2l+6)!} }{ C^{(0)}_{22;{\cal D}_4} }  \frac{1+(-1)^l}{2} + 2 C^{(1)}_{22; {\cal D}_4}
\end{equation}

The authors \cite{Aprile:2017bgs} also obtained expressions for 
 \begin{align}
C^{(0)}_{22;{\cal D}_4} & = \sqrt{\frac{(l+1)(l+6)}{3}}\frac{(l+3)!}{\sqrt{(2l+6)!}} \frac{1+(-1)^l}{2} \\ 
C^{(1)}_{22;{\cal D}_4} &= \frac{ \frac{2}{3} - 8(1+2\psi^{(0)}(l+4)-2\psi^{(0)}(2l+7) ) }{
 \sqrt{\frac{(l+1)(l+6)}{3}}}\frac{(l+3)!}{\sqrt{(2l+6)!}} \frac{1+(-1)^l}{2}
\end{align}
and therefore we can find $C^{(1)}_{[2^2][2^2] ;{\cal D}_4}$ explicitly, 
 \begin{gather}\label{C1twist4explicit}
 C^{(1)}_{[2^2][2^2] ;{\cal D}_4}=  \frac{ \frac{4}{3} (l+2)(l+5)+\frac{4}{3}-16(1+ 2\psi^{(0)}(l+4)-2\psi^{(0)}(2l+7) ) }{ \sqrt{\frac{(l+1)(l+6)}{3}} } \frac{(l+3)!}{\sqrt{(2l+6)!}} \frac{1+(-1)^l}{2}
 \end{gather}

In section \ref{OPEsect12}, given the knowledge of \eqref{appFC1} we deduced a prediction 
for the $\log(U)$ of ${\cal H}^t_{[2^2][2^2]22}$, based on the known expression for the anomalous dimension $\eta_{ {\cal D}_4}$ in \cite{Aprile:2017bgs}.

\section{Comments on triple particle operators}\label{apptriplep}

In this section we extract novel data about triple particle operators:

1. We will start with ${\cal H}_{[2^2][2^2]22}$ and the triple 
particle operators in $[000]$. Focusing on twist six first, we will infer the existence of such triple traces operators 
from the fact that a CFT analysis based only on double particle operators 
fails to make a prediction for ${\cal H}^t_{[2^2][2^2]22}$ at twist six! (Instead
it did it for twist four, see section \ref{OPEsect12}.) The argument will be similar 
to that of \cite[sec.~4.7]{Drummond:2022dxw} where the authors used the conjectured 
result  for ${\cal H}_{2222}$ at two loops. Now we can reach the same conclusion quite 
explicitly already at one-loop in ${\cal H}^t_{[2^2][2^2]22}$, from a full-fledged gravity result. 
A similar argument shows the existence of higher twist triple trace operators. 

2. We will give formulae for the block coefficients of the leading logs of ${\cal H}_{2[2^2]2[2^2]}$. 
Here the triple particle operators in $[020]$ give a leading contribution, since $C_{2[2^2]{\cal T}}\sim 1$, 
while the double particle operators are suppressed $C_{2[2^2]{\cal D}}\sim \frac{1}{N}$.
But let us note that triple particle operators are degenerate in spin thus we will only be able 
to extract mixed quantities. It would be interesting to consider some special cases, 
like spin $0,1$, but we leave this for a future work.

\subsection{Proof that $C^{}_{22;{\cal T}_6}\neq 0$ and $C^{}_{[2^2][2^2]; {\cal T}_6} -2 C^{}_{22 ;{\cal T}_6}\neq 0$ }
\label{apptriplepsec1}

Even though the CFT data of the double particle operators ${\cal D}_{\tau}$ is know from
\cite{Aprile:2017bgs,Aprile:2017xsp}, the higher than twist four OPE data entering 
${\cal H}^t_{[2^2][2^2]22}$ involves long triple operators, starting from twist six.
 For example, 
 \begin{equation}
 C^{}_{[2^2][2^2];{\cal O}_2^3}\sim \frac{1}{N}\qquad;\qquad C^{}_{22;{\cal O}_2^3}\sim \frac{1}{N}
 \end{equation}
 This implies that 
 \begin{align}\label{newprediction1_app}
\!\!\!\!\!\!\! \hat{\cal C}_{[2^2][2^2]22}\Big|_{\frac{1}{a}}  
-2\, \hat{\cal C}_{2222} \Big|_{\frac{1}{a}}  & = \sum_{l} \Big[ C^{(1)}_{[2^2][2^2]; {\cal D}_4} -2 C^{(1)}_{22 ;{\cal D}_4}\Big] C^{(0)}_{22;{\cal D}_4} {B}^{\cal L}_{4,l,[000]}  +  \\
&\ \ \ \ \sum_{ l,{\cal D}_6 } \Big[ C^{(1)}_{[2^2][2^2]; {\cal D}_6} -2 C^{(1)}_{22 ;{\cal D}_6}\Big] C^{(0)}_{22;{\cal D}_6} {B}^{\cal L}_{6,l,[000]} + 
\sum_{ l,{\cal T}_6 } \Big[ C^{(\frac{1}{2})}_{[2^2][2^2]; {\cal T}_6} -2 C^{(\frac{1}{2})}_{22 ;{\cal T}_6}\Big] C^{(\frac{1}{2})}_{22;{\cal T}_6} {B}^{\cal L}_{6,l,[000]} \notag  \\
&\ \ \ \  +{\rm higher\ twist} \notag
\end{align}
where in the second line we have both double and triple particle operators of twist six. Then we will also find
 \begin{align}\label{newprediction2_app}
\!\!\!\!\!\!\! \hat{\cal C}_{[2^2][2^2]22}\Big|_{\frac{\log(U)}{a^2}}  
-2\, &\hat{\cal C}_{2222} \Big|_{\frac{\log(U)}{a^2}}   = \sum_{l} \Big[ C^{(1)}_{[2^2][2^2]; {\cal D}_4} -2 C^{(1)}_{22 ;{\cal D}_4}\Big] \eta_{{\cal D}_4}C^{(0)}_{22;{\cal D}_4} {B}^{\cal L}_{4,l,[000]}  +  \\
&\ \sum_{ l,{\cal D}_6 } \Big[ C^{(1)}_{[2^2][2^2]; {\cal D}_6} -2 C^{(1)}_{22 ;{\cal D}_6}\Big] \eta_{{\cal D}_6}C^{(0)}_{22;{\cal D}_6} {B}^{\cal L}_{6,l,[000]} + 
\sum_{ l,{\cal T}_6 } \Big[ C^{(\frac{1}{2})}_{[2^2][2^2]; {\cal T}_6} -2 C^{(\frac{1}{2})}_{22 ;{\cal T}_6}\Big] \eta_{{\cal T}_6}C^{(\frac{1}{2})}_{22;{\cal T}_6} {B}^{\cal L}_{6,l,[000]} \notag  \\
&\ \ \ \  +{\rm higher\ twist} \notag
\end{align}

We will now show that the 
\begin{equation}\label{000tripletrace}
C^{(\frac{1}{2})}_{22;{\cal T}_6}\neq 0\qquad{\rm and}\qquad C^{(\frac{1}{2})}_{[2^2][2^2]; {\cal T}_6} -2 C^{(\frac{1}{2})}_{22 ;{\cal T}_6}\neq 0 \,.
\end{equation}
Reasoning by contradiction let us remove by hand the triple particle contribution 
from \eqref{newprediction1_app}-\eqref{newprediction2_app}. If this is consistent then 
we can use the knowledge about the double particle spectrum from \cite{Aprile:2017xsp}, 
to predict  \eqref{newprediction2_app}, generalising the procedure in section \ref{OPEsect12}. 
This is now a prediction for the twist six $\log(U)$ contribution of 
${\cal H}^t_{[2^2][2^2]22}$ that we can check to be true or false. 
The relevant block decomposition is
\begin{align}\label{resulttwist6trp000}
{\cal H}^t_{[2^2][2^2]22}\Big|_{\frac{\log(U)}{a^2} } =&
\sum_{l} \Bigg[ - 64 \frac{(l+2)(l+5)}{(l+1)(l+6)}\frac{(l+3)!^2}{(2l+6)!} \frac{1+(-1)^l}{2}\Bigg] {H}^{}_{4,l,[000]}+  \\
&\sum_{l} \Bigg[  -224 \frac{(l+4)(l+5)}{(l+1)(l+8)}\frac{(l+4)!^2}{(2l+8)!} \frac{1+(-1)^l}{2}\Bigg] {H}^{}_{6,l,[000]}+  {\rm higher\ twist} \ldots
\end{align}
where we quoted also the result \eqref{etapred5.2sec} in the first line.  
It will turn out that the prediction from double particles at twist six \emph{does not} match 
the second line, thus we infer the existence of triple particle contributions, 
and in particular \eqref{000tripletrace} holds.

The double particle operators ${\cal D}_6$, of twist six in $[000]$, are schematically 
$O_1\equiv {\cal O}_2\partial^l \Box{\cal O}_2$ and $O_2\equiv {\cal O}_3\partial^l{\cal O}_3$. 
Their CFT data is described by a matrix of leading order three-point couplings 
\begin{equation}
C_6^{(0)}=\left(\begin{array}{cc} C^{(0)}_{22;O_1}  & C^{(0)}_{22;O_2} \\[.2cm]  C^{(0)}_{33;O_1}  & C^{(0)}_{33;O_2} \end{array}\right)\,,
\end{equation}
and then, for any external operator with charge higher than six, a vector of subleading three-point couplings
\begin{equation}
C^{(1)}_{pp,6}=\left[\begin{array}{cc} C^{(1)}_{pp;O_1} \,\,,&  C^{(1)}_{pp;O_2} \end{array}\right]^t\qquad;\qquad 2p>6\,.
\end{equation}
We will now need to consider the combination of three-point functions
\begin{equation}
V_{[2^2][2^2],6}= 
C_6^{(0)}. \Big[ C^{(1)}_{[2^2][2^2],6}- 2 C^{(1)}_{22,6}\Big]
\end{equation}
whose explicit components are
\begin{align}
V^{(1)}_{[2^2][2^2],6}=& \sum_{{\cal D}_6} C^{(0)}_{22;{\cal D}_6} \Big[ C^{(1)}_{[2^2][2^2]; {\cal D}_6} -2 C^{(1)}_{22 ;{\cal D}_6}  \Big]     \\
V^{(2)}_{[2^2][2^2],6}=& \sum_{{\cal D}_6}  C^{(0)}_{33;{\cal D}_6} \Big[C^{(1)}_{[2^2][2^2]; {\cal D}_6} -2 C^{(1)}_{22 ;{\cal D}_6}\Big]\,.
\end{align}
Under the assumption that there are no triple particle contributions, we can
obtain $V$ from the block decomposition at twist six of the (difference of) correlators,
\begin{equation}
\hat{\cal C}_{[2^2][2^2]22}\Big|_{\frac{1}{a}}  
-2\, \hat{\cal C}_{2222} \Big|_{\frac{1}{a}} \qquad;\qquad 
\hat{\cal C}_{[2^2][2^2]33}\Big|_{\frac{1}{a}} 
-2\, \hat{\cal C}_{2233} \Big|_{\frac{1}{a}} \,.
\end{equation}
Then, from the twist six disconnected contributions to the single-particle 
correlators (first) and from the tree level ${\cal H}_{2222}$, ${\cal H}_{2233}$, 
${\cal H}_{3333}$ (second), we can compute
\begin{equation}
A_6=\big(C^{(0)}\big).\big(C^{(0)}\big)^t \qquad;\qquad M_6= \big(C^{(0)}\big). \left(\begin{array}{cc} \eta_1 & 0 \\ 0 & \eta_2 \end{array}\right)\!.\big(C^{(0)}\big)^t\,,
\end{equation}
as done already in \cite{Aprile:2017xsp}. Finally, we compute
\begin{equation}
\big(M_6\big).\big(A_6\big)^{\!-1}\!\!\!.\,\,V_{[2^2][2^2],6} =\left\{ 
\begin{array}{c} \sum \Big[ C^{(1)}_{[2^2][2^2]; {\cal D}_6} -2 C^{(1)}_{22 ;{\cal D}_6}  \Big]\eta_{ {\cal D}_6}  C^{(0)}_{22;{\cal D}_6} \\[.3cm]    
\sum \Big[C^{(1)}_{[2^2][2^2]; {\cal D}_6} -2 C^{(1)}_{22 ;{\cal D}_6}\Big] \eta_{ {\cal D}_6}  C^{(0)}_{33;{\cal D}_6} \end{array} \right.
\end{equation}
The lhs of the above relations gives us the prediction we were looking for. This is matrix 
generalization of what was done in section \ref{OPEsect12} for the twist four 
long operators ${\cal O}_2\partial^l{\cal O}_2$. Let us collect all relevant formulae here below. 

\begin{itemize}
\item[$\bullet$]
The normalized free theory correlators that we need are 
\end{itemize}
\begin{equation}\label{usedisconconn}
{\cal G}_{pppp}\Big|_{\rm disco}=1+ U^p\sigma^p + \frac{U^p\tau^p}{V^p}\qquad;\qquad {\cal G}_{2233}= 1+ \frac{6}{a} \Bigg[ U\sigma +\frac{U\tau}{V}\Bigg] + \frac{12}{a} \frac{U^2\sigma\tau}{V}
\end{equation}
\begin{itemize}
\item[$\bullet$]
The normalized dynamical functions are
\end{itemize}
\begin{equation}
{\cal H}_{22pp}=-\frac{2p U^p\overline{D}_{p,p+2,2,2}}{(p-2)!}\qquad;\qquad {\cal H}_{3333}\Big|_{[000]}= -\frac{3}{2}U^p \Big[ \overline{D}_{2,5,2,3}+\overline{D}_{2,5,3,2}+6 \overline{D}_{3,5,2,2}+8\overline{D}_{3,5,3,3}\Big]
\end{equation}
\begin{itemize}
\item[$\bullet$]
A simple computation gives the matrices
\end{itemize}
\begin{align}
A_6&=+\left(\begin{array}{cc} \frac{1}{10} & 0 \\ 0 & \frac{1}{40}(l+2)(l+7) \end{array}\right)(l+1)(l+8)\frac{(l+4)!^2}{(2l+8)!}\\[.2cm] 
M_6&=-6\left( \begin{array}{cc} 4 & 12 \\ 12 & 44+9l +l^2 \end{array}\right) \frac{(l+4)!^2}{(2l+8)!}
\end{align}
At this point we need to compute the vector $V$. For the first component we need to 
consider ${\cal C}_{[2^2][2^2]22}$ as in  \eqref{twist4prediction2d2d22free1}, and we 
will do the twist 6 computation shortly. For the second component, $V^{(2)}_{[2^2][2^2],6}$, 
we need to consider the correlator ${\cal C}_{[2^2][2^2]33}$ at order $\frac{1}{a}$. 
This particular contribution is still disconnected, and therefore it is 
fully determined by free theory and by ${\cal C}_{2233}$ at order $\frac{1}{a}$. In fact, 
\begin{equation}
{\cal C}_{[2^2][2^2]33}\Big|_{\frac{1}{a}}=2{\cal C}_{2233} \Big|_{\frac{1}{a}}
\end{equation} 
Thus $V^{(2)}_{[2^2][2^2],6}=0$. For completeness we quote the free theory computed by Wick contractions  
\begin{align}
{\cal P}_{[2^2][2^2]33}&=g_{12}^4g_{34}^3 {\cal N}_{[2^2]}{\cal N}_3\\[.2cm]
{\cal G}_{[2^2][2^2]33}&= 1+ \frac{12}{a}\Bigg[ U\sigma+\frac{U\tau}{V}\Bigg] + \frac{24}{a}\frac{U^2\sigma\tau}{V} + \frac{24}{a(a+2)}\Bigg[ U^2\sigma^2 +\frac{U^2\tau^2}{V^2}+\frac{4U^2\sigma\tau}{V} + 2\Bigg[\frac{U^3\sigma^2\tau}{V}+\frac{U^3\sigma\tau^2}{V^2}\Bigg]\Bigg] \notag
\end{align} 
where the two point function normalization is ${\cal N}_{[2^2]}{\cal N}_3= \frac{24(N^2-1)(N^2-4)(N^4-1)}{N}$. 
It is then obvious from \eqref{usedisconconn} that ${\cal C}_{[2^2][2^2]33}|_{\frac{1}{a}}=2{\cal C}_{2233} |_{\frac{1}{a}}$.
This should be contrasted with the case
\begin{equation}\label{app2d2d22freetripletrace}
\hat{\cal C}_{[2^2][2^2]22}\Big|_{\frac{1}{a}}  
-2\, \hat{\cal C}_{2222} \Big|_{\frac{1}{a}}= \frac{4}{a} U^2\sigma^2 + \frac{4}{a} \frac{U^2\tau^2}{V^2} + \frac{16}{a} \frac{U^2\sigma\tau}{V} \,,
\end{equation}
already used in \eqref{twist4prediction2d2d22free1}.\footnote{The crucial point of this mismatch is the fact that even 
though a diagram of $\langle{\cal O}^2_2{\cal O}^2_2{\cal O}_2{\cal O}_2\rangle$ is decomposable into 
factors of $\langle {\cal O}_2{\cal O}_2\rangle$  and $\langle{\cal O}_2{\cal O}_2{\cal O}_2{\cal O}_2\rangle$, the latter
has full crossing symmetry and so the color factors, however $\langle{\cal O}_2{\cal O}_2{\cal O}_2{\cal O}_2\rangle$ is not full crossing. 
This implies that generically $\hat{\cal C}_{[2^2][2^2]22}|_{\frac{1}{a}} \neq  2\, \hat{\cal C}_{2222}|_{\frac{1}{a}}$. In fact, upon 
explicit computation the equality holds only in certain cases: for the $\gamma=2$ diagrams, and for the $\gamma=4$ disconnected diagrams but only $O(1)$ contributions.}
From \eqref{app2d2d22freetripletrace} the computation of $V^{(1)}_{[2^2][2^2],6}$  
follows from a simple $\gamma=4$ free theory block expansion. The final result is
\begin{equation}
V_{[2^2][2^2],6}=\left[ \begin{array}{cc} \displaystyle \frac{4+9l+l^2}{5} \frac{(l+4)!^2}{(2l+8)!}\,\,, & 0 \end{array}\right] \frac{1+(-1)^l}{2}\,.
\end{equation}
Therefore we get the prediction
\begin{equation}
\sum \Big[ C^{(1)}_{[2^2][2^2]; {\cal D}_6} -2 C^{(1)}_{22 ;{\cal D}_6}  \Big]\eta_{ {\cal D}_6}  C^{(0)}_{22;{\cal D}_6} =-96 \frac{(l^2+9l+4)}{(l+1)(l+8)}\frac{(l+4)!^2}{(2l+8)!}\frac{1+(-1)^l}{2}
\end{equation}
which \emph{does not match} the result in \eqref{resulttwist6trp000}. This proves 
the existence of long triple traces contributions at twist six in $[0,0,0]$. 

Finally, note that the our same argument here can be generalised to higher twists showing the existence of triple traces contributions at higher twist in $[0,0,0]$.

\subsection{Data for the long sector in $\langle {\cal O}_2 {\cal O}_2^2 {\cal O}_2 {\cal O}_2^2\rangle$}\label{apptriplep2}

The block decomposition of the $\frac{1}{a^n}\log^n(U)$ discontinuities 
of $\langle {\cal O}_2{\cal O}_2^2 {\cal O}_2 {\cal O}_2^2\rangle$
contain new CFT data. 
In particular, CFT data for triple particle operators in the $[0,2,0]$ rep. 
Let us recall that given a function ${\cal H}$, its block decomposition will be given by
\begin{equation}
{\cal H}=\sum_{\tau,l,[aba]} A_{\tau,l,[aba]} {H}_{\tau,l,[aba]}
\end{equation}
where the functions ${H}_{\tau,l,[aba]}$ have been given in \eqref{allLongs}.  We are interested in computing 
the $A_{\tau,l,[020]}$ as function of twist and spin, for the
function ${\cal H}_{2[2^2]2[2^2]}$. Since $[0,2,0]$ is the only available rep, in the following we shall use $A_{\tau,l}\equiv A_{\tau,l,[020]}$.

To describe the block coefficients, $A_{\tau,l}$, it will be useful to introduce some definitions,
\begin{align}
\mathbb{F}_{\tau,l}=\frac{ ( \frac{\tau}{2}+1 )!^2 ( l+\frac{\tau}{2} )!^2 }{48 \tau! (2l+\tau+2)!} \qquad;\qquad
T=\frac{\tau+1}{2}\qquad;\qquad L=l+\frac{\tau+3}{2} 
\end{align}
and write the block coefficients in the factorized form $A_{\tau,l}=\mathbb{F}_{\tau,l} \times \mathbb{R}_{\tau,l}$\footnote{Using blocks with $(-1)^l$.}\\

{\bf [n=0]} The relevant disconnected free theory is ${\cal P}_{2[2^2]2[2^2]}( U^2\sigma^2+\frac{2 U^2\tau^2}{V^2} )$. We find
\begin{equation}
 \mathbb{R}_{\tau,l}=
(l+1)(2l+\tau+4)(2l+\tau+2)(\tau+2+l) \left( (-1)^l \tfrac{96}{\tau(\tau+2) } -\tfrac{1}{4} + L^2\right)
\end{equation}
Note that  $\mathbb{R}_{\tau,l}$  is invariant under $l\rightarrow -l-\tau-3$ at fixed $\tau$, 
and  under $\{\tau\rightarrow-\tau-2,l\rightarrow -l-2\} $. For illustration, the twist six line is
\begin{equation}\label{citediscussion1}
{A}_{6,l}=\frac{2}{15}(l+1)(l+4)(l+5)(l+8)\Big[\tfrac{1}{2}(l+3)(l+6)+ 2\tfrac{1+(-1)^l}{2}  \Big]\frac{(l+3)!^2}{(2l+8)!}
\end{equation}\\

{\bf [n=1]} As we pointed out in the main text, the dynamical function is $-8U^2\overline{D}_{2422}$ where
 the $\overline{D}_{2422}$ is (very) familiar from the stress tensor correlator. However here we have to 
 perform a non standard decomposition of the $\log(U)$ sector,  done with respect to blocks in 
 which $p_{12}=-2, p_{43}=2$ and the minimum twist is six. The result is new, and reads
\begin{equation}
 \mathbb{R}_{\tau,l}=  \mathbb{R}^-_{\tau,l}\frac{1-(-1)^l}{2} +\mathbb{R}^+_{\tau,l}\frac{1+(-1)^l}{2}
\end{equation}
where
\begin{align}
 \mathbb{R}^-_{\tau,l}&=-\tfrac{(-4+\tau)(-2+\tau)(\tau+4)(\tau+6)}{2}\left[(l+1)(l+\tau+2)( L^2-\tfrac{96}{\tau(\tau+2)}-\tfrac{3}{5}T^2 +\tfrac{71}{10} )\right] \\[.2cm]
 \mathbb{R}^+_{\tau,l}&= \mathbb{R}^-_{\tau,l} -24\tfrac{(-4+\tau)(-2+\tau)(\tau+4)(\tau+6)}{\tau(\tau+2)}(2+2l+\tau)(4+2l+\tau)
\end{align}
For illustration, 
the twist six line is
\begin{equation}\label{citediscussion2}
{A}_{6,l}=-8\Bigg[ (l+2)(l+7)(16+9l+l^2) \tfrac{1+(-1)^l}{2}+(l+1)(l+3)(l+6)(l+8) \tfrac{1-(-1)^l}{2}\Bigg]\frac{(l+3)!^2}{(2l+8)!}\
\end{equation}\\

{\bf [n=2,\ ${\cal H}^t$]}
The dynamical function here is  ${\cal H}^c_{ [2^2][2^2]22 }|_\frac{1}{a^2}- 4 {\cal H}_{2222}|_{\frac{1}{a}}$. The first line at twist six is
\begin{equation}\label{citediscussion3}
{A}_{6,l}=32(-1)^{l+1}\Bigg[\frac{3(l+1)(l+3)(l+6)(l+8)}{(l+4)(l+5)} -( (2l+9)^2-37) \tfrac{1+(-1)^l}{2} \Bigg] \frac{(l+3)!^2}{(2l+8)!}\
\end{equation}
The block coefficients at higher twists are a bit more complicated, 
but we manage to write them in the following form
\begin{equation}
\mathbb{R}_{\tau,l}=32\frac{(l+1)(l+\tau+2)}{\tau^2(\tau+2)^2} (-1)^{l+1}\Big[ 2 (\tfrac{\tau}{2}-2)_6 \Big[ 96 +\tfrac{8}{15}(\tfrac{\tau}{2}-2)_6\big(\tfrac{1}{2}-\tfrac{1+(-1)^l}{2}(1+ \tfrac{3}{(l+1)(l+\tau+2)})\big) \Big]+ \Sigma_{\tau,l} \Big] 
\end{equation}
where
\begin{equation}
\Sigma_{\tau,l}=  \sum_{i=0}^{ \frac{\tau-6}{2} } \frac{ (-1)^i (T-I-2)_5(T+I-2)_5\times  I\times \sum_{j=0}^3 B_j (T-I)^{2j} }{ L^2-I^2 }
\end{equation}
where $I=i-T+3$ and
\begin{align}
B_0&= \frac{ (-5+ I+T)(-3+I+T)(3+I+T)(5+I+T)(1191-71(I+T)^2)}{92400}\\
B_1&=\frac{-508221+100379(I+T)^2 -2979(I+T)^4-59 (I+T)^6}{831600}\\
B_2&=\frac{32445-2979(I+T)^2-1330(I+T)^4+64(I+T)^6}{831600}\\
B_3&=\frac{-639-59(I+T)^2+64(I+T)^4+4(I+T)^6}{831600}
\end{align}
The symmetry is $I\rightarrow -I$ and $T\rightarrow -T$ which correspond to $i\rightarrow -i-6$ and $\tau\rightarrow -\tau-2$.\\

{\bf [n=2,\ ${\cal H}_{2222}$]}
The above results  add up to the contributions from $ 2 {\cal H}_{2222}|_{\frac{1}{a^2}}$, with the latter computed in \cite{Aprile:2017bgs}.  
We will only give the twist six line
\begin{align}\label{citediscussion4}
&
{A}_{6,l}=-192\frac{(l+3)!^2}{(2l+8)!}\Bigg[   \\ 
&\rule{1cm}{0pt} \Big[ (82+45l+5l^2)-\frac{16}{(l+3)}+\frac{16}{(l+6)}-4(l+2)(l+7) {\rm HarmonicNumber}(l+4)\Big] \tfrac{1+(-1)^l}{2} + \notag\\
& \rule{1cm}{0pt}\Big[ (86+45l+5l^2)-\frac{24}{(l+4)}+\frac{24}{(l+5)}- 4(l+2)(l+7) {\rm HarmonicNumber}(l+4)\Big] \tfrac{1-(-1)^l}{2} \Bigg] \notag
\end{align}
Again, this is a non standard decomposition compared to that in \cite{Aprile:2017bgs}. 
Let us also note the presence of harmonic sums in spin. Each polynomial coefficient is 
invariant under the symmetry mentioned above, so here $l\rightarrow -l-9$.

\paragraph{A quick comparison with $\langle {\cal O}_2 {\cal O}_4 {\cal O}_2 {\cal O}_4\rangle$.}
The CFT data presented above is not available from single-particle correlator.
With similar conventions to the previous sections, we quote twist six results for $\langle {\cal O}_2 {\cal O}_4 {\cal O}_2 {\cal O}_4\rangle$, that are implicit in \cite{Aprile:2018efk}, 
\begin{equation}
\begin{array}{c|cr}
\hline
\rule{0pt}{.8cm}\textrm{\bf [n=0]} & \qquad & \frac{1}{15}(l+1)(l+4)^2(l+5)^2(l+8)  \frac{(l+3)!^2}{(2l+8)!}\\[2ex]
\hline
\rule{0pt}{.8cm}\textrm{\bf [n=1]} & \qquad & -8(l+4)(l+5)\big( (l+1)(l+8)+24\frac{1+(-1)^l}{2} \big) \frac{(l+3)!^2}{(2l+8)!}\\[2ex]
\hline
\rule{0pt}{.8cm}\textrm{\bf [n=2]} & \qquad & 480(l+1)(l+8)\big( 84 +\big( 84+\frac{1008}{(l+1)(l+8)}\big)\frac{1+(-1)^l}{2} \big) \frac{(l+3)!^2}{(2l+8)!}
\end{array}
\end{equation}
The leading asymptotics quoted in Table \eqref{table1} is readily obtained.

\section{Mellin space, ladders and holographic correlators}\label{appLadder}

In the main text we used Mellin space as the natural framework to understand features of the 
multi-particle correlators in both AdS$_5$ and AdS$_3$. Here we give additional details on their Mellin amplitudes. 

Let us first recall some basics.
In both AdS$_5$ and AdS$_3$, superconformal symmetry implies that a correlator of four $\frac{1}{2}$-BPS 
operators is given by a free (or generalized free) contribution, and a dynamical function.
The latter has a well defined Mellin tranform. We shall focus on next-to-next-to extremal correlators, since these are the
multi-particle correlators that we discussed in the main text.
\paragraph{AdS$_3$.} By following the conventions of~\eqref{fa_4ptfunc} with ${\cal P}=g_{12}^{n} g_{34}$, we have
\begin{equation}\label{conventionsMellinads3}
 \hat{\cal C}_{[1^n][1^n]11}=  {free}+ \frac{(x_1-y_1)(x_2- y_2)}{y_1 y_2}  {\cal H}_{[1^n][1^n]11}(U,V)
\end{equation}
where
\begin{equation}
{\cal H}_{[1^n][1^n]11}= \oint\!\!\frac{ds dt}{(2\pi i)^2}\, \Gamma[-s]^2\Gamma[-t]^2\Gamma[s+t+2]^2 U^{1+s} V^t {\cal M}_{[1^n][1^n]11}(s,t) \\[.3cm]
\end{equation}
The natural $u$ variable here is $u=-s-t-2$. 

\paragraph{AdS$_5$.} By following the conventions of~\eqref{fa_4ptfunc} with ${\cal P}=g_{12}^{2n} g_{34}^{2}$, we have
\begin{equation}
 \hat{\cal C}_{[2^n][2^n]22}=  {free}+ \frac{(x_1-y_1)(x_1-y_2)(x_2-y_1)(x_2-y_2)}{(y_1 y_2)^2}  {\cal H}_{[2^n][2^n]22}(U,V)
\end{equation}
where
\begin{equation}
{\cal H}_{[2^n][2^n]22}= \oint\!\!\frac{ds dt}{(2\pi i)^2}\, \Gamma[-s]^2\Gamma[-t]^2\Gamma[s+t+4]^2 U^{2+s} V^t {\cal M}_{[2^n][2^n]22}(s,t)
\end{equation}
The natural $u$ variable here is $u=-s-t-4$.

\subsection{Ladders in Mellin space}\label{appladderexplicit}

The series of 4pts ladder integrals ${\cal P}_{L}(z,\bar{z})$  is based on the well known formula \cite{Isaev:2003tk}
\begin{equation}\label{ladderdef}
{\cal P}_{L}(z,\bar{z})=\sum_{r=0}^{L} \frac{(-1)^r}{r!} B\!\Big[\,^{2L-r}_{\ \ L}\Big] \log^r(z\bar{z}) \Big( {\rm Li}_{2L-r}(z)-{\rm Li}_{2L-r}(\bar{z})\Big)
\end{equation}
where $B[\,^a_b]=\frac{a!}{b!(a-b)!}$ is Newton's binomial.
We quote the first few integrals for reference, 
\begin{align}
{\cal P}_{0}(z,\bar{z})&=-\frac{z}{z-1}+\frac{\bar{z}}{\bar{z}-1} \\
{\cal P}_{1}(z,\bar{z})&=\ \log(z,\bar{z}) \big( \log(1-z) -\log(1-\bar{z}) ) + 2 \big( {\rm Li}_{2}(z) - {\rm Li}_2(\bar{z})\big) \notag \\
{\cal P}_{2}(z,\bar{z})&=\frac{1}{2}\log^2(z,\bar{z}) \big( {\rm Li}_{2}(z) - {\rm Li}_2(\bar{z})\big) - 3 \log(z,\bar{z})\big( {\rm Li}_{3}(z) - {\rm Li}_3(\bar{z})\big) + 6 \big( {\rm Li}_{4}(z) - {\rm Li}_4(\bar{z})\big)  \notag
\end{align}
The ${\cal P}_{L}$ have a symmetry,
\begin{equation}
{\cal P}_{L}(\tfrac{1}{z},\tfrac{1}{z})=-{\cal P}_{L}(z,\bar{z})\,,
\end{equation}
and therefore 
under crossing transformations there are three (instead of six) independent orientations, namely
\begin{equation}
{\cal P}_{L}(z,\bar{z})\qquad;\qquad  {\cal P}_{L}\!\left(\frac{z}{z-1},\frac{\bar{z}}{\bar{z}-1}\right) \qquad;\qquad {\cal P}_{L}(1-z,1-\bar{z})\,.
\end{equation}
The case of ${\cal P}_{L=1}$ is special since it is singlet under crossing.

In this section we give explicit formulae for the Mellin transform of
\begin{equation}\label{flippled_ladder}
\Phi^{(L)}(z,\bar{z})=-\frac{1}{z-\bar{z}} \, {\cal P}_{L}\!\left(\frac{z}{z-1},\frac{z}{\bar{z}-1}\right)
\end{equation}
The reason why we prefer this orientation will become soon clear. For the moment let us note that  the 
differential-recurrence relations for the ${\cal P}_{L}$ gives corresponding relations for the $\Phi^L(z,\bar{z})$:
\begin{equation}\label{diff_ladder}
\frac{U V}{z-\bar{z}} \partial_{z}\partial_{\bar z}\Big[ (z-\bar{z}) \Phi^L(z,\bar{z}) \Big] = -\Phi^{(L-1)}(z,\bar{z})\qquad;\qquad \Phi^{(0)}=1
\end{equation}
where we introduced
\begin{equation}
U=(1-z)(1-\bar{z}) \qquad;\qquad V=z\bar{z}\,,
\end{equation}
and let us also note that the symmetry of ${\cal P}_{L}$ translates into 
\begin{equation}
\Phi^{(L)}(1-z,1-\bar{z})=\Phi^{(L)}(z,\bar{z}), 
\end{equation}
which is the symmetry $U\leftrightarrow V$, also manifest in the differential equation. 

We focus on the orientation $\Phi^{(L)}$ because it manifests the maximal logs in $U$ and $V$, 
and this is a property that will turn out to be useful when we consider the Mellin transform. In particular,
\begin{equation}\label{bd_ladder}
\Phi^{(L)}(z,\bar{z})\Big|_{\log^L(U)\log^L(V) } = \frac{1}{(L!)^2}\frac{1}{z-\bar{z}}
\end{equation}
where a ${1}/{L!}$ is explicit in \eqref{ladderdef} with $r=L$, while the other  ${1}/{L!}$ comes from 
expanding $ {\rm Li}_{L}(z)-{\rm Li}_{L}(\bar{z})$. 

More precisely, $\Phi^{(L)}(z,\bar{z})$ has an expansion in small $U$ and small $V$ that follows from 
expanding the polylogs \eqref{ladderdef} in the small $\bar{z}$ and small $\zeta$ variables, where 
\begin{equation}
z=1-\zeta\,,
\end{equation}
using the identities,
\begin{align}
\zeta=&\ U+\sum_{n,m\ge1} \frac{U^n}{n!}\frac{V^m}{m!} \frac{(m+n-1)!(m+n-2)!}{ (n-1)!(m-1)!}\,,  \\
\bar{z}=&\ V+\sum_{n,m\ge1} \frac{U^n}{n!}\frac{V^m}{m!} \frac{(m+n-1)!(m+n-2)!}{ (n-1)!(m-1)!}\,.
\end{align}
For example, in \eqref{bd_ladder} it is simple to find that,
\begin{equation}\label{bd_laddermore}
\Phi^{(L)}(U,V)\Big|_{\log^L(U)\log^L(V) } = \frac{1}{(L!)^2}\,\frac{1}{1-\zeta-\bar{z}} =  \frac{1}{(L!)^2} \sum_{m,n,\ge 0} \frac{ U^n V^m }{n!m!} \times\frac{(m+n)!^2}{n!m!}
\end{equation}
The more general formula for $(1-\zeta-\bar{z})^{-p}$ with $p\in\mathbb{N}$ is given in \eqref{more_generalx1mx2den}.

Turning to Mellin space, we consider 
$$
\Phi^{(L)} =\oint \frac{U^s V^t}{(2\pi i)^2} \,\Gamma[-s]^2\Gamma[-t]^2\Gamma[-u]^2\, {\cal K}^{(L)}(s,t)\qquad;\qquad u=-s-t-1 \qquad;\qquad L\ge 1
$$
The $U\leftrightarrow V$ symmetry implies that $K(s,t)=K(t,s)$. 
The differential operator \eqref{diff_ladder} turns into the following difference operator
\begin{equation}
\Delta_{st}[{\cal K}^{(L)}]:= - st {\cal K}^{(L)}[s,t]+ \frac{s^2 t}{s+t} {\cal K}^{(L)}[s-1,t] + \frac{t^2 s}{s+t} {\cal K}^{(L)}[s,t-1]
\end{equation}
and the differential-recurrence equation becomes
\begin{equation}\label{difference_eq}
\Delta_{st}[{\cal K}^{(L)}]= -{\cal K}^{(L-1)}(s,t)\,.
\end{equation}
In Mellin space we have to start the recurrence from the $L=1$ integral, since $\Phi^{(0)}=1$ has vanishing Mellin transform. 
The reason is that $\Gamma[-s]^2\Gamma[-t]^2\Gamma[-u]^2$ can be assigned 
transcendentals weight $2$, thus  ${\cal K}^{(1)}$ is a constant and becomes the starting point of the recursion.  
In particular
\begin{equation}
{\cal K}^{(1)}=1;
\end{equation}
and one can check that $\Delta_{st}[1]=0$.

For $L>1$, we expect ${\cal K}$ to have transcendental weight $2L-2$, 
in order to make manifest the uniform transcendentality of the ladders, that equals $2L$. 
Moreover, the condition \eqref{bd_laddermore} translates into a condition on the maximal poles of ${\cal K}^{(L)}$, which can only be of the form, 
\begin{equation}\label{bd_mellin}
{\cal K}^{(L)}(s,t) = \frac{1}{(s-n)^{L-1} (t-m)^{L-1}}\ +\ {\rm lower\ order\ poles}\qquad n,m\in\mathbb{N}\;.
\end{equation}
Note that the $\frac{1}{(L!)^2}$ in \eqref{bd_laddermore} comes from integration by parts. 

\subsection*{Solutions}
The first few integrals are
\begin{align}\label{listLadderMellin}
 {\cal K}^{(2)}(s,t)&=\frac{2}{(2!)^2}\bigg[ -\Psi_1^2 + \Psi_2 -\pi^2 \bigg] \\[.2cm]
 {\cal K}^{(3)}(s,t)&=\frac{3}{(3!)^2}\bigg[ +\Psi_1^4 + 3 \Psi_2^2-2\Psi_1 \Psi_3 -2 \pi^2 \Psi_2 - 4 \pi^2 {\cal K}^{(2)} -\pi^4 \bigg]\\[.2cm]
 {\cal K}^{(4)}(s,t)&=\frac{4}{(4!)^2}\bigg[ -\Psi_1^6 - 3\Psi_1^4\Psi_2 - 9\Psi_1^2\Psi_2^2+ 9 \Psi_2^3+4\Psi_1^3\Psi_3-12\Psi_1\Psi_2\Psi_3-4\Psi_3^2 +3 \Psi_1^2\Psi_4 + 3\Psi_2\Psi_4 + \notag\\
 & \rule{1.7cm}{0pt} +\pi^2 ( 6 \Psi_1^2\Psi_2 -12\Psi_2^2 + 6 \Psi_1\Psi_3 - \Psi_4) - 9\times 4\times \pi^2 {\cal K}^{(3)} - 6 \pi^4 {\cal K}^{(2)} -\pi^6 \bigg]
\end{align}
where we defined
\begin{equation}
\Psi_{n}(s,t)= {\tt PolyGamma}[n-1,-s] +(-1)^{n} {\tt PolyGamma}[n-1,-t]\,.
\end{equation}

In the main text we used $K_2(s,t)=2{\cal K}^{(2)}$ in  \eqref{doubleboxwith2}, then we used $K_4(s,t)=12{\cal K}^{(3)}$ in  \eqref{tripleboxmellinamp}.

\subsection{Maximal log trick for sugra amplitudes}

Our AdS correlators ${\cal C}(z,\bar{z})$ have the form of a sum over ladder integrals, and derivatives of ladder 
integrals, each one multiplied by a certain rational function. 
Focus on the orientation of the correlator that manifest the max logs $\log^L(U)\log^L(V)$ and 
take the following projection
\begin{equation}\label{maxlog_mellin1}
{\cal C}(z,\bar{z})\Big|_{\frac{ \log^L(U)\log^L(V) }{ (L!)^2}}=\frac{P(U,V)}{\ \ (z-\bar{z})^{2\gamma+1} }\;,
\end{equation}
where the polynomial $P(U,V)$ will be written as
\begin{equation}
P(U,V)=\sum_{a,b} c_{ab} U^aV^b \qquad ;\qquad c_{ab}\in\mathbb{R}
\end{equation}
We now show that quite straightforwardly we can reconstruct 
the top weight contribution of the Mellin amplitude ${\cal M}$, i.e.~the rational function 
in $s,t$ that will multiply ${\cal K}^{(L)}$. To this end we only need to 
consider the following identity
\begin{equation}\label{more_generalx1mx2den}
\frac{{U}^a\,{V}^b }{(z-\bar{z})^{2\gamma+1}}=\frac{\gamma!}{(2\gamma)!} \oint  (-{U})^s (-{V})^t \ \frac{\Gamma[-s]\Gamma[-t]}{\Gamma[s+1]\Gamma[t+1]}\Gamma[s+t+X]^2\times  m_{ab,\gamma} 
\end{equation}
where
\begin{equation}
m_{ab,\gamma}=
\left[ (-)^{a+b} \frac{ (-s)_a (-t)_b}{ (s+1)_{\gamma-a} (t+1)_{\gamma-b}} (s+t+X)_{2\gamma+1-a-b-X} (s+t+X)_{\gamma+1-a-b-X}\right]\qquad 
\label{Mellin-resum1}
\end{equation}
with $X$ corresponding to the definition of the $u$ variable in Mellin space, and apply it repeatedly to all terms of $P(U,V)$. Then, 
\begin{equation}
A(s,t) = \sum_{a,b} c_{ab} m_{ab,\gamma} 
\end{equation}
because
\begin{equation}
\oint \frac{U^s V^t}{(2\pi i)^2} \,\Gamma[-s]^2\Gamma[-t]^2\Gamma[-u]^2\, \Bigg[ {\cal K}^{(L)}(s,t) A(s,t) +\ldots \Bigg]= \frac{\log^L(U)\log^L(V) }{(L!)^2} \frac{P(U,V)}{\ \ (z-\bar{z})^{2\gamma+1} }+\ldots 
\end{equation}
The non maximal log contributions to the Mellin amplitude cannot be found this way, it is nevertheless important to have this as a starting point.

\subsection{Transcendental basis for $\langle {\cal O}_2^3 {\cal O}_2^3 {\cal O}_2{\cal O}_2\rangle$}\label{appF3}

The dynamical function ${\cal H}^t_{[2^3][2^3]22}$ 
has maximum transcendental weight six given by 
the ladder integral ${\cal P}_{L=3}$.  
In this section we describe its lower weight 
completion in the basis of svhpl functions, 
from weight five (antisymmetric) to weight zero. 
This basis is closely related to (a subset of) 
the basis that is used for ${\cal H}_{2222}$ at two-loops, studied in \cite{Drummond:2022dxw,Huang:2021xws}. 
However, the $\log^3(U)$ discontinuity of 
the two-loop function ${\cal H}_{2222}$ 
does not have a contribution from ${\cal P}_{L=3}$. 
Instead, its top weight six is given by the zig-zag 
integrals and therefore the authors of 
\cite{Drummond:2022dxw} tailored the lower weights basis 
to the weight six zig-zag integrals. Here we 
will give a version of the basis tailored to ${\cal P}_{L=3}$. 
Before doing so, let us also observe that 
from the structure of BMN respresentation 
of our HHLL correlator, see \eqref{eq:psiexpads5}-\eqref{expagravityCFT3}, 
we expect to find even a restricted basis, 
simply made from ladders, derivatives of the ladders and simple logs.

The simplest base elements are those at weight 0,1,2,3. 
This is because the OPE predicts a $\log^3(U)$ 
discontinuity at order $\frac{1}{a^3}$, 
as the maximal $\log$ discontinuity. 
Therefore, all svhpl functions up to weight 
$3$ have to be included. We summarize them in the following table,
\begin{equation}
\begin{array}{c|ll}
\textrm{weight} & \textrm{anti-symmetric functions} &;\ \ \, \textrm{symmetric functions}  \\
\hline\\
0 & &;\ \ \,1 \\[2ex]
\hline\\
1&\ &;\ \ \,\log(V)\quad,\quad \log(U)\\[2ex]
\hline\\
2 &\ {\cal P}_1 &;\quad \log^2(V) \quad,\quad  \log(U)\log(V) \quad,\quad \log^2(U) \\[2ex]
\hline\\
3 &\ {\cal P}_1 \log(V)\quad,\quad {\cal P}_1\log(U) &;\quad   \log^3(V) \quad,\quad \log(U)\log(V)\log(U V) \quad,\quad \log^3(U) \\[2ex] 
& & \ \ \ \ \,\log(U)\log(V)\log(\frac{U}{V})\\[2ex]
& & \ \ \ \ \, \partial_-{\cal P}_2(1-z)\quad,\quad \partial_-{\cal P}_2(z)\quad,\quad \partial_-{\cal P}_2(\frac{z}{z-1})
\end{array}\vspace{2mm}
\end{equation}\vspace{2mm}
For the same reason, all weight $4^-$ svhpl antisymmetric functions have to be included
\begin{equation}
\begin{array}{c|lll}
\textrm{weight} & \textrm{functions}  \\
\hline\\
4^- 
&\ {\cal P}_1 \log^2(V)\quad&,\quad {\cal P}_1\log^2(U)\quad&,\quad {\cal P}_1\log^2(\frac{U}{V}) \\[2ex]
&\ {\cal P}_2(1-z)\quad&,\quad{\cal P}_2(z)\quad&,\quad {\cal P}_2(\frac{z}{z-1})
\end{array}
\end{equation}

Starting from {weight} $4^+$, there will be restrictions. The 
complete {weight} $4^+$ basis goes up to $\log^4$ terms, eg.~it contains all products $\log^m(U)\log^n(V)$ such that $m+n=4$. 
But for our purpose we need to truncate it such that at most we find $\log^3$.
To do so let us start by writing the complete basis of {weight} $4^+$ symmetric svhpl functions, which consists of the following 12 elements,
\begin{equation}
\begin{array}{c|l}
\textrm{weight} & \textrm{functions}  \\
\hline\\
4^+ 
&\  {\cal P}_1^2\quad,\quad \cancel{\log(U)\log(V)(\log^2(U) +\log^2(V) )}\quad,\quad \cancel{\log^4(V)}\quad,\quad \cancel{\log^4(U)}\\[2ex]
 &\rule{1.6cm}{0pt} \log(U)\log^2(V)\log(\frac{U}{V}) \quad,\quad \log^2(U) \log(V) \log(\frac{U}{V}) \\[2ex]
 &\ \zeta_3 \log(V) \quad,\quad \zeta_3\log(U)\quad,\quad \log(U V) \partial_{-} {\cal P}_2(z)  \\[2ex]
 &\ \partial_-\partial_+{\cal P}_3(1-z)\quad,\quad \partial_-\partial_+{\cal P}_{3}(z)\quad,\quad \partial_-\partial_+{\cal P}_{3}(\frac{z}{z-1})
\end{array}
\end{equation}
A way to truncate the basis to $\log^3$ elements is to take a 
linear combination  $v=\sum_i a_i b_i$, where $b_i$ are all the basis elements, impose that there are no $\log^4$ terms, and repeat the process 
for all the other crossing images of $v$.  By doing so the \cancel{elements} are excluded.

Proceeding in a similarly way at weight $5^-$, we list the complete basis of weight $5^-$ antisymmetric svhpl functions first, which consists of
\begin{equation}
\begin{array}{c|ll}
\textrm{weight} & \textrm{functions}  \\
\hline\\
5^- 
&\ \cancel{{\cal P}_1\log^3(V)}\quad,\quad \cancel{{\cal P}_1\log(U)\log(V)\log(UV)}\quad,\quad\cancel{{\cal P}_1\log^3(U)} \\[2ex]
&\ {\cal P}_1\log(U)\log(V)\log(\frac{U}{V})\\[2ex]
&\ {\cal P}_1\partial_-{\cal P}_2(1-z)\quad,\quad {\cal P}_1\partial_-{\cal P}_2(z)\quad,\quad {\cal P}_1\partial_-{\cal P}_2(\frac{z}{z-1})\\[2ex]
&\ \log(U){\cal P}_2(1-z)\quad,\quad \log(V){\cal P}_2(z)\quad,\quad\log(U V)  {\cal P}_2(\frac{z}{z-1})\\[2ex]
&\ \partial_+{\cal P}_3(1-z)\quad,\quad \partial_+{\cal P}_3(z)\quad,\quad \partial_+{\cal P}_3(\frac{z}{z-1})
\end{array}
\end{equation}
and as in the case of $4^+$, the \cancel{terms} are excluded. 

It turns out that out the position space result for the correlator ${\cal H}_{[2^3][2^3]22}^t$ is spanned 
by an even restrict set of basis elements, namely 
\begin{equation}\label{basis3dappendix}
{\cal H}_{[2^3][2^3]22}^t={\rm span}\Big\{ {\cal P}_3\,,\, \partial_-{\cal P}_3\,,\, \partial_+{\cal P}_3\,,\, \partial_+\partial_- {\cal P}_3\Big\}\cup\Big\{\,4^-,3,2,1,0\Big\} 
\end{equation}
where the first parenthesis $\{\ldots\}$ contain weight $6,5^{\pm},4^{+}$, but only in the form of  ${\cal P}_3$ and its derivatives.  

In order to conclude our discussion here we need to understand how to map \eqref{basis3dappendix} to Mellin space.
We derived already in \eqref{listLadderMellin} the amplitudes corresponding to ${\cal P}_3$ and ${\cal P}_2$, 
thus we only need to understand $\partial_+{\cal P}_3$ and ${\cal P}_1\log^2(U)$ in the weight $4^-$ 
basis, the rest of the basis will follow from the one-loop basis. For the function 
${\cal P}_1\log^2(U)$ we can make a preliminary comment. Recall that that at 
one-loop the term ${\cal P}_1\log(U)$ turned out to imply the presence of the 
amplitude $\Psi_1(s,t)$. The justification in this case is simply that ${\cal P}_1$ has a constant 
Mellin amplitude, unity, thus in order to capture $\log^2U\log V$ one has to add weight 
one polygammas, namely $\Psi_1(s,t)$.  Now, ${\cal P}_1\log^2(U)$ has $\log^3(U)\log(V)$, 
and by reasoning in a similar way we are led to think that  ${\cal P}_1\log^2(U)$ implies the 
presence of weight two polygammas, and a natural guess is $\Psi_2(s,t)$.\footnote{Recall that $-\Psi_1^2+\Psi_2$, 
i.e.the double box has at most $\log^2 U\log^2 V$.}

In order to map the basis element $\partial_+{\cal P}_3$ and ${\cal P}_1\log^2(U)$ to Mellin space, we will do the following. 
First let us write $\partial_+{\cal P}_3(\frac{z}{z-1},\frac{\bar z}{{\bar z}-1})$ in Mellin space by considering the expression
\begin{equation}
\frac{1}{(z-\bar{z})} \Bigg[z(1-z)\frac{\partial}{\partial z} +{\bar z}(1-{\bar z})\frac{\partial}{\partial {\bar z}}\Bigg] \Bigg[ (z-\bar{z}) \oint \frac{U^s V^t}{(2\pi i)^2}\Gamma[-s]^2\Gamma[-t]^2\Gamma[s+t+1]^2 {\cal K}^{(3)}(s,t)\Bigg]\;.
\end{equation} 
This gives the result
\begin{equation}
\oint \frac{1}{(2\pi i)^2}\Gamma[-s]^2\Gamma[-t]^2\Gamma[s+t+1]^2 \Bigg[ (s-t)U^sV^{t} +(1+s+t)(U^sV^{t+1}-U^{s+1}V^{t}) \Bigg] {\cal K}^{(3)}(s,t)
\end{equation}
which leads to
\begin{equation}
\oint \frac{U^s V^t}{(2\pi i)^2}\Gamma[-s]^2\Gamma[-t]^2\Gamma[s+t+1]^2 \Bigg[(s-t)+\frac{t^2{\cal K}^{(3)}(s,t-1) -s^2 {\cal K}^{(3)}(s-1,t)}{s+t} \Bigg]
\end{equation}
The Mellin amplitude simplifies to weight $3$ polygammas, giving the result, 
\begin{equation}
\frac{1}{3}\Bigg[ \Psi^3_1(u,t) -\frac{1}{2} \Psi_3(u,t)  + \pi^2 \Psi_1 \Bigg]\,.
\end{equation}
This is the amplitude we used to define $K_3(s,t)$ in \eqref{tripleboxmellinamp}. Finally, to get ${\cal P}_1\log^2(U)$ we consider another 
equation on ${\cal P}_3$ such that, once decomposed on the basis, has a component along ${\cal P}_1\log^2(U)$.
For example 
\begin{equation}
\frac{1}{(z-\bar{z})} \Bigg[z\frac{\partial}{\partial z} +\bar{z}\frac{\partial}{\partial {\bar z}}\Bigg] \Bigg[ (z-\bar{z}) \oint \frac{U^s V^t}{(2\pi i)^2}\Gamma[-s]^2\Gamma[-t]^2\Gamma[s+t+1]^2 {\cal K}^{(3)}(s,t)\Bigg] \end{equation}
(Note this is not $\partial_+{\cal P}_3(z,\bar{z})$.) With manipulation similar to the one above we find that the new 
element that is needed to expand this function in a basis, is proportional to 
\begin{equation}
\Psi_2-\frac{2}{3} \pi^2
\end{equation}
As expected we have found a weight 2 combination whose leading term is just $\Psi_2$. This is the amplitude we used to define ${\tilde K}_2(s,t)$ in \eqref{tripleboxmellinamp}.

\providecommand{\href}[2]{#2}\begingroup\raggedright\endgroup

\end{document}